\documentclass[]{aa}

\usepackage{txfonts}
\usepackage{float}
\usepackage{graphicx}
\usepackage{subfigure}
\usepackage{tabularx}
\usepackage{natbib}
\usepackage{epsfig}
\usepackage{longtable}
\usepackage{txfonts}
\usepackage{pdflscape} 

\begin{document}

\title{Structure and substructure analysis of DAFT/FADA galaxy clusters
in the [0.4-0.9] redshift range~\thanks{Based on
    XMM-Newton archive data and on data retrieved from the NASA/IPAC
    Extragalactic Database (NED), which is operated by the Jet
    Propulsion Laboratory, California Institute of Technology, under
    contract with the National Aeronautics and Space
    Administration. The scientific results reported in this article are 
also based in part on data obtained from the Chandra Data Archive.
Based on observations made with the FORS2 
multi-object spectrograph mounted on the Antu VLT telescope at ESO-Paranal 
Observatory (programme 085.A-0016; PI: C. Adami). 
Also based on observations 
obtained at the Gemini Observatory, which is operated by the
Association of Universities for Research in Astronomy, Inc., under a 
cooperative agreement
with the NSF on behalf of the Gemini partnership: the National Science 
Foundation (United
States), the Science and Technology Facilities Council (United Kingdom), 
the National Research Council (Canada), CONICYT (Chile), the Australian 
Research Council (Australia), Minis\'erio da Ci\^encia, Tecnologia e 
Inova\c c\~ao (Brazil), and Ministerio de Ciencia, Tecnolog\'ia e Innovaci\'on 
Productiva  (Argentina). 
Also based on observations made with the Italian Telescopio Nazionale Galileo 
(TNG) operated on the island of La Palma by the Fundaci\'on Galileo Galilei of 
the INAF (Istituto Nazionale di Astrofisica) at the Spanish Observatorio del 
Roque de los Muchachos of the Instituto de Astrof\'isica de Canarias.
Also based on service observations made with the WHT operated on the island of 
La Palma by the Isaac Newton Group in the Spanish Observatorio del Roque de 
los Muchachos of the Instituto de Astrof\'isica de Canarias.
Also based on observations collected at the German-Spanish Astronomical 
Center, Calar Alto, jointly operated by the Max-Planck-Institut fur 
Astronomie Heidelberg and the Instituto de Astrof\'isica de Andalucía (CSIC).
Based on observations obtained with MegaPrime/MegaCam, a joint project of CFHT 
and CEA/IRFU, at the Canada-France-Hawaii Telescope (CFHT) which is operated 
by the National Research Council (NRC) of Canada, the Institut National des 
Science de l'Univers of the Centre National de la Recherche Scientifique 
(CNRS) of France, and the University of Hawaii. This work is based in part 
on data products produced at Terapix available at the Canadian Astronomy Data 
Centre as part of the Canada-France-Hawaii Telescope Legacy Survey, a 
collaborative project of NRC and CNRS. 
Also based on observations obtained at the WIYN telescope (KNPO). The WIYN 
Observatory is a joint facility of the University of Wisconsin-Madison, 
Indiana University, Yale University, and the National Optical Astronomy 
Observatory. Kitt Peak National Observatory, National Optical Astronomy
Observatory, is operated by the Association of Universities for Research in 
Astronomy (AURA) under cooperative agreement with the National Science 
Foundation.
Also based on observations obtained at the MDM observatory (2.4m telescope).
MDM consortium partners are Columbia University Department of Astronomy 
and Astrophysics, Dartmouth College Department of Physics and Astronomy, 
University of Michigan Astronomy Department, The Ohio State University 
Astronomy Department, Ohio University Dept. of Physics and Astronomy. 
Also based on observations obtained at the Southern Astrophysical Research 
(SOAR) telescope, which is a joint project of the Minist\'{e}rio da 
Ci\^{e}ncia, Tecnologia, e Inova\c{c}\~{a}o (MCTI) da Rep\'{u}blica Federativa 
do Brasil, the U.S. National Optical Astronomy Observatory (NOAO), the 
University of North Carolina at Chapel Hill (UNC), and Michigan State 
University (MSU).
Also based on observations obtained at the Cerro Tololo Inter-American 
Observatory, National Optical Astronomy Observatory, which are operated by the 
Association of Universities for Research in Astronomy, under contract with the 
National Science Foundation.
Finally, this research has made use of the VizieR 
catalogue access tool, CDS, Strasbourg, France.
}}

\author{
L.~Guennou\inst{1,2} \and
C.~Adami\inst{1} \and
F.~Durret\inst{3} \and
G.B.~Lima Neto\inst{4} \and
M.P.~Ulmer\inst{5} \and  
D.~Clowe\inst{6} \and
V.~LeBrun\inst{1} \and
N.~Martinet\inst{3} \and \\
S.~Allam\inst{7,8} \and
J.~Annis\inst{7} \and
S.~Basa\inst{1} \and
C.~Benoist\inst{9} \and
A.~Biviano\inst{3,10} \and
A.~Cappi\inst{9,11} \and
E.S. Cypriano\inst{4} \and
R.~Gavazzi\inst{3} \and
C.~Halliday\inst{12} \and
O.~Ilbert\inst{1} \and
E.~Jullo\inst{1} \and
D.~Just\inst{13,14} \and
M. Limousin\inst{1} \and
I.~M\'arquez\inst{15}\and
A.~Mazure\inst{1} \and
K.J.~Murphy\inst{6} \and
H.~Plana\inst{16} \and
F.~Rostagni\inst{9} \and
D.~Russeil\inst{1} \and
M.~Schirmer \inst{17,18} \and 
E.~Slezak\inst{9} \and
D.~Tucker\inst{7} \and 
D.~Zaritsky\inst{13}\and 
B.~Ziegler \inst{19}
}

\offprints{L.~Guennou \email{guennou@ukzn.ac.za}}

\institute{
Aix Marseille Universit\'e, CNRS, LAM (Laboratoire d'Astrophysique de Marseille) 
UMR 7326, 13388, Marseille, France 
\and
Astrophysics and Cosmology Research Unit, University of KwaZulu-Natal, Durban, 4041, SA 
\and
UPMC-CNRS, UMR7095, Institut d'Astrophysique de Paris, F-75014, Paris, France 
\and
Departamento de Astronomia, Instituto de Astronomia Geofisica e Ci\^encias Atmosf\`ericas, Universidade de S\~ao Paulo, 
Rua do Mat\~ao 1226, 05508-900 S\~ao Paulo, Brazil
\and
Dept of Physics and Astronomy \& Center for Interdisciplinary Exploration 
and Research in Astrophysics (CIERA), Evanston, IL 60208-2900, USA 
\and
Department of Physics and Astronomy, Ohio University, 251B Clippinger
Lab, Athens, OH 45701, USA 
\and
Fermi National Accelerator Laboratory, P.O. Box 500, Batavia, IL 60510, USA 
\and
CSC/STScI, 3700 San Martin Dr., Baltimore, MD 21218 , USA 
\and
OCA, Cassiop\'ee, Boulevard de l'Observatoire, BP 4229, 06304 Nice Cedex 4, France 
\and
INAF/Osservatorio Astronomico di Trieste, via Tiepolo 11, 34143 Trieste,
Italy
\and
INAF - Osservatorio Astronomico di Bologna, via Ranzani 1, 40127 Bologna, Italy 
\and
Osservatorio Astrofisico di Arcetri, Largo Enrico Fermi 5, 50125 Firenze, Italy, 
\and
Steward Observatory, University of Arizona, 933 N. Cherry Ave. Tucson, 
AZ 85721, USA 
\and
Department of Astronomy \& Astrophysics, University of Toronto, 50 St
George Street Toronto, Ontario M5S 3H4, Canada 
\and
Instituto de Astrof\'\i sica de Andaluc\'\i a, CSIC, Glorieta de la Astronom\'\i a s/n, 18008, Granada, Spain  
\and
Laborat\'orio de Astrof\'\i sica Te\'orica e Observacional, Universidade Estadual de Santa Cruz, Ilh\'eus, Brazil 
\and
Gemini Observatory, Casilla 603, La Serena, Chile 
\and
Argelander-Institut f\"ur Astronomie, Universit\"et Bonn, Auf dem H\"ugel 71, 53121, Bonn, Germany 
\and
University of Vienna, Department of Astronomy, T\"urkenschanzstrasse 17, 1180 
Vienna, Austria  
}

\date{Accepted . Received ; Draft printed: \today}

\authorrunning{Guennou et al.}

\titlerunning{Substructures in 0.4$<$z$<$0.9 clusters}

\abstract 
{.}
{We analyse the structures of all the clusters in the DAFT/FADA survey
  for which XMM-Newton and/or a sufficient number of galaxy redshifts
  in the cluster range is available, with the aim of detecting
  substructures and evidence for merging events.  These properties are
  discussed in the framework of standard cold dark matter
  ($\Lambda$CDM) cosmology.}
{.}
{XMM-Newton data were available for 32 clusters, for which we derive
  the X-ray luminosity and a global X-ray temperature for 25 of
  them. For 23 clusters we were able to fit the X-ray emissivity with
  a $\beta -$model and subtract it to detect substructures in the
  X-ray gas. A dynamical analysis based on the SG method was applied
  to the clusters having at least 15 spectroscopic galaxy redshifts in
  the cluster range: 18 X-ray clusters and 11 clusters with no X-ray
  data. The choice of a minimum number of 15 redshifts implies that
  only major substructures will be detected.  Ten substructures were
  detected both in X-rays and by the SG method.  Most of the
  substructures detected both in X-rays and with the SG method are
  probably at their first cluster pericentre approach and are
  relatively recent infalls.  We also find hints of a decreasing X-ray
  gas density profile core radius with redshift. }
{The percentage of mass included in substructures was found to be
  roughly constant with redshift with values of 5-15$\%$, in
  agreement both with the general CDM framework and with the results
  of numerical simulations. Galaxies in substructures show the same
    general behaviour as regular cluster galaxies; however, in substructures,
    there is a deficiency of both late type and old stellar population 
    galaxies. Late type galaxies with recent bursts
    of star formation seem to be missing in the substructures close to the 
    bottom of the host cluster potential well. However, our sample would need 
  to be increased to allow a more robust analysis. }


\maketitle

\section{Introduction}

The DAFT/FADA survey\footnote{PIs: M. Ulmer, C. Adami, and D. Clowe, see
Guennou et al. 2010 and http://cencos.oamp.fr/DAFT/ for a full
description of the project} is based on the study of $\sim$90 rich
(masses found in the literature $>2 \times 10^{14} M_{\odot}$) and
moderately distant ($0.4<z<0.9$) galaxy clusters, all with HST
imaging data available. This survey has two main
objectives. The first one is to constrain dark energy (DE) using weak
lensing tomography on galaxy clusters. The second one is to build a
database of rich distant clusters to study their properties. The
requirement of obtaining photometric redshifts for the DAFT/FADA
survey fields has indeed allowed us to build a rich multi-band imaging
database for these clusters. For a number of them, we have also
obtained spectroscopic data for several tens of galaxies in the
cluster redshift range, either from our own observations, or from
public databases.

DAFT/FADA is a nice complement to other cluster surveys such as for
example the Local Cluster Substructure Survey (LoCuSS, X-ray selected,
around z$\sim$0.2, Smith et al. 2010), the MAssive Cluster Survey
(MACS, X-ray selected, z$>$0.3, but with only 12 clusters above z=0.5,
Ebeling et al. 2001a, 2007, 2010), or the Cluster Lensing And
Supernova survey with Hubble (CLASH, Postman et al. 2012) which
analyses the high mass end of the cluster population, with only 14
clusters in the redshift range of our survey.

The redshift range chosen for the DAFT/FADA survey is an important one
in terms of studying cluster evolution. First, because clusters have
achieved nearly full growth in terms of mass by redshift of about 1
(see review by Kravtsov \& Borgani 2012, hereafter KB12, and
references therein) and the dark energy (DE) density becomes the
dominant form of energy density just below redshift 0.4 (under the
standard cold dark matter, hereafter $\Lambda$CDM, cosmology).
Second, clusters have acquired a hot and dense enough ICM to become
detectable at $z < 2$, and yet, clusters continue to evolve, with
infalling substructures present to current day, such as the Coma
cluster and the NGC~4839 subgroup (Neumann et al. 2003). The [0.4,0.9]
redshift range is also interesting because it spans a time frame of
about 3~Gyr, giving substructures time to infall, and thus allowing
comparison of the younger systems with the older ones on a meaningful
time scale.  With a typical infall speed of 1000~km~s$^{-1}$,
substructures have enough time to cross the cluster about three times
between $z$ of 0.9 (age of universe 6.3~Gyr) and 0.4 (9.4~Gyr).
Understanding how the three major components (the ICM, the galaxies,
and the dark matter concentrations that are the seeds of clusters)
form to grow into massive clusters is still a work in progress though
(KB12).

The cluster formation simulations all involve an assumption about
initial density perturbations and must include interplay between (at
least) non-interacting cold dark matter (CDM) and the evolution of the
baryon content in the cluster, including the ICM and galaxies.

While on the galactic scale there is a possible disagreement between
the number of subhaloes found and predicted (e.g. Strigari et
al. 2010), there have not been enough observations yet at the cluster
and substructure scales to require any adjustments or re-examination of
the $\Lambda$CDM paradigm.

Tonnesen \& Bryan (2008) offer a useful review of cluster substructure
observations.  Among the numerous papers dealing with the observations
of cluster substructures, we can also quote for X-rays B\"ohringer et
al. (1994), Dupke \& Bregman (2001), Furusho et al. (2001), Shibata et
al. (2001), Churazov et al. (2003), B\"ohringer et al. (2010), and
Weissmann et al. (2013).  And in the optical: Adami et al. (2005),
Ulmer et al. (2009), Einasto et al. (2012), and Wen \& Han (2013).
However, there has been little coupling of X-ray and optical data,
especially in the redshift range [0.4,0.9] of the DAFT/FADA
sample. Over this time frame of about 3~Gyr, some galaxy groups are
infalling for the first or second time (e.g. Poole et al. 2007 and the
present paper). As noted above, clusters have already accreted enough
material to become detectable by $z\sim 1$, and the time scale from
$z$=0.9 to 0.4 is just sufficient for a substructure to move in and
out of a cluster.

Thus, we can compare our work with the predictions of simulations such
as those by Poole et al. (2007) and the observations of subgroups that
are in the field.  Further work done by Tonnesen and Bryan (2008) also
shows that the existence and properties of subclusters affect the
evolution of ram pressure stripping of galaxies due to the local
relative velocities between the substructures and the cluster
galaxies.  In related work on the ICM in subgroups, Takizawa (2005),
for example, has shown that subclusters do not lose all their mass via
ram pressure stripping, say, in the first passage, which agrees with
the simulations of Poole et al. (2007). Thus, one would expect to see
(as is observed) substructure in some of the cluster X-ray emission
images over all redshift ranges below about 1. In complementary work
by Gao et al.  (2012: the Phoenix project), they and others (e.g.
Springel et al. 2008a,b, Navarro et al. 2010, the Aquarius Project)
predict the fraction of mass clusters made up of subclusters (at $z =
0$, however), and we can compare our results (albeit at higher
redshift) with those simulations as a first step in linking these
cluster formation codes with our observations.

Our data could therefore lead to improvements in the simulations to
describe the substructure mass fraction growth and evolution over the
$z = 1-0$ time frame in a more quantitative way than results obtained,
for example, by simulation movies (e.g. Diemand,
http://krone.physik.unizh.ch/~diemand/clusters/ and Hydro-ski,
http://astro-staff.uibk.ac.at/~hydroskiteam/) or images (e.g.  Borgani
\& Kravtsov (2009). 
Perhaps the present paper and future observational papers will
encourage more quantitative simulations and their analysis.
 
From an observational point of view, the substructure forming groups
that clusters continue to accrete at later epochs than $z \simeq 1$
have smaller scales than do the clusters themselves or those of
larger groups accreted above about $z \simeq 2$ (e.g. Adami et
al. 2012, Connelly et al. 2012). The existence of such groups of
galaxies has been confirmed up to z$\sim$1.3 (e.g. Gerke et al. 2007
or Lilly et al. 2009) and groups are very common at lower redshifts
(e.g. Carlberg et al. 2001).  This also implies different mechanisms
in the group accretion (e.g. simulated in Poole et al. 2007, see also
KB12) than during the initial cluster formation and should have direct
consequences on the cluster dynamical state, which can be probed by
detecting substructures. The search for substructures in the [0.4,0.9]
redshift range thus allows us to search for traces of this accretion
mechanism inside galaxy clusters. This search can be made either in
the galaxy distribution or in the intracluster medium (ICM) through
X-ray data.

We  primarily used data from the XMM-Newton archive to detect
substructures in all the clusters of the DAFT/FADA survey for which
such data were available (about half of the sample) and then we
carried out follow-up Chandra analysis.  A review of methods available to
search for substructures in X-rays can be found in Andrade-Santos et
al. (2012).
In the optical, we used the Serna \& Gerbal (1996) hierarchical code
(based on spectroscopic redshifts and optical magnitudes) to detect
optical substructures.  A number of better known methods are available
to search for substructures in the optical, such as the $\Delta -$test
(Dressler \& Schechtman 1988), which searches for deviations in the
local mean velocity and velocity dispersion from the overall
values. However, the SG method is quite powerful for showing evidence
of substructuring, as illustrated by the results for a number of
different clusters, at low redshift (Abell~496: Durret et al. 2000;
Coma: Adami et al. 2005, 2009; Abell~780: Durret et al. 2009;
Abell~85: Bou\'e et al. 2008), moderate redshift (Abell~222/223:
Durret et al. 2010), and high redshift (RX~J1257.2+4738: Ulmer et
al. 2009).  The SG method has also been extensively tested on
simulations by Guennou (2012), in particular on the effect of
undersampling on mass determinations.

 Our aim in this paper is to investigate the
structure of the DAFT/FADA clusters for which X-ray and/or optical
spectroscopic+imaging data are available. This will improve our
knowledge of clusters (analysed in a homogeneous way) in the [0.4,0.9]
redshift range.

As we show below, at least in a general way, there is agreement
between theory (simulations) and our observations of substructure in
rich clusters, but further work is needed on both fronts to determine
if the standard $\Lambda$CDM model of the Universe needs any
modification with regards to the effects of the CDM on the growth of
large scale--structures in the Universe.

The paper is presented in the following way. In Section~2, we present
the X-ray data, analyses, and results. Section~3 is dedicated to
optical data and Serna \& Gerbal analyses. Our results are presented
in Section~4 and summarized in Section~5.  The majority of the figures
(X-ray images and X-ray residuals over an azimuthally averaged $\beta
-$model, spectroscopic redshift histograms), except for a few
illustrative ones, are grouped in the Appendix (available in
electronic form).

We adopted the Dunkley et al. (2009) concordance cosmological model
(H$_0$=71.9~km~s$^{-1}$~Mpc$^{-1}$, $\Omega _\Lambda = 0.742$,
$\Omega_M=0.258$).

\section{X-ray data and data analysis}

We retrieved XMM-Newton data from the public archive and only kept the
clusters with data of sufficient quality and depth (typically those
for which the relative error on the MOS1 count rate, hence the X-ray
luminosity, is less than 20\%). The XMM-Newton data obtained with the
EPIC-MOS (Metal Oxide Semi-conductor) instruments were reduced using
the SAS (Science Analysis System developed by the XMM-Newton team)
tool from the Heasarc package. After this we applied the code created
by Andy Read to remove flares, using a 3$\sigma$ clipping technique,
and we calibrated the images.

\subsection{Basic parameters} 

We analysed the XMM-Newton data available for 42 clusters to
derive their X-ray temperatures and luminosities and search for
substructures. A spatial analysis was possible for 32 of these
clusters, but only 23 had deep enough X-ray data for a really robust
spatial analysis (i.e. the $\beta$-model ftting process converged for
those 23 clusters).

The information on the 32 clusters in our sample that have usable
XMM-Newton data is given in Table~\ref{tab:Xandz}. X-ray luminosities
in the [0.5,8.0] keV interval were computed for all of them, but in
seven cases the X-ray emission was not sufficient to estimate the
temperature of the X-ray gas. For 17 clusters that we have in common
with Baldi et al. (2012), we compared our X-ray gas temperatures and
find good agreement (mean difference of $-0.27$~keV with a
dispersion of 1.34~keV).

Though the main aim of this paper is to study the substructures in the
DAFT/FADA survey, we computed the gas masses and total masses for the
25 clusters with measurable X-ray temperatures, and these masses can
give interesting information on cluster properties.  We estimated the
X-ray gas masses and total masses in the $r_{500}$ radius for the
clusters with reliable X-ray temperatures using the proxy calculated
by Kravtsov et al. (2006), based on simulations with cosmological
parameters close to ours.  The parameters of the proxy determined by
Kravtsov et al. (2006) are given for relaxed and unrelaxed clusters,
and for z=0 and z=0.6, with small differences from one category to
another. Because our X-ray clusters cover a redshift range between 0.4
and 0.9, and some are relaxed and some are not (and in a number of
cases we cannot classify our clusters as relaxed or unrelaxed), we
took average values in the Kravtsov et al. (2006) Table~2
(log$_{10}$C=14.4 and $\alpha = 1.500$) to obtain the following
formula (which give the gas and total masses in solar masses as a
function of the X-ray gas temperature kT in keV):

\begin{equation}
M_{gas,500} = \frac{2.5 \times 10^{14}}{11.2} (\frac{kT_{{\rm keV}}}{3})^{1.5}\ M_\odot.
\end{equation}

To compute the total mass, we decided to take one of the best 
determinants, also given by Kravtsov et al. (2006):
\begin{equation}
Y_x = kT \times M_{gas,500}.
\end{equation}
\noindent
Using the average values of the parameters (for all redshifts and all clusters: 
log$_{10}$C=14.27 and $\alpha = 0.581$), we computed the total mass with the relation:

\begin{equation}
 M_{tot,500} = 10^{14.27} \times ({\frac{Y_x}{4.0 \times 10^{13}}})^{0.581}\ M_\odot.
\end{equation}

Total masses may be slightly underestimated here, since the stellar
contribution (stars in galaxies and intracluster light) has not been
taken into account (though it has been shown not to be negligible, see
e.g. Gonzalez et al. 2007).  The corresponding masses are given in
Table~\ref{tab:Xandz}.  In the case of unrelaxed clusters, we may
expect the total masses derived from X-rays to be overestimated
(e.g. Mamon 2000, Chon et al. 2012 and references therein).

\begin{landscape}
\begin{table}[t!]
  \caption{Data for the clusters with usable XMM-Newton data. Notes:
    (1)~name given by NED (LCDCS clusters come from Gonzalez et al. 2001), 
    (2)~right ascension in degrees (J2000.0), (3)~declination in degrees (J2000.0), 
    (4)~redshift, (5)~number of galaxies with redshifts in the cluster range, 
    (6)~useful XMM-Newton exposure time (in seconds), 
    (7)~temperature of the X-ray gas, (8)~X-ray luminosity in the [0.5,8.0]~keV band, (9)~X-ray gas mass
    in the $r_{500}$ radius, (10)~total mass in the $r_{500}$ radius, (11)~subtructure inside the cluster (1 means yes, 
    -1 means no, 0 means not detectable with data in hand).
    ** The position given for LCDCS 504 comes from Guennou et al. (2013) where we determined the centre as the position of the cD, 
    whereas the position given in parentheses was obtained from NED.
    *** The cluster MACS~J1423.8+2404 did not have any XMM-Newton public data, but we collected Chandra data with enough
    depth to be able to subtract a $\beta -$model and search for substructures.}

\begin{tabular}{lrrrrrcrrrr}
\hline
\hline
Name                 & RA       & DEC       & z~~    & Nz & $\Delta$t (s)& kT (keV) & L$_{\rm X}$ (erg~s$^{-1}$) & M$_{gas,r500}$ (M$_\odot$) & M$_{tot,r500}$ (M$_\odot$) & Substructure\\
\hline
CL~0016+1609         &  4.63888 &  16.44329 & 0.5455 & 173& 30320 & 8.97$\pm$0.40  & 2.23$\pm 0.06$e+45 & 1.17$\pm 0.03$e+14 & 1.24$\pm 0.04$e+15 & 1 \\ 
CL~J0152.7-1357      & 28.17083 & -13.96250 & 0.8310 & 115& 51150 & 7.55$\pm$0.65  & 7.91$\pm 0.74$e+44 & 7.72$\pm 0.28$e+13 & 8.83$\pm 0.38$e+14 & 1 \\ 
MS~0302.5+1717       & 46.32911 &  17.47729 & 0.4250 &  1 & 12440 & 6.23$\pm$1.13  & 1.41$\pm 0.35$e+44 & 6.02$\pm 0.72$e+13 & 6.84$\pm 0.67$e+14 & 0/-1\\ 
XDCS cm J032903.1+025640 & 52.26175 & 2.94033 & 0.4122& 13& 47520 & 3.26$\pm$0.38 & 1.40$\pm 0.89$e+43 & 2.60$\pm 0.30$e+13 & 2.88$\pm 0.40$e+14 & -1 \\
RX~J0337.6-2522      & 54.43812 & -25.37669 & 0.5850 &  5 & 11820 & 2.40$\pm$0.78  & 5.73$\pm 4.98$e+43 & 1.65$\pm 0.25$e+13 & 1.85$\pm 0.36$e+14 & 0 \\ 
MACS J0454.1-0300    & 73.54552 &  -3.01865 & 0.5377 & 194& 25420 & 9.97$\pm$0.59  & 2.04$\pm 0.06$e+45 & 1.29$\pm 0.04$e+14 & 1.39$\pm 0.04$e+15 & 1 \\ 
BMW-HRI~J052215.8-362452 & 80.55917 & -36.41778 & 0.4720 & 1 & 17920 & 5.64$\pm$0.66 & 1.70$\pm 0.27$e+44 & 4.99$\pm 0.32$e+13 & 5.78$\pm 0.41$e+14 & 0/-1 \\ 
MACS~J0647.7+7015    & 101.94125 & 70.25083 & 0.5907 &  1 & 32030 & 7.74$\pm$0.35  & 1.61$\pm 0.07$e+45 & 9.76$\pm 0.32$e+13 &  1.02$\pm 0.04$e+15 & 1 \\ 
MACS~J0744.9+3927    & 116.21583 & 39.45917 & 0.6860 &  2 & 71260 & 7.87$\pm$0.28  & 1.87$\pm 0.06$e+45  & 8.96$\pm 0.19$e+13 & 9.86$\pm 0.30$e+14 & 1 \\ 
RX~J0847.1+3449      & 131.79708 & 34.82111 & 0.5600 &  1 &  6908 &                & 7.11$\pm 0.49$e+44 & & & 0 \\ 
MACS J0913.7+4056    & 138.40277 & 40.94315 & 0.4420 &   2& 12430 & 5.39$\pm$0.19  & 1.58$\pm 0.06$e+45 & 5.29$\pm 0.09$e+13 & 5.83$\pm 0.20$e+14 & 0 \\ 
Abell 851            & 145.73601 & 46.98942 & 0.4069 & 213& 40970 &  5.17$\pm$0.16 & 6.13$\pm 0.20$e+44 & 5.04$\pm 0.10$e+13 & 5.53$\pm 0.21$e+14 & 1 \\ 
MS~1054-03           & 164.25093 & -3.62428 & 0.8231 & 326& 25680 & 8.64$\pm$ 0.66 & 1.49$\pm 0.13$e+45 & & & 1 \\ 
UM 425 Cluster       & 170.83542 &  1.62944 & 0.7685 &   8& 26090 & 14.5$\pm$4.2   & 4.90$\pm 0.34$e+44  & 1.17$\pm 0.06$e+14 & 1.64$\pm 0.06$e+15 & 0 \\ 
MS 1137.5+6624       & 175.09696 & 66.14485 & 0.7820 &  17& 15560 &  7.43$\pm$0.90 & 7.41$\pm 1.05$e+44  & 8.49$\pm 0.58$e+13 & 9.25$\pm 0.59$e+14 & 0 \\ 
CLG J1205+4429       & 181.46410 & 44.48600 & 0.5915 &  10& 25520 &                & 6.77$\pm 3.32$e+43  & & & 0 \\ 
RXC~J1206.2-0848     & 181.54991 & -8.80001 & 0.4400 &  53& 10200 & 9.36$\pm$0.59  & 2.32$\pm 0.08$e+45  & 1.23$\pm 0.05$e+14 & 1.31$\pm 0.05$e+15 & 1 \\ 
LCDCS~0504**           & 184.18845 & -12.02147 & 0.7943 &  65& 23460 & 5.49$\pm$0.64  & 3.91$\pm 0.50$e+44  & 9.00$\pm 0.93$e+13 & 8.02$\pm 0.77$e+14 & 1 \\ 
                     & (184.18792) & (-12.02139) &        &            &                &           &                    &                    &   \\
BMW-HRI~J122657.3+333253 & 186.74167 & 33.54836 & 0.8900 & 35 & 65350 &  8.74$\pm$0.42 & 2.02$\pm 0.10$e+45  & 1.16$\pm 0.03$e+14 & 1.21$\pm 0.04$e+15 & -1 \\ 
GHO 1322+3027        & 201.20091 & 30.19276 & 0.7550 &  38& 36390 & 5.98$\pm$1.33 & 8.91$\pm 3.64$e+43  & 5.17$\pm 0.63$e+13 & 6.11$\pm 0.62$e+14 & 0 \\ 
ZwCl~1332.8+5043     & 203.58333 & 50.51506 & 0.6200 &  1 & 27860 & 5.08$\pm$0.59  & 2.40$\pm 0.51$e+44  & 4.39$\pm 0.33$e+13 & 5.05$\pm 0.42$e+14 & 0 \\ 
LCDCS 0829           & 206.88333 & -11.76167 & 0.4510 &  50& 32370 & 11.31$\pm$0.24 & 7.75$\pm 0.06$e+45  & 1.58$\pm 0.02$e+14 & 1.69$\pm 0.03$e+15 & 1 \\ 
LCDCS~0853           & 208.53958 & -12.51639 & 0.7627 &  18& 27420 &                & 3.65$\pm 0.38$e+44  & & & 0 \\ 
RX~J1354.2-0221      & 208.57042 & -2.36306 & 0.5460 &  2 &  18390 & 2.18$\pm$0.98  & 2.04$\pm 0.62$e+44  & 1.38$\pm 0.29$e+13 & 1.58$\pm 0.29$e+14 & 0 \\ 
MACS~J1423.8+2404***    & 215.95125 & 24.07972 & 0.5450 &   9  &  113400     & 5.3$\pm$0.1  & 1.71$\pm 0.05$e+45 & 5.24$\pm 0.10$e+13 & 5.74$\pm 0.10$e+14 &1 (Chandra)\\ 
GHO~1602+4312        & 241.10483 & 43.08131 & 0.8950 &  26& 12000 &               & 8.96$\pm 8.39$e+43  & & & 0 \\ 
MS~1621.5+2640       & 245.89863 & 26.56378 & 0.4260 & 104&  2210 &               & 9.13$\pm 3.46$e+43  & & & 0 \\ 
CXOU J205617.1-044155& 314.07150 & -4.69864 & 0.6002 &   1& 16240 & 2.46$\pm$1.02 & 1.98$\pm 0.42$e+44  & 5.02$\pm 0.71$e+13 & 3.59$\pm 0.66$e+14 & 0 \\ 
MS~2053.7-0449       & 314.09321 & -4.62873 & 0.5830 &  30& 16240 & 4.81$\pm$1.17  & 2.13$\pm 0.56$e+44  & 4.42$\pm 0.63$e+13 & 4.91$\pm 0.62$e+14 & 0 \\ 
GHO~2143+0408        & 326.52000 &  4.38861 & 0.5310 &  1 & 20170 & 4.39$\pm$0.43  & 1.51$\pm 0.30$e+44  & 4.60$\pm 0.29$e+13 & 4.77$\pm 0.39$e+14 & 0 \\ 
RX~J2202.7-1902      & 330.68708 & -19.03611 & 0.4380 & 8 & 26060 & 4.97$\pm$1.34  & 4.56$\pm 2.07$e+43  & 4.41$\pm 0.57$e+13 &  5.00$\pm 0.58$e+14 & -1 \\ 
RX~J2328.8+1453      & 352.20792 & 14.88667 & 0.4970 &  1 & 25510 & 2.63$\pm$0.53  & 4.28$\pm 1.78$e+43  & 2.03$\pm 0.26$e+13 & 2.20$\pm 0.37$e+14 & 0/-1 \\ 
\hline
\end{tabular}
\label{tab:Xandz}
\end{table}
\end{landscape}

\begin{table*}
  \caption{Detected substructures. Notes: (1)~cluster name, (2)~substructure number,
    (3)~substructure number of galaxies, (4)~substructure mean redshift,
    (5)~substructure to total cluster mass ratio estimated with
    the SG method and given in 10$\%$ wide intervals (the asterisk 
means that we only detected the main structure),
    (6)~substructure to total cluster mass ratio estimated with
    the method based on a scaling relation described in
Section~\ref{subsec:massnico} and given in 10$\%$ wide intervals,
    (8)~substructure velocity dispersion estimated
    with SG for the substructures also detected in X-rays ,
    (9)~substructure X-ray luminosity, (10)~merging stage (see
text). }
\begin{tabular}{lrrrrrrrr}
\hline
\hline
Name & \#  & N$_{gal}$ & z & ${\rm (M_{SS}/M_{tot})_{SG}}$ & ${\rm
(M_{SS}/M_{tot})_{sc}}$   & Vel. Disp. & L$_X$ & merging  \\
     &     &          &   & (\%)&(\%) & (km/s)     & (erg/s)& stage \\
\hline
CL~0016+1609      & 1 & 64& 0.5418 & [10;20] & [10;20]&  & &\\ 
                  & 2 & 17& 0.5505 & [0;10] & [0;10]&  & &\\
                  & 3 & 13& 0.5530 & [0;10] & [0;10]&  & &\\
                  & 4 & 24& 0.5597 & [0;10] & [0;10]&  200 & 1.56 10$^{44}$&t1\\
CL~J0152.7-1357   & 1 & 49& 0.8382 & [20;30] & [40;50]&  680 & 3.28
10$^{44}$&t1\\        
                  & 2 & 17& 0.8323 & [0;10] & [0;10]&  & &\\
                  & 3 & 29& 0.8279 & [0;10] & [0;10]&  & &\\
                  & 4 & 34& 0.8458 & [20;30] & [0;10]&  320 & 2.27 10$^{44}$ &t1\\
XDCS cm J032903.1+025640  & 1 & 7& 0.4115 & [0;10] & [50;60] &  & &\\ 
                  & 2 &  4& 0.4095 & [0;10] & [0;10] &  & &\\
MACS J0454.1-0300 & 1 & 31& 0.5365 & [10;20] & [0;10]&  & &\\  
                  & 2 &  5& 0.5407 & [0;10] & [0;10]&  & &\\
                  & 3 & 18& 0.5434 & [0;10] & [0;10]&  320 & 3.24 10$^{44}$&t1\\
                  & 4 &  6& 0.5376 & [0;10] & [0;10]&  & &\\
                  & 5 &  3& 0.5320 & [0;10] & [0;10]& & &\\
                  & 6 & 12& 0.5309 & [0;10] & [0;10]&  & &\\
                  & 7 &  3& 0.5390 & [0;10] & [0;10]&  & &\\
                  & 8 &  4& 0.5457 & [0;10] & [0;10]&  & &\\
                  & 9 &  4& 0.5287 & [0;10] & [0;10]&  & &\\
Abell 851         & 1 &  6& 0.4070 & [0;10] & [0;10]&  & &\\ 
                  & 2 &  3& 0.4100 & [0;10] &  [0;10]& &  &\\
                  & 3 &  4& 0.4036 & [0;10] & [0;10]& 1300 & 5.63 10$^{43}$&t0
(t1 or t2)\\
                  & 4 &  3& 0.4059 & [0;10] &  [0;10]& & &\\
                  & 5 &  4& 0.4100 & [0;10] &  [0;10]& & &\\
                  & 6 &  8& 0.4142 & [0;10] & [0;10]& & &\\
                  & 7 &  3& 0.4100 & [0;10] &   [0;10]& &\\
                  & 8 &  3& 0.4163 & [0;10] & [0;10]& & &\\
MS~1054-03        & 1 &  7& 0.8218 & [0;10]&  [0;10]& 1250 & 1.94 10$^{44}$
&t1 (t0 or t2)\\ 
                  & 2 & 26& 0.8261 & [0;10] & [0;10]&  & &\\
                  & 3 &  5& 0.8267 & [0;10] & [0;10]& && \\                 
                  & 4 &  5& 0.8270 &  [0;10]       & [0;10]& & &\\
CLG~J1205+4429    & 1 & 11& 0.5948 & [90;100]* & [90;100]& & &\\ 
BMW-HRI~J122657.3+333253 & 1 & 10 &0.8816 & [20;30] & [0;10]& & &\\ 
                  & 2 &    4 &0.8910 & [0;10] & [0;10] & & &\\
                  & 3 &    5 &0.8920 & [0;10] & [0;10] & & &\\
                  & 4 &    5 &0.8930 & [0;10] & [0;10] & & &\\
                  & 5 &    4 &0.8960 & [0;10] & [0;10] & & &\\
                  & 6 &    4 &0.8970 & [0;10] & [0;10] & & &\\
RXC~J1206.2-0848  & 1 &    5 &0.4255 & [0;10] & [0;10]& & &\\ 
                  & 2 &    3 &0.4336 & [0;10] & [0;10]& & &\\
                  & 3 &    6 &0.4409 & [0;10] & [0;10]& & &\\
                  & 4 &    4 &0.4373 & [0;10] & [0;10]& 690 & 5.10 10$^{43}$& t1
(t0 or t2)\\
LCDCS~0504        & 1 &    7 &0.8036 & [0;10] & [0;10]& 110 & 3.10
10$^{43}$&t1 (t0 or t2)\\
                  & 2 &   10 &0.7996 & [0;10] & [0;10]& & &\\
                  & 3 &   17 &0.7858 & [0;10] & [0;10] & & &\\
                  & 4 &    5 &0.7913 & [0;10] & [0;10] & & &\\
                  & 5 &    6 &0.7966 & [0;10] & [0;10] & & &\\
                  & 6 &    7 &0.7953 & [0;10] & [0;10] & & &\\
                  & 7 &    6 &0.7940 & [0;10] & [0;10]& & &\\
GHO 1322+3027     & 1 &   44 &0.7562 & [90;100]* & [90;100]& & &\\ 
LCDCS 0829        & 1 &   22 &0.4503 & [0;10] & [40;50]& 230 & 5.79
10$^{44}$&t1\\
                  & 2 &   14 &0.4529 & [0;10] &[20;30] & & &\\
                  & 3 &   15 &0.4465 & [0;10] &[0;10] & & &\\
                  & 4 &   12 &0.4553 & [0;10] &[0;10] & 180 & 1.82 10$^{45}$&t1\\
LCDCS~0853        & 1 &    3 &0.7648 & [90;100]* & [20;30]&  & &\\  
MACS~J1423.8+2404 & 1 &    3 &0.5445 & [90;100]*      & [30;40]&  & &\\  
GHO~1602+4312     & 1 &   29 &0.8941 & [90;100]* & [90;100]&  & &\\  
MS 1621.5+2640    & 1 &   24 &0.4245 & [0;10] & [10;20]&  & &\\  
                  & 2 &   19 &0.4264 & [0;10] & [0;10]&  & &\\
                  & 3 &   43 &0.4307 & [10;20] & [20;30]&  & &\\
                  & 4 &   20 &0.4211 & [0;10] & [0;10]&  & &\\
MS~2053.7-0449    & 1 &   28& 0.5837 & [90;100]* & [90;100]& & &\\  
GHO~2143+0408     & 1 &    4  &0.5205  & [90;100]*  & [80;90]& & &\\  
\hline
\end{tabular}
\label{tab:SG}
\end{table*}

\subsection{Model subtraction to search for substructures}

The X-ray images, with a pixel size of $3.2\times 3.2$~arcsec$^2$,
were then fit with an azimuthally symmetric elliptical $\beta -$model
(as given by Sherpa, see
http://cxc.cfa.harvard.edu/sherpa4.4/ahelp/beta2d.html):

$$\Sigma (r) = \Sigma _0 [1+(\frac{r}{r_c})^2]^{-3\beta +0.5} +b $$
\noindent
where $\Sigma (r)$ is the surface brightness as a function of radius
$r$, $\Sigma _0$ is the central surface brightness, $r_c$ the core
radius, $\beta$ the shape parameter, and $b$ accounts for the
background, which is assumed to be constant throughout the image.

To analyse the best quality data possible to search for substructures,
we had to make a compromise between having a high number of photons to
improve our detections and avoiding artifacts due in particular to
the superposition of images obtained with different detectors (and
thus summing up their defects). In this context, to limit the number
of artefacts, we ignored the observations that were contaminated by
bad pixels and/or had CCD gaps passing through the cluster image,
mainly the PN and sometimes the MOS-2 data.

We opted to model our clusters with a simple beta-model rather than
with more complex ones, such as the ``modified $\beta$-model''
(e.g. Vikhlinin et al. 2006) or double $\beta$-model (e.g. Eckmiller
et al. 2011). The reason is that many of our clusters are very faint
and have a small angular size, and their cores are hardly resolved by
XMM. This makes it very difficult to give meaningful constraints for
models with 12 or more free parameters (remembering that we are also
fitting the ellipticity, position angle, and coordinates of the
centre). Therefore, to compare all clusters uniformly we used
the standard 2D $\beta$-model described above.

This model represents a  relaxed cluster with a homogeneous
gravitational potential, simulated with the
Sherpa tool\footnote{http://cxc.harvard.edu/sherpa4.4/index.html} from 
CIAO\footnote{Chandra Interactive Analysis of Observations, see
  http://asc.harvard.edu/ciao}. 
The residuals were computed as the
difference between the image and the fit, allowing us to bring out any
perturbation from a homogeneous gravitational potential due to the
substructures that are not completely merged with the cluster
yet. This is a classical technique used, for example, to study the Coma
cluster with XMM-Newton (Neumann et al. 2003).

 Results for each
cluster are shown in the Appendix and a summary of the properties of
the X-ray detected substructures is given in Table~\ref{tab:SG}, together
with other quantities described in Section~\ref{subsec:stages}. We only give
the SG-estimated velocity dispersions for substructures also detected
in X-rays.

\subsection{Assessing substructure detections with simulations}

\begin{figure}
\centering
\caption[]{A few examples of simulated images showing the situation
  where the subcluster can be easily detected (low redshift, well
  separated) and the case where the subcluster is hardly seen (high
  redshift, small separation). See text for more details.}
\label{fig:exemplo}
\end{figure}

\begin{figure}
\centering
\includegraphics*[width=6.8cm,angle=270]{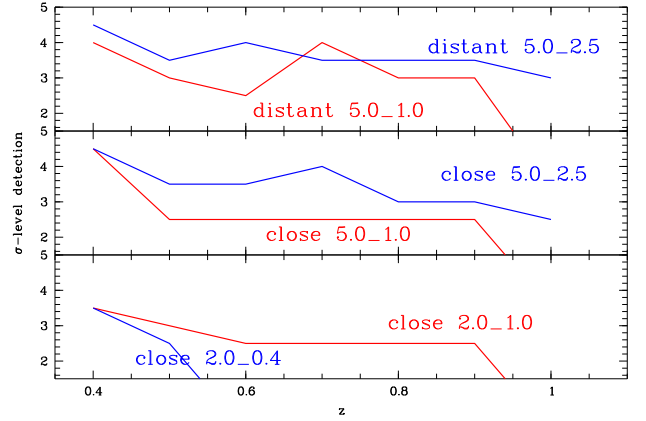}
\caption[]{Detection level (in numbers of $\sigma$ ) of the X-ray simulated 
substructures (50 ksec exposures) versus redshift. Lower figure: 2.0 
10$^{44}$ erg/s cluster + 
0.4 10$^{44}$ erg/s subcluster (close 2.0\_0.4) and 2.0 10$^{44}$ erg/s cluster 
+ 1.0 10$^{44}$ erg/s subcluster (close 2.0\_1.0). Separated by 6 px in both 
cases. Middle figure: 5.0 10$^{44}$ erg/s cluster + 
1.0 10$^{44}$ erg/s subcluster (close 5.0\_1.0) and 5.0 10$^{44}$ erg/s cluster 
+ 2.5 10$^{44}$ erg/s subcluster (close 5.0\_2.5). Separated by 6 px in both 
cases. Upper figure: 5.0 10$^{44}$ erg/s cluster + 
1.0 10$^{44}$ erg/s subcluster (distant 5.0\_1.0) and 5.0 10$^{44}$ erg/s 
cluster + 2.5 10$^{44}$ erg/s subcluster (distant 5.0\_2.5). Separated by 12 
px in both cases. When the line stops, it means that the subcluster is
no longer detected. The close 2.0\_0.4 configuration, for example, provides
detections of the subcluster only up to z=0.5 at the 2.5$\sigma$ level.}
\label{fig:simures}
\end{figure}

To test our method of identifying substructures in X-rays, we
have generated a series of synthetic X-ray images. These images
consist of a primary luminous cluster and a fainter subcluster, both
represented by a projected $\beta -$model, with $\beta = 0.6$ (a
typical value, see e.g. Jones \& Forman 1984). The surface brightness
per pixel is converted in counts/s taking the MOS
response, the cluster temperature, and the $K-$correction, which is
important given the redshift range we are covering ($0.4 \le z \le
1.0$) into account. The simulated cluster temperature is fixed using a $L_X - T_X$
scaling relation (Xue \& Wu 2000). The cluster images are then
generated assuming a Poisson distribution for each pixel.

We assumed the equivalent of 50~ks, 25~ks, and 10~ks exposures. This
optimally covers our range of exposure durations (see
Table~\ref{tab:Xandz}).  The synthetic images have a scale of
3.2~arcsec/pixel, corresponding to the binning we used.  We added
a flat Poissonian background with a count rate of $6.6\times 10^{-7}$
counts~s$^{-1}$~arcsec$^{-2}$, corresponding to the typical on-axis
MOS background in the [0.3--8.0] keV band observed in our (real)
images.

We generated synthetic images in a coarse grid (see
Fig.~\ref{fig:exemplo}) where the cluster and subcluster were
separated by either 6 or 12 pixels (19.2~arcsec and 38.4~arcsec, or
$\sim 140$ and 280~kpc at $z = 0.7$). For both separations, we fixed
the luminosity ratio at 1/5 and 1/2, with the luminosity of the
primary cluster fixed to be either $L_X = 5\times 10^{44}$ or $2
\times 10^{44}\,$erg~s$^{-1}$. Images were generated between redshifts
0.4 and 1.0 in steps of 0.1. In this way, we generated 42 simulated
images per exposure time (126 in total), to which we can apply our
X-ray substructure detection procedure.  The analysis of these
simulations is given in Fig.~\ref{fig:simures} for 50~ks exposures
(small bumps in the curves are due to hot pixels coinciding with a
substructure, which artificially increase the S/N).

\begin{figure}
\centering
\includegraphics*[width=6.cm,angle=270]{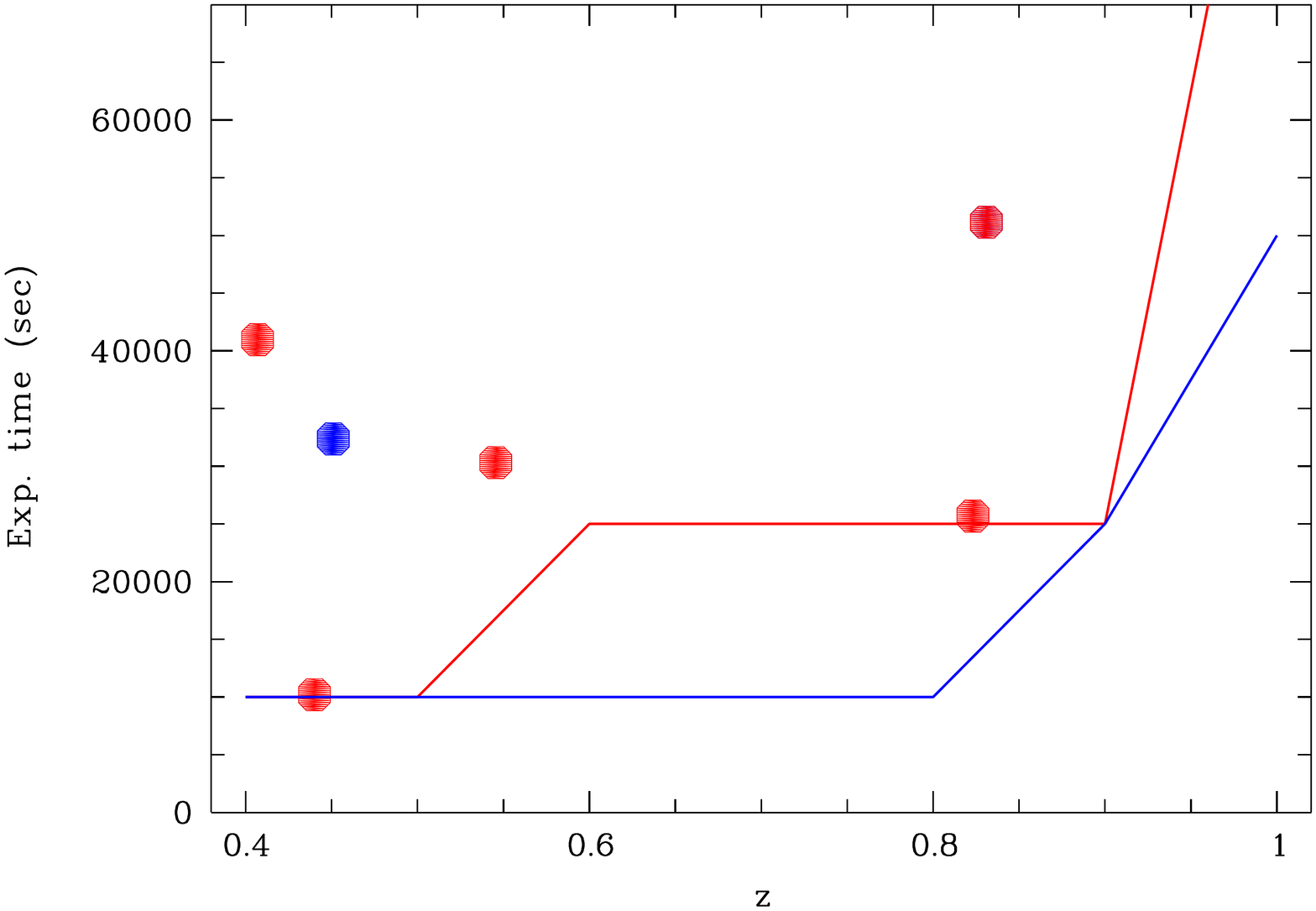}
\includegraphics*[width=6.cm,angle=270]{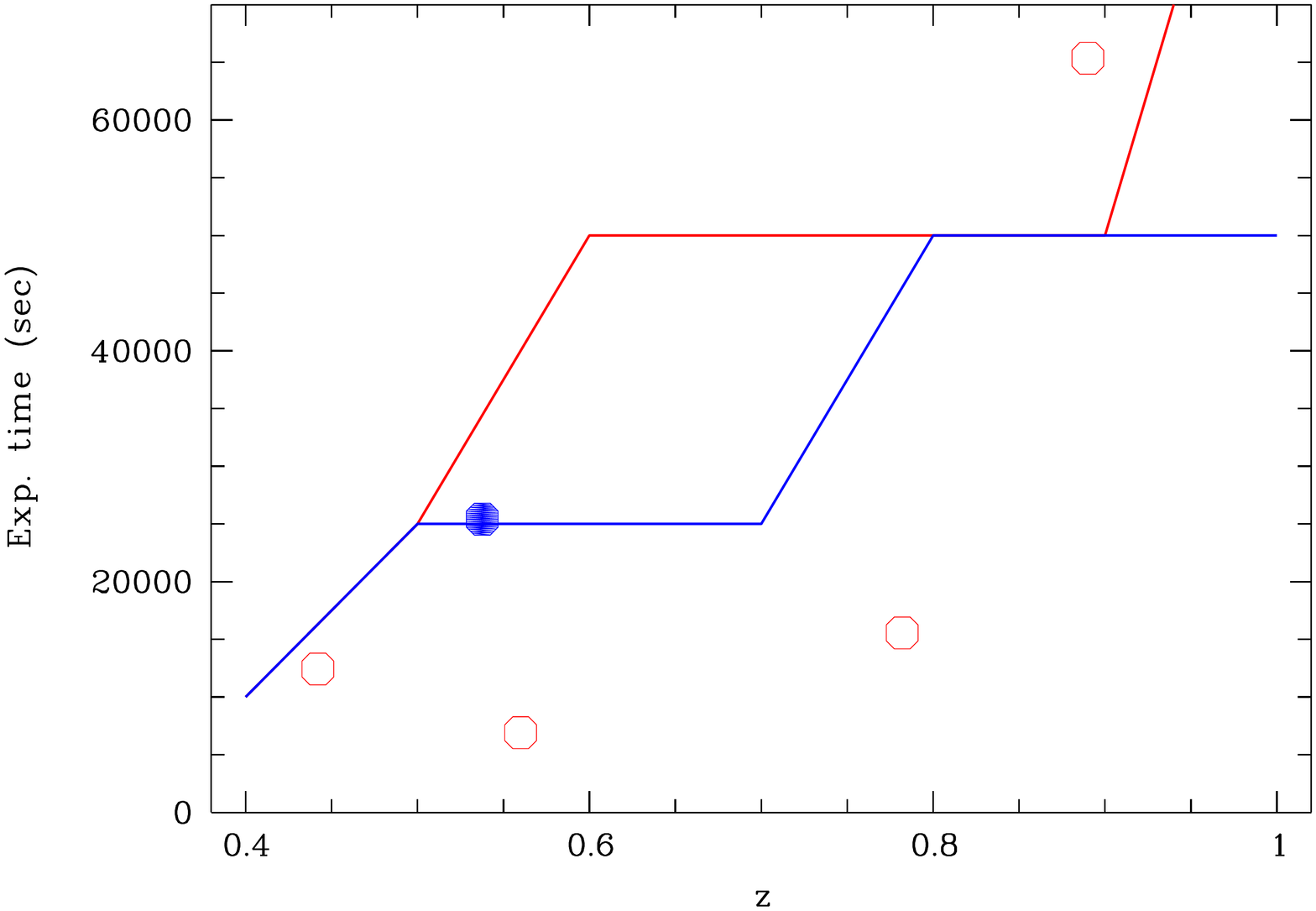}
\includegraphics*[width=6.cm,angle=270]{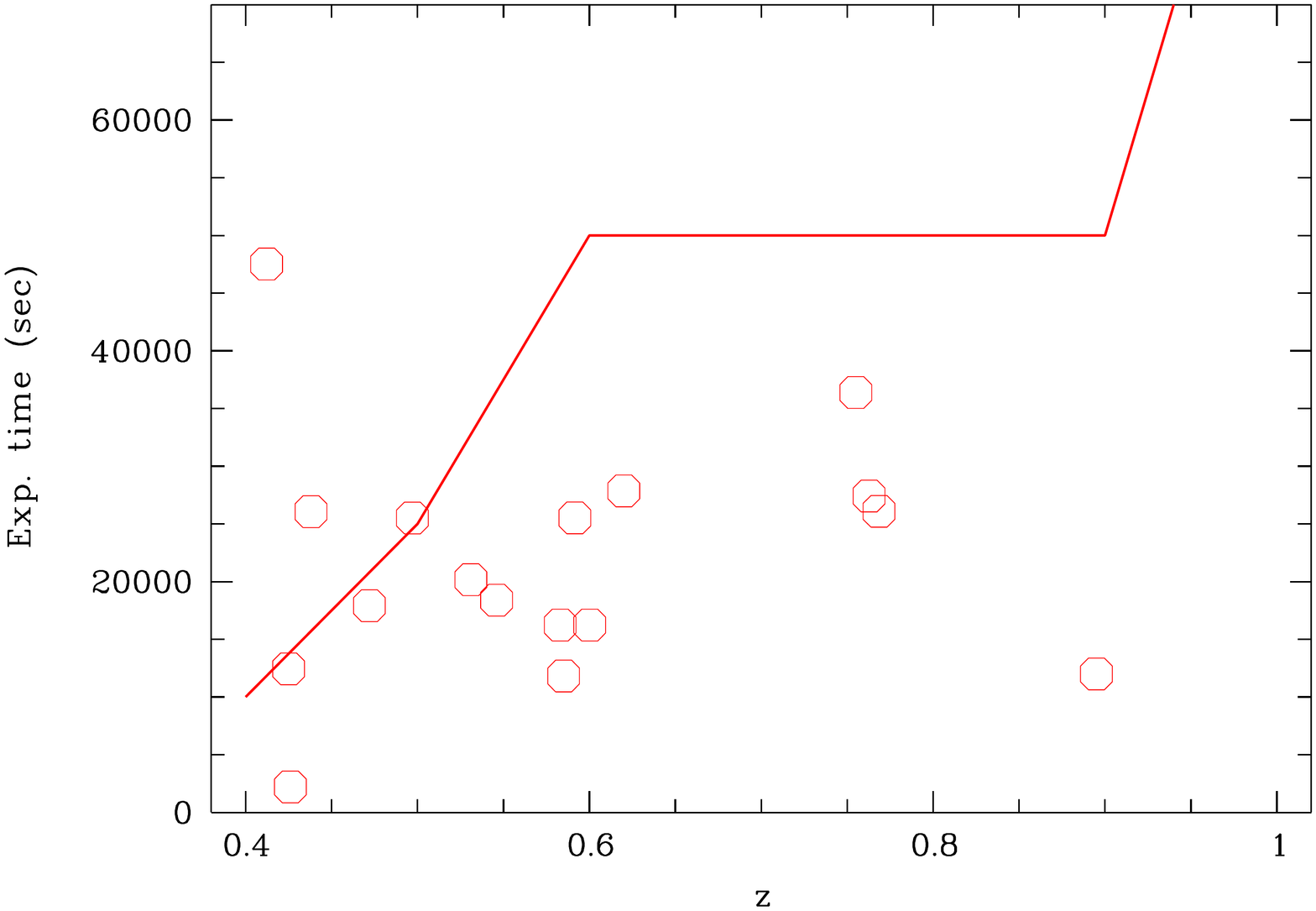}
\caption[]{Clusters in our sample in diagrams of
useful XMM-Newton exposure time versus
redshift. Filled symbols are clusters with one or more detected
substructures, and open symbols clusters without detected substructures.
Solid lines are susbstructure detection limits below which it is not possible 
to detect a substructure.

Upper figure:
Red symbols are clusters brighter than 5.0 10$^{44}$ erg/s with detected 
substructures fainter than 2.5 10$^{44}$ erg/s and more distant than 7 pixels
from the cluster centre (or no detected substructures).
Blue symbols are clusters brighter than 5.0 10$^{44}$ erg/s with detected 
substructures brighter than 2.5 10$^{44}$ erg/s and more distant than 7 pixels
from the cluster centre.

Middle figure:
Red symbols are clusters brighter than 5.0 10$^{44}$ erg/s with detected 
substructures fainter than 2.5 10$^{44}$ erg/s and less distant than 7 pixels
from the cluster centre (or no detected substructures).
Blue symbols are clusters brighter than 5.0 10$^{44}$ erg/s with detected 
substructures brighter than 2.5 10$^{44}$ erg/s and less distant than 7 pixels
from the cluster centre.

Lower figure: Red symbols are clusters fainter than 5.0 10$^{44}$
erg/s with detected substructures fainter than 0.4 10$^{44}$ erg/s and
less distant than 7 pixels from the cluster centre (or no detected
substructures).  }
\label{fig:simures2}
\end{figure}

We also show in Fig.~\ref{fig:simures2} the detection limits of the
50~ks, 25~ks, and 10~ks runs overlaid on our cluster distributions in
a plot of useful XMM-Newton exposure time as a function of redshift.
The results can be summarized as follows. There are indeed no clusters
without detected substructures in Fig.~\ref{fig:simures2} located
inside the area where we theoretically cannot detect substructures.
The detection of a substructure strongly depends on its X-ray
luminosity (more than on the ratio of the luminosity of the
substructure to that of the main structure). The more luminous the
substructure, the better it is detected at high redshift.  The
distance between the substructure and the cluster centre does not
change the maximum redshift of the detection, but has an influence on
the significance level of the detection. It is easier to detect a
substructure far from the cluster centre.

To be more quantitative, simulations predict the detection of
substructures brighter than $1.0 \times 10^{44}$~erg/s in our redshift
interval, while fainter substructures should be detected only for
$z<0.5$.  This is consistent with our results: all the substructures
that we detect are brighter than $10^{44}$~erg/s except three
(Abell~851, RXC~J1206, and LCDCS~0504), out of which the first two
clusters are at $z<0.45$.  In three quarters of the cases, we also
detect residuals of the cluster itself at a $2.5\sigma$
level. Therefore, potential substructures detected very close to the
cluster centre are probably only residuals from the cluster itself,
and not real substructures.

The results of these simulations allow us to indicate in Table~1
whether a substructure was detected or not. If it was not detected,
these simulations allow confirming that there is indeed no
substructure above the solid lines in Fig.~\ref{fig:simures2}, or if
substructures, if any, are below our detection limit (the solid lines
in Fig.~\ref{fig:simures2}). Open symbols touching the solid lines
were flagged 0/-1 in Table~\ref{tab:Xandz}.

\subsection{Contamination by point sources}

In some of the residual images it was difficult to distinguish
substructures from point sources owing to the limited spatial resolution
of XMM-Newton.  To check this point, we considered Chandra
images when available.  Thanks to their very good angular resolution,
these data allowed us to locate bright point sources, which were
usually bright enough not to be plagued by the relatively poor Chandra
collecting power.

Second, we also searched for all X-ray point sources known as AGN and
QSOs in the fields covered by XMM-Newton but not by Chandra. This was
done via the NED and Vizier databases, as well as with our own
spectroscopic observations (see below). Most of the time, public
databases referred to the work by Gilmour et al. (2009).

\subsection{Validation of X-ray luminosities and temperatures}

\begin{figure}[!ht]
  \begin{center}
    \includegraphics[angle=0,width=3.2in,bb=100 44 600 500]{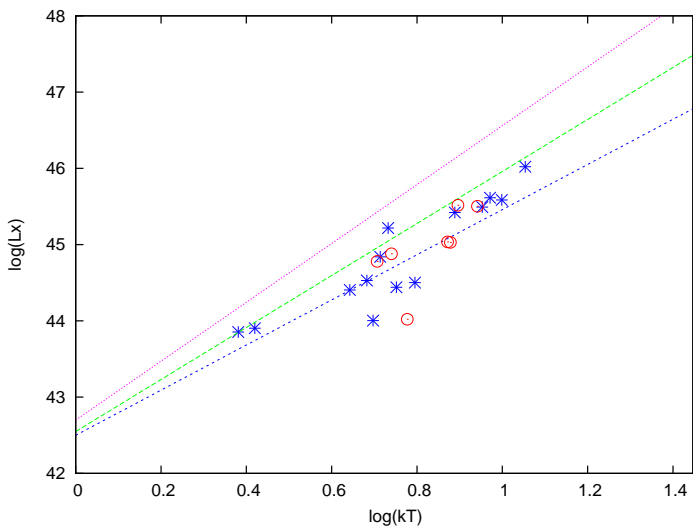}
    \caption{X-ray luminosity as a function of X-ray temperature for
      our clusters on a logarithmic scale.  The blue crosses and red
      circles represent clusters with redshifts respectively lower
      and higher than 0.6.  The green central dashed line represents 
      the ${\rm L_X-T_X}$ relation found by Takey et al. (2011), and the
      blue and pink dashed lines indicate the uncertainty on this relation.}
  \label{fig:kT_logLX}
  \end{center}
\end{figure}

Even though this was not one of the primary goals of the present
study, it was important to validate our X-ray measurements. We plot
the X-ray luminosities as a function of the X-ray temperatures in
Fig.~\ref{fig:kT_logLX}, with different symbols for the clusters at
redshifts lower and higher than 0.6 (the redshift limit of Takey et
al. 2011), and we plot on the same figure the Takey et al. (2011)
relation.  We can see that our X-ray data agree reasonably well with
the Takey et al. (2011) ${\rm L_X - T_X}$ relation.  If we try to
separate clusters with and without substructures, we find no obvious
dependence in the relation between ${\rm L_X}$ and ${\rm T_X}$ and the
level of substructuring, perhaps because of the relatively small size
of our sample.


\section{Optical data and Serna \& Gerbal analysis}

\subsection{Optical imaging and spectroscopic data}

Our survey was initiated for clusters with HST images available,
generally at least in the F814W band. We obtained deep optical imaging
for most of the clusters in several bands, with various telescopes
(Blanco, Calar Alto, CFHT, Gemini, GTC, SOAR, TNG, VLT, WIYN, WHT). Part of
the images were also taken from observatory archives (e.g. CFHT, Subaru,
ESO).  We are also presently in the process of acquiring infrared J
band images  to have better constraints on the photometric
redshifts of distant galaxies (for weak lensing tomography), see
http://cencos.oamp.fr/DAFT/. 

The galaxy magnitudes used here were measured in the V photometric
band images, calibrated and extinction--corrected in the usual way (see
Guennou et al. 2010). Some archive images were observed in other
bands, and in this case they were converted into the V band following
Fukugita et al. (1995).

We retrieved all the galaxy spectroscopic redshifts available in NED,
Vizier, and in the literature in a region of radius 5~arcmin around
each cluster centre. Such a zone corresponds to a radius of 1.59~Mpc
at z=0.4 and 2.24~Mpc at z=0.8, and therefore covers the entire
cluster.  We also added redshifts that we obtained during several
observing runs with 8m telescopes (42 redshifts obtained with GMOS on
the Gemini telescope and 60 obtained with FORS2 on the VLT).  Typical
errors on the velocities measured with FORS2 are $\pm 180$~km~s$^{-1}$
(see Guennou et al. 2013, in preparation). For redshifts taken from NED, the
uncertainties are not given, so we assume them to be comparable
to those of our FORS2 data.

Gemini/GMOS spectroscopy has a resolution of 7~\AA /px (see Guennou et
al. 2013 for details). The targeted objects were
brighter than $I_{AB} \sim 24$, and the reduction was made with the IRAF
package and the Gemini/Gmos environment.

VLT/FORS2 spectroscopy was obtained during  ESO period 85
(programme: 085.A-0016) with a resolution of 3~\AA/px. The targeted
objects were brighter than $I_{AB} \sim 23$. We applied the standard ESO
reduction since it proved to be good enough to measure redshifts for most
of our targets.

In order not to eliminate a priori galaxies that could be close to the
clusters on the line of sight, we applied the Serna \& Gerbal
(1996, hereafter SG, see below) method to all the galaxies with
measured redshifts.
For each cluster, the number of galaxies in the cluster
redshift range is given in Tables~\ref{tab:Xandz} and
\ref{tab:onlyzSG}. The cluster redshift range is defined as the range
within $\pm 0.025$ of the mean cluster redshift (which corresponds to
$\pm 3$ times the maximum velocity dispersion observed in clusters: 
$\sim$1500~km~s$^{-1}$).

Galaxy spectroscopic redshift histograms are given in the Appendix.
In most cases we only show a zoom around the mean cluster redshift.
When necessary, a full redshift histogram is also given (e.g. when
several structures are detected along the line of sight and/or when
the cluster redshift given in NED could be wrong).

\subsection{Searching for substructures by applying a Serna \& Gerbal
  analysis: the method} 

For each cluster, the catalogue of galaxies for which positions,
spectroscopic redshifts and magnitudes were available was analysed
with the SG method, in a region comparable to the X-ray image (when
available). We limited our sample to the clusters having at least 15
spectroscopic galaxy redshifts in the cluster range. Assuming a mean
number of three SG detected substructures per cluster, this allows us
to have statistically about 5 galaxies per substructure, close to the
number required to avoid being too affected by incompleteness (see
below).

The SG hierarchical method calculates the potential binding energy
between pairs of galaxies and detects substructures taking positions
and redshifts into account. The output is a file containing lists of
galaxies distributed in a hierarchical way. For example, structure~1
will be the cluster and structure~2 will be galaxies outside the
cluster.  Structure 1 will then be divided into 11 and 12, etc.

Assuming a value of 100 (in solar units) for the total
mass--to--stellar luminosity ratio (but results do not strongly depend
on this quantity, e.g. Adami et al. 2005), galaxy magnitudes can be
transformed into masses, and approximate values can be estimated for
the total masses of the various substructures detected by the SG
method.  The comparison with other mass estimates (see Sections~3.4
and 4.2) strongly suggests that the absolute values of the optical
masses estimated by the SG method cannot be considered as fully
reliable.  However, mass ratios (i.e. when trying to determine if a
substructure is more massive than another one) can be considered as
robust, keeping in mind, however, that assuming a constant M/L for all
galaxies is probably an oversimplification. Figure~\ref{fig:compl}
allows us to estimate the typical uncertainties on these mass
ratios. Considering only completeness levels lower than 80$\%$ (we
never reach higher completeness levels in spectroscopy), the typical
dispersion in mass estimates is of the order of 18$\%$. The mass
ratios of two such masses therefore must have uncertainties of the
order of 25$\%$ (the quadratic sum of the relative uncertainties on
cluster and substructure masses, both taken to be 18\%). By 25\%, we
mean that if a mass ratio is 40\%, the uncertainty on this value will
be $0.25\times 40\% = 10\%$.  Typical uncertainties on the mass ratios
computed with the scaling law method (see below) are estimated to be
of the same order, however we neglect here the uncertainty
on the M/L ratio assumed for the galaxies.  This, by the way, makes
 any comparison between the stellar and total masses of the
substructures difficult, in particular because by assuming the same M/L ratio
for all galaxies, mass ratios are in fact just optical luminosity
ratios.  Given the typical uncertainties estimated above, we chose not
to give the actual values of the mass ratios in Tables~\ref{tab:SG}
and~\ref{tab:SG2}, but rather to give estimates of these values in
10$\%$ wide intervals.

\subsection{Results of the Serna \& Gerbal analysis}

A summary of the substructures (if any) found by the SG method for
each cluster is given in Table~\ref{tab:SG}.

\begin{table}
  \caption{Clusters with no usable X-ray data but with more than 15 galaxy 
    redshifts in the cluster range. Notes: (1)~name (as in NED), 
    (2)~right ascension in degrees (J2000.0), (3)~declination in degrees (J2000.0), 
(4)~redshift, (5)~number of galaxies with redshifts in the cluster range. }
\begin{tabular}{lrrrr}
\hline
\hline
Name                     & RA & DEC & z      & Nz   \\ 
\hline
Cl J0023+0423B           &   5.96587 &   4.37797 & 0.8453 &  23 \\ 
CXOMP~J    &   6.70917 &  17.32658 &  0.4907 &  29 \\ 
002650.2+171935   &    &   &  &   \\ 
CXOMP~J   & 137.86083 &   5.83681 & 0.7682 &  18 \\ 
091126.6+055012   &  &   &  &   \\ 
LCDCS 0110               & 159.46917 & -12.72889 & 0.5789 &  18 \\  
LCDCS 0130               & 160.17333 & -11.93083 & 0.7043 &  30 \\  
LCDCS 0172               & 163.60083 & -11.77167 & 0.6972 &  48 \\ 
LCDCS 0173               & 163.68125 & -12.76389 & 0.7498 &  37 \\  
CL~J1103.7-1245a         & 165.89542 & -12.77944 & 0.6300 &  19 \\ 
CXOMP~J   & 169.35875 &   7.72639 & 0.4770 &  36 \\  
111726.1+074335   &  &    &  &   \\  
LCDCS 0340               & 174.54292 & -11.56639 & 0.4798 &  51 \\  
LCDCS~0531               & 186.97458 & -11.63889 & 0.6355 &  24 \\  
LCDCS 0541               & 188.12708 & -12.84250 & 0.5414 &  80 \\  
ClG J1236+6215           & 189.16500 &  62.26500 & 0.8500 &  40 \\  
XDCS mf~J  & 197.50792 &  32.35278 & 0.4370 &  19 \\  
131001.9+322110 &  &   &  &   \\  
NSCS~J       & 200.91500 &  30.37600 & 0.461  &  19 \\  
132336+302223      &  &   &  &   \\  
MJM98  034              & 203.80742 &  37.81564 & 0.383  &  16 \\  
3C 295 Cluster           & 212.83396 &  52.20251 & 0.4600 &  66 \\  
GHO~1601+4253            & 240.80762 &  42.76005 & 0.5391 &  50 \\  
RX J1716.4+6708          & 259.20667 &  67.14167 & 0.809 &   37 \\ 
\hline
\end{tabular}
\label{tab:onlyzSG}
\end{table}

\begin{table*}
  \caption{Optical structures found in the clusters with no X-ray data.
    Notes: (1)~cluster name, (2)~number of the substructure,
    (3)~number of galaxies with redshift in the substructure, (4)~mean 
    redshift of the substructure, (5)~substructure to total cluster mass ratio estimated 
    with the SG method and given in 10$\%$ wide intervals,
    (6)~substructure to total cluster mass ratio estimated with
    the method based on a scaling relation described in
    Section~\ref{subsec:massnico} and given in 10$\%$ wide intervals. }
\begin{tabular}{lrrrrrr}
\hline
\hline
Name              & \# & N$_{gal}$ &z & ${\rm (M_{SS}/M_{tot})_{SG}}$ &
${\rm (M_{SS}/M_{tot})_{sc}}$ \\
                  &    &          &  & (\%)                      & (\%)\\
\hline
CXOMP~J091126.6+055012 & 1 & 11 &0.7687 & [20;30]  & [30;40]\\  
                  & 2 & 7&0.7623 & [0;10]    & [10;20]\\
                  & 3 &  6&0.7748 & [0;10]   & [0;10]\\
LCDCS 0110        & 1 &  4 & 0.5807 & [0;10] & [10;20]\\ 
                  & 2 &  9 & 0.5777 & [0;10] & [40;50]\\
LCDCS 0130        & 1 & 7 & 0.7041 & [0;10]  & [0;10]\\ 
                  & 2 & 5 & 0.7028 & [0;10]  & [0;10]\\
                  & 3 & 8 & 0.7059 & [0;10]  & [0;10]\\
LCDCS 0172        & 1 & 24 & 0.6977 & [0;10] & [30;40]\\ 
                  & 2 & 6 & 0.6979 & [0;10]  & [0;10]\\
                  & 3 & 7 & 0.6944 & [0;10]  & [0;10]\\
LCDCS 0173        & 1 & 12 & 0.7498 & [0;10] & [10;20]\\ 
                  & 2 & 11 & 0.7477 & [0;10] & [10;20]\\
                  & 3 &  8 & 0.7523 & [0;10] & [0;10]\\
                  & 4 &  5 & 0.7573 & [10;20] & [0;10]\\
CXOMP~J111726.1+074335       & 1 & 22 & 0.4833 & [0;10] & [30;40]\\  
                  & 2 & 16 & 0.4790 & [0;10] & [10;20]\\
LCDCS~340         & 1  & 9 & 0.4852 & [0;10] & [0;10]\\ 
                  & 2  & 5 & 0.4765 & [0;10] & [0;10]\\
                  & 3 & 5 &0.4818 & [0;10]   & [0;10]\\
                  & 4 & 4 &0.4801 & [0;10]   & [0;10]\\
                  & 5 & 4 &0.4787 & [0;10]   & [0;10]\\
                  & 6 & 5 &0.4796 & [0;10]   & [0;10]\\
LCDCS~0541        & 1 & 9  &0.5420 & [0;10]  & [0;10]\\
                  & 2 & 11 &0.5447 & [0;10]  &[0;10]\\
                  & 3 & 10 &0.5395 & [0;10]  &[0;10]\\
                  & 4 & 13 &0.5347 & [20;30]  &[0;10]\\
                  & 5 & 7  &0.5432 & [0;10]  &[0;10]\\
                  & 6 & 9  &0.5408 & [0;10]  &[0;10]\\
                  & 7 & 4  &0.5379 & [0;10]  &[0;10]\\
                  & 8 & 12 &0.5492 & [0;10]  &[0;10]\\
ClG J1236+6215    & 1 & 23 & 0.8521 & [0;10] & [30;40]\\ 
                  & 2 & 17 & 0.8494 & [0;10] & [0;10]\\
                  & 3 & 13 & 0.8495 & [0;10] & [0;10]\\
                  & 4 & 16 & 0.8462 & [10;20] & [10;20]\\
3C 295 Cluster    & 1 & 10 & 0.4560 & [0;10] & [10;20]\\ 
                  & 2 & 17 & 0.4618 & [0;10] & [20;30]\\
GHO~1601+4253     & 1 & 37 & 0.5392 & [0;10] & [50;60]\\ 
                  & 2 & 14 & 0.5439 & [0;10] & [0;10]\\
\hline
\end{tabular}
\label{tab:SG2}
\end{table*}

\subsection{Influence of the undersampling of redshift catalogues}

\begin{figure}[!ht]
  \begin{center}
    \includegraphics[angle=270,width=3.2in]{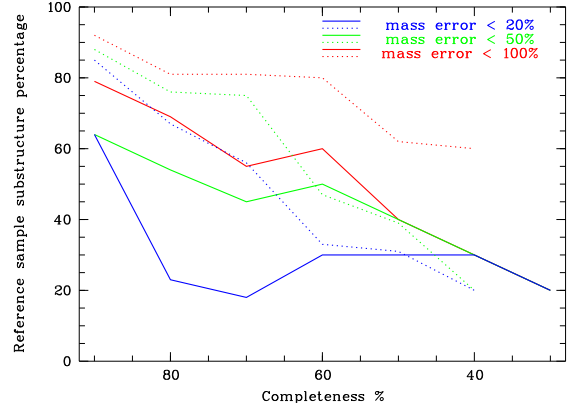}
    \caption{Effects of the undersampling of the input spectroscopic
      catalogue on the results of the Serna \& Gerbal (1996) method.
      The figure shows the percentages of substructures in the total
      reference cluster sample (versus the spectroscopic catalogue
      completeness percentage) for which the mass estimates are
      plagued by an error of less than X$\%$ (compared with the
      100$\%$ complete spectroscopic catalogue). Blue lines are for
      X=20$\%$, green lines for X=50$\%$, and red lines for
      X=100$\%$. Continuous and dashed lines respectively correspond
      to substructures that merged and which did not merge with other
      substructures during the spectroscopic catalogue resampling
      process. }
  \label{fig:compl}
  \end{center}
\end{figure}

An important question concerning the application of the SG method is
that of estimating how the undersampling in the optical spectroscopy
of galaxies can affect our results. To estimate the importance of this
effect, we considered six reference clusters from the literature that
have more than one hundred spectroscopic redshifts in the cluster
redshift range and in various dynamical states (relaxed, with minor,
or with major ongoing mergers). We observed the changes in the results
depending on the completeness of the input catalogue.  The clusters
that were considered (and total numbers of galaxy redshifts in
parentheses) were Abell~85 (815), Abell~168 (695), Abell~496 (499),
Abell~851 (211), Abell~2744 (131), and Coma (595).

To estimate the effects of undersampling, we considered between 100\%
and 10\% of the complete catalogue for each of the six clusters. At each
step we randomly took out 10\% of the galaxies, reapplied the SG, and
checked the results to detect differences with growing
incompleteness. This is what we call in the following the
resampling process of the spectroscopic catalogue. In this way, we can
observe the impact of the undersampling on substructure detection and
its effects on the results (numbers of substructures found and
corresponding masses) depending on the richness of the substructures.

As expected, substructures with many members tend 
to be detected down to lower completeness levels, while substructures with
few members disappear faster when the  undersampling increases. Typically,
substructures detected with more than six members in the original spectroscopic
catalogues will remain detected by the SG analysis down to completenesses of
$\sim$50\% to 30$\%$.

The precision on the mass of the substructures detected by the SG
analysis varies with completeness level, as seen in
Fig.~\ref{fig:compl}.  We distinguished two cases: (1) a given
substructure is artificially merging with another one during the
spectroscopic catalogue resampling process (this process will naturally
tend to overestimate the substructure masses) and (2) a given
substructure is not polluted by other substructures during the
spectroscopic catalogue resampling process. Figure~\ref{fig:compl} 
shows that the SG precision on the mass remains better than
50$\%$ for about half of the sample down to incompletenesses of about
60$\%$, while the SG analysis is not able to estimate  the
mass of most of the substructures precisely for completeness levels lower than
$\sim$50$\%$. 

As a further test of the influence of undersampling on our results, we
also applied the SG method to a halo from the Millennium simulation
(halo \#51037100000000). This halo has a theoretical mass of
$4.4\times 10^{14}$~M$_\odot$ and is at a redshift $z=0.37$. It has 23
subhaloes with more than three galaxies.

Considering the mass resolution of the Millennium simulation, if we
use the same semi--analytical models as those applied to the CFHTLS
clusters (Adami et al. 2010) we estimate that the
completeness limit in this halo corresponds to an absolute magnitude
${\rm M_I \sim -17.5}$, which roughly corresponds to the completeness
limit of the DAFT/FADA survey.
In this simulated cluster, if we consider 100\% of the galaxies, the
SG method detects the five most massive substructures (numbered from \#1
to \#5), the mass of the least massive one (\#5) being $3.5\
10^{12}$~M$_\odot$.  Two smaller structures are also
detected with masses of $7.9\ 10^{11}$~M$_\odot$ and $5.2\ 10^{11}$~M$_\odot$.

If we now randomly remove 10\% of the galaxies from the initial
simulated galaxy catalogue, then 20\%, etc., we start to lose some of
the  initially detected substructures. The percentage of undersampling
inducing the loss of a given substructure is the result of a complex
interplay between the intrinsic richness of the substructure (a poor
substructure will obviously be easier to lose when undersampling the
catalogue) and the galaxy mass distribution in the substructure (a
cD-dominated substructure will be easier to lose if the cD is
removed).  For example, because substructure \#1 is both very rich (451
galaxies) and not strongly cD-dominated (only 19$\%$ of the total mass
of the substructure is associated with the cD), we are able to detect
it down to a sampling rate of only 10$\%$. The other substructures
disappear between sampling rates 90\% and 10\%, but as a general
statement we can say that we are able to detect some of the most
massive substructures down to about 30\% of the original sampling.
The mass estimate remains within a factor of 2 down to 40\% sampling.

We  also see below that in most cases massive substructures are
detected both in X-rays and with the SG analysis, when both types of
data are available. However, since the SG mass estimate is sometimes
inaccurate, we describe in the next section an alternative way
to estimate the mass of a substructure based on optical data.

\subsection{Alternative determination of the masses of the substructures 
detected  by the Serna \& Gerbal method}
\label{subsec:massnico}

As mentioned above, the SG algorithm allows the total structure masses
to be estimated in a rather crude way, since the mass--to--luminosity
ratio (set to M/L=100) is assumed to be the same for all the galaxies.
However, the exact value of M/L does not strongly affect the
substructure content since it mainly relies on the spectroscopic
redshifts of galaxies, as shown for example by Adami et al. (2005).
We have then developed a new method to estimate masses of the
substructures detected with the SG algorithm based on more physical
arguments. It is well known that the stellar--to--luminosity ratio of a
galaxy depends on its spectrum, hence on its magnitude (e.g. Bell
et al. 2003, Tremonti et al. 2004, Cappellari et al. 2006). Following
this, we propose to apply a different stellar mass--to--light ratio to
each galaxy, depending on its luminosity, and then to convert the
stellar mass of the substructure to its total mass (i.e. including
X-ray gas and dark matter) using the cluster scaling relation between
those two quantities.

We first compute absolute magnitudes by calculating the distance
modulus for each galaxy. This is made possible by our only using
galaxies with an accurate spectroscopic redshift. We then convert our
V-band magnitudes to the g-band, applying a colour correction of
$g-V=0.28$, as found in Fukugita et al. (1995). This allows the
stellar mass--to--light ratio to be estimated, as measured for SDSS
data for an ``average type'' galaxy by Kauffmann et al. (2003, see
their Fig. 14). Each galaxy is assigned its own stellar
mass--to--light ratio and luminosity, taking an absolute magnitude of
5.11 for the Sun in the g-band. Summing the stellar masses of the
galaxies belonging to the substructure provides us with the estimated
stellar mass of the substructure.

The last step is to convert stellar masses to total masses using the
following scaling relation defined by Giodini et al. (2009) for
clusters within r$_{500}$:
 
\begin{equation}
 \frac{M_{star}}{M_{tot}}=(5.0 \pm 0.1) \times 10^{-2}\ ({\frac{M_{tot}}{5\ 10^{13}M_{\odot}}})^{-0.37\pm0.04}.
\end{equation} 
\noindent
One must note that the $\pm$0.04 uncertainty on the exponent in the
previous equation results in large error bars on the total
substructure mass (about 85\%).

We calculated substructure masses following this method and compared
them to those calculated with the Serna \& Gerbal method for the 18
clusters with optical substructures and X-ray data
(e.g. Table~\ref{tab:SG}) and we have a rough overall agreement.  We
are aware that none of these methods gives the exact mass of
substructures, so we present both to cross check our mass ratios. When
the results obtained with both methods agree, we have a good chance of
having a reasonable estimate.

We also calculated substructure masses for the clusters to
which we applied the SG method but for which no X-ray analysis was
possible, and give results in Table~\ref{tab:SG2}.

\section{X-ray gas distribution}

The results for the invidual clusters with X-rays and/or a SG analysis
are described in the Appendix. In Appendix~A, we give a full X-ray
(and optical when enough galaxy redshifts are available) analysis for
the clusters with usable XMM-Newton data. In Appendix~B, we present
the SG analysis for the clusters with no usable X-ray data but with at
least 15 galaxy redshifts in the cluster range. In Appendix~C we give
brief notes on the clusters for which we have little spatial
information, but which are worth mentioning in particular because
several of them seem to have redshifts differing from those given by
NED.

\subsection{Cluster core radius versus redshift}

\begin{figure}
\begin{center}
\includegraphics[width=7cm,angle=270]{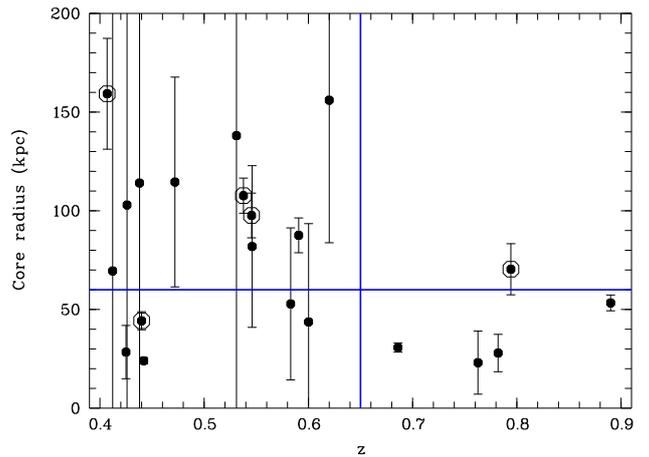}
\caption{$\beta$-model core radius as a function of redshift for the considered
clusters (see text). The circled disks are the clusters with detected X-ray
substructures. The two  lines symbolize the z=0.65 limit and the maximal
value of the core radii for z$\geq$0.65 not significantly substructured clusters 
(from an X-ray point of view).}
\label{fig:CRz}
\end{center}
\end{figure}

We first investigate the possible variations in the cluster X-ray gas
distribution (modelled by a $\beta$-model) versus redshift. There is
no significant correlation between the redshift and the $\beta$-model
slope. We may, however, detect a tendency between the $\beta$-model
core radius and the redshift (see Fig.~\ref{fig:CRz}). To produce this
figure, we first only selected clusters for which the $\beta$-model
fitting provides a converging solution (true for $\sim$80\% of the
clusters). Then, we eliminated double clusters as CL~J0152.7-1357.
Figure~\ref{fig:CRz} shows that clusters at redshifts lower than 0.65
tend to have larger core radii.  

If we now eliminate the clusters with X--ray--detected substructures
for which the core radius may be biased the tendency is even more
pronounced. The respective mean core radii are 76$\pm$42~kpc,
93$\pm$45~kpc, and 34$\pm$13~kpc for the clusters in the redshift bins
[0.4;0.5], [0.5;0.65], and [0.65;0.9]. Even though our sample is
limited (only 16 clusters in total after all the selections), it
suggests that the highest redshift clusters are younger structures
that have not yet accreted large amounts of matter and therefore have
smaller core radii.  However, this could be due to several selection
effects, since our cluster sample is not homogeneous in terms of
numbers of detected clusters versus redshift: it is a collection of
clusters known in the literature to which we applied simple criteria
on mass and available data (see section~1).

The first question is to know if the XMM-Newton data we have in hand
are able to measure large core radii at z$\geq$0.65. The collected
XMM-Newton data are of two types: (1) pointed observations for which
the cluster was the main target or (2) observations made for other
purposes and where clusters were observed by chance.

We can hope that type (1) XMM-Newton pointings will not be too
affected by the inability to measure core radii, because exposure
times have been selected by the original PIs to specifically study the
clusters and modelling a gas distribution is one of the most basic
tasks. To test this point, we computed that $\sim$80\% of core radius
measurements in the type (1) pointings were successful.  The
calculation of the core radius with Sherpa converged for 80\% of the
clusters. Among the $\sim$20$\%$ unsuccessful measurements, half were
at z$\geq$0.65 and half at z$\leq$0.65. Type~(1) observations
therefore do not seem to be affected by a variable (with redshift)
ability to measure core radii.  Similarly, we also checked that
z$\geq$0.65 and z$\leq$0.65 did not provide X-ray photon count rates
that are too different (the basic parameter for modelling the gas
distribution). Clusters at z$\leq$0.65 are sometimes very bright, but
if these bright objects are excluded, we have somewhat lower count
rates at low redshifts than at high reshifts. Almost 40$\%$ of the
z$\leq$0.65 clusters have count rates that are higher than 0.04
counts/sec, while 50$\%$ of the z$\geq$0.65 clusters have count rates
higher than 0.04 counts/sec. We therefore do not find a strong
variation in the cluster count rates with redshift in our sample.

Type (2) observations could  present a variable core radius
measurement efficiency because their exposure times were not
specifically selected to study the clusters. However, these observations
represent  only 17$\%$ of the sample, and only one third of them
corresponds to z$\geq$0.65 clusters. The effect is therefore limited.

A second bias could be due to the fact that X-ray clusters of galaxies
often exhibit relations between size, luminosity, and temperature (the
well known fundamental plane of galaxy clusters, e.g. Adami et
al. 1998). Since distant clusters in the literature are more easily
detected when they are luminous (and therefore have high luminosity or
temperature), we may imagine a tendency to select clusters with small
core radii due to the selection of high--luminosity or
high--temperature clusters. To test this point, we checked that
relations between the core radius and the luminosity or temperature
were visible in our sample. We did not detect any clear correlations,
so even a luminosity or temperature selection effect in our sample
would not induce a core radius effect, such as the one seen in
Fig.~\ref{fig:CRz}.

However, that the high redshift clusters could systematically have
smaller core radii will have to be confirmed on larger samples in the
coming years.  We detect no correlation between the core radius and
the X-ray temperature (and therefore the X-ray mass), and we detect no
clear relation between the velocity dispersion and the X-ray
luminosity either.

\subsection{X-ray versus SG substructures}

\begin{figure}
\begin{center}
\includegraphics[width=7cm,angle=-90]{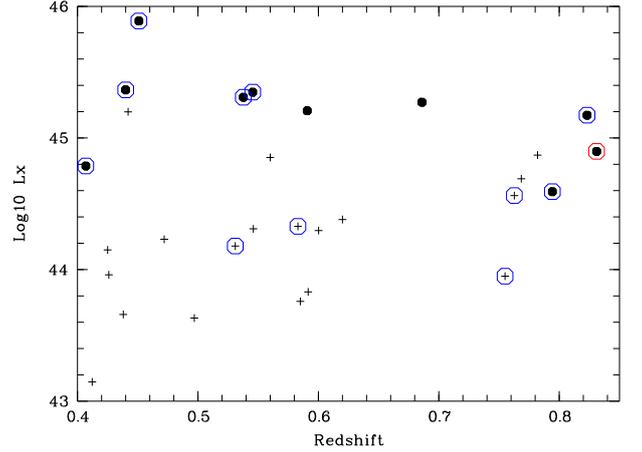}
\caption{Cluster total X-ray luminosity as a function of
  redshift. Filled disks are clusters for which substructures were
  detected with the X-ray method. Crosses indicate clusters with no
  X-ray detected substructures. The blue and red circles correspond to
  clusters for which substructures respectively representing less and
  more than 15$\%$ of the total cluster mass (estimated with the SG
  method) were detected with the SG method. }
\label{fig:logLXtot}
\end{center}
\end{figure}

\begin{figure}
\begin{center}
\includegraphics[width=7cm,angle=-90]{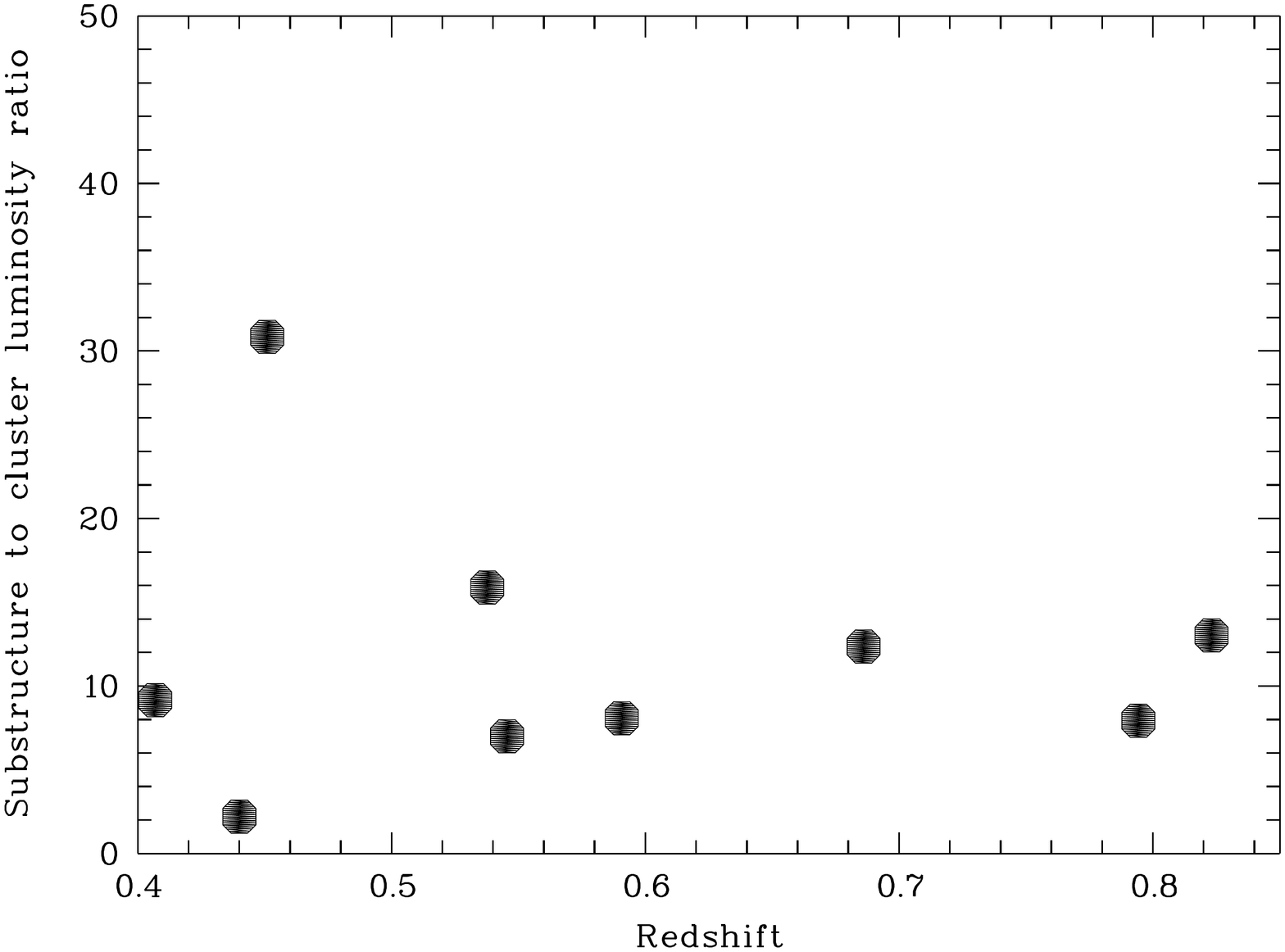}
\caption{Substructure to cluster luminosity ratios from the X-ray
  analysis (in \%) versus redshift.  }
\label{fig:percX}
\end{center}
\end{figure}

Figure~\ref{fig:logLXtot} shows that substructures were detected with
the X-ray method at all redshifts. However, it is mainly the most
X-ray luminous clusters ($\geq$ 4~10$^{44}$ erg/s) that provided such
detections. To check that these are instrumental effects (only
luminous clusters could provide substructure detections due to their
higher S/N), we put the substructure information coming from the SG
analysis on the same figure. We see that we detected substructures
with the SG method even for clusters with undetected X-ray
substructures, and even for clusters with low X-ray luminosities.
Therefore, X-ray selection effects must indeed be at work.  However,
nearly all the substructures undetected in X-rays but detected by the
SG method seem to be minor (less than 15$\%$ of the total cluster
mass). It is therefore tempting to conclude that all the major
substructures (or at least a large percentage of these) were
effectively detected with the X-ray data in hand.

We can now examine the variations in the X-ray luminosity ratio
(substructure versus cluster) as a function of redshift. We find that
this percentage remains more or less constant with redshift in a 5\%
-- 15\% interval (see Fig.~\ref{fig:percX}). Even if such percentages
are probably an underestimate of the mass fraction of clusters that is
in the form of substructures (because we do not detect them all, see
Section~2.3), they are in good agreement with the Millennium
simulation (see Appendix of Adami et al. 2013 or Gao et al. 2012)
predicting that clusters below z$\sim$1 primarily undergo minor
mergers.

\subsection{X-ray substructure merging stages}
\label{subsec:stages}

We concentrate in this part on substructures detected both in X-rays
and with the SG method. We consider the X-ray substructures that we
detected as relics of more or less recently infalling groups of
galaxies. We were able to measure physical parameters for these groups
based on their X-ray gas phase and on their galaxy phase. Gas and
galaxies have different time--scales in galaxy structures in response to
gravitational perturbations such as mergers. We are therefore
theoretically in the position of deducing the merging stage of the
considered infalling groups with the corresponding clusters.

More precisely, we know the position of isolated groups in $L_X$
versus galaxy velocity dispersion diagrams (e.g. Connelly et
al. 2012). We compared the substructures that we have detected in the
present paper to these isolated groups.  This was done after applying
a 1.73 correction factor to translate the Connelly et al. (2012) X-ray
luminosities (measured in the 0.1$-$2.4~keV energy range (see their
section 3.1 for details) to our 0.5$-$8~keV energy range.

Four of our substructures exhibit velocity dispersions over
500~km/s. This may appear large for dynamically relaxed groups. For
SG1 in CL J0152.7-1357, it is not surprising because this cluster is
undergoing a major merger, so SG1 is already a pretty massive
structure. It is more difficult to explain the high values for Abell
851 (SG3), MS 1054-03 (SG1), and RXC J1206.2-0848 (SG4) if they are
relaxed. We may then deal with highly unrelaxed groups in an already
quite advanced fusion stage.

\begin{figure}
\begin{center}
\includegraphics[width=7cm,angle=270]{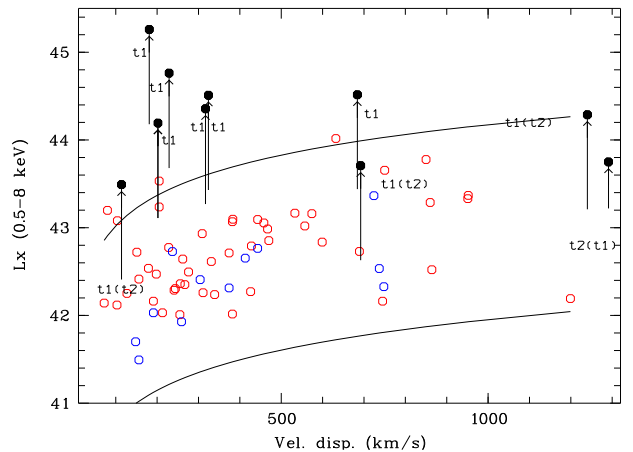}
\caption{X-ray luminosity as a function of galaxy velocity
  dispersion. Open circles are from Connelly et al. (2012), red: X-ray
  selected groups, blue: optically selected groups. The two black
  curves show the 3$\sigma$ envelope of the Connelly et al. (2012)
  X-ray selected sample. The black filled disks correspond to our detected
  substructures. The lower extremities of the vertical  lines show
  the places of the infalling structures prior to their dynamical
  capture assuming the epoch written next to the lines. Other epochs
  written in parentheses are also possible to place the infalling
  structures inside the Connelly et al. (2012) envelope. }
\label{fig:age}
\end{center}
\end{figure}

We clearly see in Fig.~\ref{fig:age} that most of our substructures
have higher X-ray luminosities than the isolated groups of Connelly et
al. (2012). This shows that, as expected, our substructures are not
classical isolated groups but have already undergone some
transformations when falling into the clusters.

Merging simulations (e.g. Poole et al. 2007) show that these
transformations include, among others, several increases in the main
cluster X-ray luminosity during the merging process at well--defined
epochs. More precisely, there is a first increase of the X-ray
luminosity at the epoch that we will call t0 in the present paper
(first virial encounter of the infalling structure with the impacted
cluster), then a more significant one at t1 (first pericentre approach
of the infalling structure), and a third and a fourth smaller
increases at epochs t2 (second pericentre approach) and t3 (relaxation
of the impacted structure).

That we still detect X-ray substructures in our clusters
naturally excludes the t3 epoch. Given the depth of our X-ray data, we
never detect the cluster X-ray contributions up to the virial radius,
and this prevents us from detecting substructures at the t0 stage. Our
clusters are therefore somewhere between the t1 and t2 epochs.

The simulations of Poole et al. (2007) quantified the increase in the
X-ray luminosity with time. By applying their predicted increasing
factors for major mergers (which the infalling groups experience when
merging with a larger cluster) and considering the t1 and t2 epochs,
we were able to estimate the original X-ray luminosities of our
infalling groups prior to their gravitational capture. We then
selected the possible epochs (t1 and/or t2), allowing the infalling
groups to have been optimally located in the $L_X$ versus velocity
dispersion diagram for isolated groups (from Connelly et al. 2012)
before gravitational capture.  These epochs are listed in
Table~\ref{tab:SG}. The increasing factors that we applied are
applicable to clusters merging with another cluster of comparable
mass. This is not the case here, but it is the best estimation at our
disposal of the increase in X-ray luminosity caused by a merger. Poole
et al. (2007) did not simulate what happens for merging ratios greater
than 1. The X-ray luminosity shifts in Fig.~\ref{fig:age} could
therefore be greater.  We see that most of the X-ray detected
substructures are probably at the t1 epoch (first pericentre approach)
and therefore are relatively recent in the clusters.

\section{Galaxy content in substructures}

As already mentioned, we collected ground--based imaging and spectroscopy at visible
and near--infrared wavelengths. Our ultimate goal is to gather at least five bands in 
the visible domain (typically B, V, R, I, z') and one in the near--infrared 
domain (J or Ks) to compute photometric redshifts with the LePhare tool (see
Guennou et al. 2010). We are in the process of completing this data collection
for our $\sim$90 lines of sight, but for now all clusters of Table 2 with 
detected substructures (except MS 1621.5+2640) have the full dataset available. 
This allowed us to compute photometric redshifts for these clusters, and we are 
therefore now ready to investigate the galaxy content of these structures
and of their detected substructures.

The LePhare tool, as for other photometric redshift tools, can be used
to compute photometric redshifts, as well as to characterize the
considered galaxies in terms of type or stellar population age
(e.g. Adami et al. 2009). If we can fix the redshift at its true value
(considering only spectroscopic catalogues), then we limit the number
of free parameters and we have even better constraints on the type and
stellar population ages. This is the approach we have presently
chosen, with the spectroscopic catalogues at our disposal. We selected
galaxies with a known spectroscopic redshift and chose Bruzual $\&$
Charlot (2003) spectral templates in LePhare, fixed the redshift to
its value, and estimated the stellar population ages and photometric
types. These photometric types are arbitrarily coded between 10 (early
galaxies) and 31 (late starbursts). We then investigated the galaxy
distribution in an age versus type space. We assume that the age can
be considered as driven by the epoch of the last burst of star
formation.

\subsection{Cluster versus field galaxies}

The first thing to check is the general behaviour of cluster versus
field galaxies. In order to limit the contamination of the cluster
sample by field galaxies as much as possible, we defined the cluster
sample as all galaxies with a redshift differing by less than 0.01
from the lower redshift and higher redshift substructures in
Table~2. This corresponds to three times the typical velocity
dispersion of a massive cluster. The field sample was defined as all
galaxies with a redshift differing by more than 0.02 from the lower
redshift and higher redshift subtructures in Table~2. This corresponds
to six times the typical velocity dispersion of a massive cluster. We
are aware that this eliminates galaxies between three and six times
the typical cluster velocity dispersion, but these galaxies are
potentially in an intermediate state, and they would have made our results
noisier.

\begin{figure}
\centering
\includegraphics*[width=7cm,angle=270]{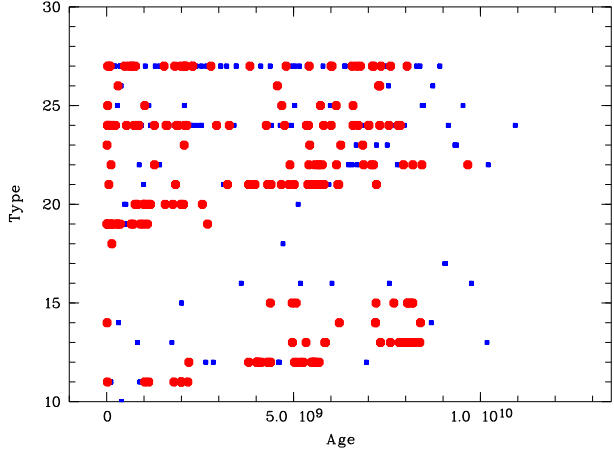}
\caption[]{Cluster (large red circles) and field (small blue squares) galaxies
in an age versus type plot.}
\label{fig:AgeMod1}
\end{figure}

Figure~\ref{fig:AgeMod1} shows the expected behaviour in both the
field and cluster samples, with a population of earlier type galaxies
more prominent in the cluster sample.  We see a clear dichotomy
between early and late types happening around type 17 for both the
cluster and the field samples, the gap being less populated in the
cluster sample.

Performing a two--dimensional Kolmogorov-Smirnov test gives a probability
greater than 99.9$\%$ to have populations (cluster and field galaxies)
coming from different parent samples.  This behaviour has been reported before,
but it confirms that our approach to compute age and type is valid. 

\subsection{General position of galaxies in substructures}

\begin{figure}
\centering
\includegraphics*[width=7cm,angle=270]{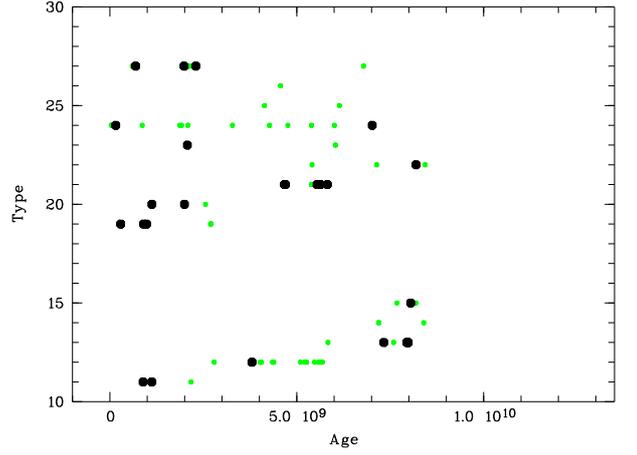}
\caption[]{Galaxies in SG detected (small green disks) and SG+X-rays detected 
(large black disks) substructures in a plot of galaxy type versus age.}
\label{fig:AgeMod2}
\end{figure}

We first note that galaxies members of substructures are very
different from field galaxies. A Kolmogorov-Smirnov test gives a
probability over 99.9$\%$ to have populations (galaxies in
substructures and field galaxies) coming from different parent
samples.  Figure~\ref{fig:AgeMod2} shows that there is perhaps a lack
of both late--type and old stellar population galaxies in
substructures detectable in X-rays. This could be explained if
galaxies in X-ray detected substructures had undergone recent bursts
of star formation induced by shocks in the hot medium. However, this
has to be confirmed on larger samples.

\subsection{Galaxies in substructures as a function of substructure characteristics}

Given the modest size of the galaxy samples in substructures detected
both by the SG and X-ray methods, we concentrate here on the
substructures detected at least by the SG method in order to maximize
our sample.

\begin{figure}
\centering
\includegraphics*[width=7cm,angle=270]{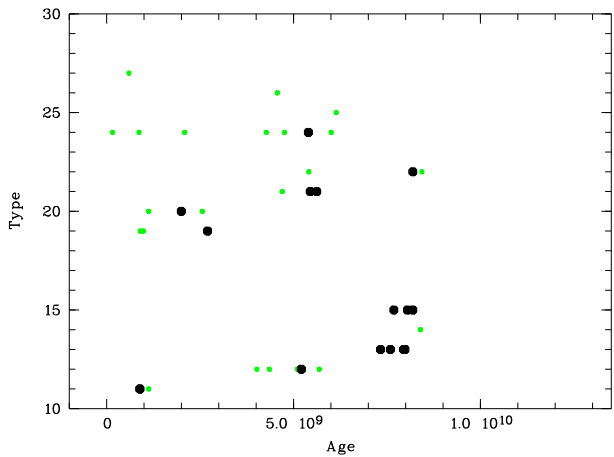}
\caption[]{Galaxies in SG detected substructures at less than 900 km/s
  (small green disks) and at less than 300 km/s (large black disks)
  from the mean cluster velocity in a plot of galaxy type versus age.}
\label{fig:AgeMod3}
\end{figure}

We chose to characterize the substructures by their velocity difference
with the  mean cluster velocity. This gives an idea of the importance
of the cluster influence on the substructure. 
For example, the difference between the substructure members closer than 
300km/s and 900km/s from the cluster mean velocity appears in 
Fig.~\ref{fig:AgeMod3} to be due to the lack of late type galaxies with recent
bursts of star formation in the  substructures closest to the cluster mean
velocity. This may indicate that substructures close to the bottom of the 
cluster potential well have  consumed a large part of their gas and are
therefore less able to form new generations of stars. 

\subsection{Summary}

In conclusion to this section, we can say that galaxy populations in
substructures have the same general behaviour as regular cluster
galaxies, but with several noticeable differences:

- a possible lack of both late type and old stellar population galaxies in 
substructures detectable in X-rays and in SG substructures contributing the 
most to the cluster mass,

- a possible lack of late type galaxies with recent bursts of star formation 
in the closest substructures to the cluster mean velocity.

These tendencies can be explained by classical expected behaviours in
clusters where galaxies close to the bottom of the cluster potential
have probably consumed a large part of their gas and are therefore
less able to form new generations of stars, and where galaxies in
important substructures would have undergone recent bursts of star
formation initiated by shocks induced in the hot medium and energy
transfer from the surrounding cluster.

\section{Discussion}

We studied a sample of 32 clusters of galaxies with usable X-ray data
and of 19 clusters of galaxies without X--ray data but with more than 15
available spectroscopic redshifts in the cluster range. Ten
substructures were detected both in X-rays and by the SG method at
optical wavelengths.

We eliminated point source contamination by using {\it Chandra} data
when available and with public catalogues of active galactic nuclei or
radio sources.  We detected substructures based on X-ray analysis via
the statistically significant detection of residuals based on a
surface--brightness $\beta$-model subtraction, or optically via an
application of the Serna \& Gerbal (1996) dynamical method to our
spectroscopic redshift catalogues.

From this work we derived a new set of substructures in rich clusters
of galaxies in the redshift range [0.4,0.9]. We have verified that
these are dynamically bound systems by combining the detection in
X-rays with a dynamical analysis based on spectroscopic redshifts.  We
now discuss these results in the context of cluster evolution, as well
as in comparison with previous work, where appropriate. A major goal
in understanding cluster formation is to be able to use clusters as a
tool for studying cosmology, we begin with a brief summary of the
relationship between clusters and cosmology.

Because astrophysicists have come to realize there is a direct link
between (1) clusters, (2) how large scale structure formed, (3) dark
matter, and (4) dark energy, clusters of galaxies are being used more
and more as cosmological tools (see for example Allen et al. 2011 and
Kravtsov \& Borgani 2012).  Those works (and references therein) show
how cluster counts and cluster-cluster correlations may shed light on
the Gaussianity or non-Gaussianity of the initial primordial
fluctuations in the early Universe.  Furthermore, these papers and
others have shown that studies of clusters can also be used to delimit
the value of $w$, if $w$ is variable with redshift, or if there are
deviations from General Relativity that are causing the apparent
acceleration of the Universe (e.g. Vikhlinin et al. 2009, Allen et
al. 2011, Kravtsov \& Borgani 2012).  On the finest cosmological scale
of galaxies, work is going on to determine if there is indeed a missing
sub-halo problem or not, but so far there is no consensus (see for
example Strigari et al. 2010).

On the in-between scales of groups of galaxies that are continually
falling into galaxy clusters, the hierarchical build up of clusters
with groups has been described both in published works, such as Poole
et al. (2007), Gao et et al. (2012; and references therein, too
many to list them all), and the Millennium project (Springel et
al. 2005), but also in simulations posted on the web.

The Poole et al. (2007) results are in partial agreement with our data,
as are those of Gao et al. (2012), in that (a)~we find that the X-ray
detected groups are in the luminous stage as if they were ``lit up''
by infall to the cluster as predicted by Poole et al. (2007); 
(b)~Gao et al. (2012) found that at redshift $z = 0$, the total masses
in substructures relative to the total cluster masses in their
simulations were about 5-15\%.  This 5-15\% value is about the same as
the one we found for clusters in our sample, but what is uncertain is
how this percentage should change (if at all) with redshift between
0.4 and 0.9. For example, a cluster initially without substructure
could grow in such a way that at higher redshifts its initial mass is
relatively low, so that added sub-clumps are a relatively high
percentage of the total cluster mass.  Conversely, it could be that as
clusters grow, many of the substructures are not dissipated enough to
disappear, and the total mass in substructures actually grows over
time.  That our mass percentage in substructures at $z = 0.4-0.9$ is
about equal to what is predicted by Poole et al. (2007) at $z = 0$ argues
that most likely events conspire to keep the detectable mass in
substructures in clusters approximately constant from $z = 0.9$ to the
current day. This would also be in good agreement with the quite
constant level of diffuse light present in clusters between z=0.4 and 
0.9 (see Guennou et al. 2012).

The work of Mann \& Ebeling (2012) states that the fraction of
``disturbed'' clusters increases with increasing redshift, implying a
higher substructure mass with higher redshift, in apparent
contradiction to our work. To be consistent with their work, we would
expect the mass in substructures to be higher at higher redshift, if
we could add more clusters to our sample and divide them into several
bins within the $z=0.4-0.9$ range.

In comparison, Baldi et al. (2012) find no change in the temperature
profile,
over the redshift range that they broke their sample into (above and
below 0.4), implying that there is little change in shape over this
redshift range, in apparent contradiction (in terms of a trend) to
what we found. However, Baldi et al. (2012) did not have enough data
points to subdivide their $z=0.4-0.9$ cluster results into smaller
redshift bins for a direct comparison to our observations.  Also, as
by-product of fitting the X-ray data with a simple beta model, we
found an increase in the physical extent (i.e. a larger core radius)
of the X-ray surface brightness with decreasing redshift. We judge
that this effect is {\it not} due to an inability to detect more
extended emission at higher redshift, based on the analysis presented
in Section~4.1.

Our findings are consistent in a general way with the hierarchical
cluster growth scenario in that the extent of clusters apparently
grows with decreasing redshift.  However, if cluster temperature is a
valid measure of the cluster mass (independent of redshift), then the
fact that we found no relationship between cluster extent and
temperature would argue for the mass  not having grown significantly over
the redshift range from 0.9 to 0.4, as also indicated by our
finding no correlation between the total mass estimated from the X-ray
temperature (given in the last column of Table~1) and the redshift.
This is plausible if (a) the number of groups infalling over this
time period of about 3.6~Gyr is relatively small, and (b) at the same
time, the infalling subgroups have caused the ICM to become more
extended but not hotter or significantly more massive.

\section{Summary and conclusions}

By the means of a comparison with Takey et al. (2011), we showed that
our X-ray luminosities and temperatures were consistant with
literature studies. We estimated the substructure detection efficiency
with simulations for the X-ray and SG methods. The X-ray detections
proved to be efficient up to z$\sim$0.9 for substructures brighter
than 1.0~10$^{44}$ erg/s and up to z$\sim$0.5 for substructures only
brighter than 0.4~10$^{44}$ erg/s.

The SG detection efficiency was tested by considering six reference
clusters outside of our sample, all very well sampled
spectroscopically. Substructures with more than six members in the
original spectroscopic catalogues remained detected by the SG analysis
down to completenesses of 50\% to 30\%. We showed that the SG
precision on the mass estimate remained better than 50\% for about
half of the sample down to incompletenesses of about 60\%, while the
SG analysis was not able to precisely estimate the mass of most of the
substructures for completeness levels lower than 50\%. SG masses were
also compared to an optically based cluster mass determination, and we
found qualitative agreement. We emphasize, however, that only {\it
  relative} SG masses should be considered as reliable.

We found that the core radius of the X-ray gas density profile may
decrease with redshift, but this needs to be confirmed with a larger
sample of clusters.  Ten substructures were detected by both methods
(X-rays and SG). These were systematically the SG most massive
substructures in each cluster. For a given cluster, the percentage of
mass included in substructures was roughly constant with redshift at
values of $\sim $5-15\%.  We also showed that most of our
substructures detected both in X-rays and with the SG method were
probably at their first cluster pericentre approach and therefore
corresponded to relatively recent infalls.

Finally, compared to regular cluster galaxies, galaxy populations in
substructures exhibit a possible lack of both late type and old
stellar population galaxies, and a possible lack of late type galaxies
with recent bursts of star formation in the substructures closest to
the mean cluster velocity.  In general, our results are consistent
with the picture of CDM hierarchical structure formation in that
substructure exists. The approximate X-ray properties and masses of
the substructures relative to the entire clusters are in the range
predicted by theory: 5-15\%, see Gao et al. (2012) for the fraction
found at $z$ = 0, and by the Millennium simulation, which predicts
that clusters below z$\sim$1 only undergo minor mergers.  On the
simulation front, the percentage of the substructure mass relative to
the total mass of the cluster would be interesting to compare with the
data we have presented here, as well as with future increased samples
produced by our DAFT/FADA collaboration and others.

\begin{acknowledgements}

 We thank the referee for useful comments.
  We gratefully acknowledge financial support from the Centre National
  d'Etudes Spatiales for many years. This project has benefitted
  from CAPES/COFECUB (programme 711/11). IM has been partially funded by
  the projects AYA2010-15169 from the Spanish Ministerio de Ciencia e
  Innovaci\'on and TIC 114 and PO08-TIC-3531 from Junta de Andaluc\'\i a.
We thank Calar Alto Observatory for allocation of director's discretionary 
time to this programme.

\end{acknowledgements}


\clearpage

\appendix

\section{Substructure analysis of the invidual clusters with usable
  XMM-Newton and/or Chandra data}

\begin{figure*}
  \begin{center}
    \includegraphics[width=2.7in,angle=0,bb=35 144 575 651,clip]{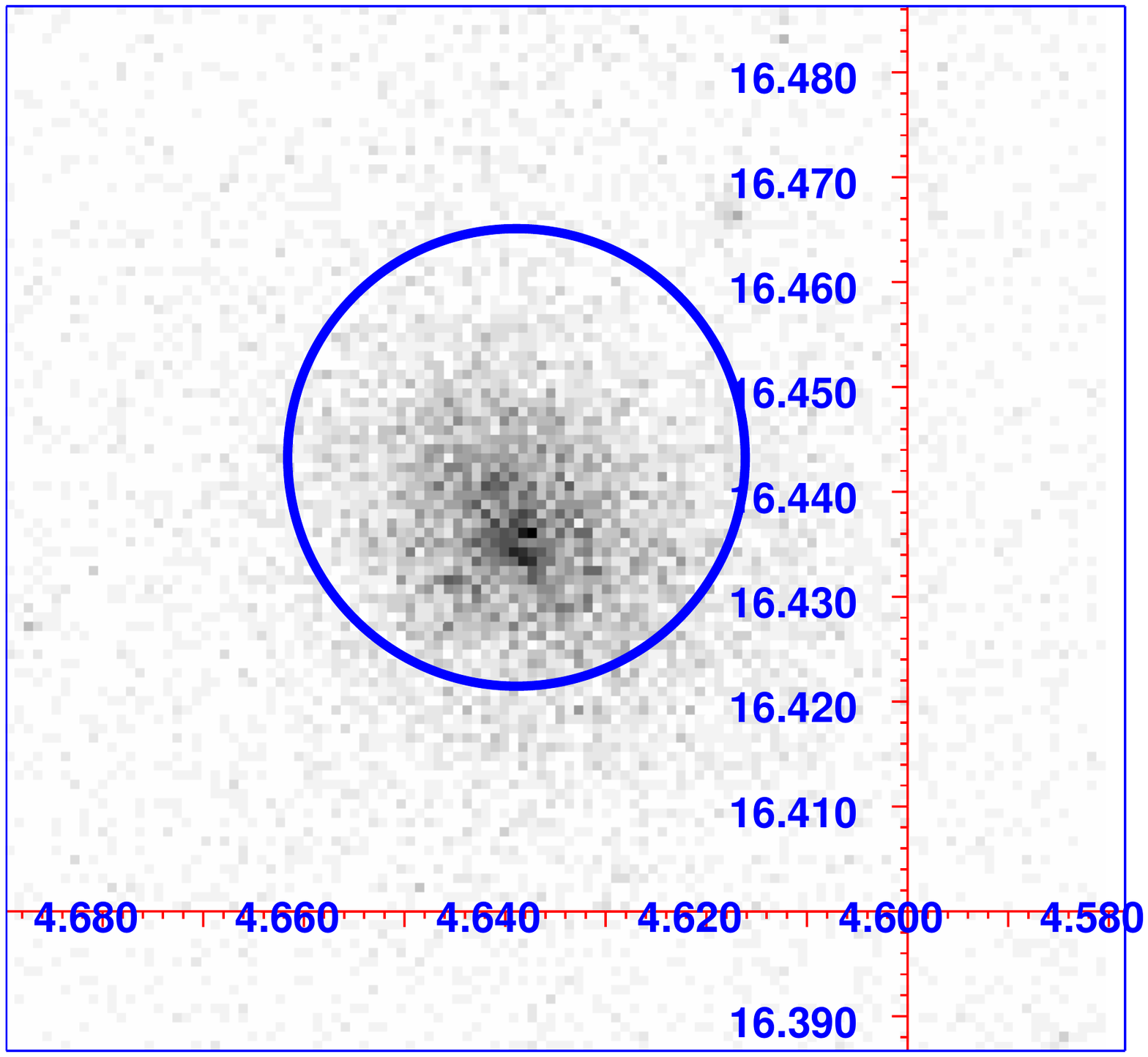}\includegraphics[width=2.7in,angle=0,bb=35 144 575 651,clip]{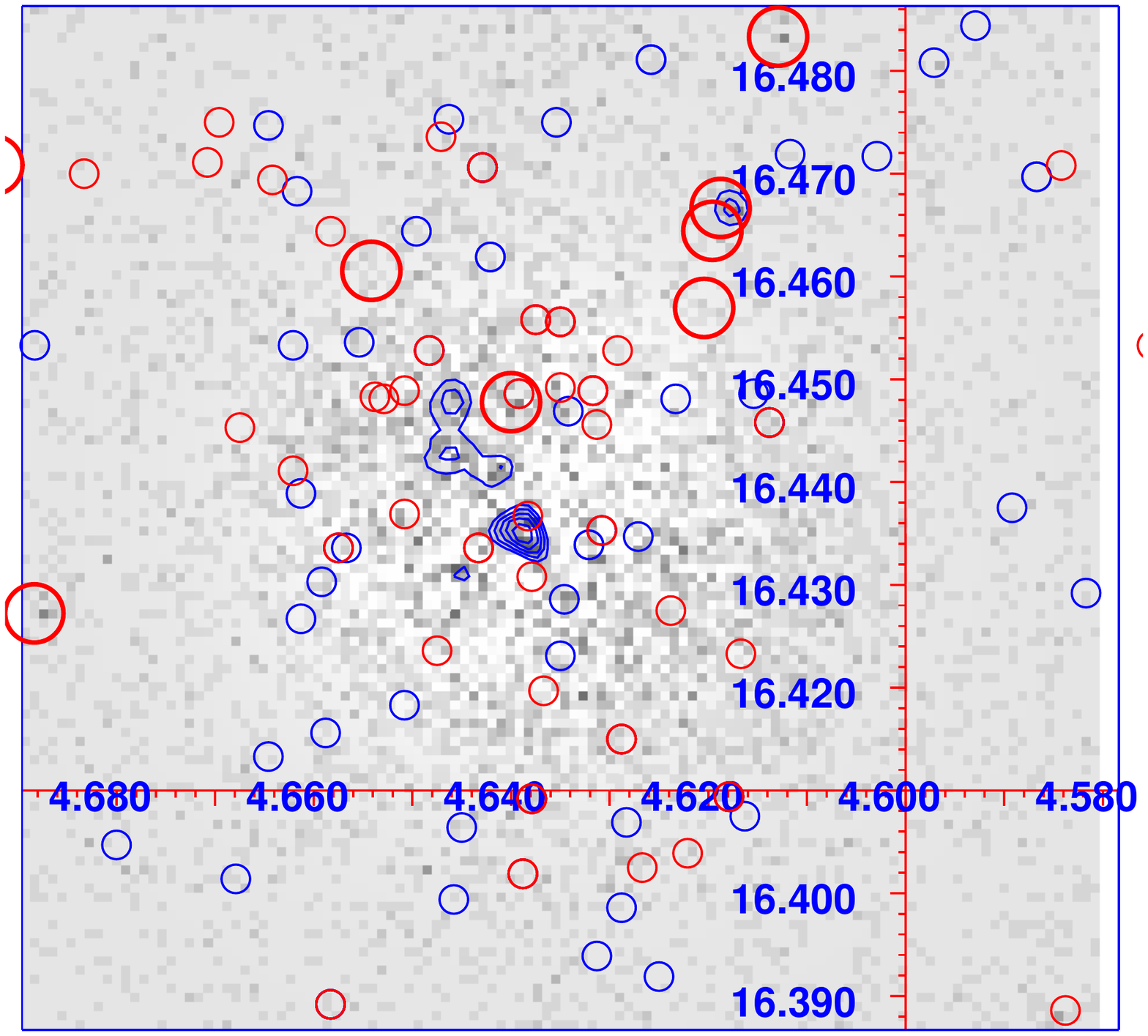}\\
    \includegraphics[width=2.7in,angle=0,bb=35 144 575 651,clip]{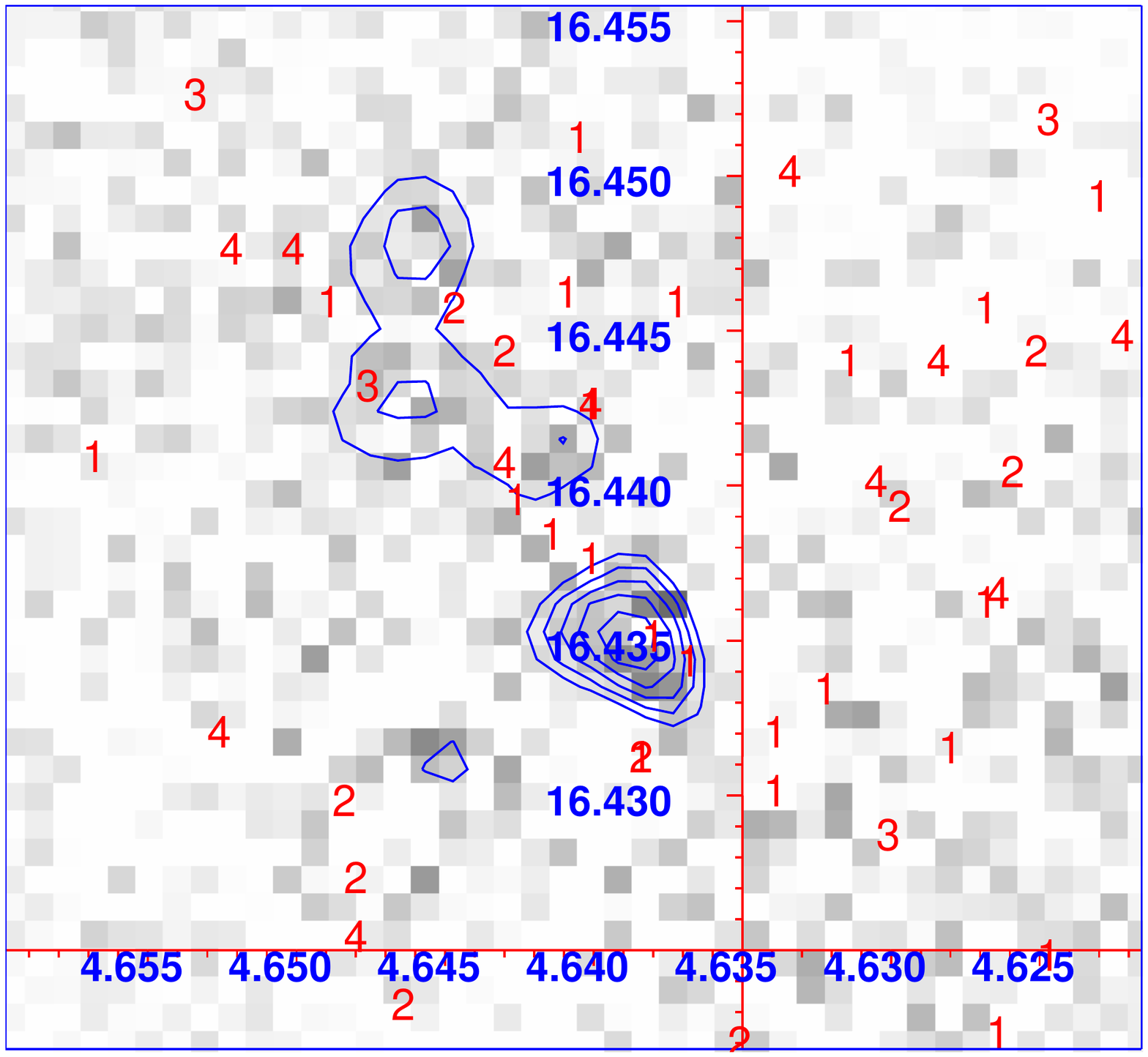}\includegraphics[width=2.7in,angle=0,bb=15 144 575 701,clip]{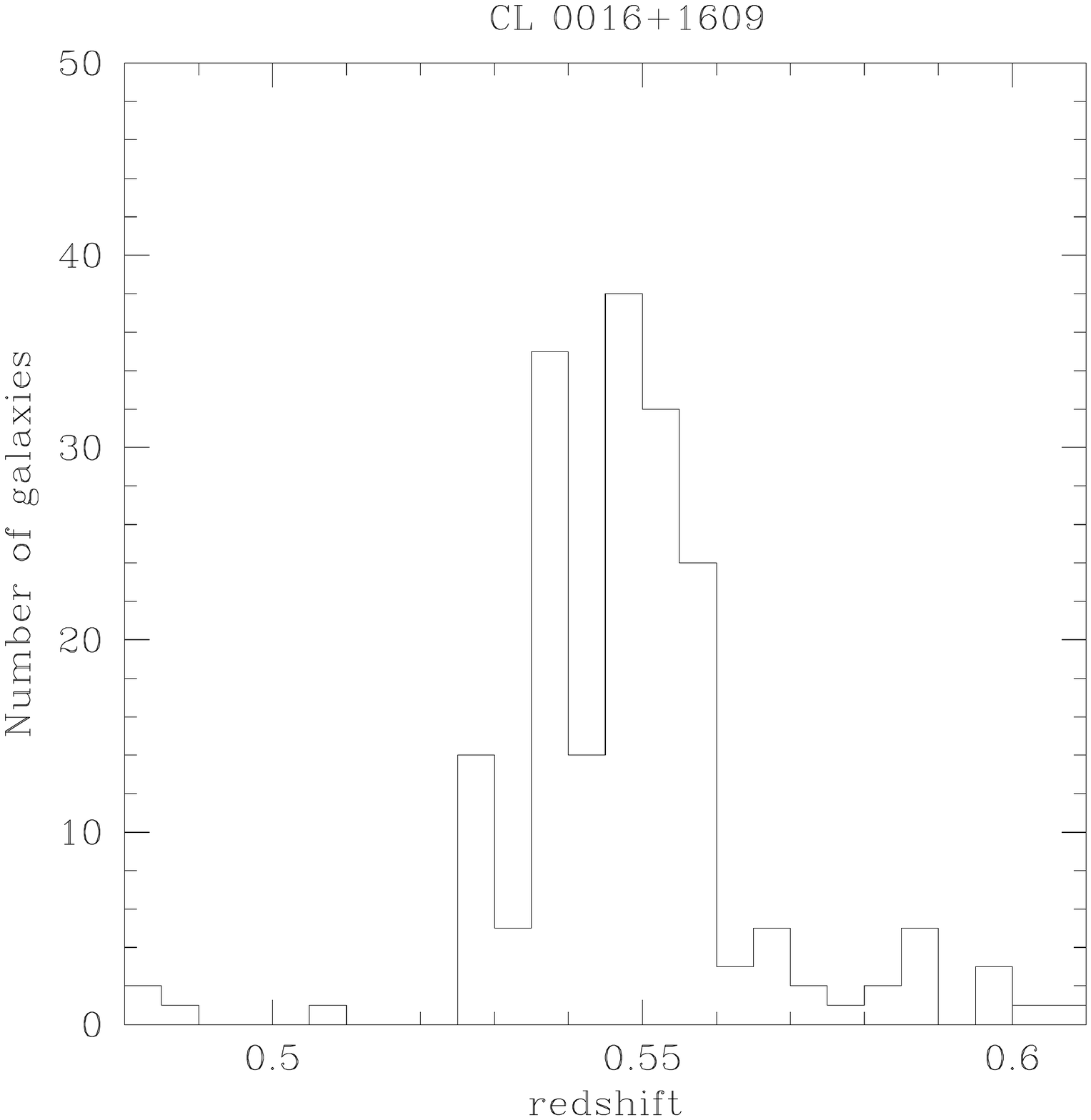}
    \caption{XMM-Newton X-ray image (upper left), residual image
      (upper right), and zoom on the residual image (lower left). The
      residual image was obtained by subtracting the model from the
      image for CL~0016+1609. The large blue circle in the upper left
      figure corresponds to a 500 kpc radius circle centred on the
      cluster position found in the literature. The large red circles
      in the upper right figure show the positions of the known active
      objects along the line of sight, the small red circles show the
      positions of the galaxies belonging to structures beyond the
      cluster from the SG analysis, and the small blue circles show
      the positions of the galaxies belonging to structures in front
      of the cluster from the SG analysis. The red numbers in the
      lower left figure show the positions of the galaxies belonging
      to substructures in the cluster from the SG analysis, the number
      being the one given in Table~\ref{tab:SG}.  Finally, blue contours
      in the lower left are the X-ray residuals, starting at the
      2.5$\sigma$ level and spaced by 1$\sigma$ intervals. Lower
      right: redshift histogram for the CL~0016+1609 area.}
  \label{fig:cl0016_X}
  \end{center}
  \end{figure*}

\begin{figure*}
  \begin{center}
    \includegraphics[width=2.7in,angle=0,bb=35 144 575 651,clip]{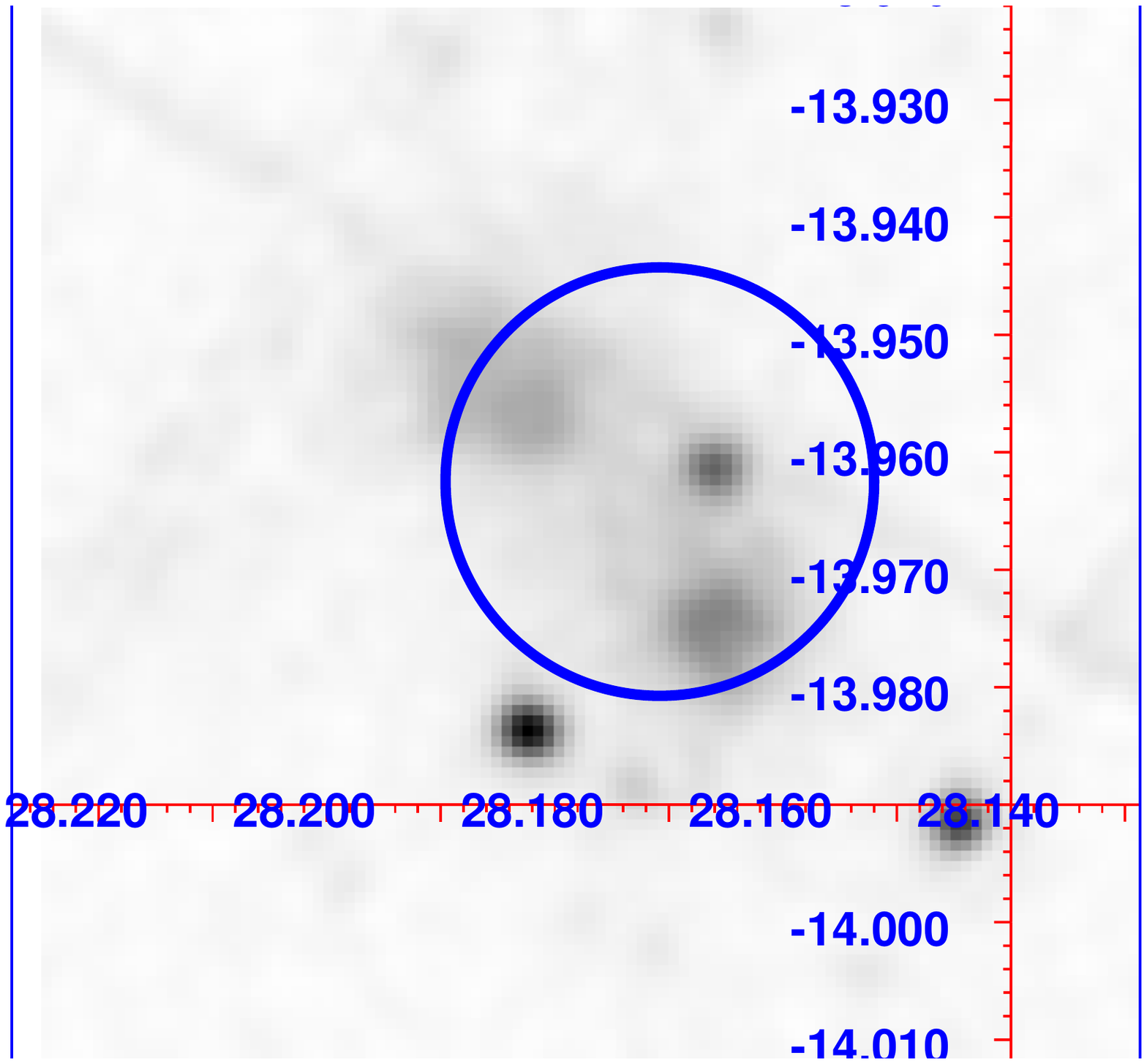}\includegraphics[width=2.7in,angle=0,bb=35 144 575 651,clip]{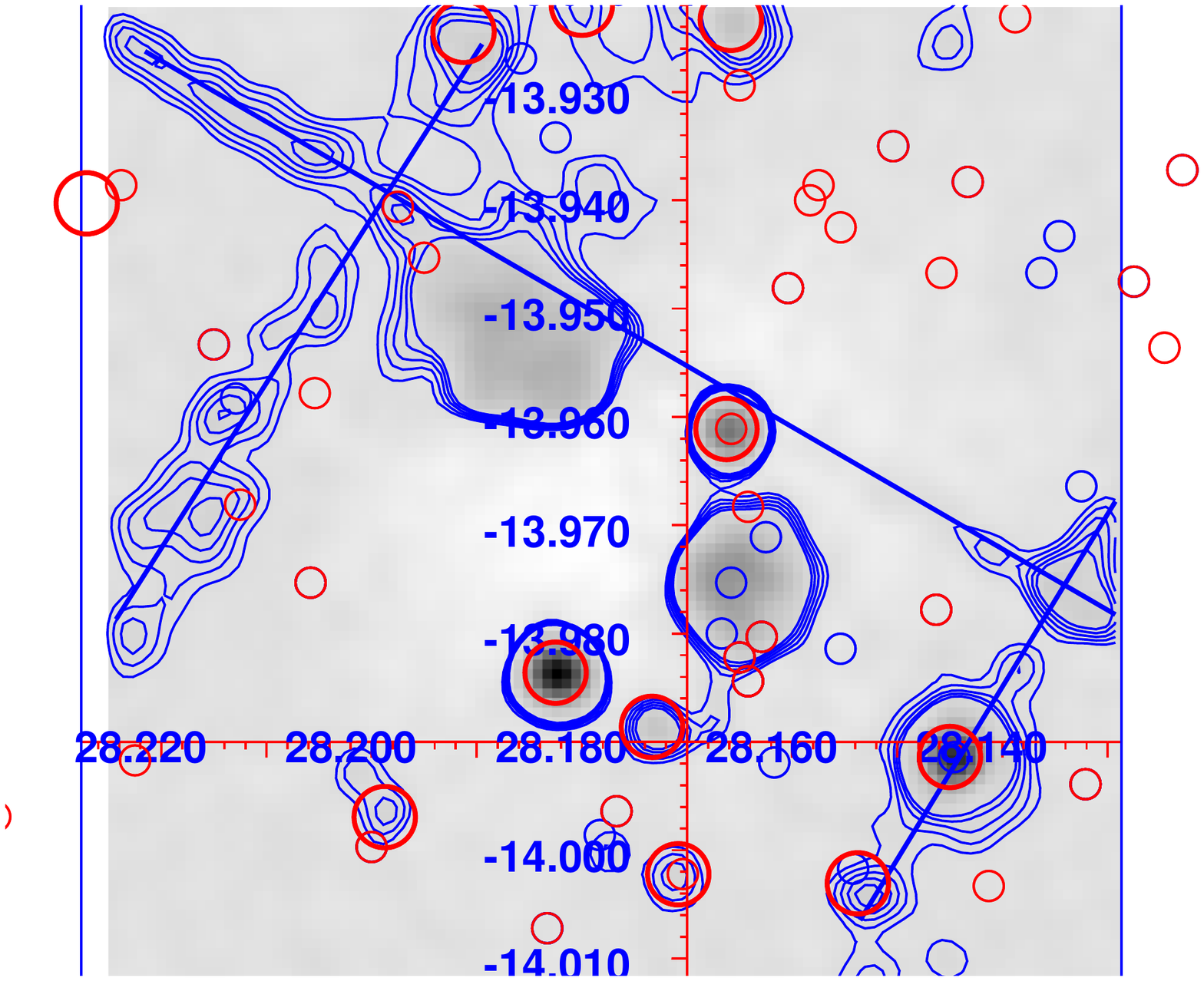}\\
    \includegraphics[width=2.7in,angle=0,bb=35 144 575 651,clip]{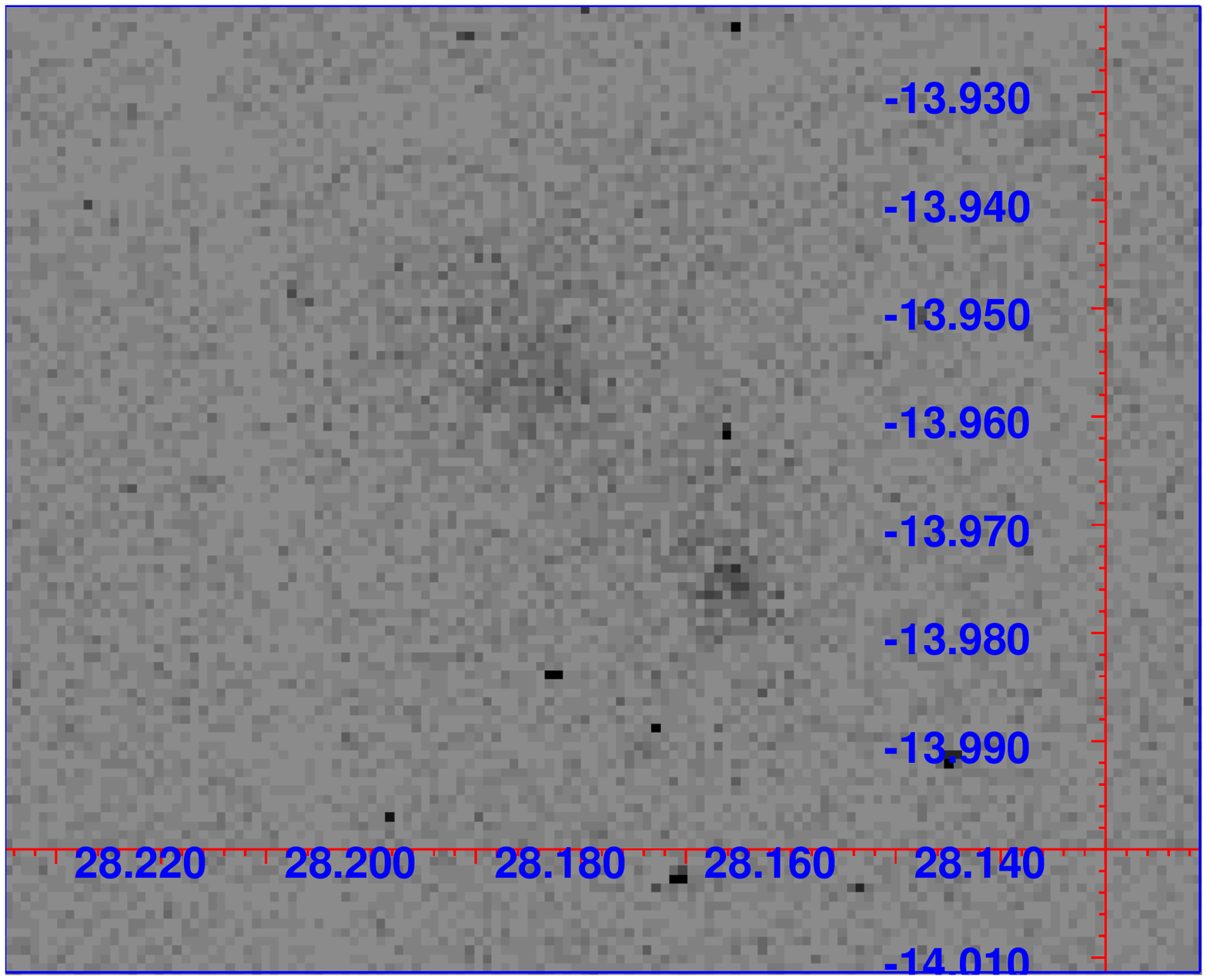}\includegraphics[width=2.7in,angle=0,bb=35 144 575 651,clip]{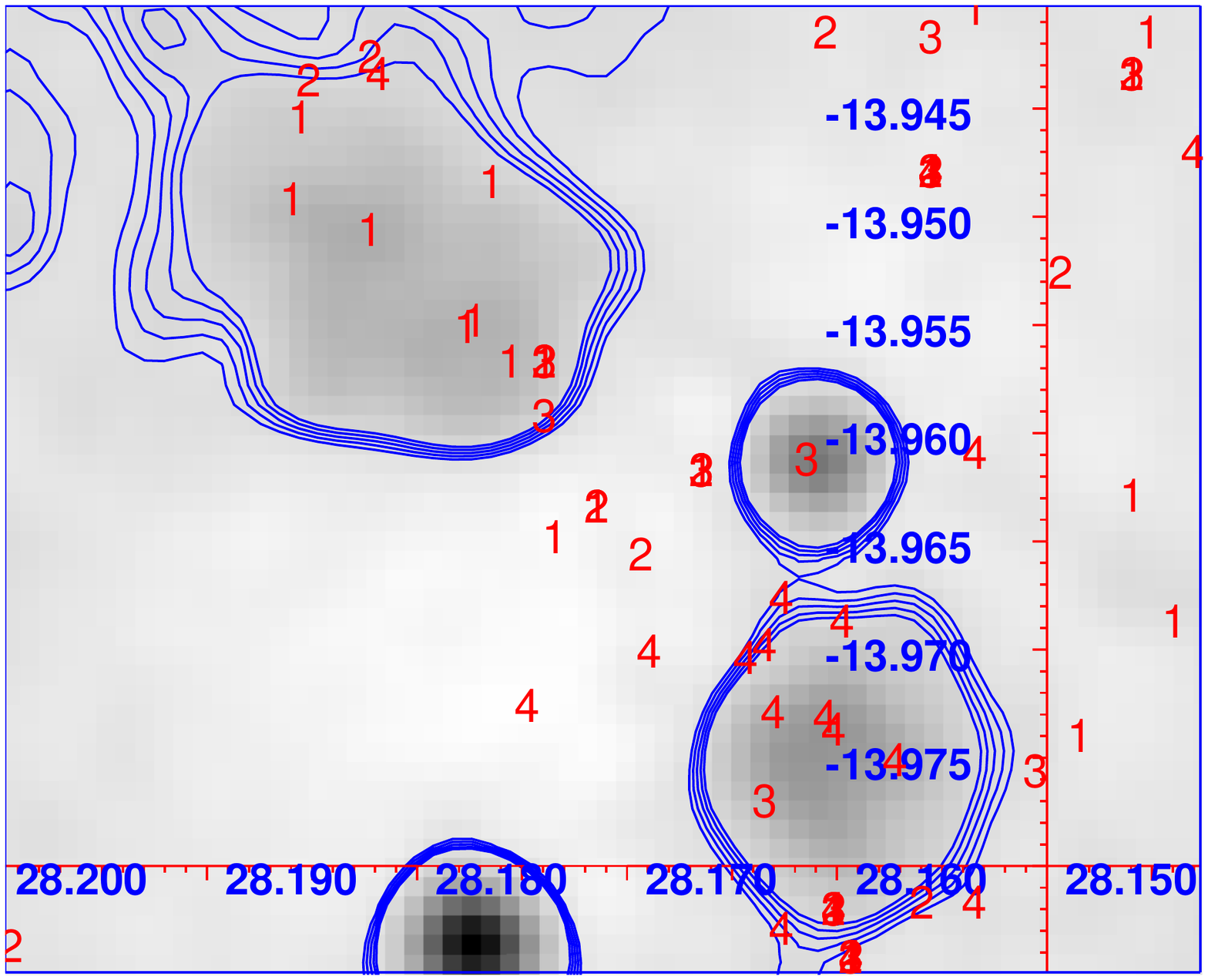}\\
    \includegraphics[width=2.7in,angle=0,bb=15 144 575 701,clip]{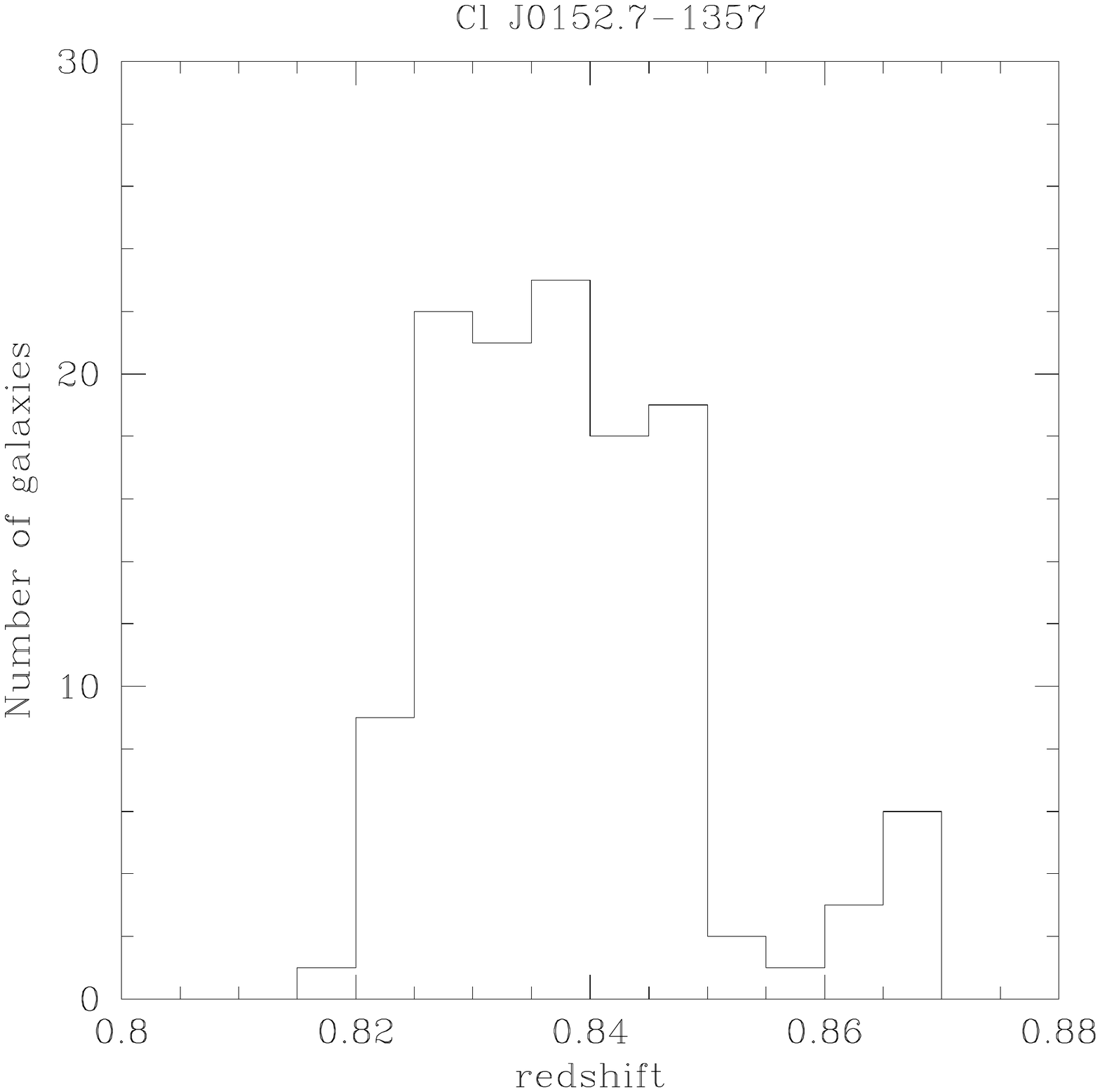}
    \caption{XMM-Newton X-ray image (upper left), residual image with
      straight blue lines showing the interchip detector gaps (upper
      right), Chandra image (middle left), and zoom on the residual
      image (middle right). The residual image was obtained by
      subtracting the model from the XMM-Newton image for
      CL~J0152.7-1357. The large blue circle in the upper left figure
      corresponds to a 500 kpc radius circle centred on the cluster
      position found in the literature. The large red circles in the
      upper right figure show the positions of the known active
      objects along the line of sight, the small red circles show the
      positions of the galaxies belonging to structures beyond the
      cluster from the SG analysis, and the small blue circles show
      the positions of the galaxies belonging to structures in front
      of the cluster from the SG analysis. The red numbers in the
      middle right figure show the positions of the galaxies belonging
      to substructures in the cluster from the SG analysis, the number
      being the one given in Table~\ref{tab:SG}.  Finally, blue contours
      are the X-ray residuals, starting at the 2.5$\sigma$ level and
      spaced by 1$\sigma$ intervals. Lower figure: redshift histogram
      for CL~J0152.7-1357.}
  \label{fig:cl0152_X}
  \end{center}
  \end{figure*}

We now describe clusters individually in order of increasing right
ascension, focussing on the X-ray and Serna \& Gerbal (1996) analyses.
Rounded coordinates (in degrees) and redshifts are given in
parentheses for each cluster (in some cases followed by the ?~symbol
when the cluster redshift is uncertain). For all similar
figures in the Appendix, the large blue circle shown in the top left
figure corresponds to a 500~kpc radius at the cluster redshift, but
the centre of this circle is the one given in the literature, and may
differ a little from the one we find from our images.
In the following, we refer to the n$^{th}$ substructure detected by
the Serna $\&$ Gerbal (1996) analysis and listed in Table~\ref{tab:SG}
as SGn.

\subsection{CL~0016+1609 (4.63888$^o$, +16.4433$^o$, z=0.5455)} 

The subtraction of a $\beta -$model in X-rays shows two X-ray sources
in the cluster area and a compact emission north--west of the cluster
(identified as an AGN by Gilmour et al., 2009). Following the results 
of our simulations, the central X-ray source is probably not a real
substructure. 
 
The other extended X-ray source (north--east of the cluster) is
probably a real substructure of CL~0016+1609. This X-ray emission does not
correspond to any galaxy structure detected on the line of sight 
(Fig.~\ref{fig:cl0016_X}).  Several cluster galaxies belonging
to substructures detected by the SG method are present inside this
X-ray emission. However, galaxies of groups SG1, SG2, and SG3 are spread
over the entire cluster area, while two thirds of the galaxies of SG4
are very close to the X-ray source. We therefore choose to relate SG4
to this extended X-ray source, putting it $\sim$5400 km/s beyond the
cluster main core (group SG1 in Table~\ref{tab:SG}).

The existence of such a substructure is reinforced by the fact that
the velocity histogram around z$\sim 0.54$ is clearly asymmetric and
seems to show at least three peaks (Fig.~\ref{fig:cl0016_X}).

\subsection{CL~J0152.7-1357 (28.17083$^o$, --13.9625$^o$, z=0.8310)} 

We can see in Fig.~\ref{fig:cl0152_X} that there are many X-ray
sources in the XMM-Newton residual image of this cluster. Part of
these sources are due to the interchip gaps (see
Fig.~\ref{fig:cl0152_X}). Both the Chandra image and Gilmour et
al. (2009) allow us to identify several point sources. Two main
extended and highly significant X-ray sources remain, which were
previously identified as two major structures in the process of
merging (Ebeling et al. 2001b), based on ROSAT data.  We are therefore
dealing with a major ongoing merger.

The 115 available redshifts in the cluster interval (see
Fig.~\ref{fig:cl0152_X}) allow us to characterize this
merger. Out of these 115 redshifts, 95 are distributed in four
substructures (see Table~\ref{tab:SG}) detected by our SG
analysis. Very clearly, substructure SG1 is identified with the
extended X-ray residual at the north east and substructure SG4 is
identified with the extended X-ray residual at the south
west. Substructure SG2 is located between the two extended X-ray
emissions and SG3 is probably infalling onto the cluster from the
foreground.

We cannot exclude a contamination by a foreground galaxy structure on
the line of sight (Fig.~\ref{fig:cl0152_X}), but given the very good
correspondence between substructure SG4 and the south--west X-ray
emission, we somewhat arbitrarily consider this contamination as
negligible. This is consistent with the detailed analysis of Girardi
et al.  (2005).

\subsection{MS~0302.5+1717  (46.32911$^o$, +17.47729$^o$, z=0.4250)  } 

\begin{figure*}
  \begin{center}
    \includegraphics[width=2.7in,angle=0,bb=35 144 575 651,clip]{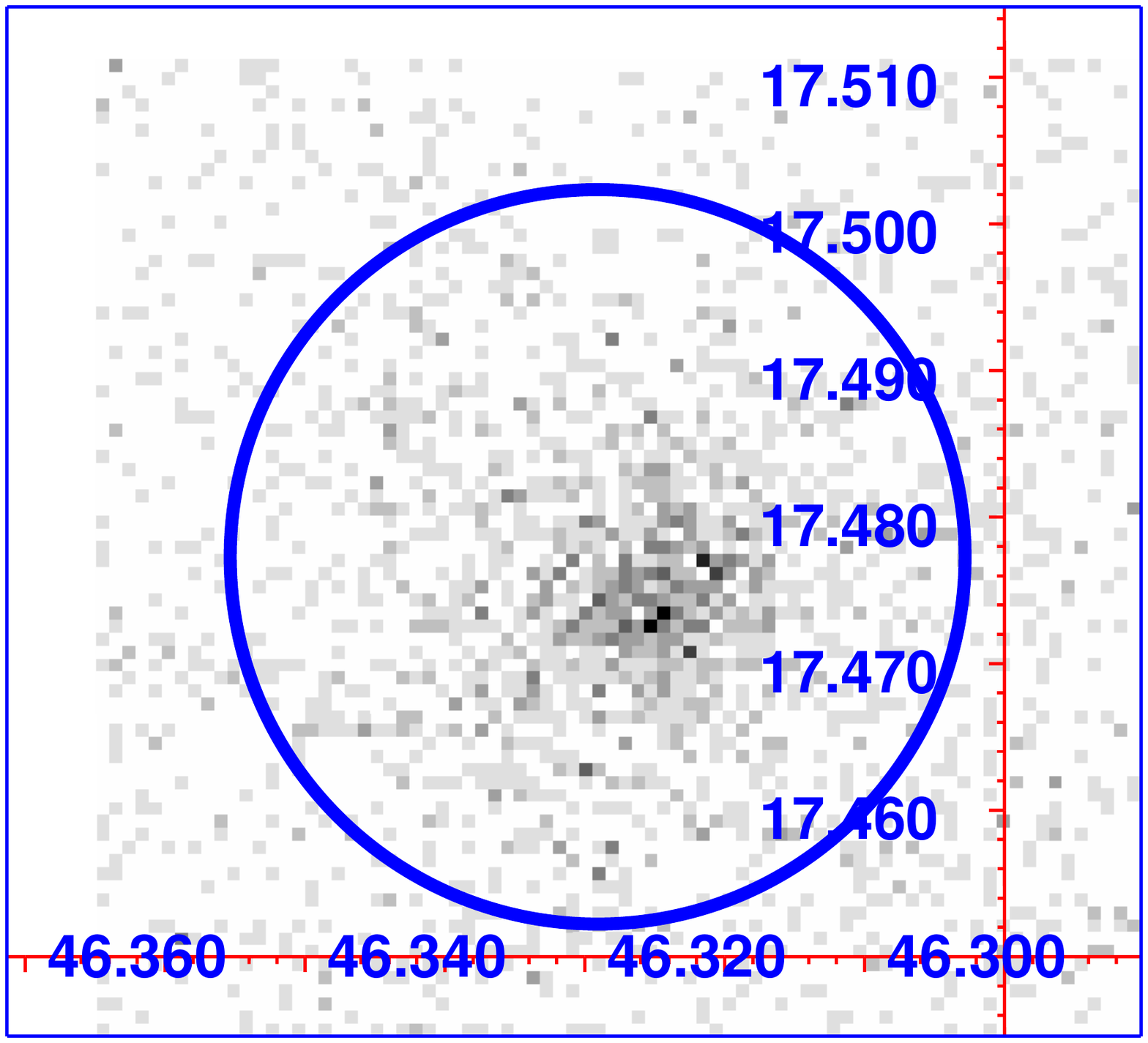}\includegraphics[width=2.7in,angle=0,bb=35 144 575 651,clip]{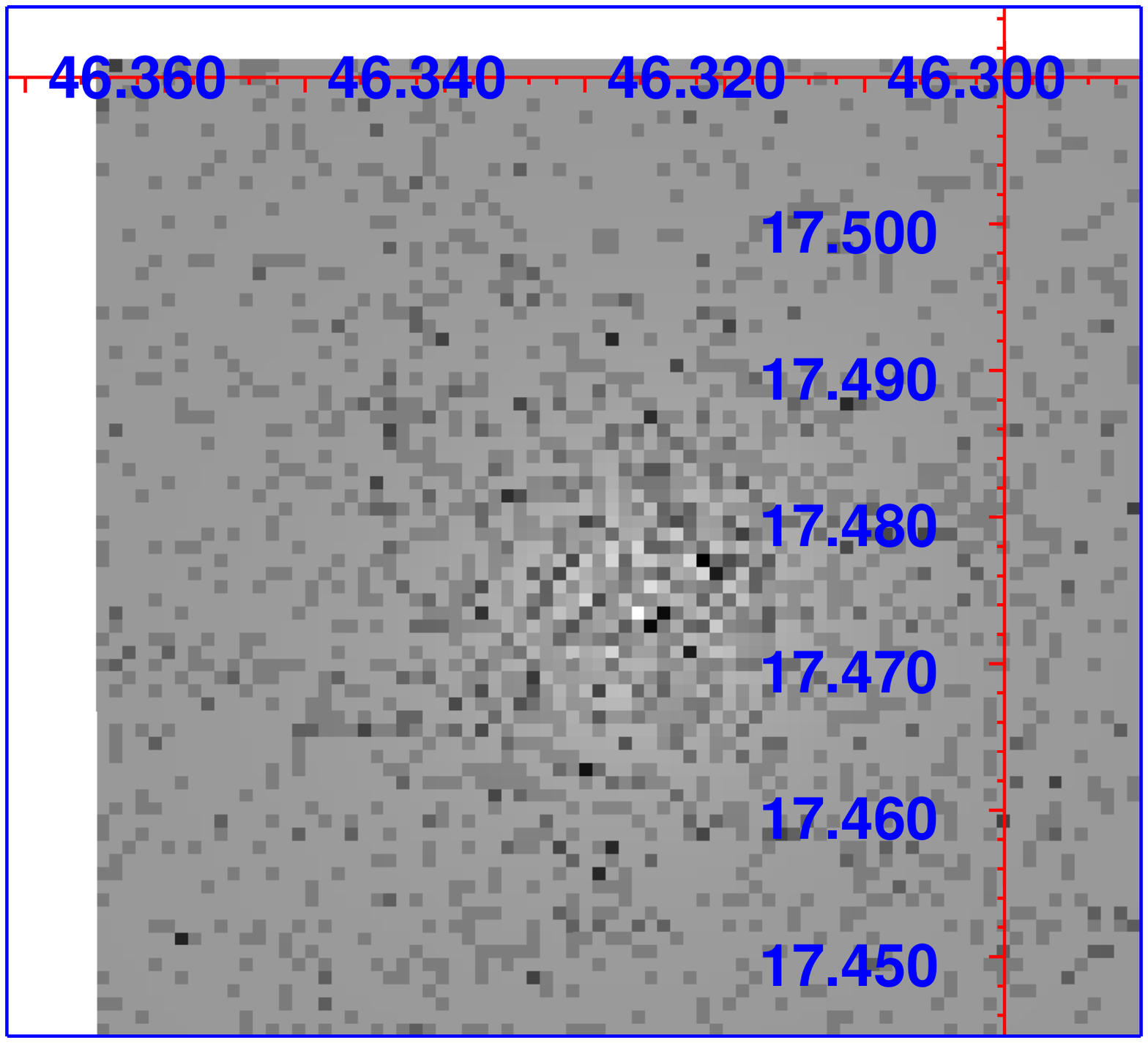}\\
    \includegraphics[width=2.7in,angle=0,bb=35 144 575 651,clip]{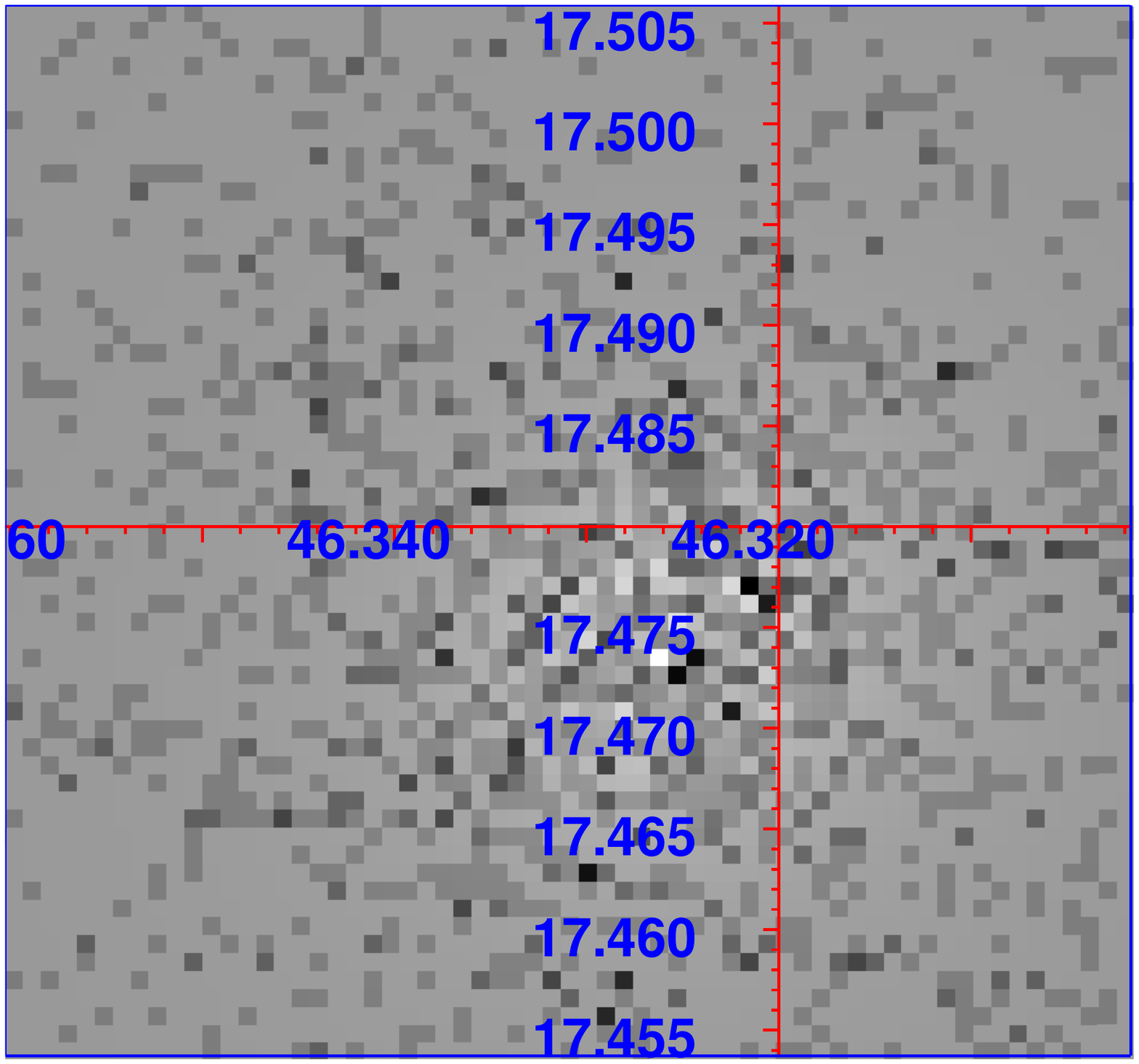}
    \caption{XMM-Newton X-ray image (upper left), residual image
      (upper right), and zoom on the residual image (lower left). The
      residual image was obtained by subtracting the model from the
      image for MS~0302.5+1717. The large blue circle in the upper
      left figure corresponds to a 500 kpc radius circle centred on
      the literature cluster position.  Finally, blue contours
      (present in some of the following figures) in the lower left are
      the X-ray residuals, starting at the 2.5$\sigma$ level and
      spaced by 1$\sigma$ intervals.}
  \label{fig:ms0302X}
  \end{center}
  \end{figure*}

  This cluster is rather weak (Gioia et al. 1990 and
  Fig.~\ref{fig:ms0302X}).  In X-rays, no substructure is detected at
  better than the 2.5$\sigma$ level when subtracting a $\beta
  -$model. A single redshift is available in NED so the SG analysis is
  not possible. X-ray data are, however, deep enough (the Fe line in the
  X-ray spectrum gives the same redshift for the cluster than the
  central galaxy: z=0.426) to be sure that this cluster is not hosting
  major substructures.

\subsection{XDCS cm J032903.1+025640 (52.26175$^o$, +2.94033$^o$, z=0.4122)} 

\begin{figure*}
  \begin{center}
    \includegraphics[width=2.7in,angle=0,bb=35 144 575 651,clip]{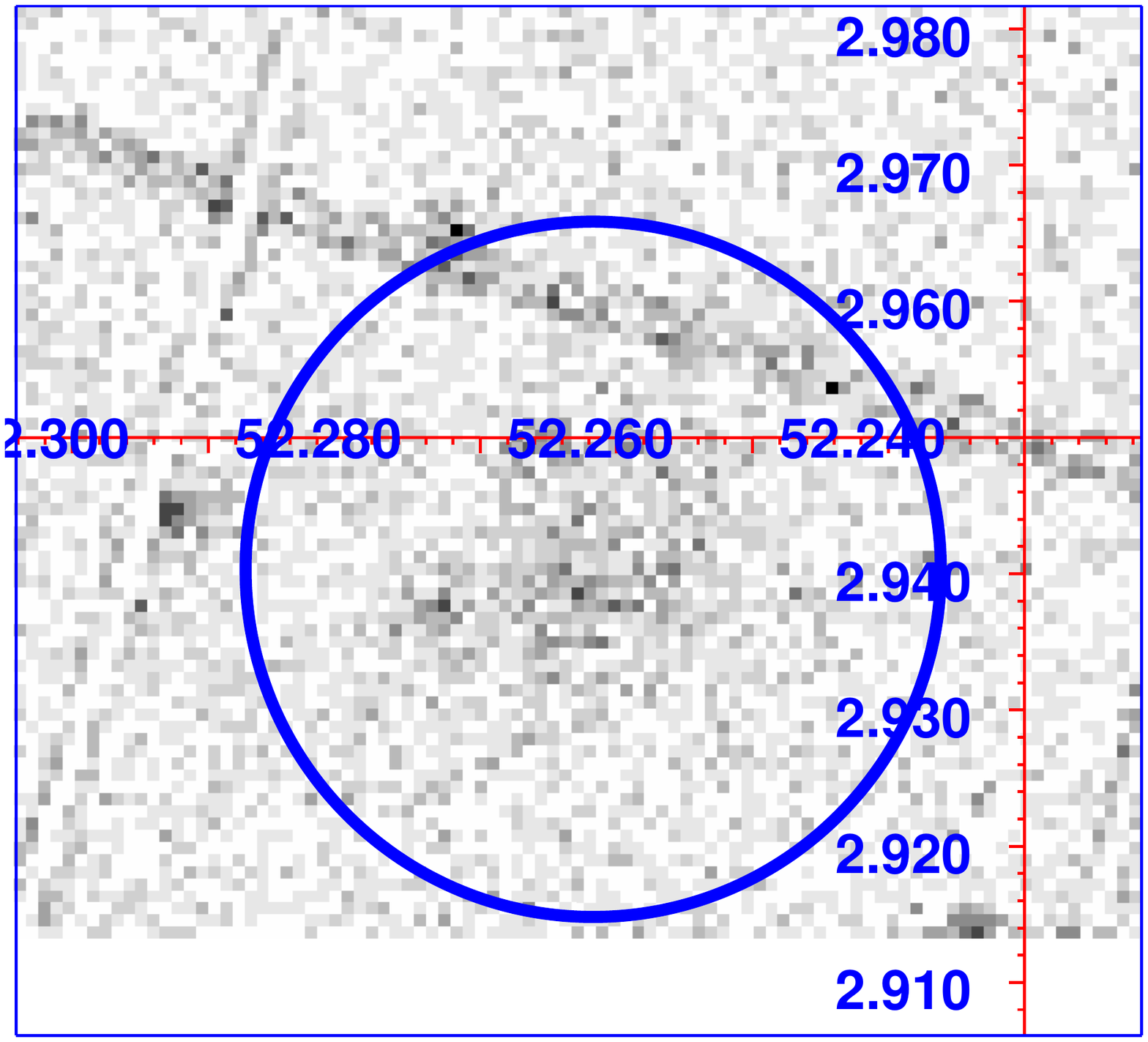}\includegraphics[width=2.7in,angle=0,bb=35 144 575 651,clip]{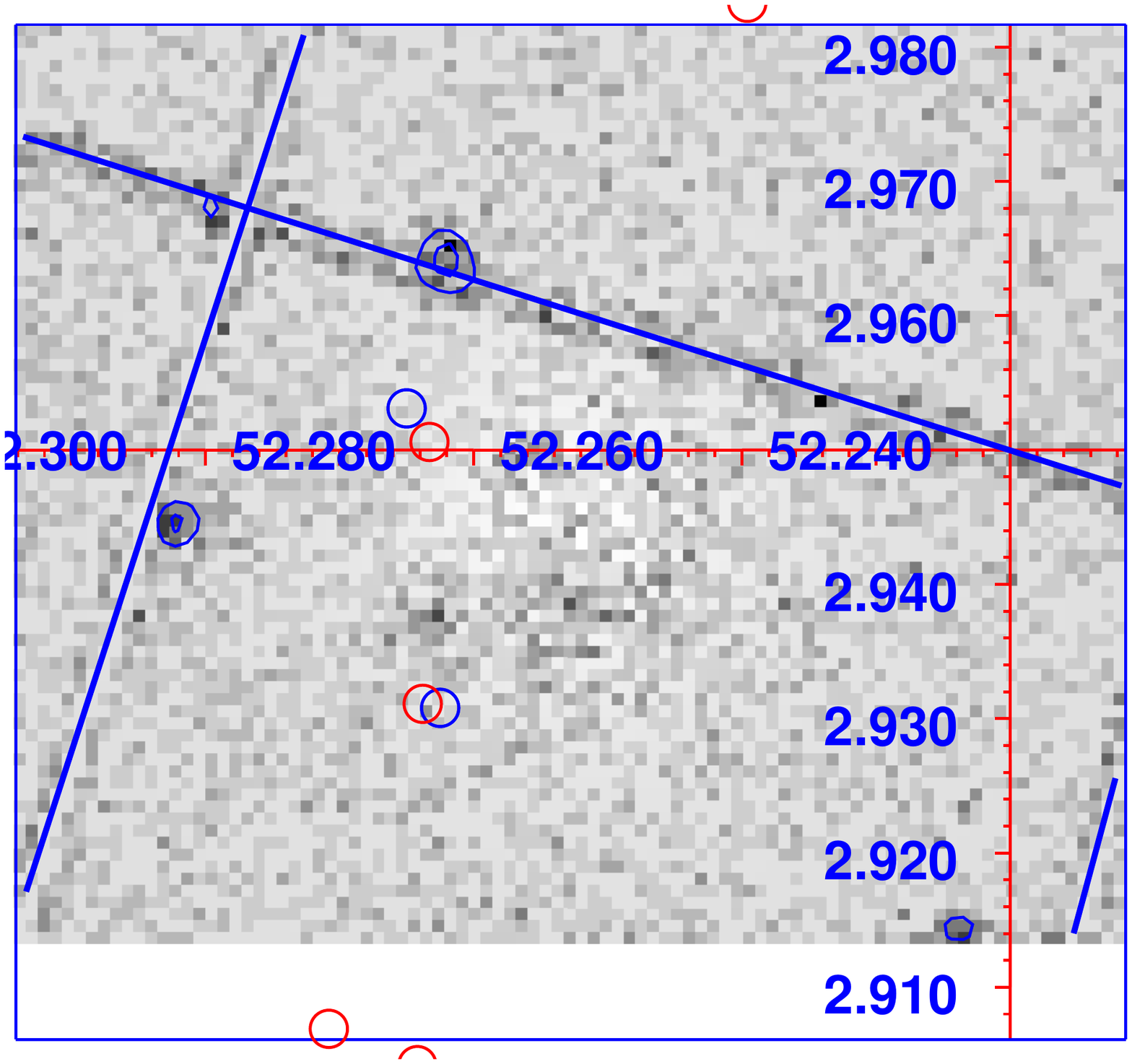}\\
    \includegraphics[width=2.7in,angle=0,bb=35 144 575 651,clip]{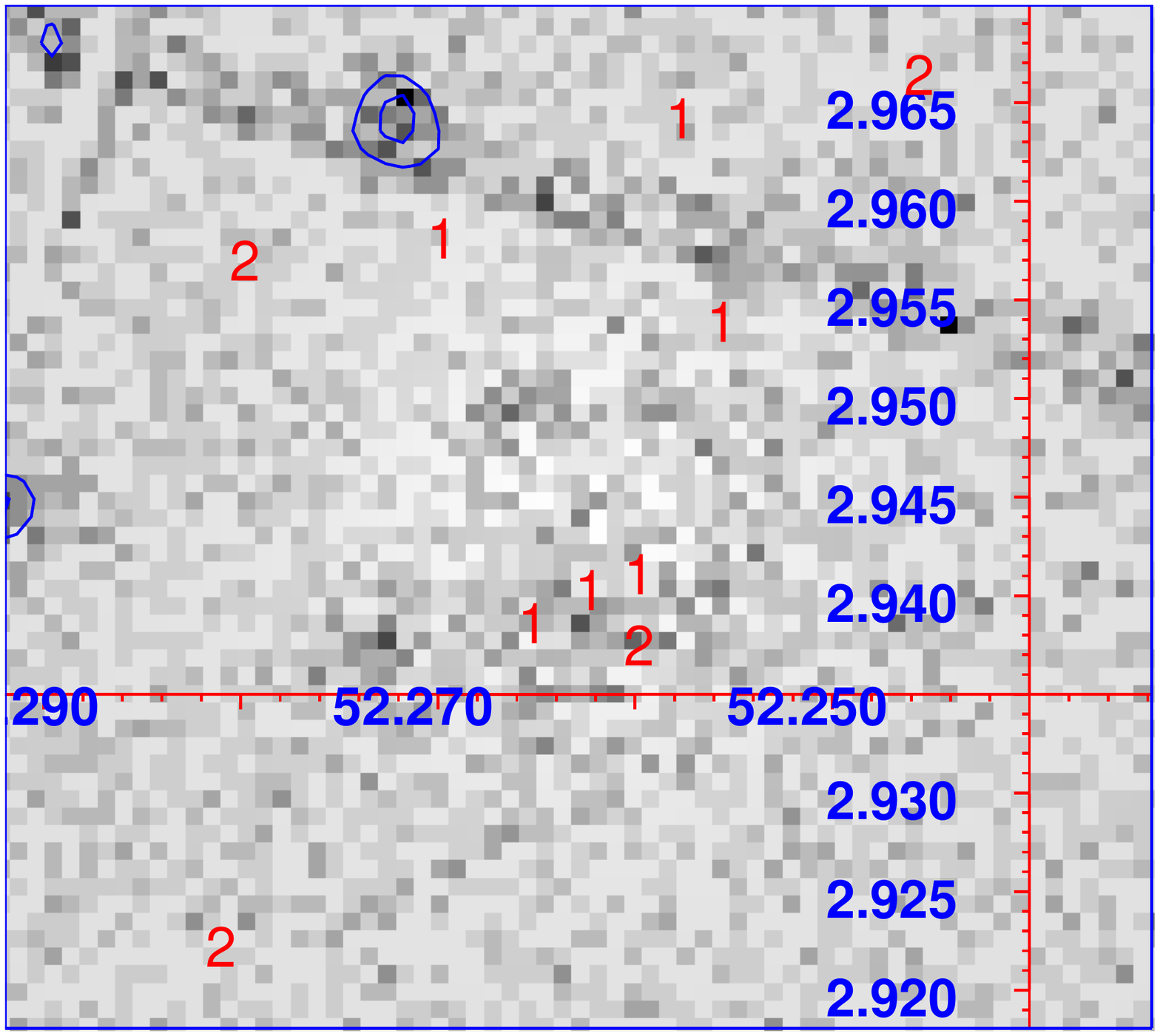}\includegraphics[width=2.7in,angle=0,bb=15 144 575 701,clip]{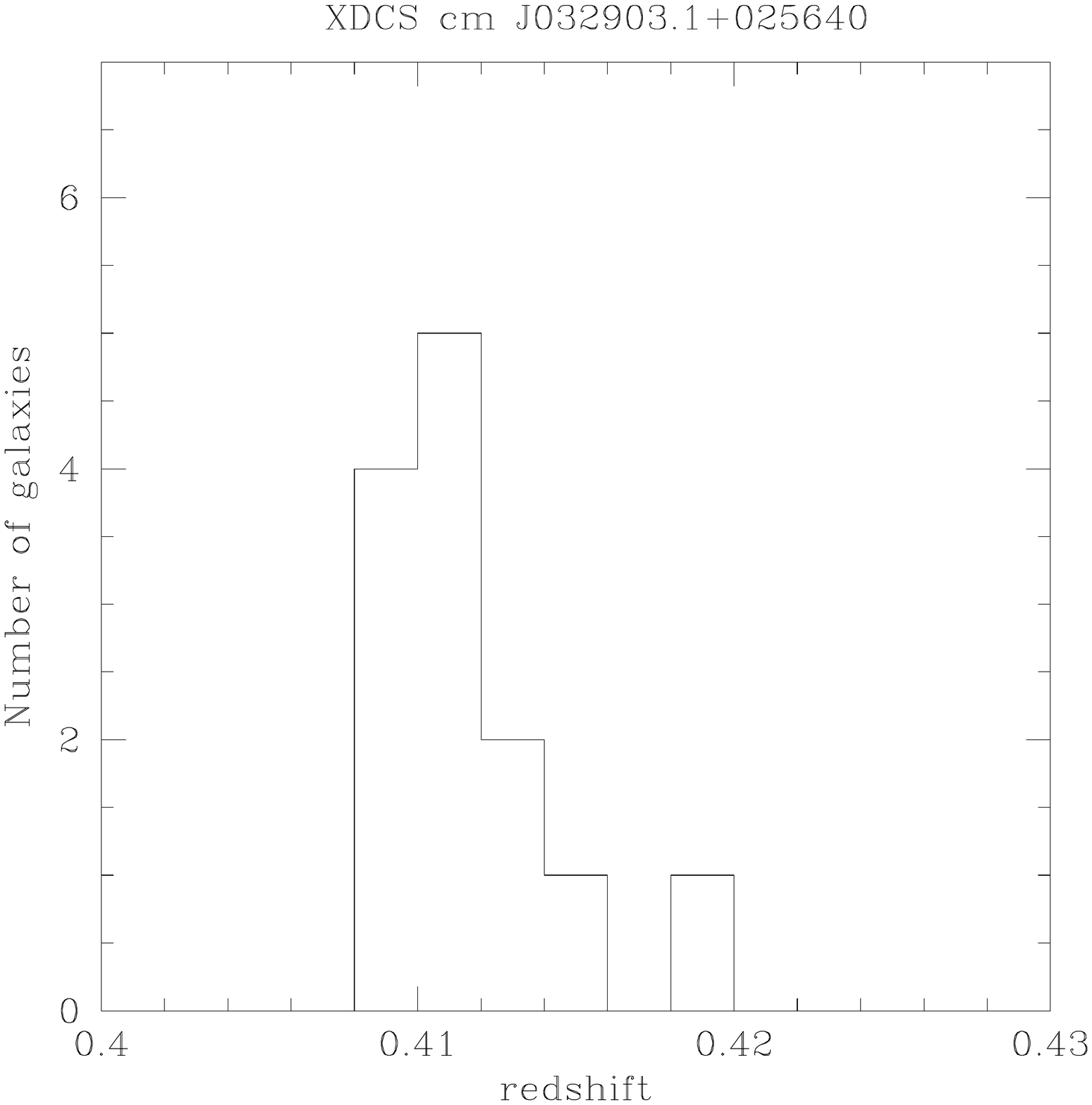}
    \caption{Same as Fig.~\ref{fig:cl0016_X} for XDCS cm J032903.1+025640 
    with straight blue lines (upper right) showing the interchip detector gaps.}
  \label{fig:XDCS_cm_X}
  \end{center}
  \end{figure*}

The XDCS cm J032903.1+025640 cluster is detected in X-rays
(Fig.~\ref{fig:XDCS_cm_X}), but it is rather weak and diffuse, as in
the ROSAT image of Mulchaey et al. (2006). After model subtraction,
there is hardly any emission left except two residuals probably due to
interchip gaps.

There are 13 galaxies in the [0.40,0.42] redshift range. The SG analysis detects
two low--mass structures (see Table~\ref{tab:SG}), SG1 probably being identified
with the cluster itself. If we consider SG1 and SG2 together, the mass computed
by the SG analysis for XDCS cm J032903.1+025640 is 7.10 10$^{13}$ M$_\odot$.

\subsection{RX~J0337.6-2522 (54.43812$^o$, --25.38$^o$, z=0.5850)  }  

\begin{figure*}
  \begin{center}
    \includegraphics[width=2.7in,angle=0,bb=35 144 575 651,clip]{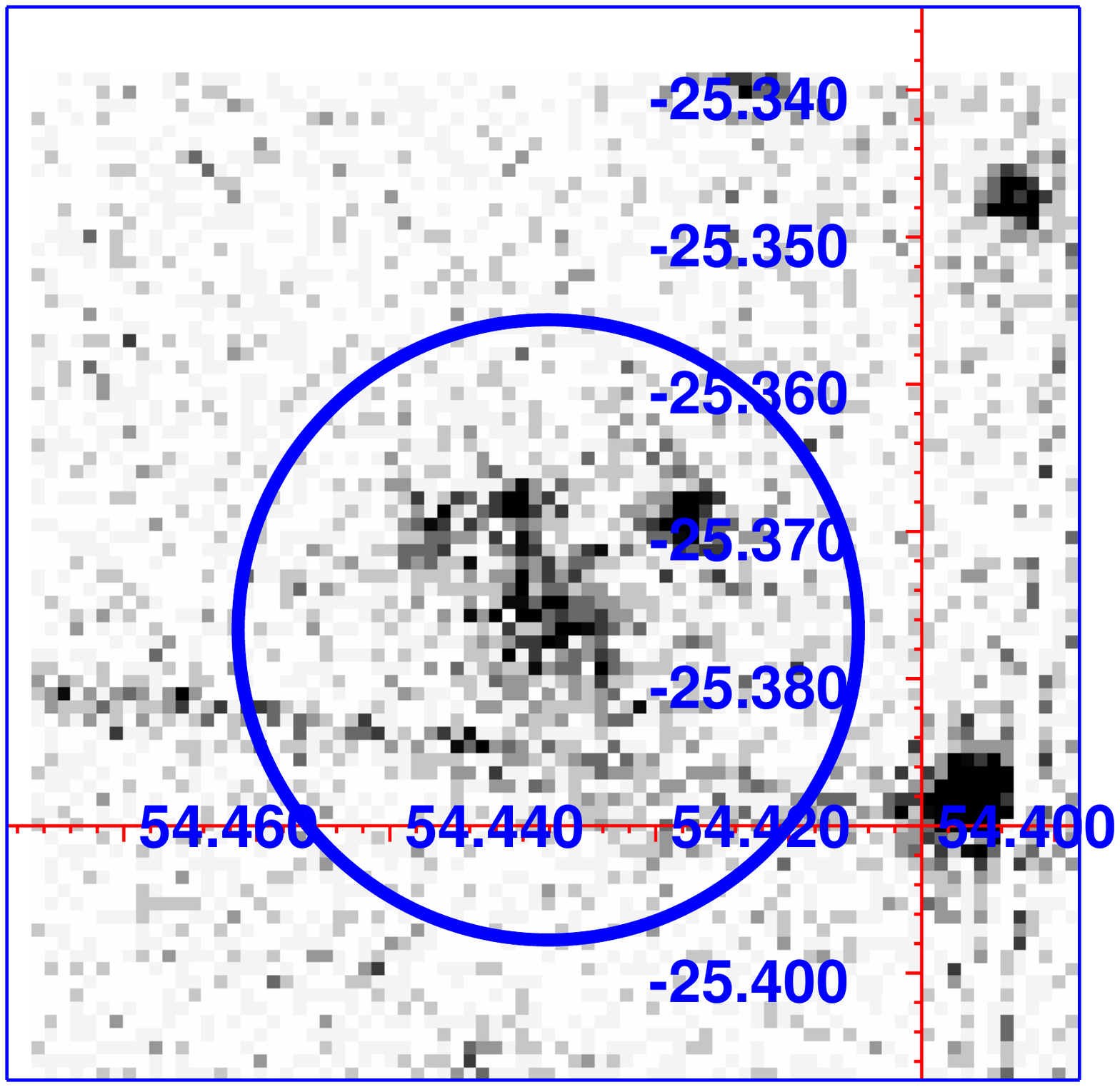}\includegraphics[width=2.7in,angle=0,bb=35 144 575 651,clip]{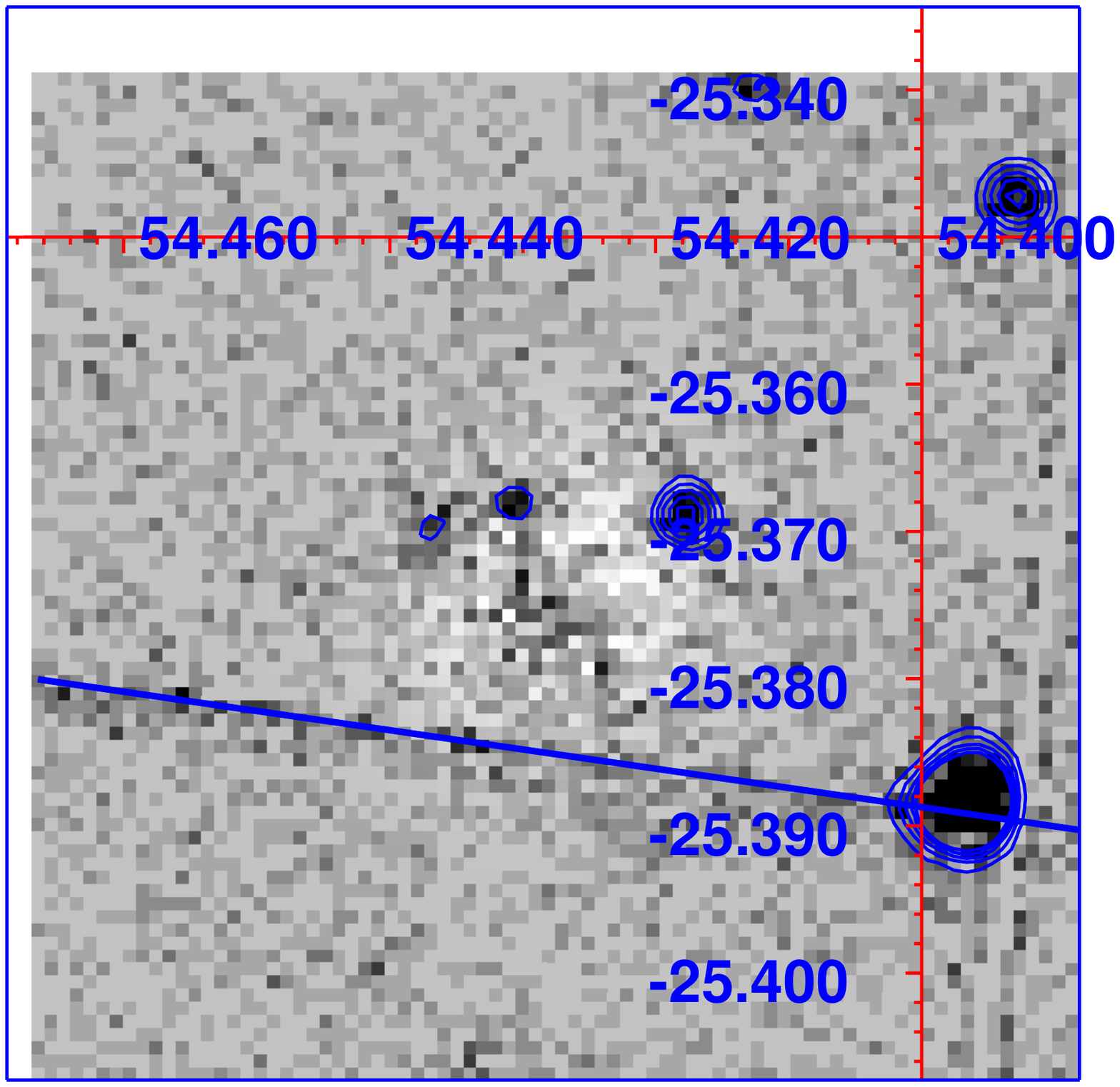}\\
    \includegraphics[width=2.7in,angle=0,bb=35 144 575 651,clip]{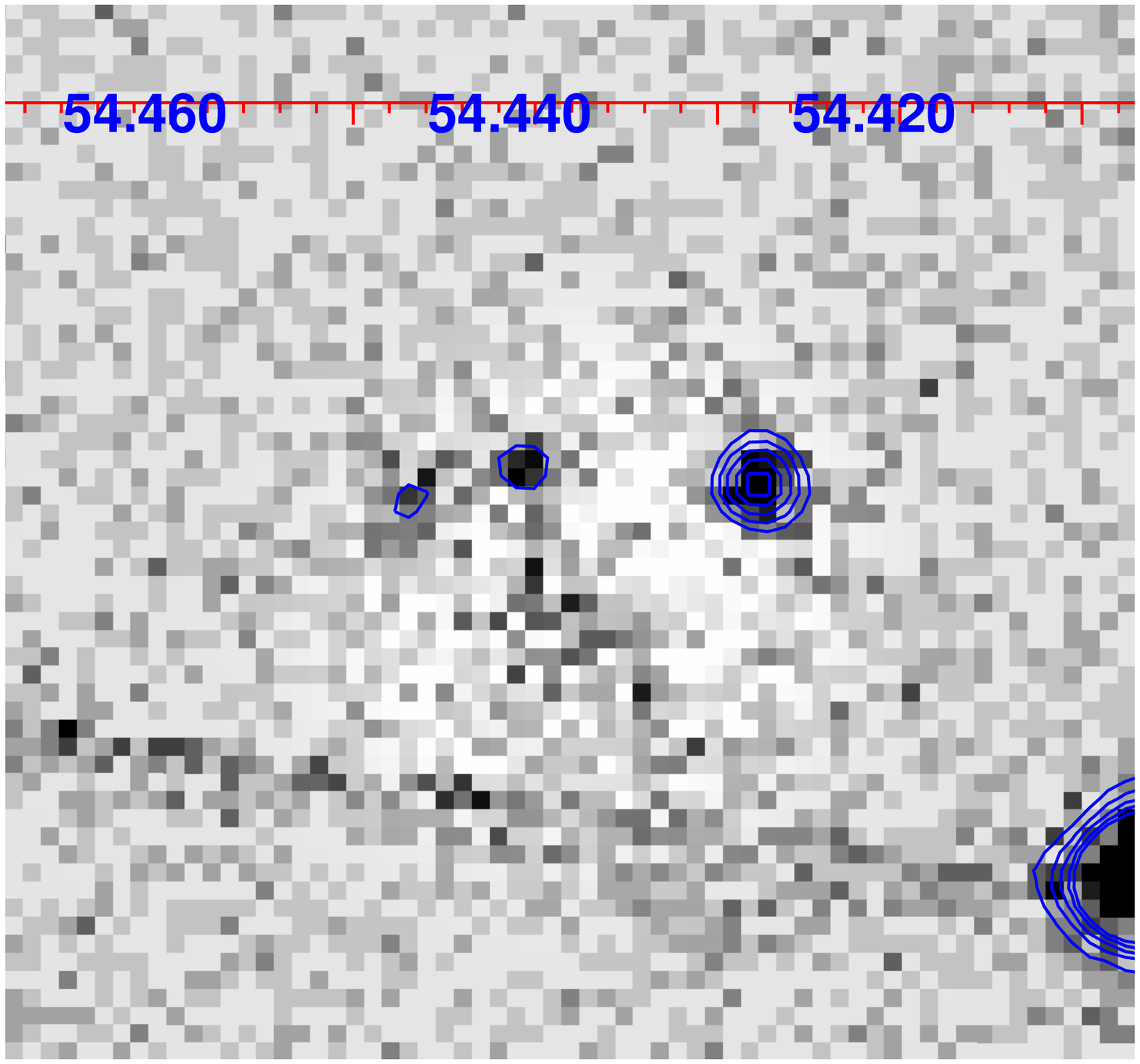}
    \caption{Same as Fig.~\ref{fig:ms0302X} for RX~J0337.6-2522.}
  \label{fig:rx0337_X}
  \end{center}
\end{figure*}

We only have limited information for this cluster, which was
discovered by Vikhlinin et al. (1998). Its XMM-Newton emission is
relatively weak, and neither Chandra data nor catalogues of known
active objects in the field are available.  There are most probably
three X-ray point sources in the field of view (south--west, west, and
north--west of the cluster centre (see Fig.~\ref{fig:rx0337_X}). We
also detect two small 2.5$\sigma$ level sources in the residual image
close to the cluster location, but even the largest one is too faint
to provide a successful luminosity measurement. We are therefore
probably dealing with a residual of the cluster X-ray emission itself
because a $\beta -$model does not perfectly fit the cluster emission.

We only have five galaxy redshifts in the cluster range.

\subsection{MACS J0454.1-0300 (73.54552$^o$, --3.0187$^o$, z=0.5377)}  

\begin{figure*}
  \begin{center}
    \includegraphics[width=2.7in,angle=0,bb=35 144 575 651,clip]{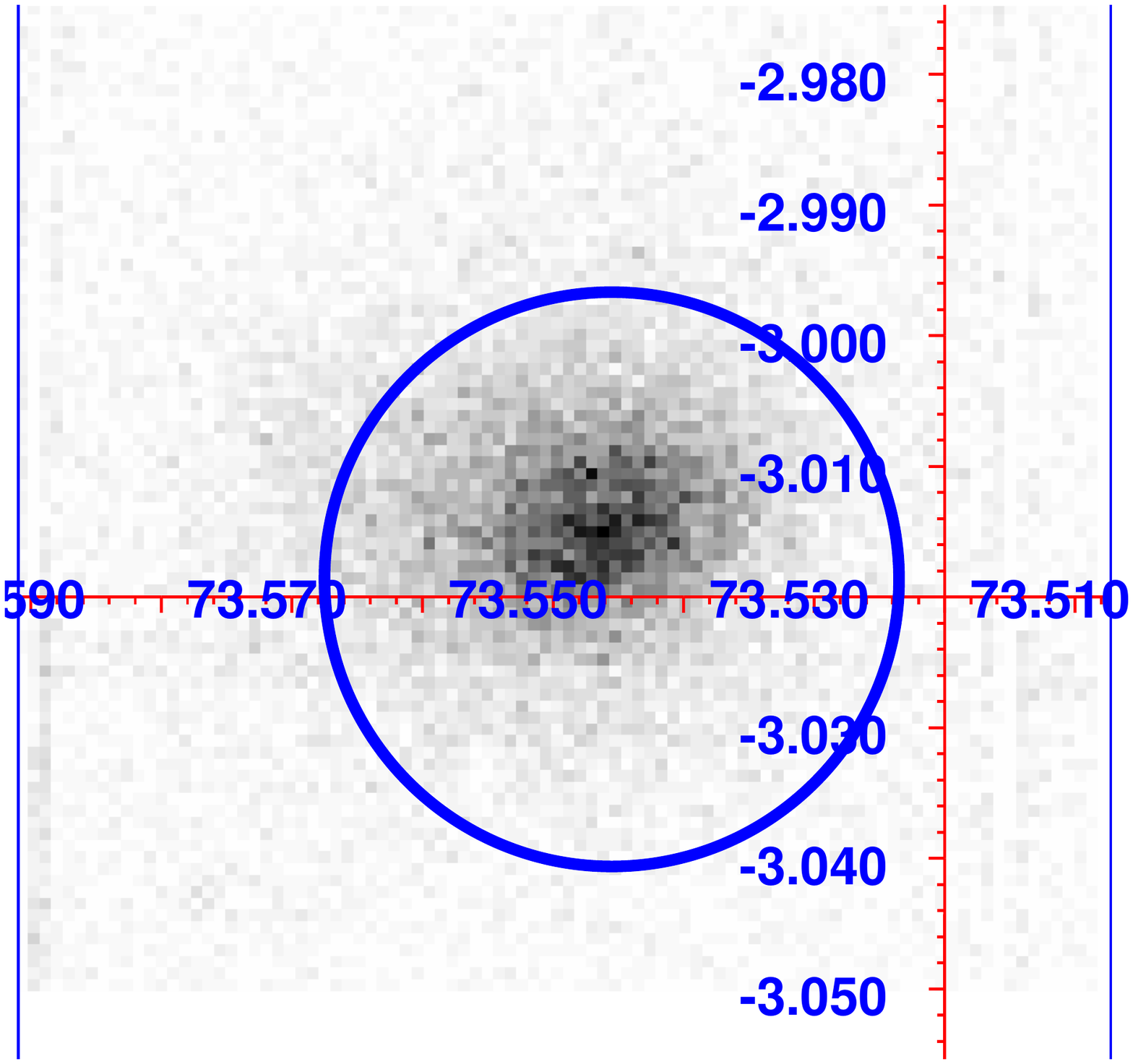}\includegraphics[width=2.7in,angle=0,bb=35 144 575 651,clip]{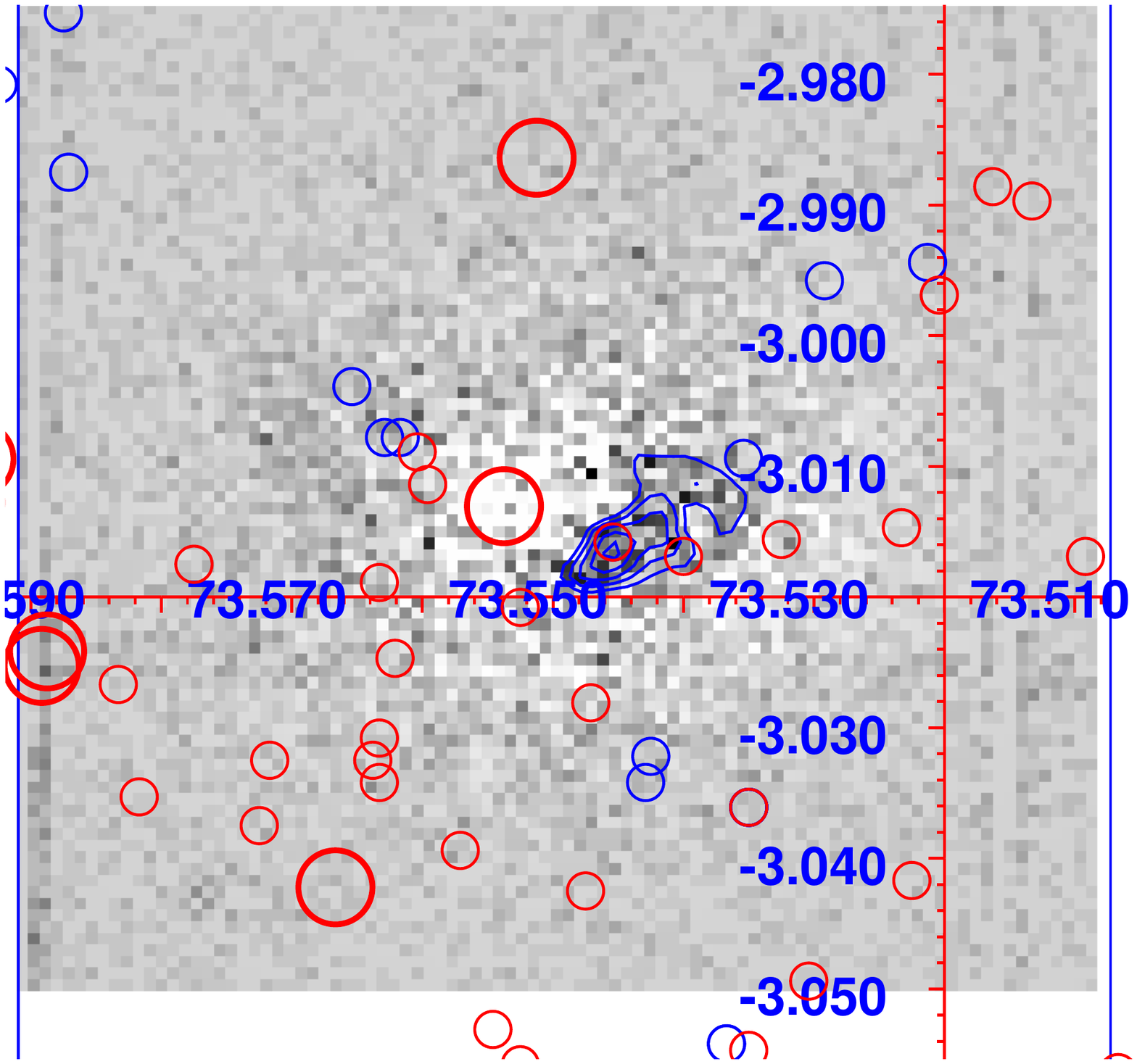}\\
    \includegraphics[width=2.7in,angle=0,bb=35 144 575 651,clip]{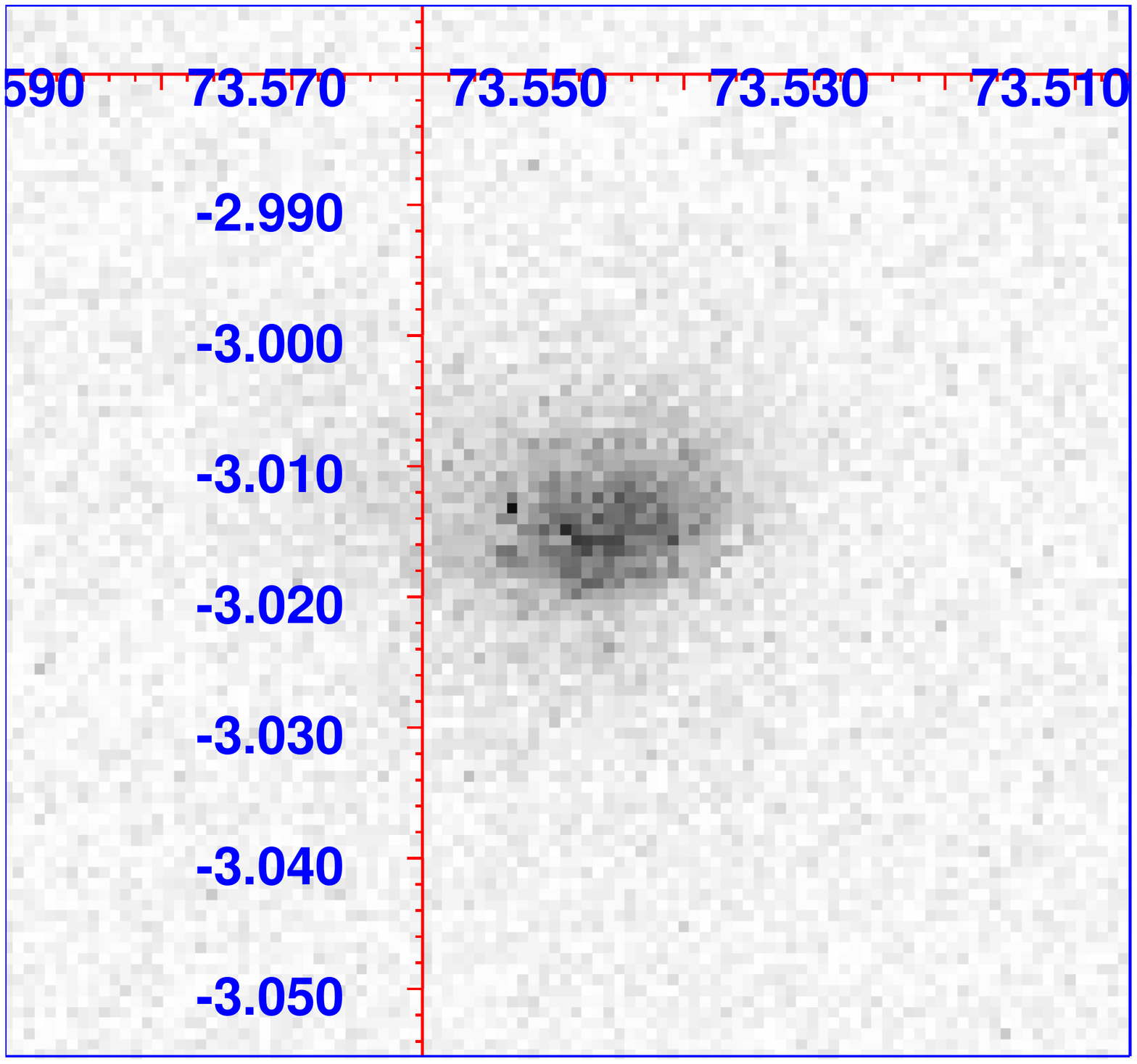}\includegraphics[width=2.7in,angle=0,bb=35 144 575 651,clip]{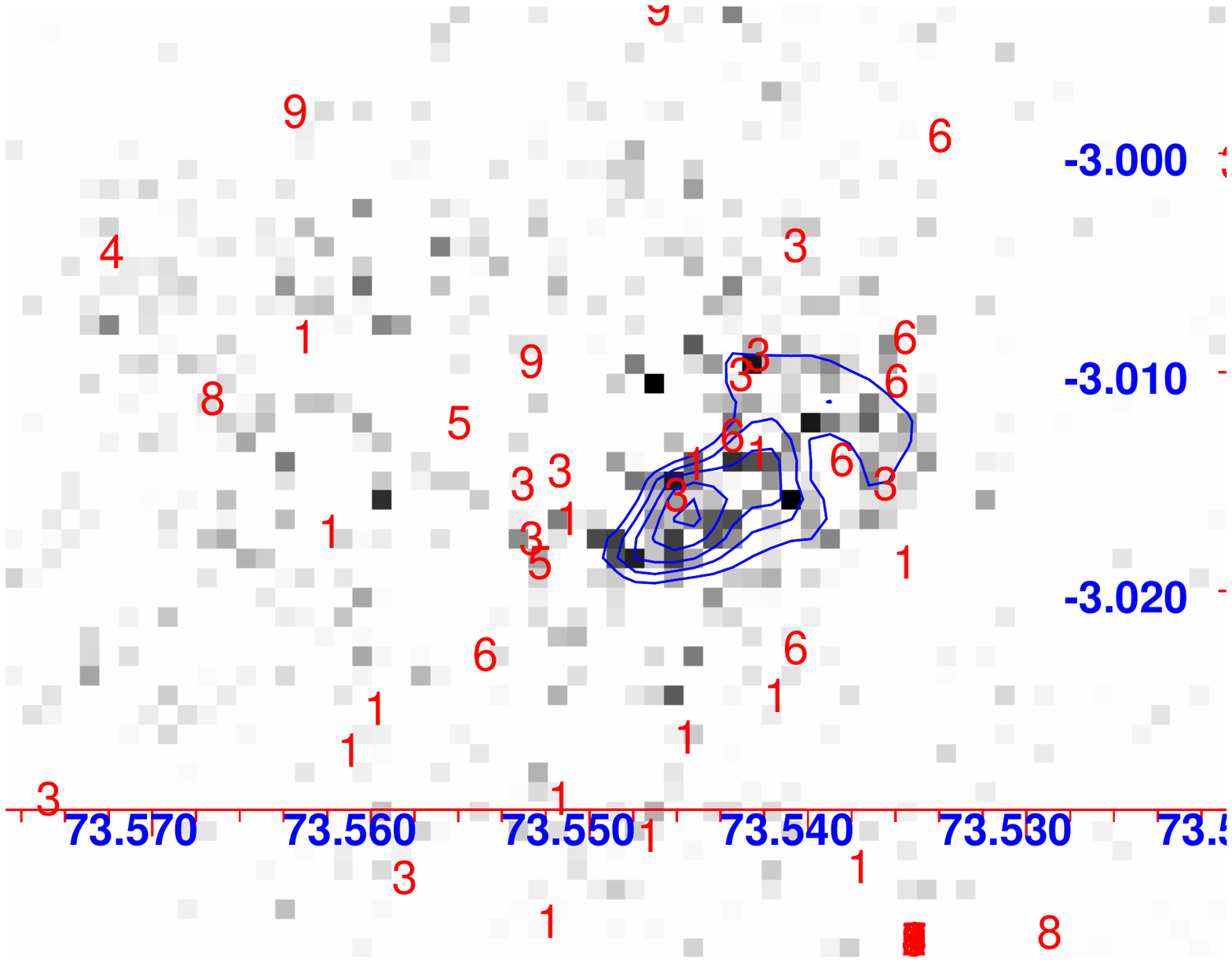}\\
    \includegraphics[width=2.7in,angle=0,bb=15 144 575 701,clip]{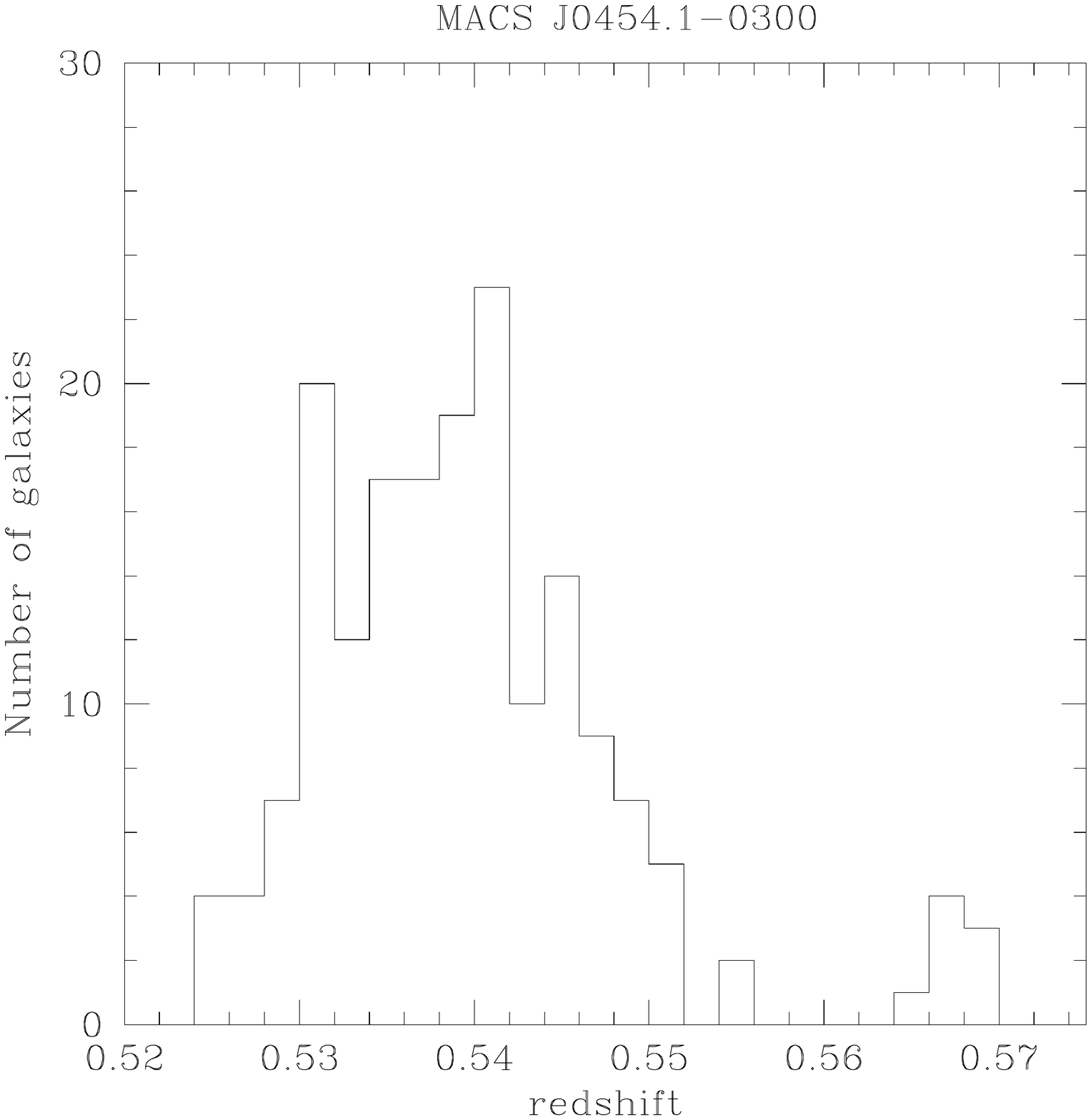}
    \caption{Same as Fig.~\ref{fig:cl0152_X} for MACS J0454.1-0300.}
  \label{fig:macs0454_X}
  \end{center}
\end{figure*}

The X-ray image of the well known X-ray cluster MACS J0454.1-0300
(Gioia et al. 1990) appears rather smooth, but after model subtraction
a significant excess emission (4.5$\sigma$ level) is detected about
0.6~arcmin west of the cluster centre (Fig.~\ref{fig:macs0454_X}). It
is too extended to be simply due to incorrect modelling by the $\beta
-$model. We also detect on a larger scale (not shown on
Fig.~\ref{fig:macs0454_X}) X-ray emission that could originate from
the fossil group J0454-0309 at z=0.26 analysed by Schirmer et
al. (2010). They found that this group was located about 8~arcmin away
from MACS J0454.1-0300.

Many redshifts are available in this region, thanks to the spectroscopic
observations of Moran et al. (2007), who analysed the transformation of
spirals into S0s. The velocity histogram around z$\sim 0.54$ is
asymmetric (see Fig.~\ref{fig:macs0454_X}), with several peaks (at
least 3); there are 194 galaxies in the [0.52,0.555] redshift range).
Nine substructures are found by the SG analysis (see
Table~\ref{tab:SG}).  SG1 is probably the one associated with the
whole cluster, while SG3 is probably associated with the X-ray residual
(900 km/s in front of the cluster itself).  We are therefore probably
observing a collection of several low--mass groups (the most massive
being SG3) in the process of merging onto the core of the structure
(MACS J0454.1-0300 itself).

\subsection{BMW-HRI~J052215.8-362452 (80.55917$^o$, --36.4178$^o$, z=0.4720)} 

\begin{figure*}
  \begin{center}
    \includegraphics[width=2.7in,angle=0,bb=35 144 575 651,clip]{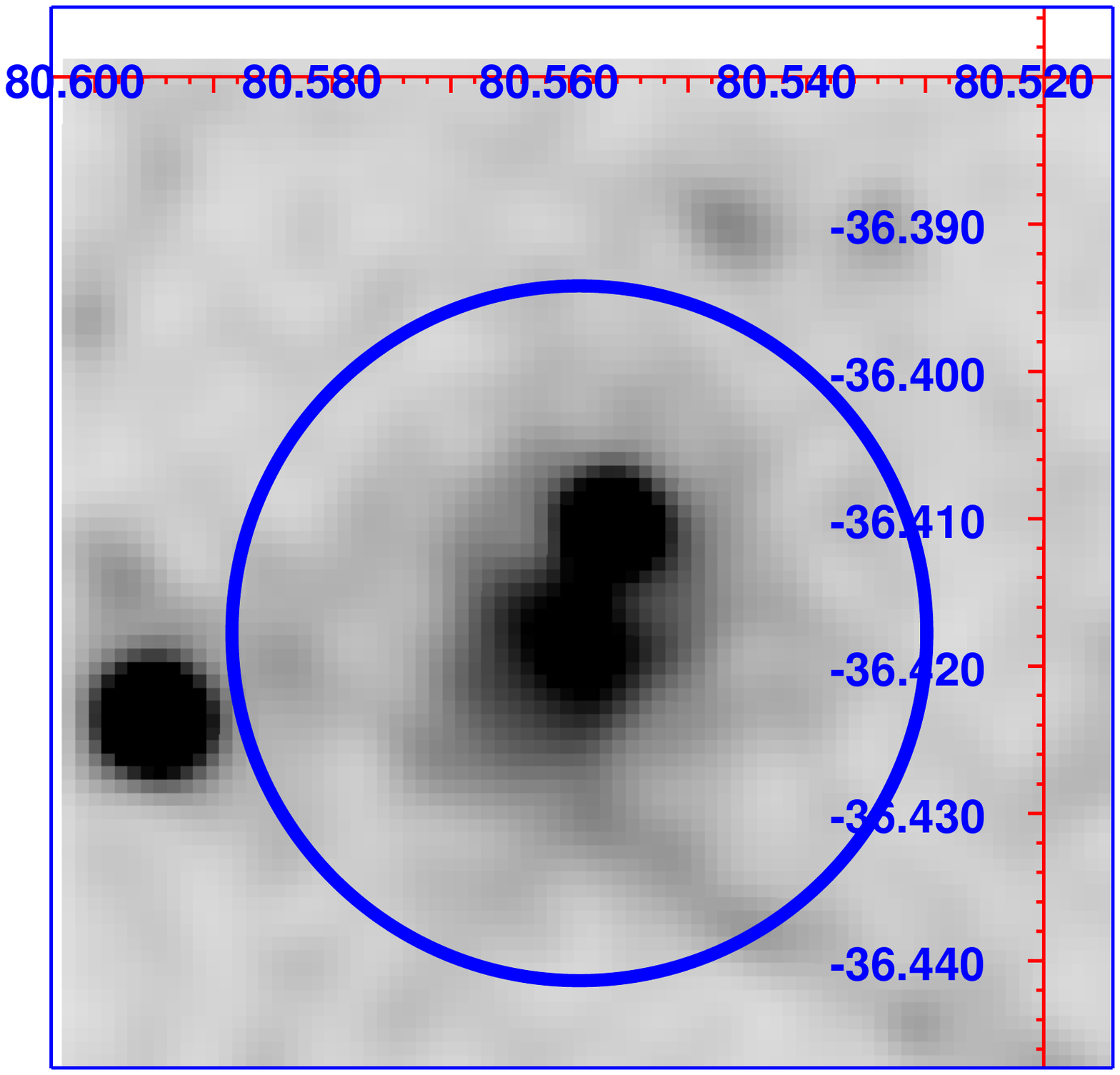}\includegraphics[width=2.7in,angle=0,bb=35 144 575 651,clip]{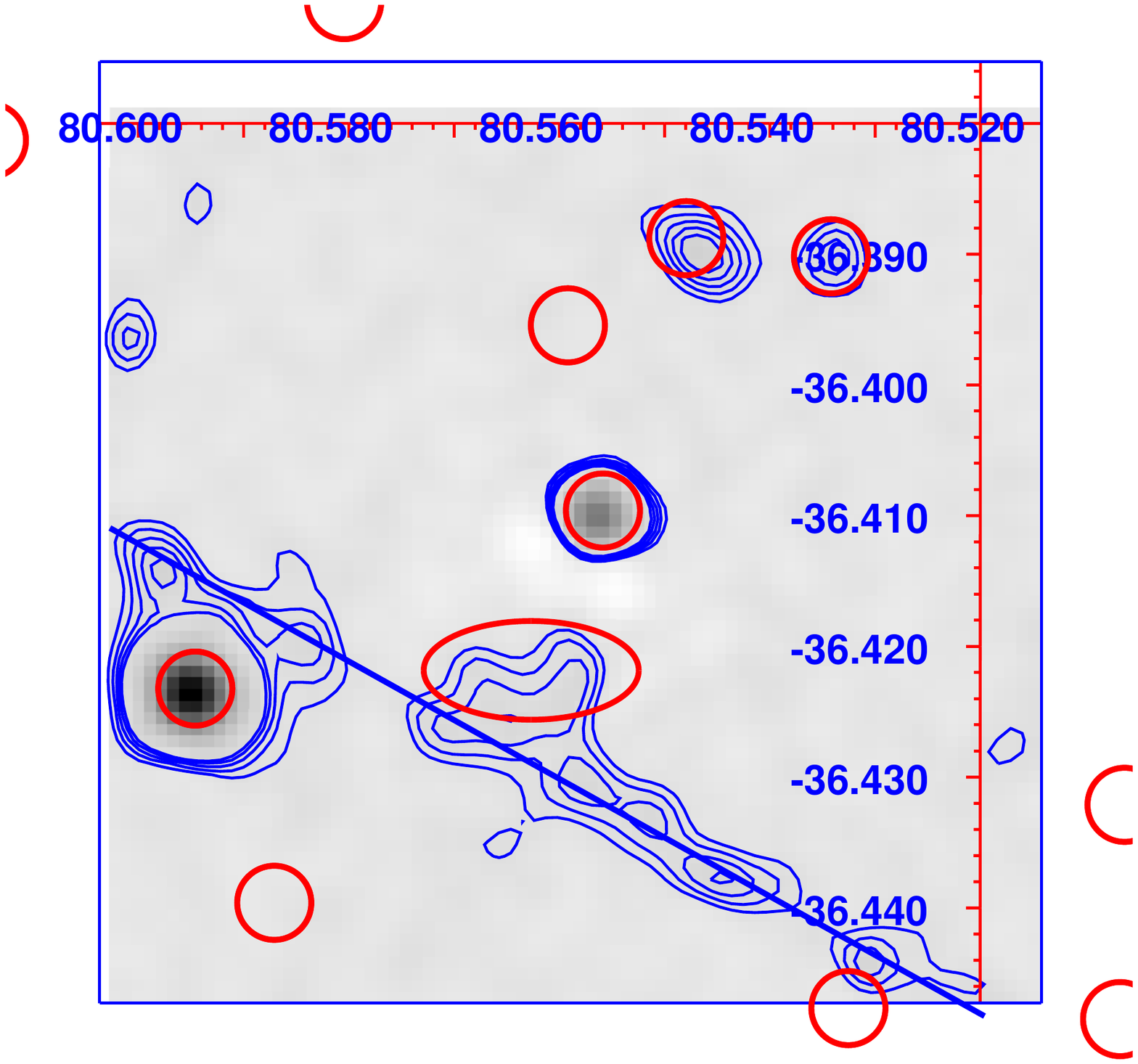}\\
    \includegraphics[width=2.7in,angle=0,bb=35 144 575 651,clip]{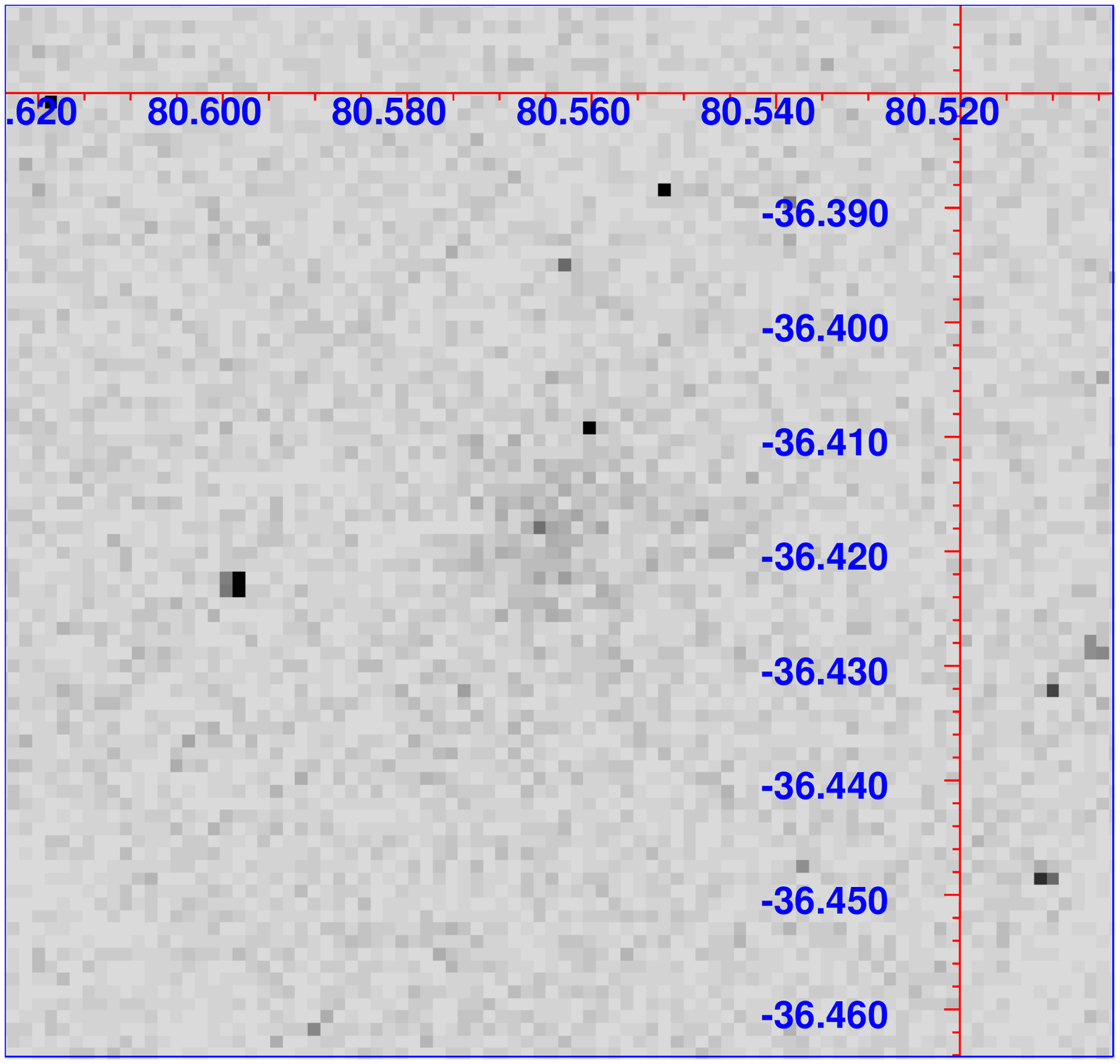}\includegraphics[width=2.7in,angle=0,bb=35 144 575 651,clip]{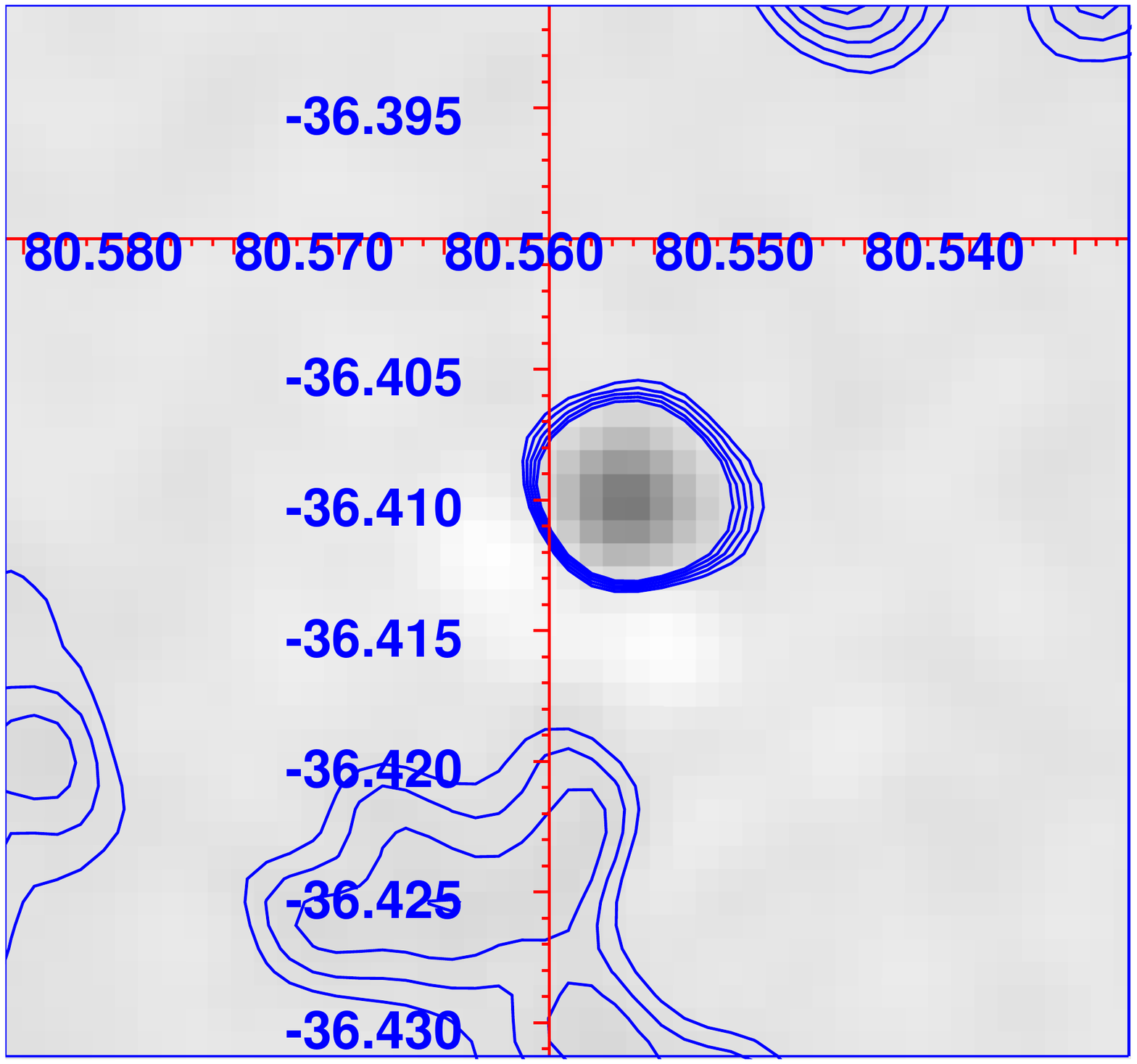}
    \caption{XMM-Newton X-ray image (upper left), residual XMM-Newton
      image (upper right), Chandra image (lower left), and zoom on the
      residual XMM-Newton image (lower left). The residual image was
      obtained by subtracting the model from the XMM-Newton image for
      BMW-HRI~J052215.8-362452. The large blue circle in the upper
      left figure corresponds to a 500 kpc radius circle centred on
      the literature cluster position. Finally, blue contours are the
      X-ray residuals, starting at the 2.5$\sigma$ level and spaced by
      1$\sigma$ intervals.}
  \label{fig:bmw0522_X}
  \end{center}
\end{figure*}

The residual X-ray image of BMW-HRI~J052215.8-362452 (serendipitously
discovered by ROSAT, Vikhlinin et al. 1998) shows the presence of
several sources superimposed on the cluster, all but one being very
compact.  The four brightest of these sources are known as AGNs
(Gilmour et al. 2009) and are visible in the Chandra image.  A fifth
one, north-east of the cluster remains unknown and not clearly
detected in the Chandra image. The MOS1 interchip separation is also
visible (Fig.~\ref{fig:bmw0522_X}).  Finally, we detect a faint X-ray
source to the south (at the 3.5$\sigma$ level), which seems extended
in the residual image (Fig.~\ref{fig:bmw0522_X}). This could be a
substructure both of the BMW-HRI~J052215.8-362452 cluster itself, even
if we cannot be sure of this, because a single redshift is available
in NED, and of the vicinity of the interchip gap.

We note that given the strong cluster contamination by AGN and by the
MOS1 interchip separation, the X-ray luminosities of the detected 
substructure are quite uncertain.

\subsection{MACS~J0647.7+7015 (101.94125$^o$, +70.2508$^o$, z=0.5907)} 

\begin{figure*}
  \begin{center}
    \includegraphics[width=2.7in,angle=0,bb=35 144 575 651,clip]{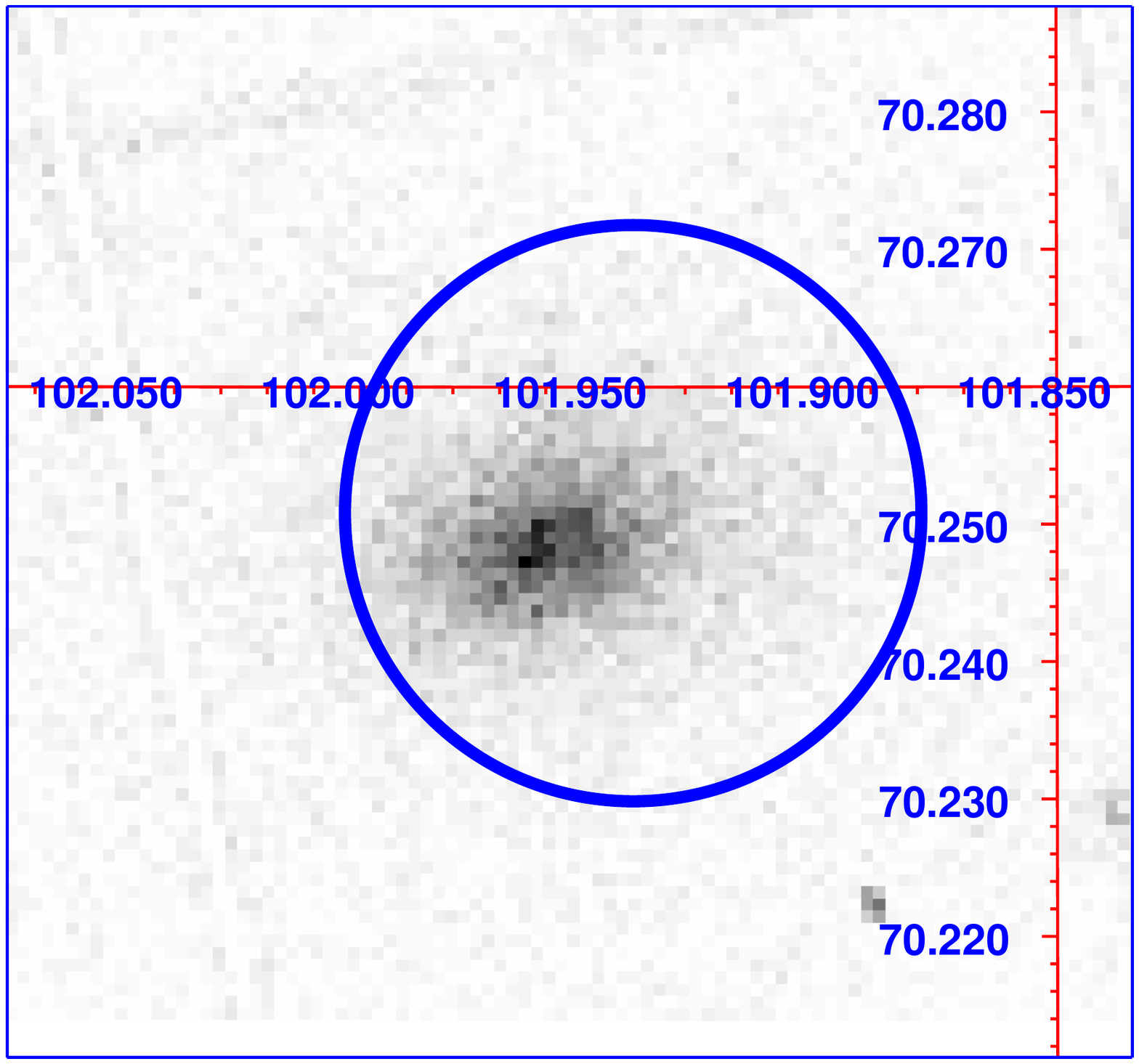}\includegraphics[width=2.7in,angle=0,bb=35 144 575 651,clip]{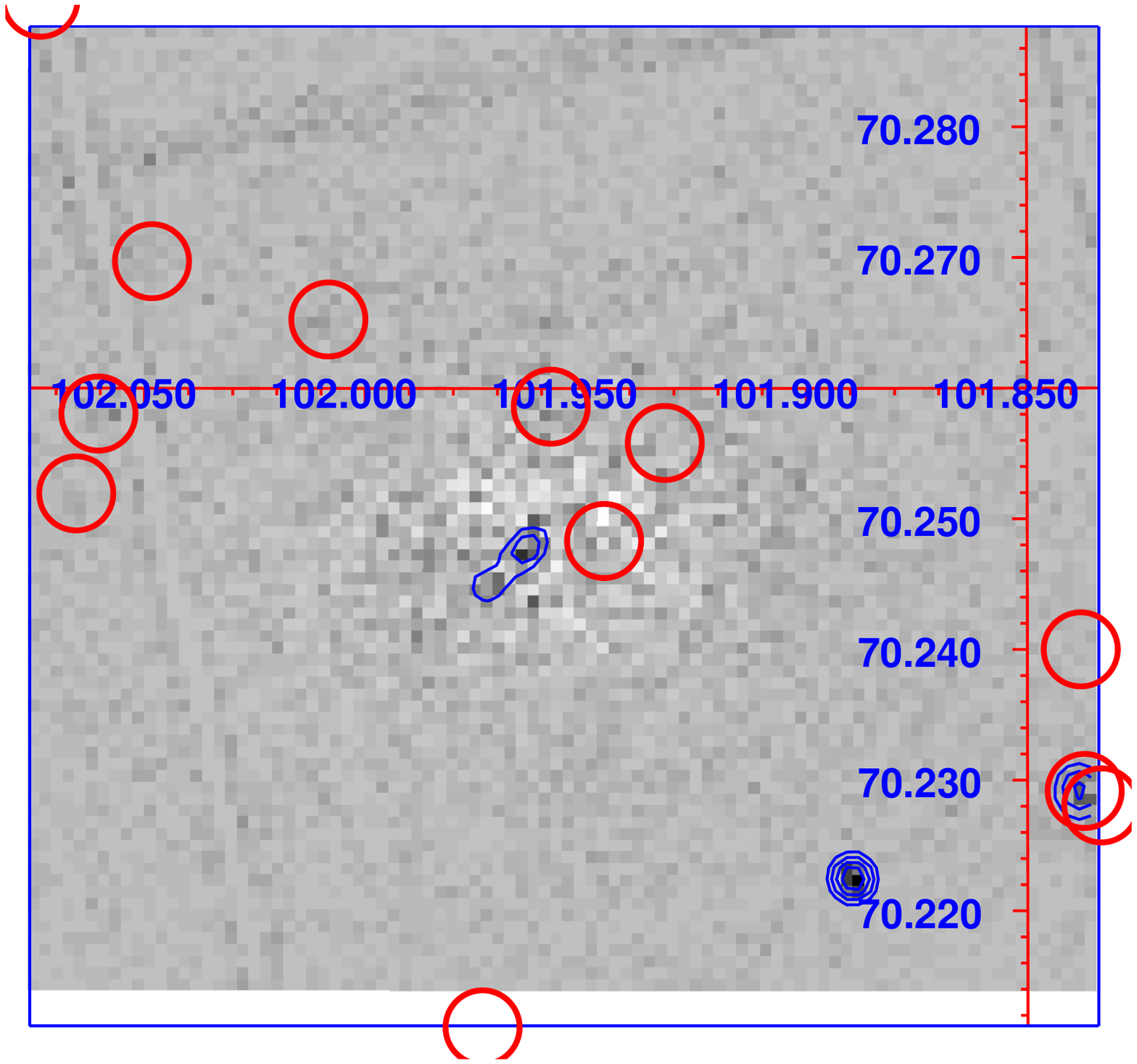}\\
    \includegraphics[width=2.7in,angle=0,bb=35 144 575 651,clip]{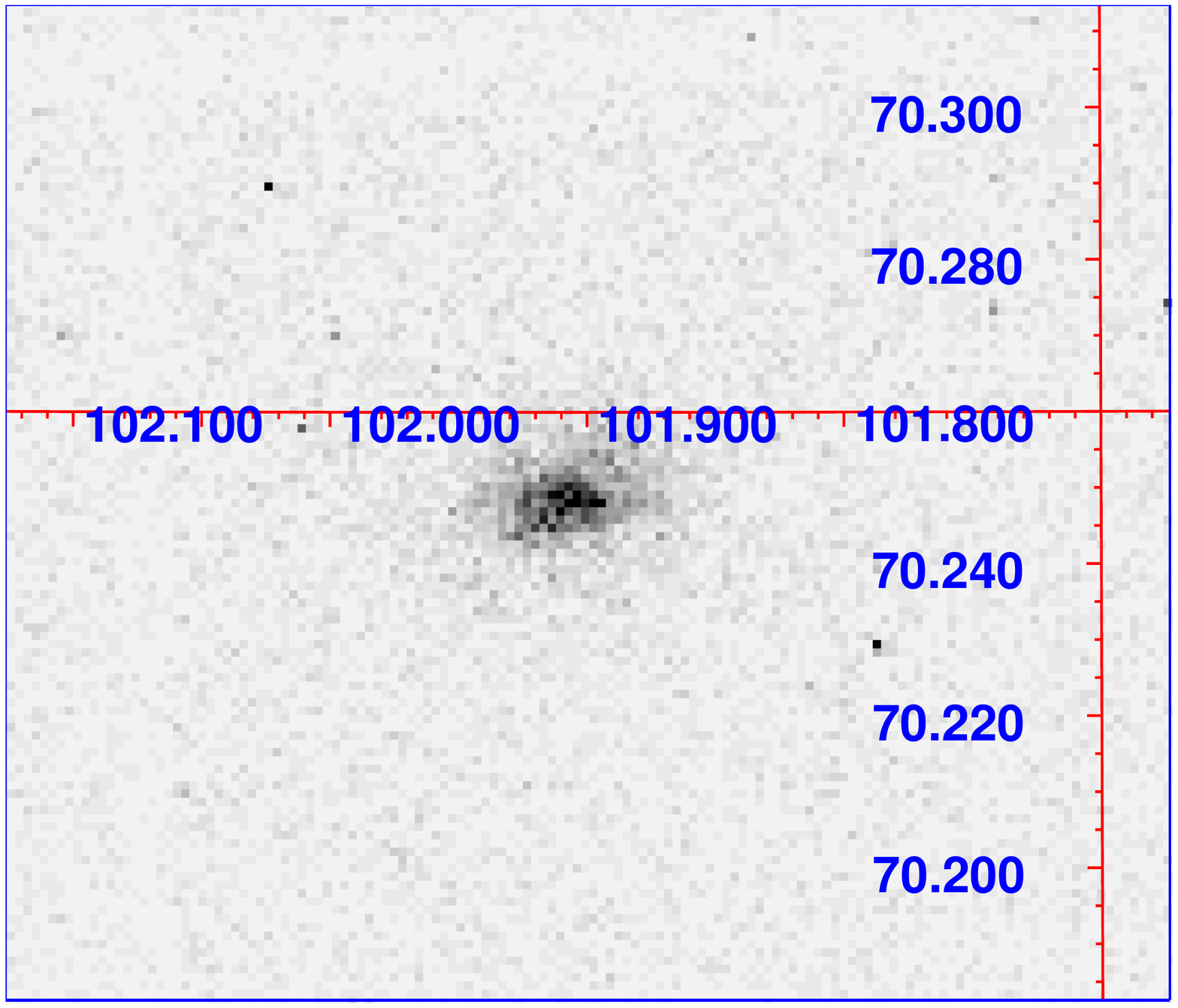}\includegraphics[width=2.7in,angle=0,bb=35 144 575 651,clip]{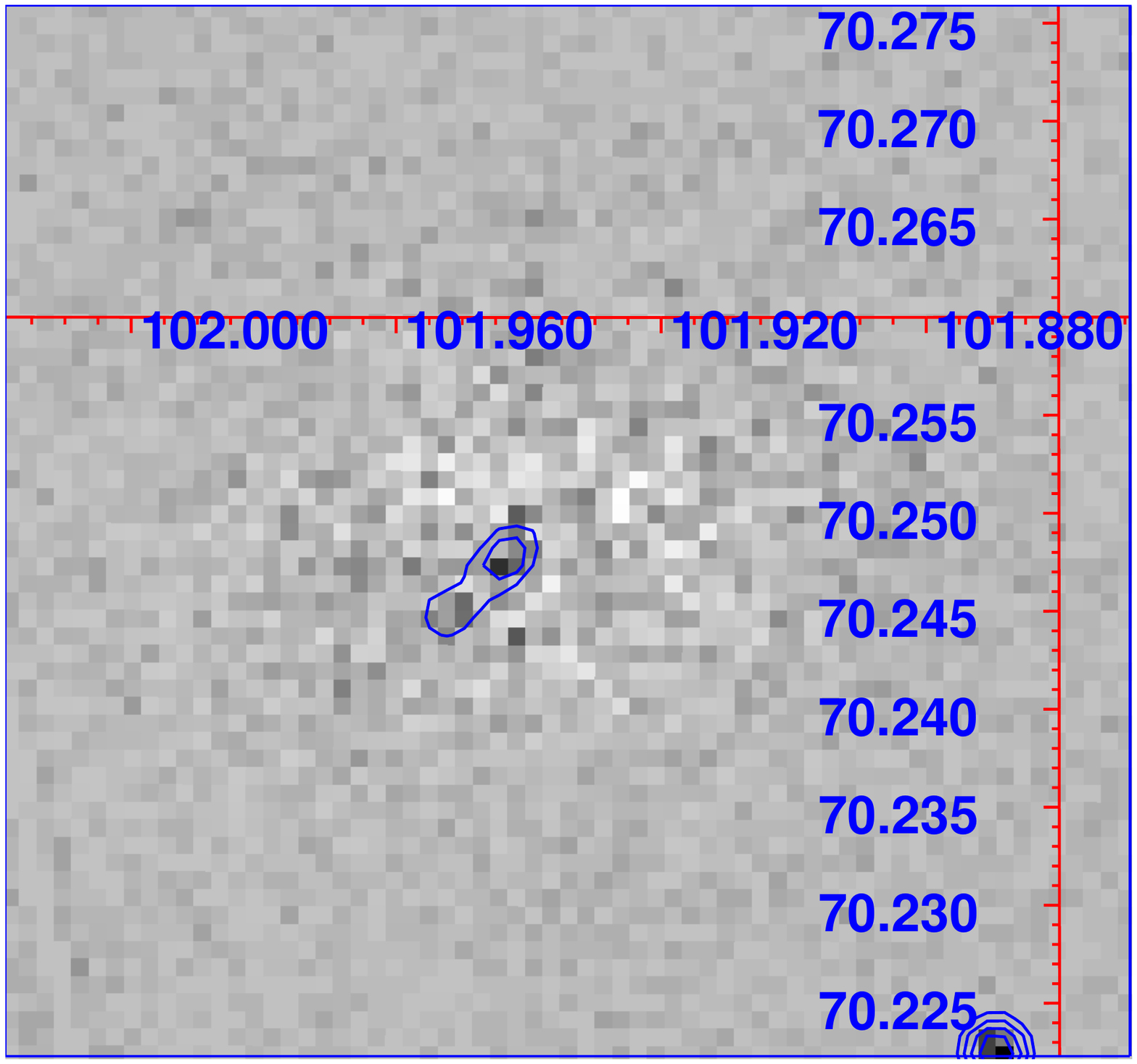}
    \caption{Same as Fig.~\ref{fig:bmw0522_X} for MACS~J0647.7+7015.}
  \label{fig:macs0647_X}
  \end{center}
\end{figure*}

The X-ray image of MACS~J0647.7+7015 (see e.g. Voges et al. 1999) is
smooth but some emission is detected after model subtraction to the
south--east (at the 3$\sigma$ level: see
Fig.~\ref{fig:macs0647_X}). The extension of this possible
substructure is 26 arcsec (equivalent to 172~kpc at z=0.5907).  There
is a single redshift available in the cluster area in NED, so the SG
analysis is impossible. Even if we cannot be sure that this
substructure is attached to the MACS~J0647.7+7015 cluster, its
elongation toward the cluster centre tends to indicate that we have
detected a substructure in this cluster. Two compact sources are
visible in the XMM-Newton data south-west of the cluster. One of them
is detected in the Chandra image and is indicated in Gilmour et
al. (2009) as a point source.

\subsection{MACS~J0744.9+3927 (116.21583$^o$, +39.4592$^o$, z=0.6860)} 

\begin{figure*}
  \begin{center}
    \includegraphics[width=2.7in,angle=0,bb=35 144 575 651,clip]{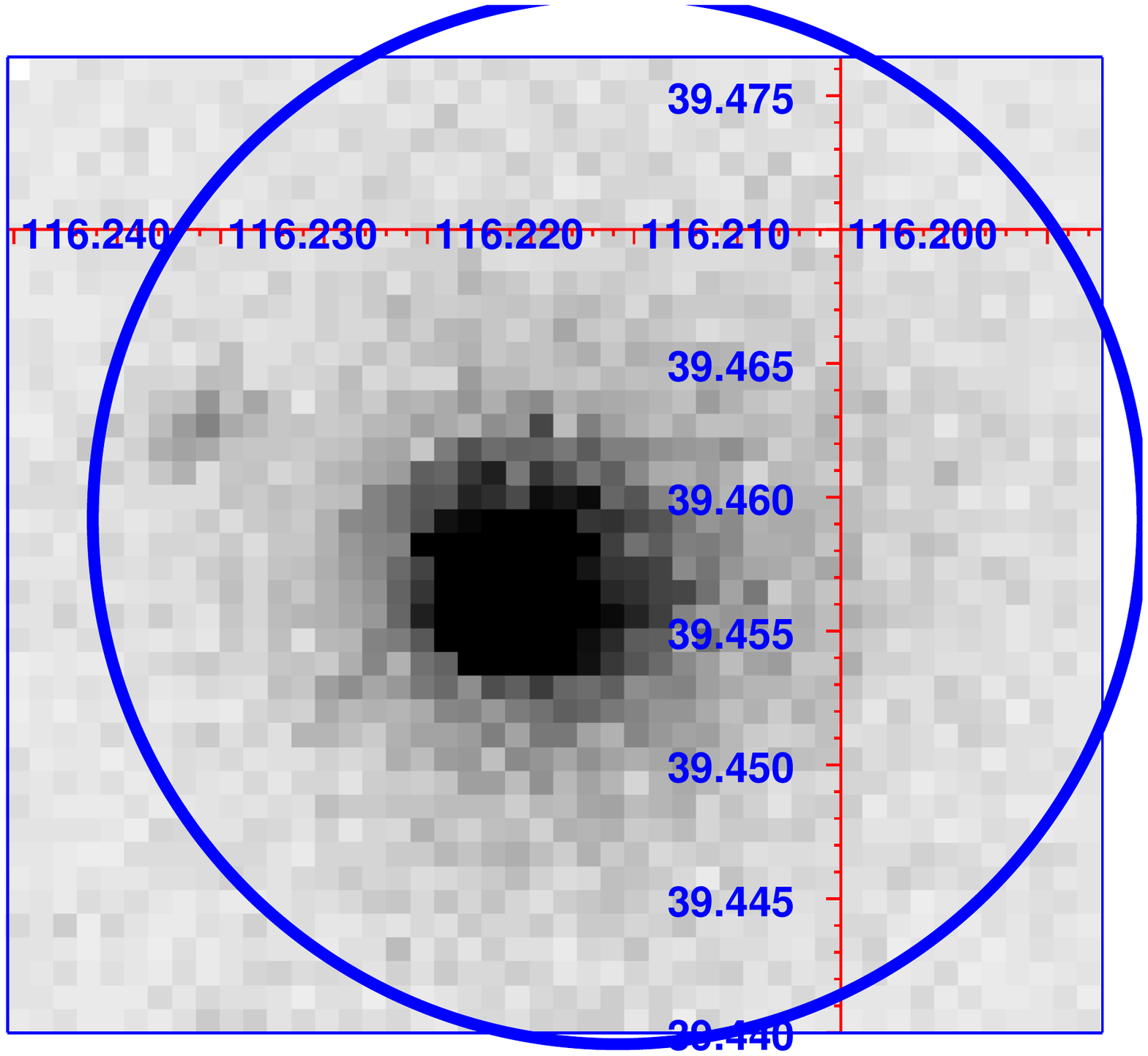}\includegraphics[width=2.7in,angle=0,bb=35 144 575 651,clip]{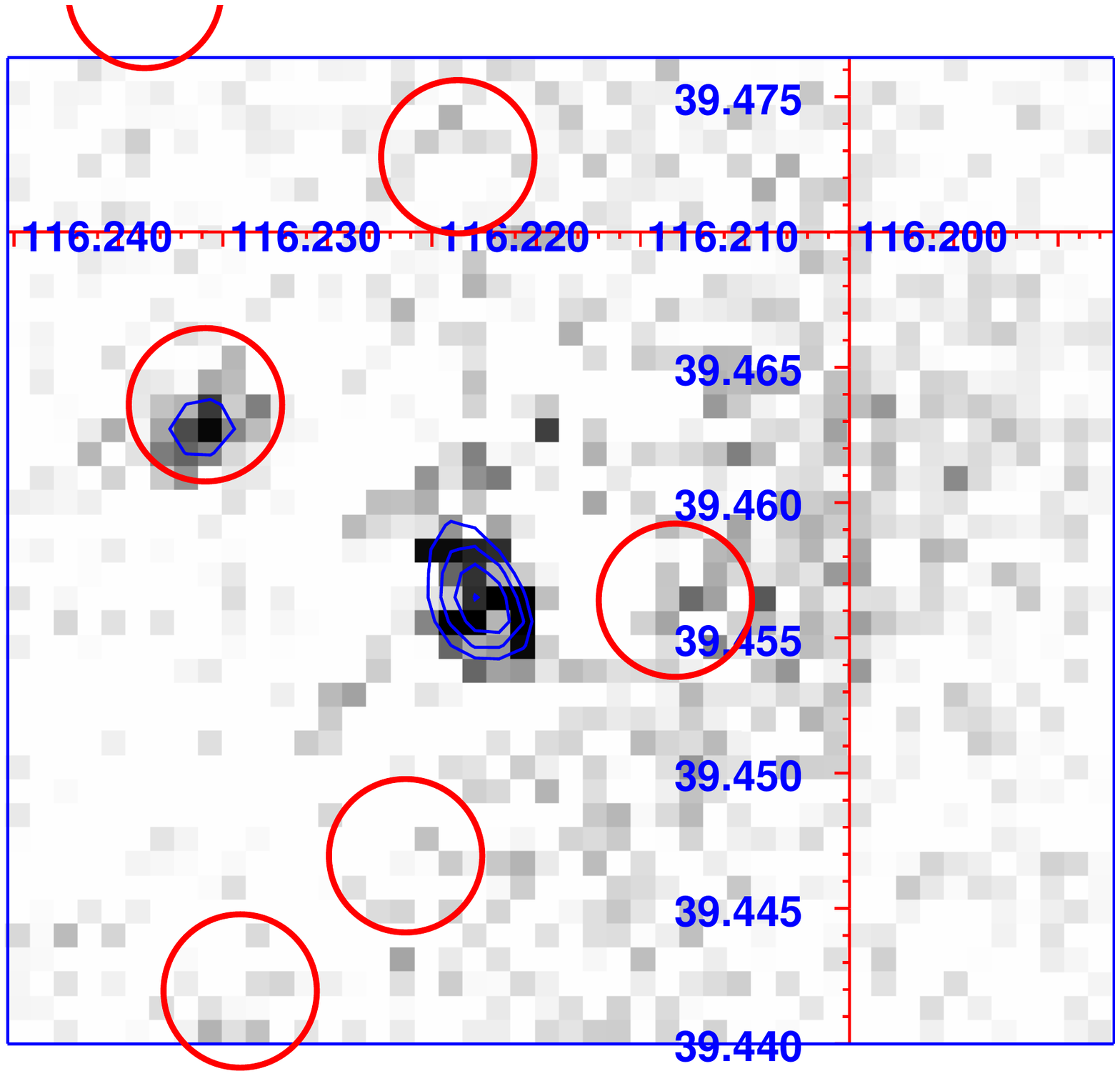}\\
    \includegraphics[width=2.7in,angle=0,bb=35 144 575 651,clip]{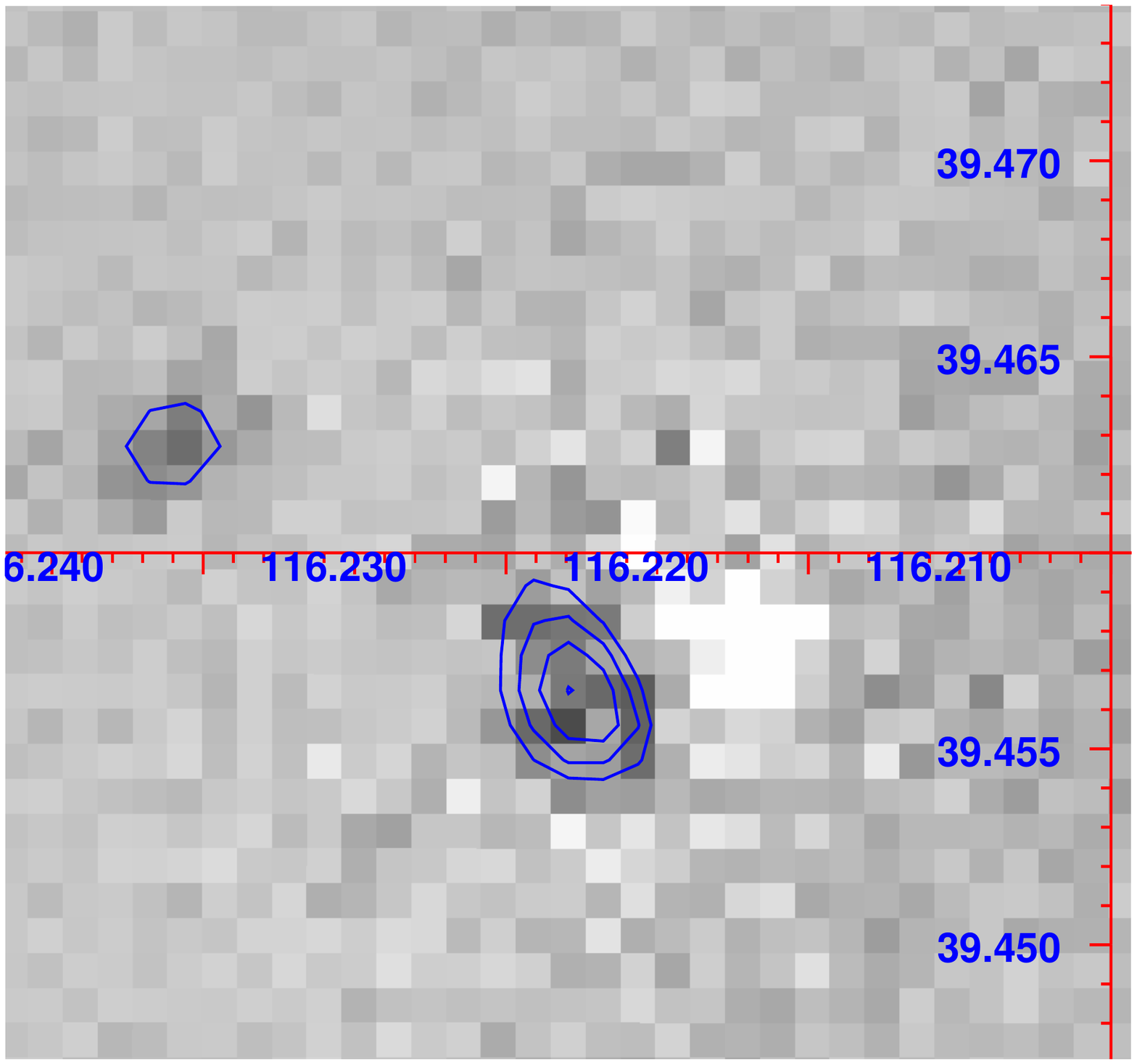}
    \caption{Same as Fig.~\ref{fig:ms0302X} for MACS~J0744.9+3927.}
  \label{fig:macs0744_X}
  \end{center}
  \end{figure*}

The X-ray image of MACS~J0744.9+3927 (Voges et al. 1999) is smooth,
but the residuals show two significant sources
(Fig.~\ref{fig:macs0744_X}). One is identified with an active galaxy
(Gilmour et al. 2009) and the other one appears extended and seems to
correspond spatially to the SZ excess detected by Korngut et
al. (2011) and attributed to a region shocked by a merger.  NED only
provides two redshifts in the cluster field of view, so
we cannot be sure we are really dealing with a substructure attached
to the MACS~J0744.9+3927 cluster. However, it is tempting to assume
that we have detected a group falling onto the considered cluster, and
we  assume this in the present work.

\subsection{RX~J0847.1+3449 (131.79708$^o$, +34.8211$^o$, z=0.5600)} 

\begin{figure*}
  \begin{center}
    \includegraphics[width=2.7in,angle=0,bb=35 144 575 651,clip]{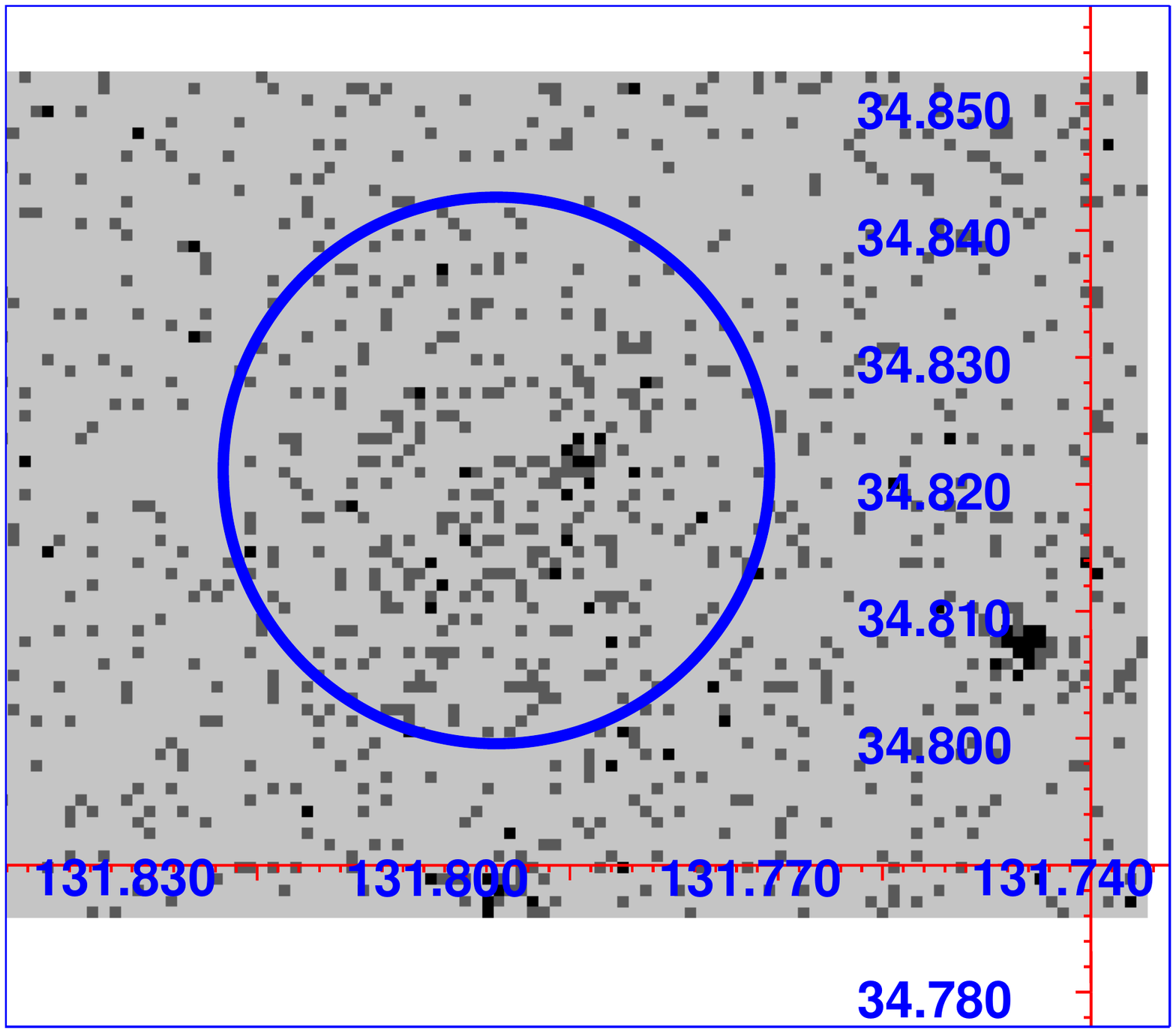}\includegraphics[width=2.7in,angle=0,bb=35 144 575 651,clip]{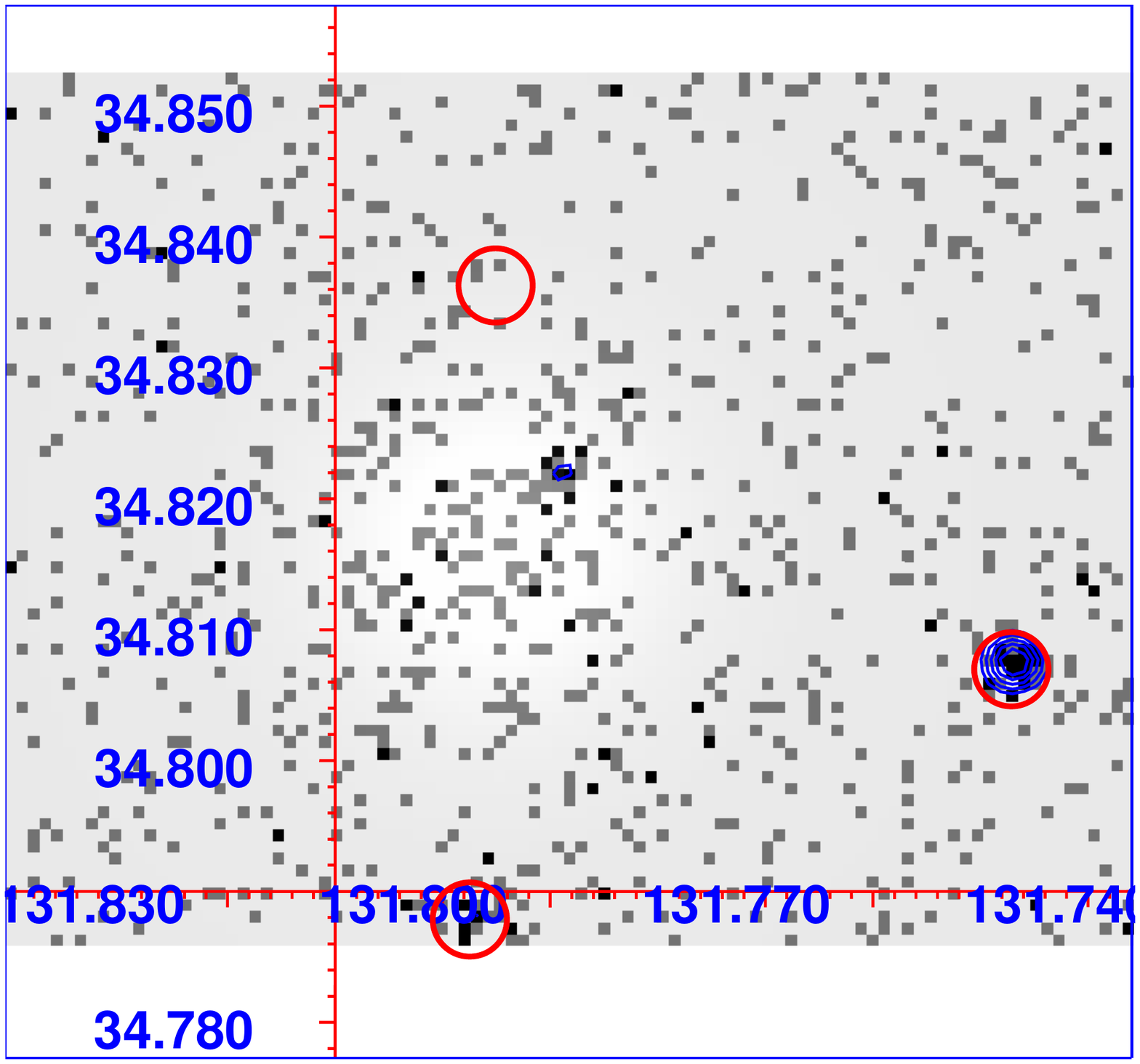}\\
    \includegraphics[width=2.7in,angle=0,bb=35 144 575 651,clip]{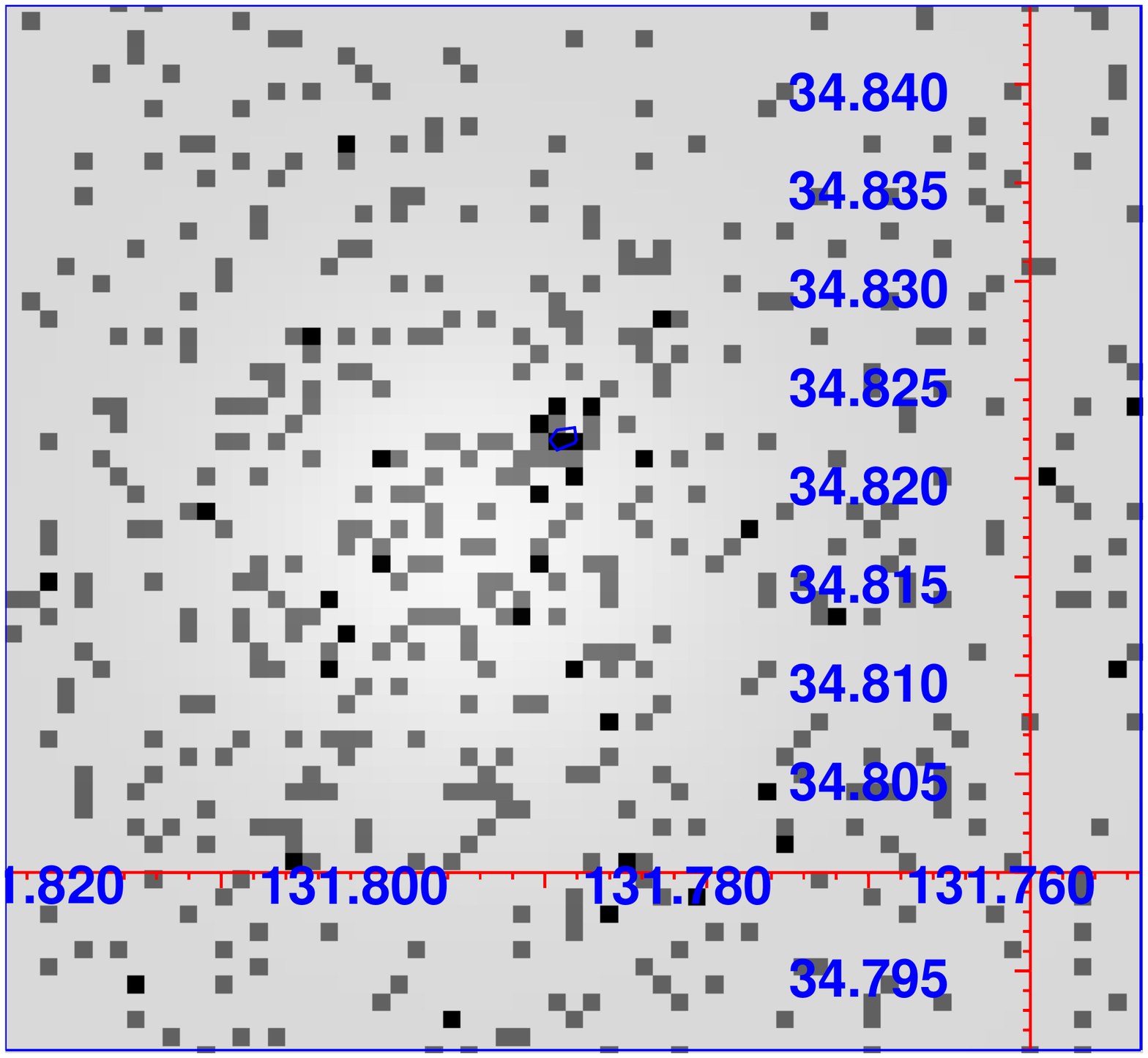}
    \caption{Same as Fig.~\ref{fig:ms0302X} for RX~J0847.1+3449.}
  \label{fig:rx0847_X}
  \end{center}
  \end{figure*}

The X-ray emission of RX~J0847.1+3449 (Vikhlinin et al. 1998) is very
faint (Fig.~\ref{fig:rx0847_X}).  Subtracting a $\beta -$model basically
only makes a point source appear south--west of the cluster, identified
as an AGN in Vizier.  NED provides a single redshift in the cluster
area.

\subsection{MACS~J0913 (138.40277$^o$, +40.94315$^o$, z=0.4420) }  

\begin{figure*}
  \begin{center}
    \includegraphics[width=3.5in,angle=0,bb=80 210 500 590,clip]{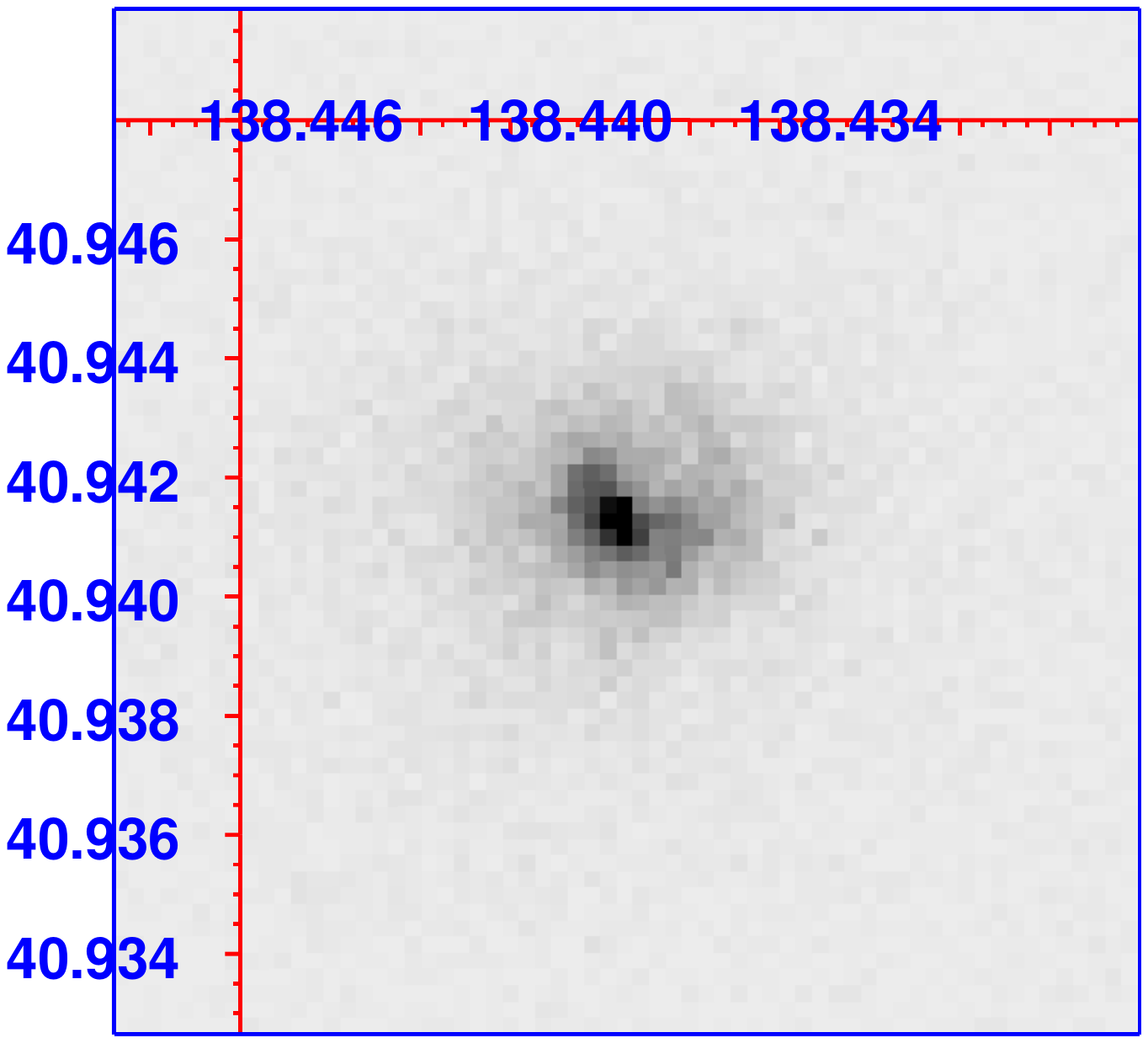}\includegraphics[width=3.5in,angle=0,bb=150 247 500 550,clip]{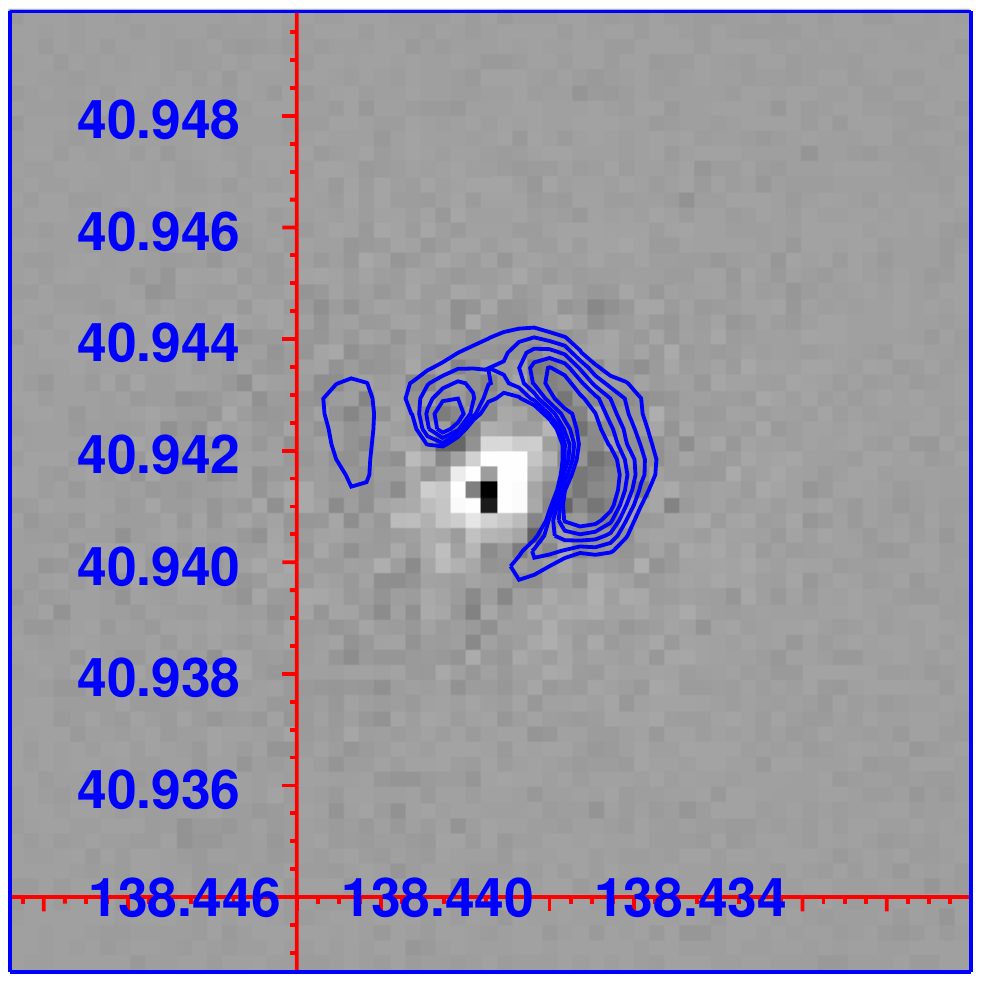}\\
    \includegraphics[width=3.5in,angle=0,bb=35 144 575 651,clip]{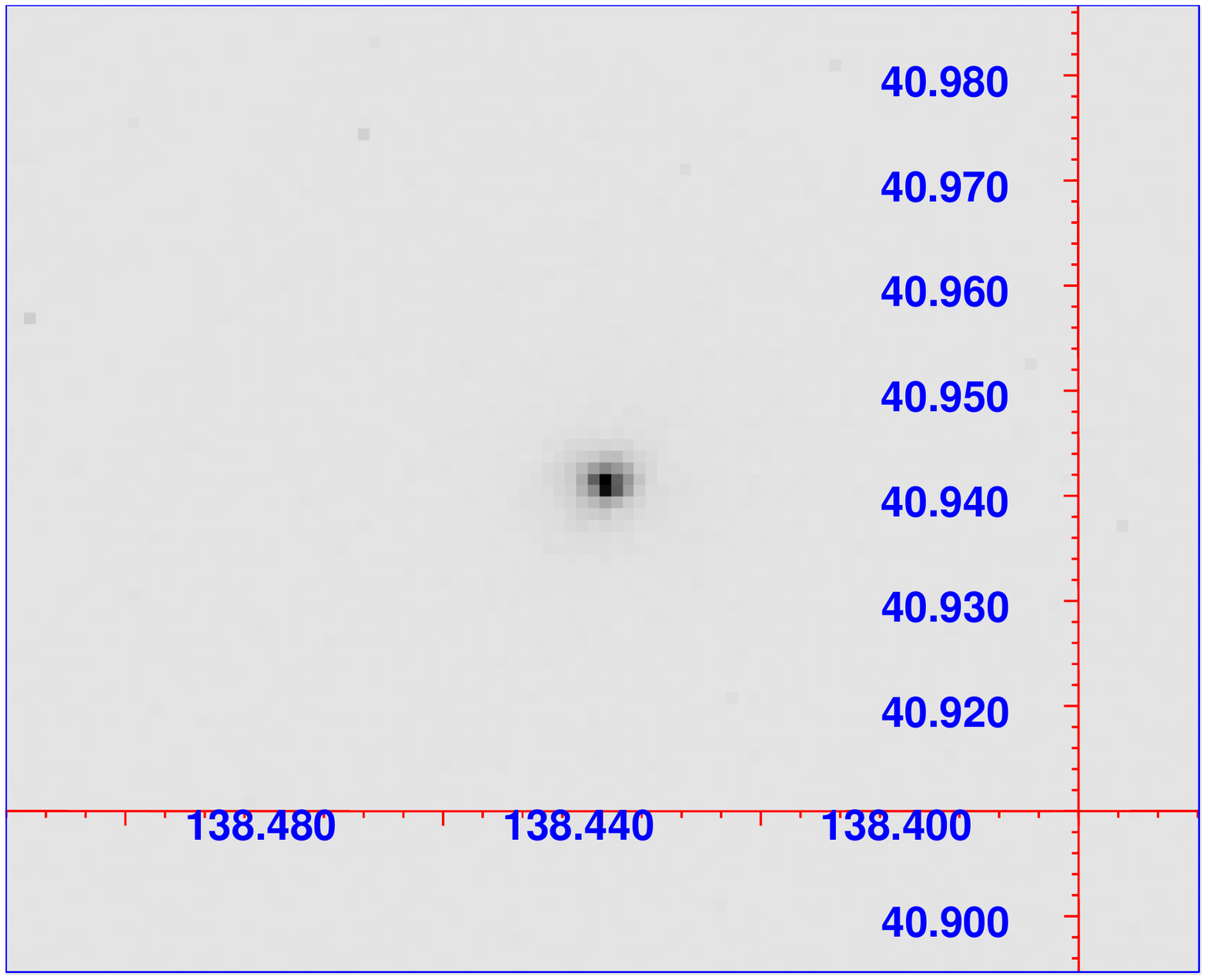}\includegraphics[width=3.5in,angle=0,bb=35 144 575 651,clip]{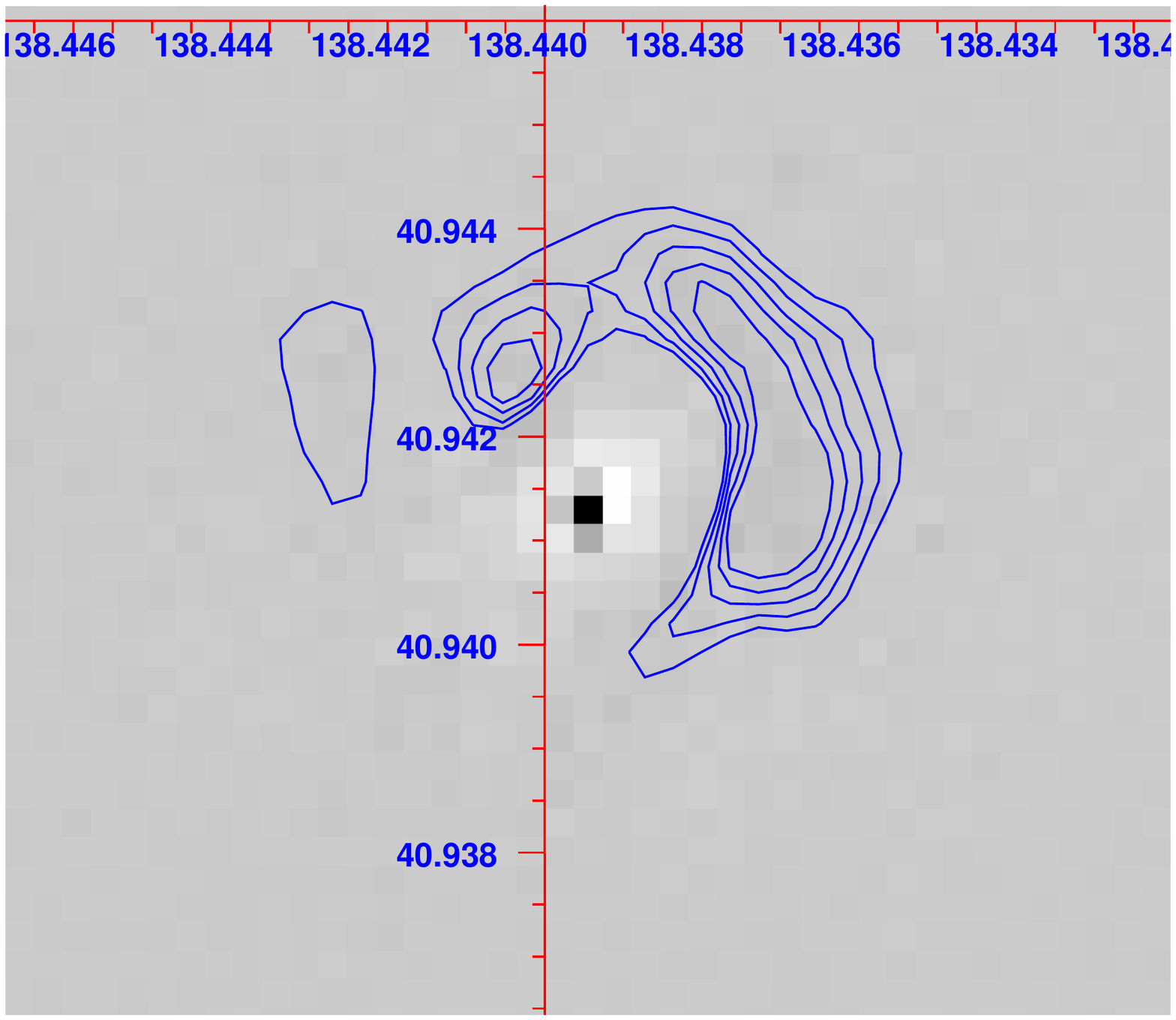}
    \caption{Same as Fig.~\ref{fig:bmw0522_X} for MACS~J0913.}
  \label{fig:cl0910_X}
  \end{center}
\end{figure*}
 
Although the X-ray emission appears quite smooth, a model subtraction
reveals the presence of a structure very close to the cluster centre,
perhaps due to an imperfect $\beta -$model subtraction.  Such an
effect is also seen in some of our simulations.  We perhaps detect a
very weak structure (significant only at the 2.5$\sigma$ level) east
of the centre (Fig.~\ref{fig:cl0910_X}).  The X-ray emission extent of
this source is, however, very small, making it doubtful. This is
confirmed by the Chandra data, pointing towards a very peaked core
without any visible substructures.

Only a few redshifts are available in the cluster area, making any SG
analysis impossible.

\subsection{Abell 851 (145.73601$^o$, +46.9894$^o$, z=0.4069) } 

\begin{figure*}
  \begin{center}
    \includegraphics[width=2.7in,angle=0,bb=35 144 575 651,clip]{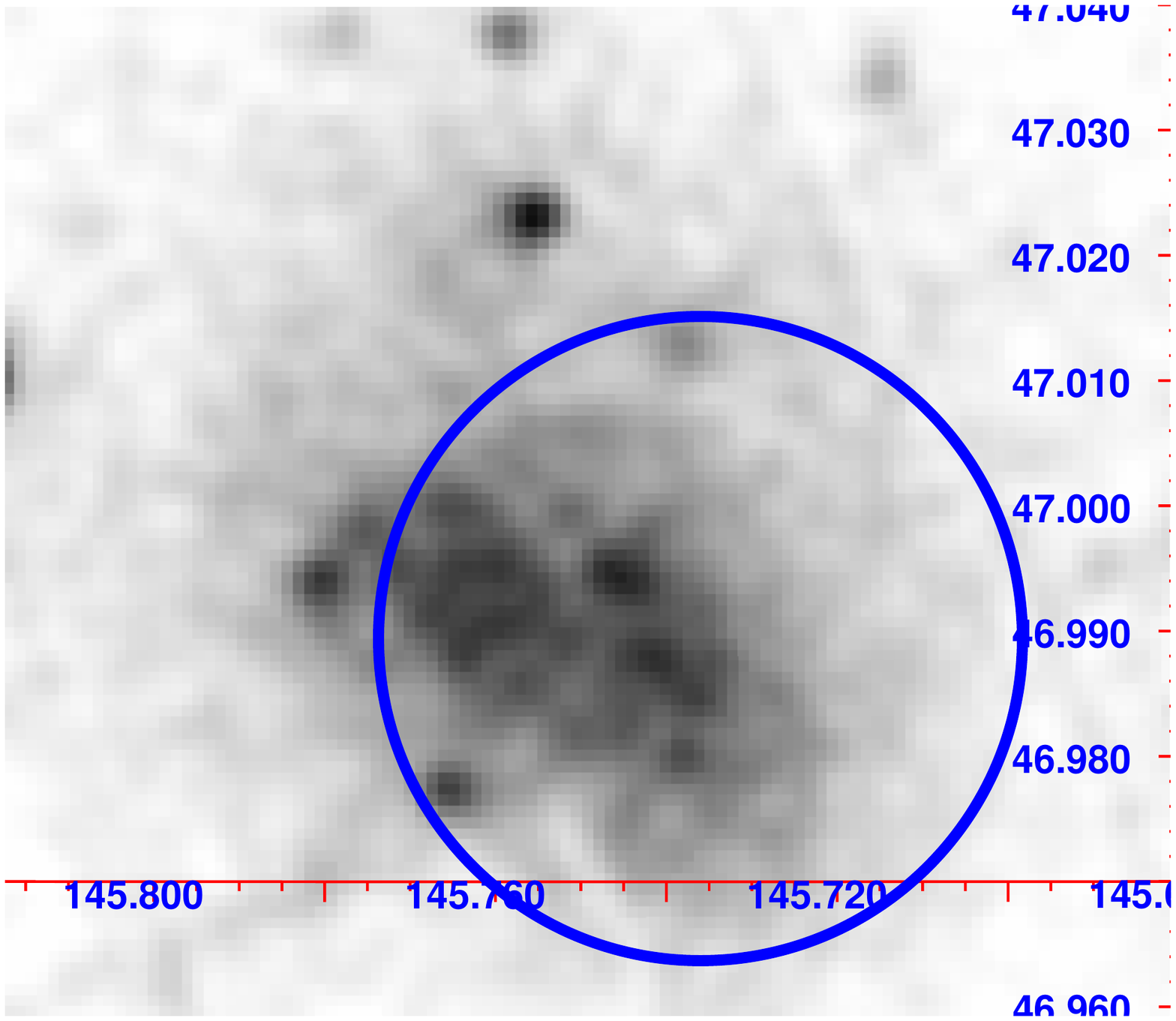}\includegraphics[width=2.7in,angle=0,bb=35 144 575 651,clip]{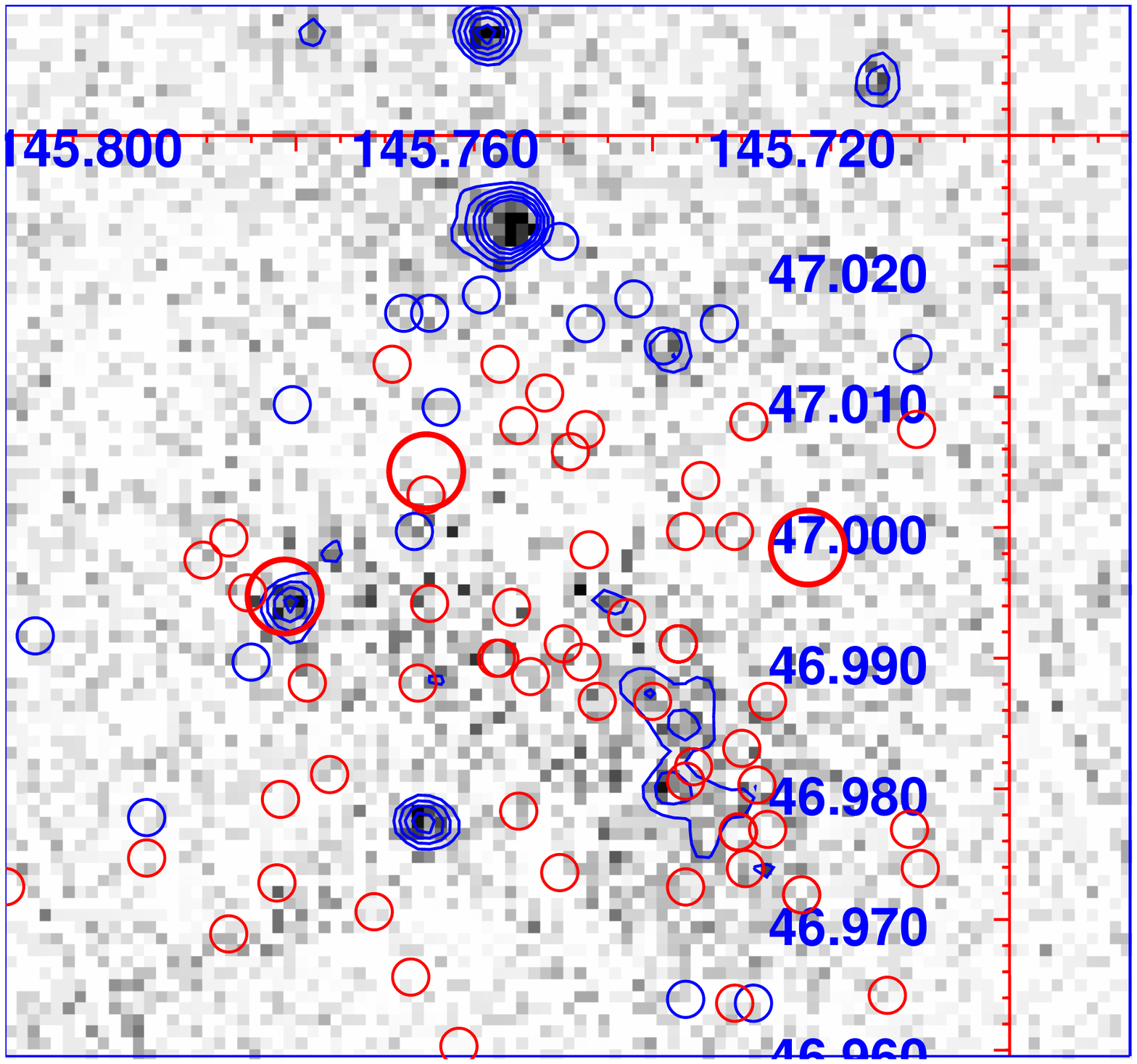}\\
    \includegraphics[width=2.7in,angle=0,bb=35 144 575 651,clip]{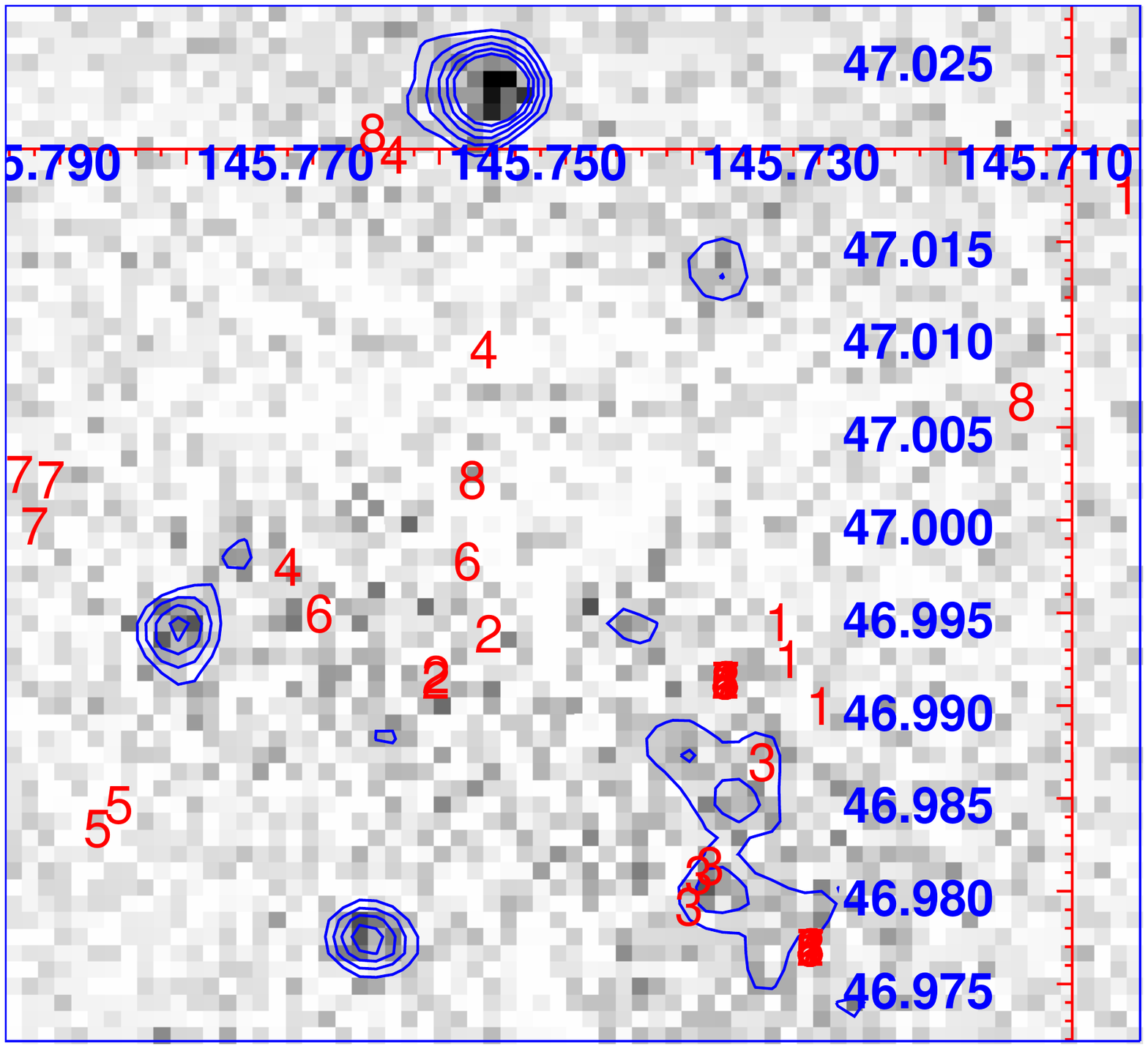}\includegraphics[width=2.7in,angle=0,bb=15 144 575 701,clip]{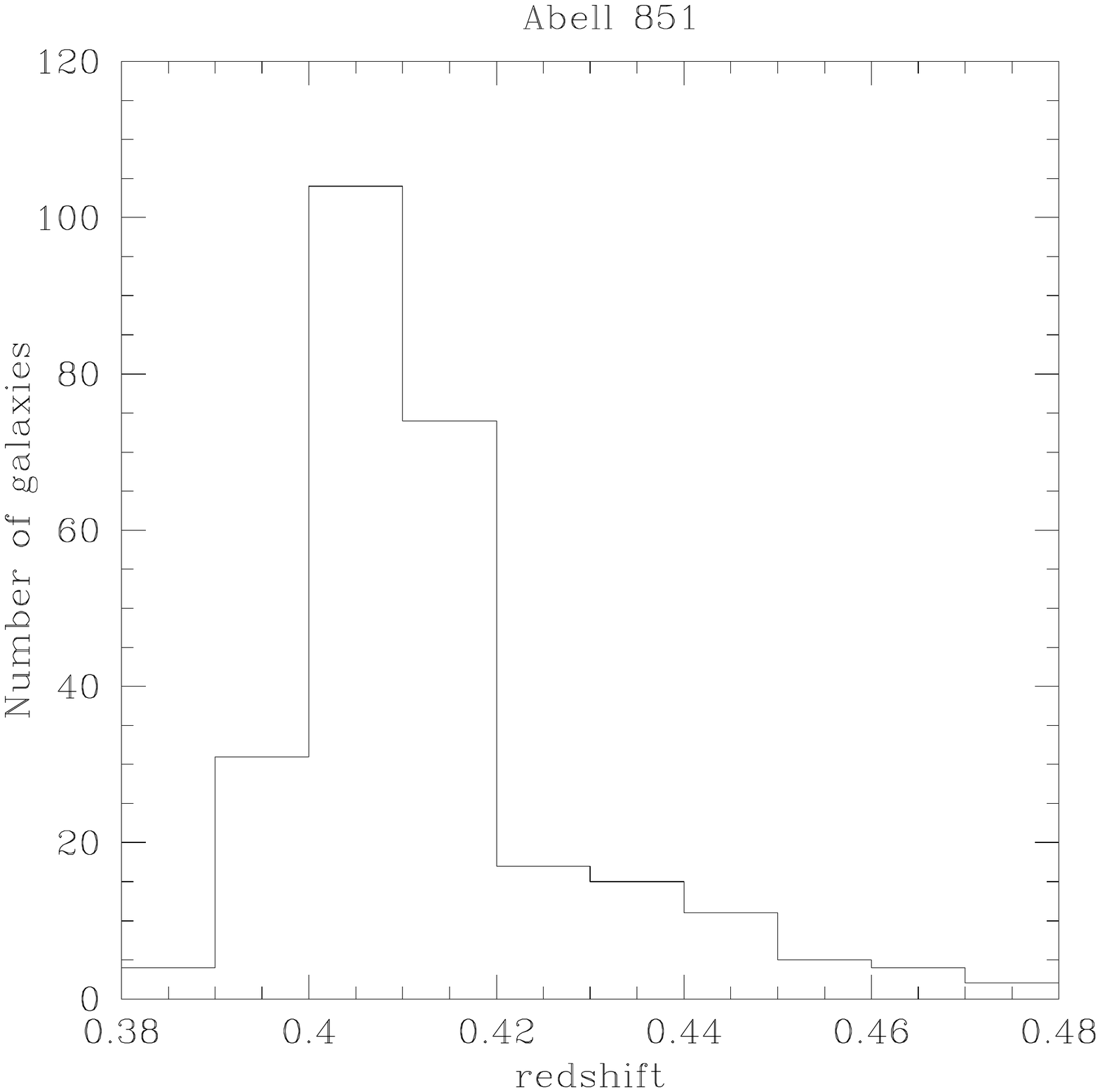}
    \caption{Same as Fig.~\ref{fig:cl0016_X} for Abell 851.}
  \label{fig:A851_X}
  \end{center}
\end{figure*}

The overall aspect of the X-ray image is very clumpy, suggesting that
one or several cluster mergers are taking place
(Fig.~\ref{fig:A851_X}). The model-subtracted XMM-Newton X-ray image
shows four main peaks that are more significant than the 3$\sigma$ level: a
large and rather faint apparently extended source to the south-west,
and three other more compact sources to the south, north, and east
(Fig.~\ref{fig:A851_X}). Our residual image agrees qualitatively with
that of De Filippis et al. (2003), who show that this cluster is
most probably formed by the merger of two clusters of comparable
masses. However, that the X-ray luminosity of the
substructure is only about 10\% that of the cluster implies that this
is not a major merger.   The comparison with spectroscopically
confirmed active objects allows us to identify the east compact source
as an AGN.  We have no Chandra image to characterize the other two
compact sources, but given their shape in the XMM-Newton data, it is
likely that they are also AGNs.

A large number of redshifts are available in NED, with 213 galaxies in
the [0.39,0.42] redshift range.  The velocity histogram around z$\sim
0.41$ is clearly asymmetric, and the SG analysis shows the presence of
several substructures. In particular, the most massive (group SG3 in
Table~\ref{tab:SG}) is clearly associated with the extended X-ray
source, all the galaxies belonging to this group being located inside
the X-ray contours. Only two galaxies that are not members of Abell 851
and are included in galaxy structures along the line of sight are located
inside the considered X-ray contours. This makes projection effects
very unlikely.

\subsection{MS~1054-03 (164.25093$^o$, --3.6243$^o$, z=0.8231)} 

\begin{figure*}
  \begin{center}
    \includegraphics[width=2.7in,angle=0,bb=35 144 575 651,clip]{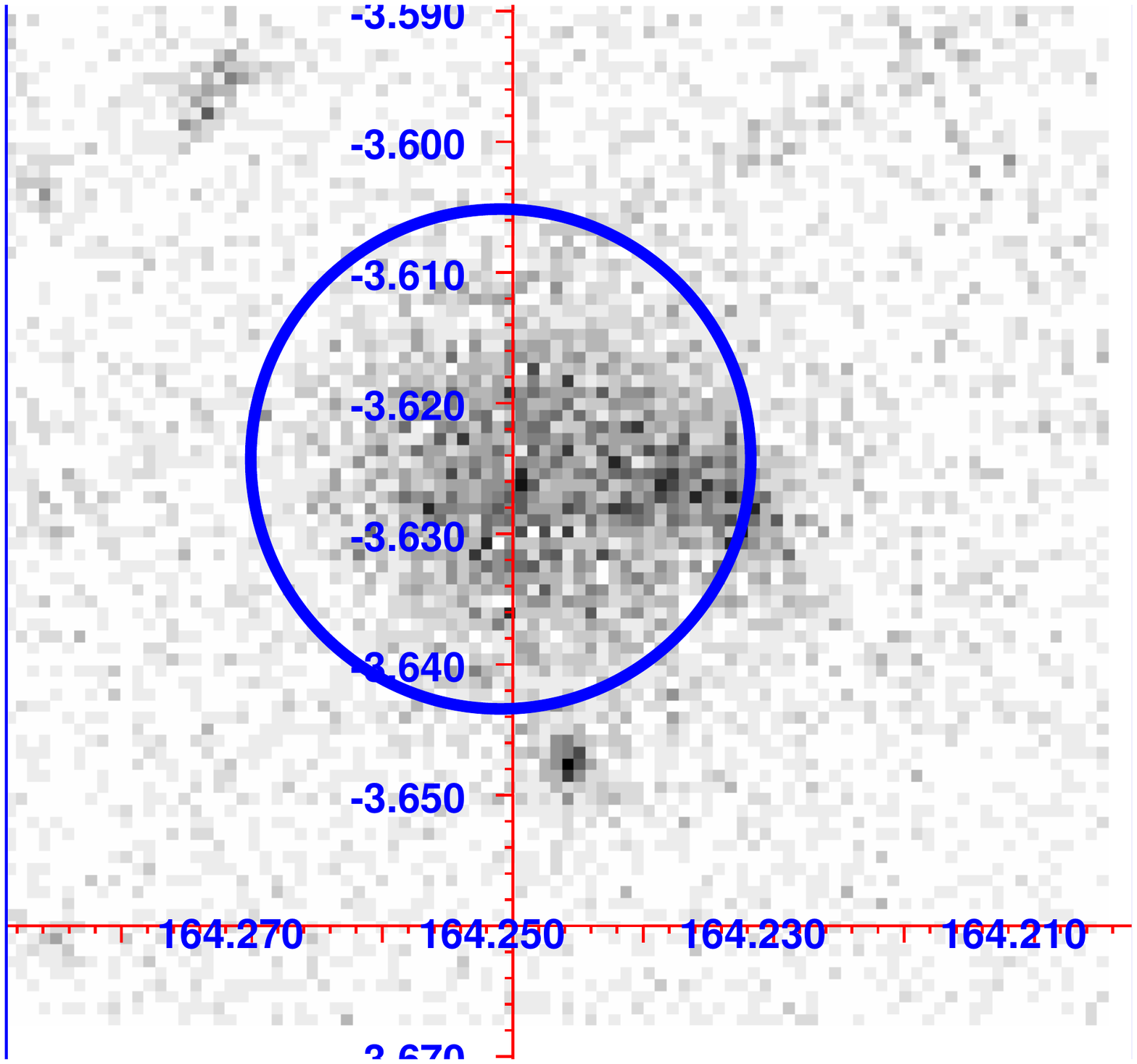}\includegraphics[width=2.7in,angle=0,bb=35 144 575 651,clip]{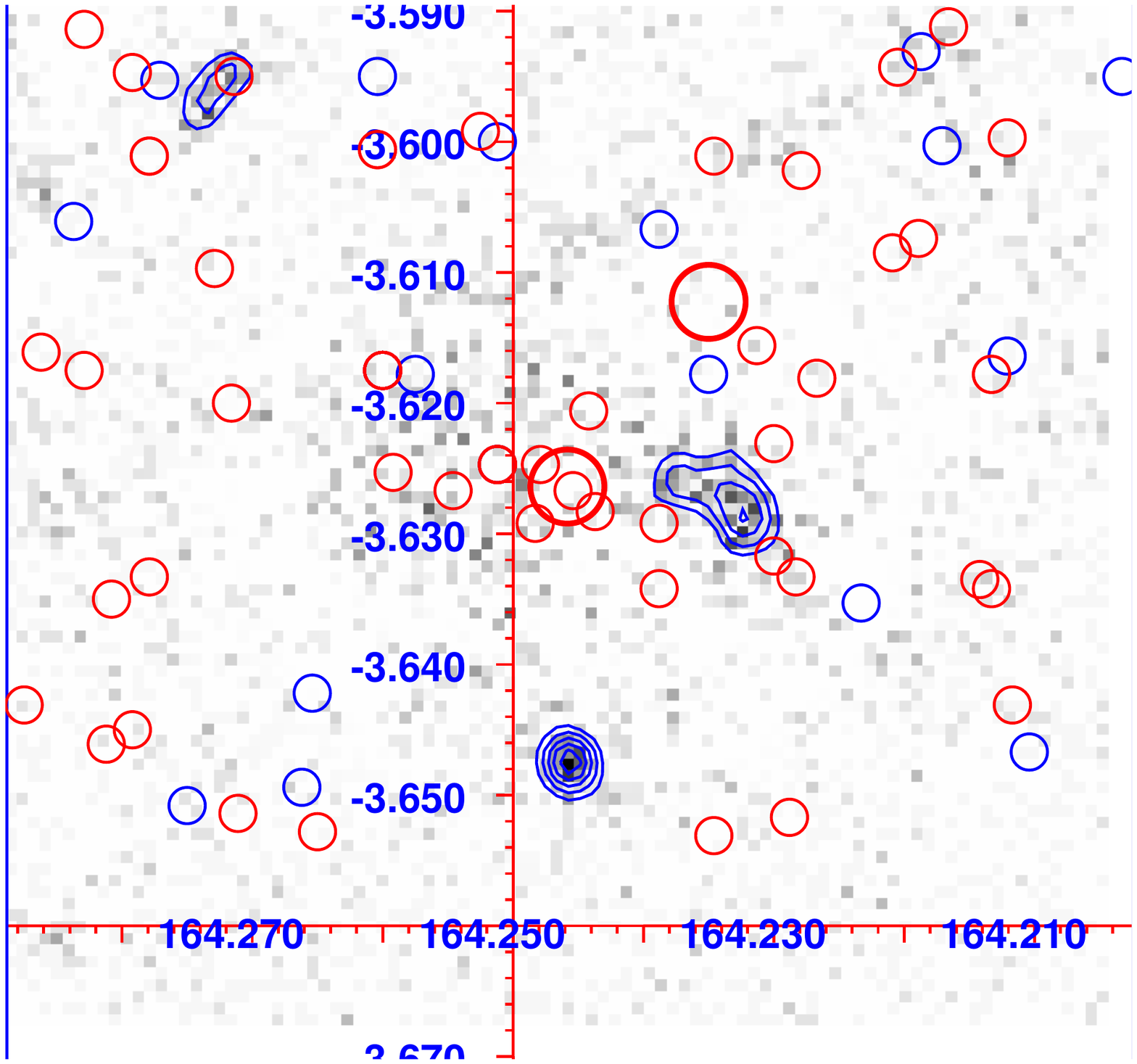}\\
    \includegraphics[width=2.7in,angle=0,bb=35 144 575 651,clip]{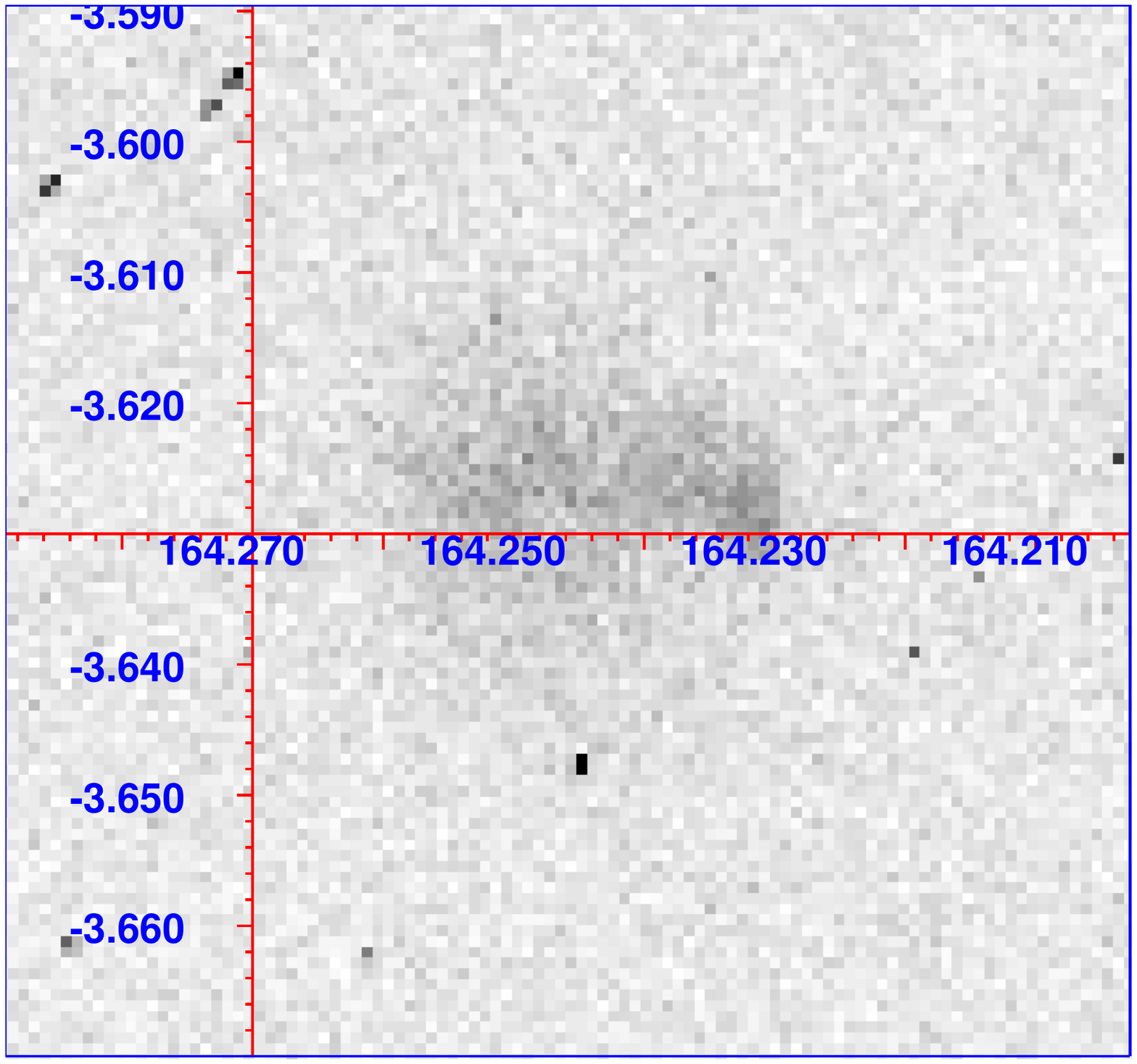}\includegraphics[width=2.7in,angle=0,bb=35 144 575 651,clip]{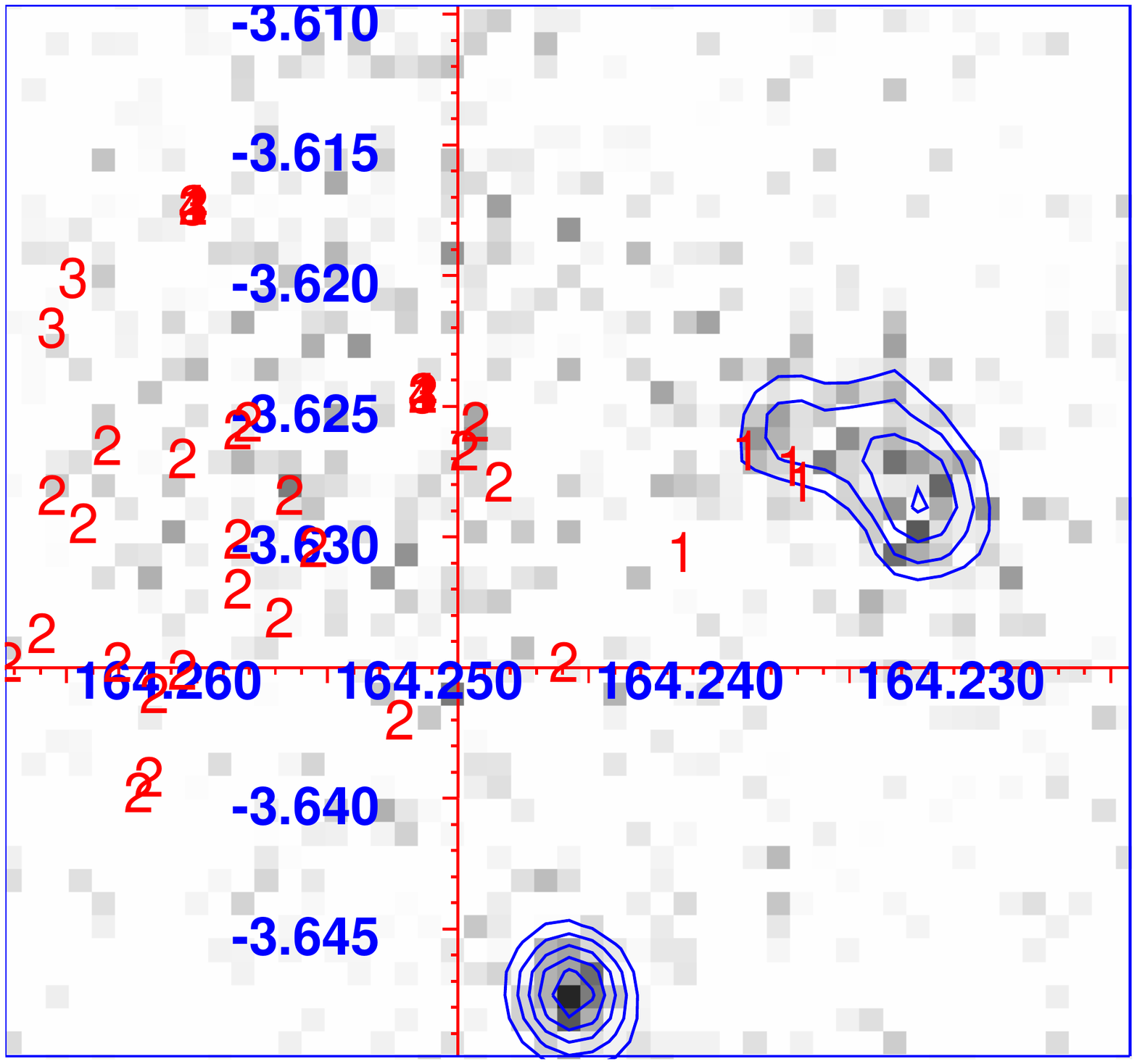}\\
    \includegraphics[width=2.7in,angle=0,bb=15 144 575 701,clip]{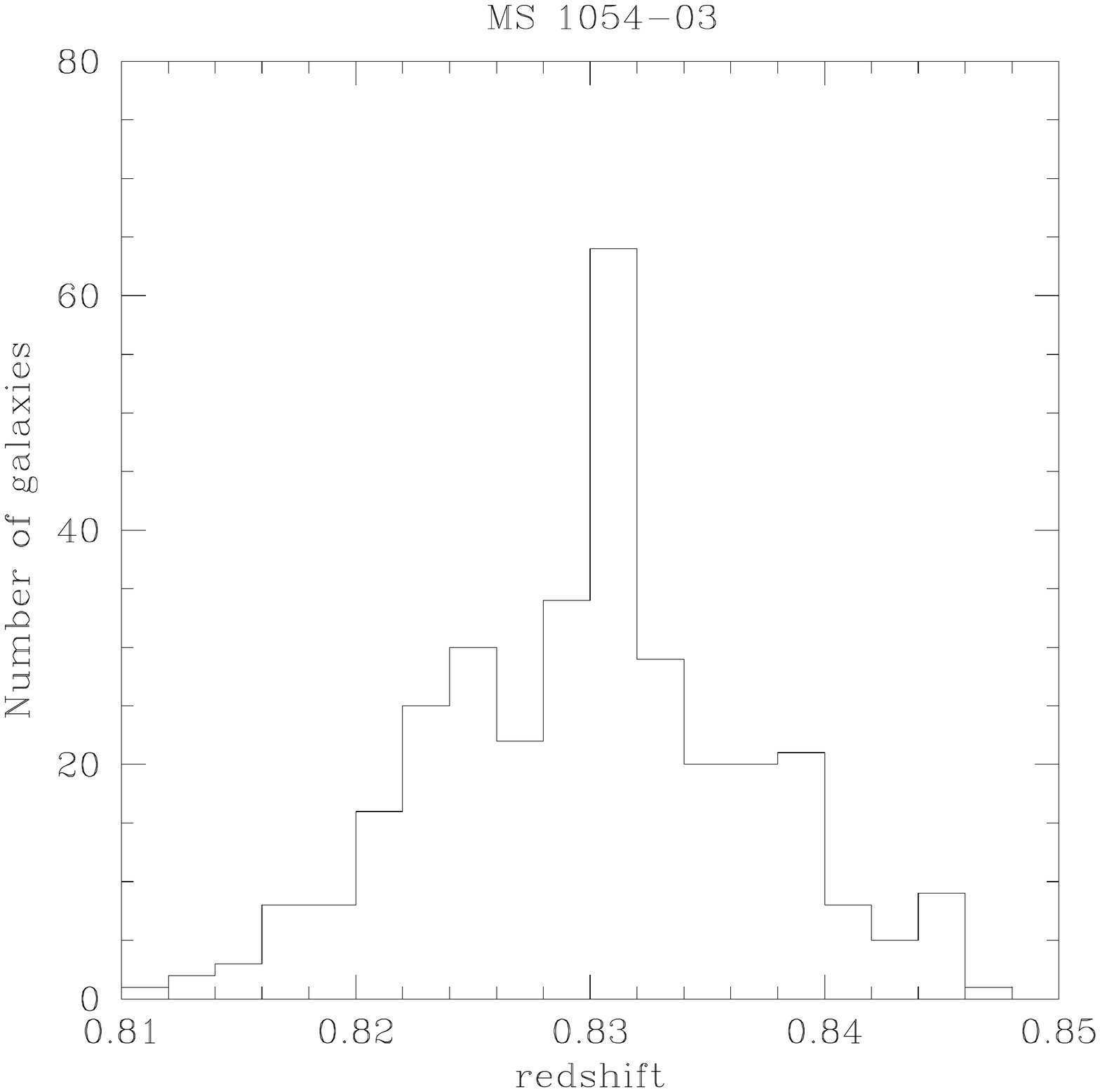}
    \caption{Same as Fig.~\ref{fig:cl0152_X} for MS~1054-03.}
  \label{fig:ms1054_03_X}
  \end{center}
  \end{figure*}

  The model-subtracted XMM-Newton image of MS~1054-03 shows three main
  peaks above the 5.5$\sigma$ level: a large and rather faint
  apparently extended source to the west (too far from the cluster
  centre to be due to a $\beta -$model subtraction that is not
  perfect), and two other more compact and somewhat brighter sources
  to the south and north east (Fig.~\ref{fig:ms1054_03_X}). This
  agrees with the statement by Gioia et al. (2004), who classify
  MS~1054-03 as a ``young, massive, highly luminous cluster with
  significant substructure''.

  The comparison with the Chandra image clearly shows that the two
  compact sources are probably due to point sources even though
  none of them corresponds to spectroscopically known active objets. The
  XMM-Newton extended residual to the west is also extended in the
  Chandra image, so we can probably conclude that we are dealing with
  extended thermal emission related to the cluster: a real
  substructure in the MS~1054-03 cluster of galaxies.

  On the optical side, the redshift histogram shows a broad
  distribution for the 326 redshifts in the [0.81,0.85] range
  (Fig.~\ref{fig:ms1054_03_X}).  The SG analysis shows the presence of
  a massive substructure of 26 galaxies (group SG2), which can probably
  be considered as the main core of MS~1054-03.  We also detect three
  other less populated substructures (Table~\ref{tab:SG}).  Among
  them, the SG1 group is the most massive and is probably associated
  with the X-ray substructure described above.

\subsection{UM 425 cluster (170.83542$^o$, +1.6294$^o$, z=0.7685)} 

\begin{figure*}
  \begin{center}
    \includegraphics[width=2.7in,angle=0,bb=35 144 575 651,clip]{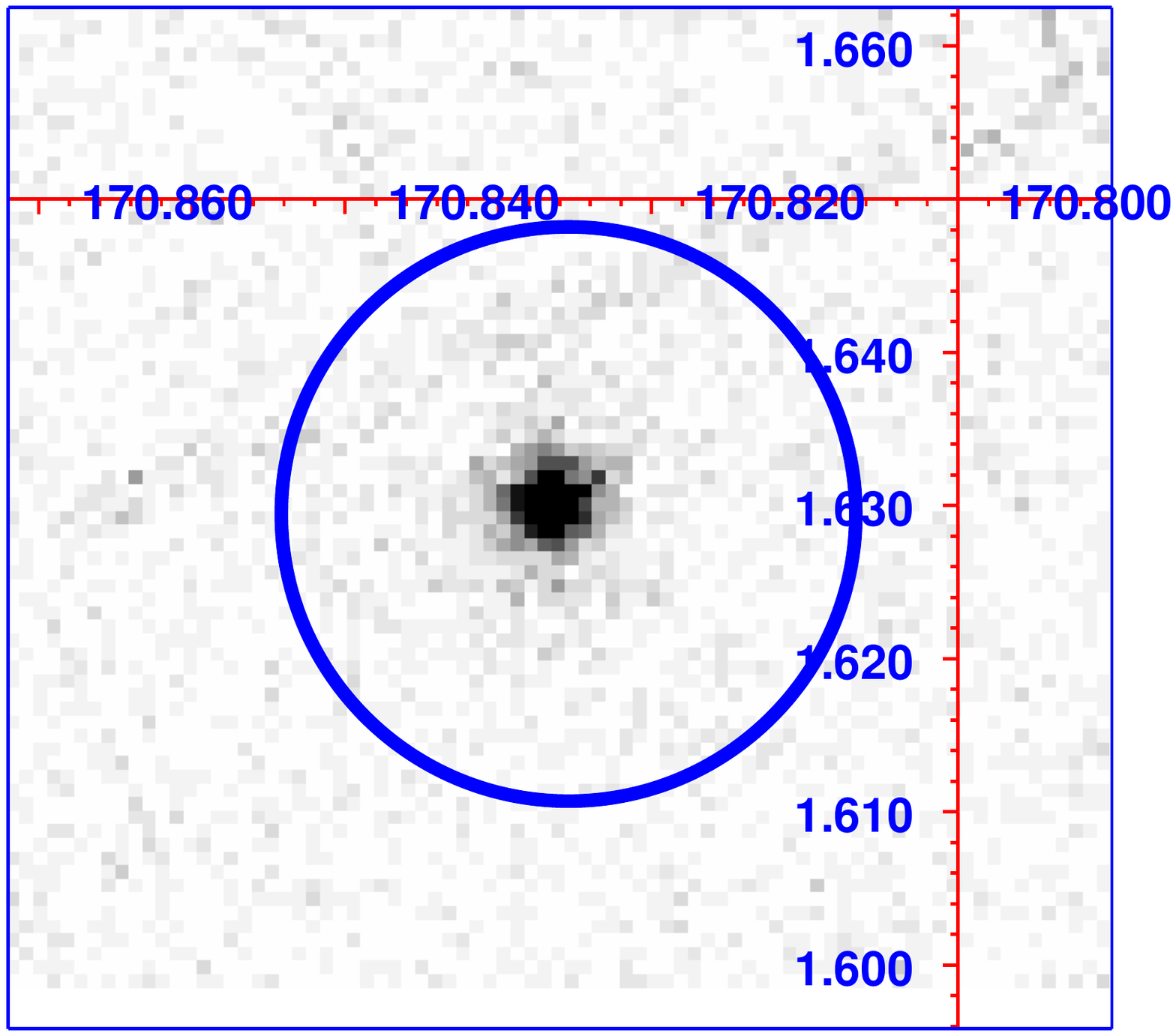}\includegraphics[width=2.7in,angle=0,bb=35 144 575 651,clip]{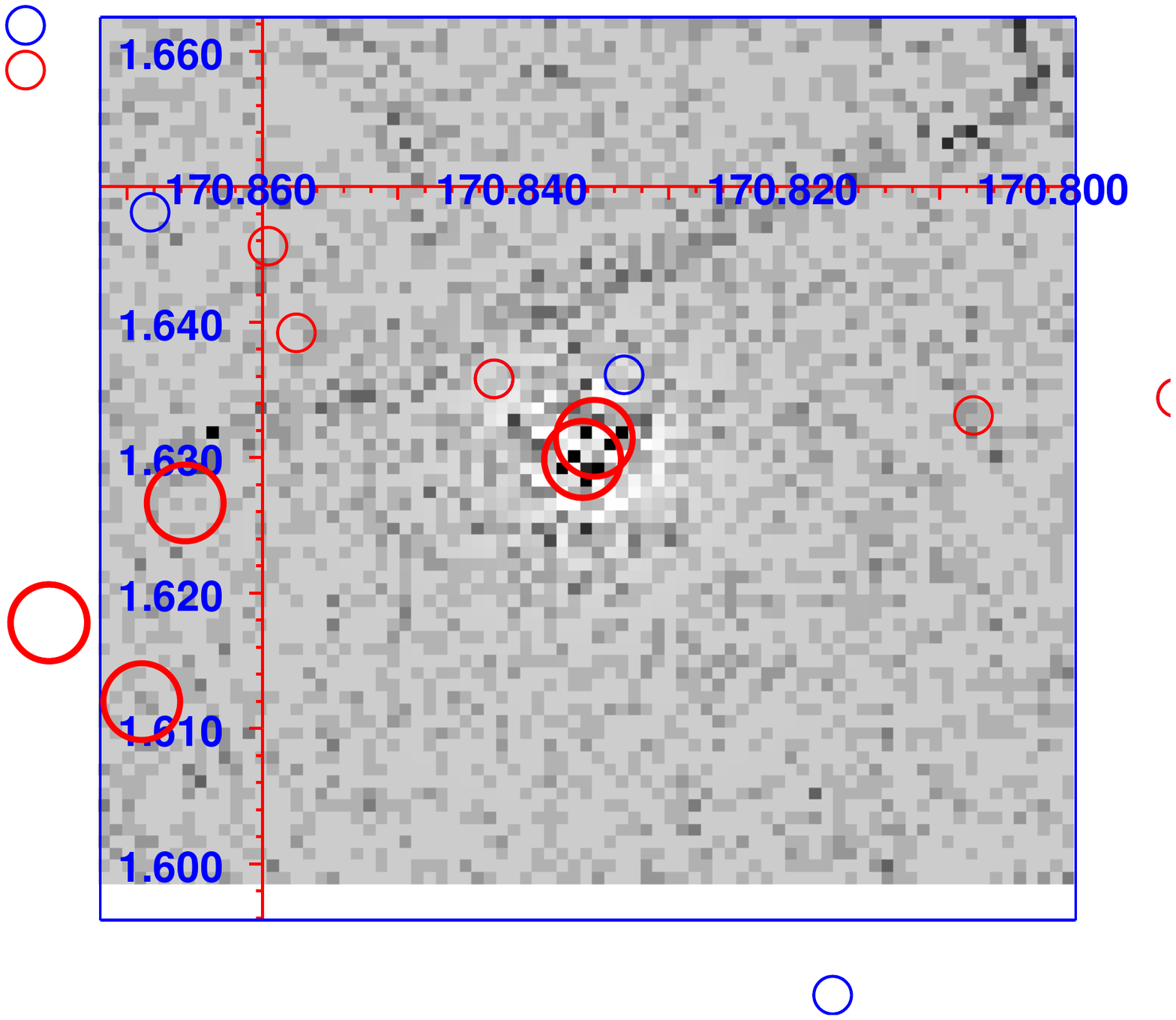}\\
    \includegraphics[width=2.7in,angle=0,bb=35 144 575 651,clip]{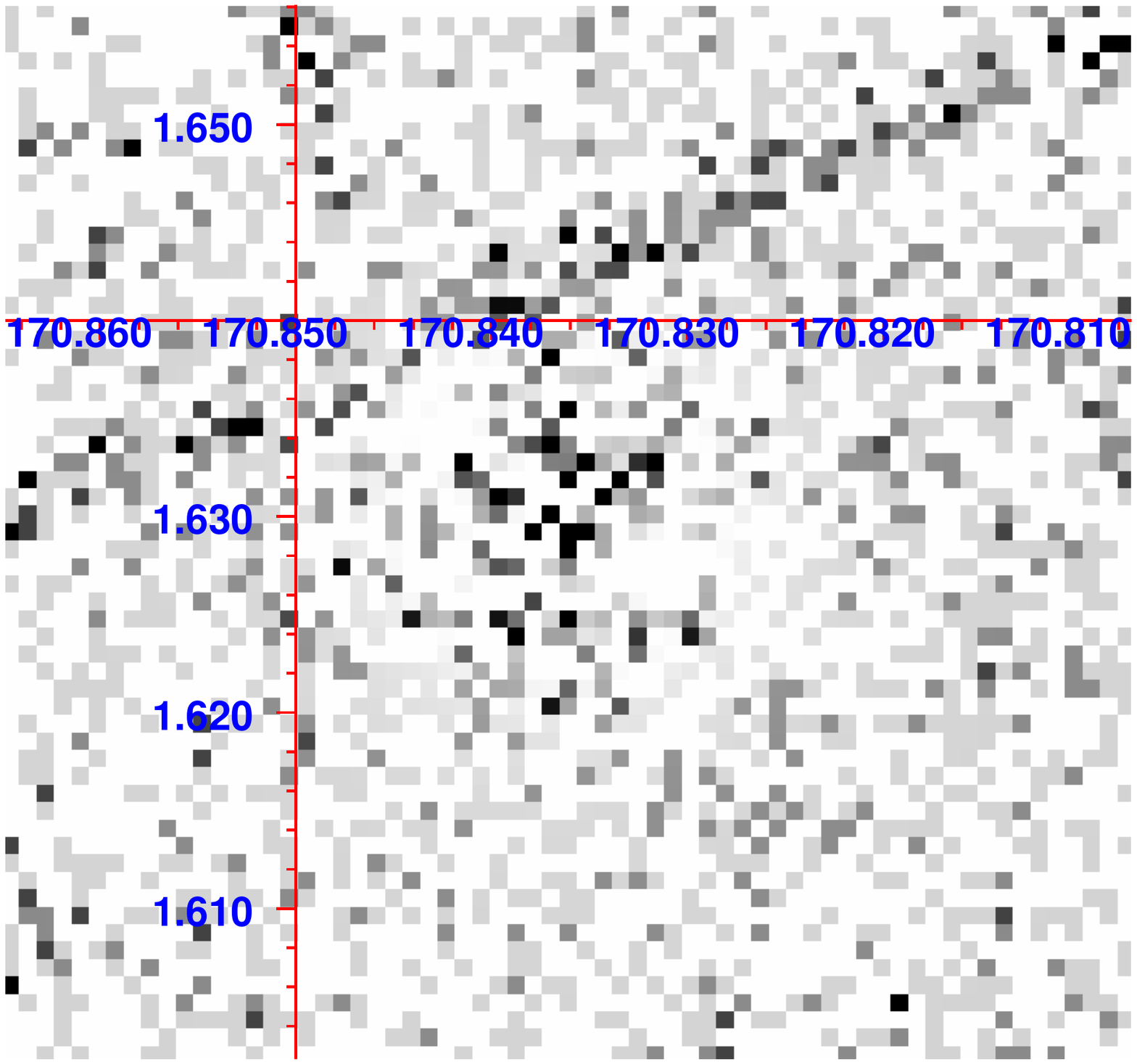}\includegraphics[width=2.7in,angle=0,bb=15 144 575 701,clip]{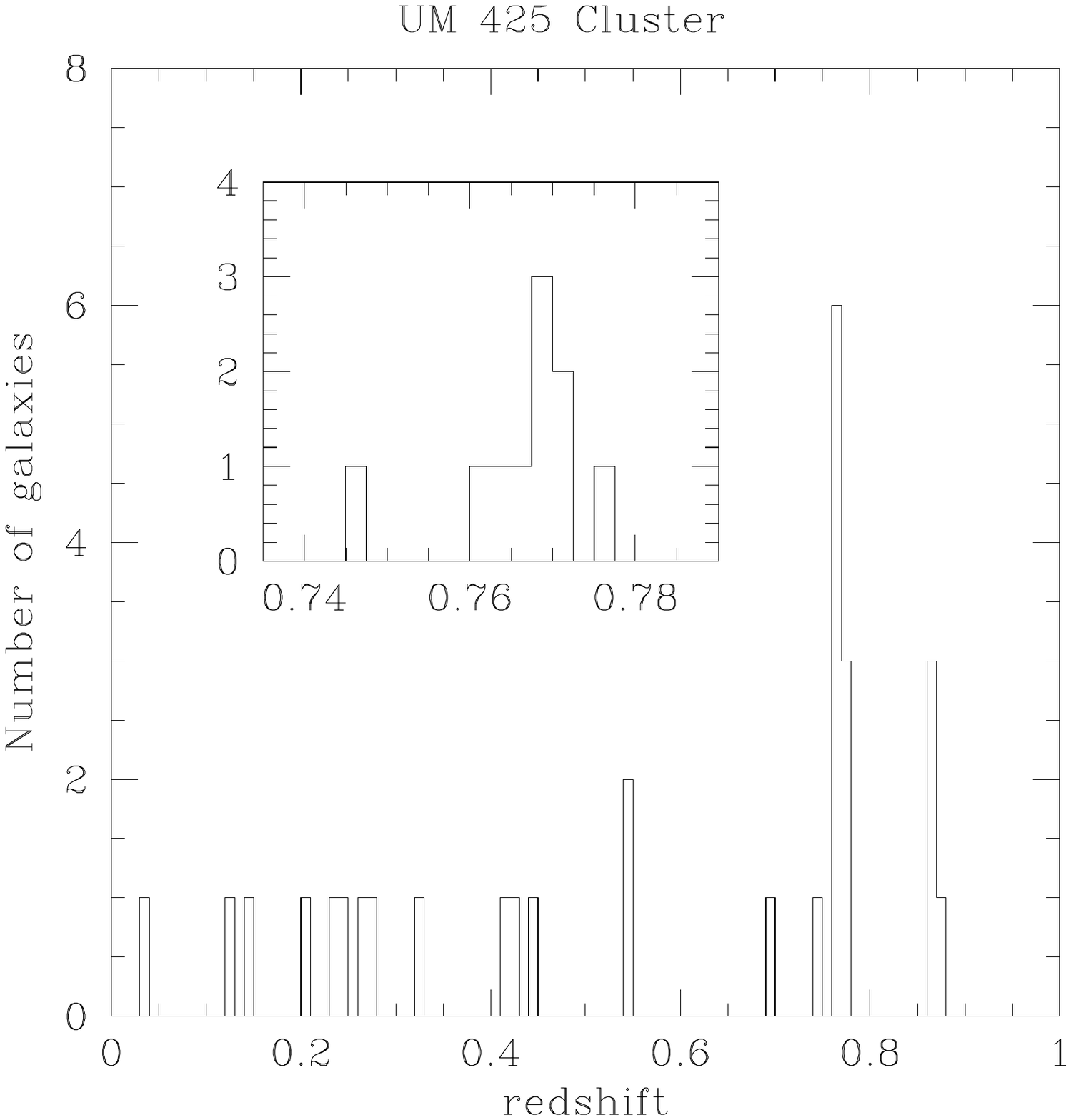}
    \caption{Same as Fig.~\ref{fig:cl0016_X} for the UM 425 cluster. The insert 
in the redshift histogram (lower right) shows a zoom around the cluster redshift.}
  \label{fig:UM425_X}
  \end{center}
  \end{figure*}

The UM~425 cluster is located along the line of sight of a known quasar pair. It was first 
supposed to be  at the same redshift than the quasars (z$\sim$1.47: Mathur \& 
Williams 2003). Later, Green et al. (2005), proposed a redshift of $\sim$0.77 based on VLT
spectroscopy.

The UM~425 candidate cluster shows no obvious substructures in X-rays
and is well fit by a simple model (Fig.~\ref{fig:UM425_X}) without any
significant X-ray residual.  The redshift histogram of the
UM~425~Cluster shows a peak at z$\sim$0.7685 (the redshift given by
NED), in agreement with Green et al. (2005), and a smaller background
peak at z$\sim$0.87 (Fig.~\ref{fig:UM425_X}). However, the SG analysis
does not detect any structure (cluster or subcluster). Even if there
are only eight galaxies in the [0.760,0.773] range, this leads to
doubts about the massive cluster nature of the UM 425 cluster. We are
perhaps intercepting a filament (explaining the peak in the redshift
histogram), including a quasar emitting in X-rays.

\subsection{MS 1137.5+6624 (175.09696$^o$, +66.1449$^o$, z=0.7820)}  

\begin{figure*}
  \begin{center}
    \includegraphics[width=2.7in,angle=0,bb=35 144 575 651,clip]{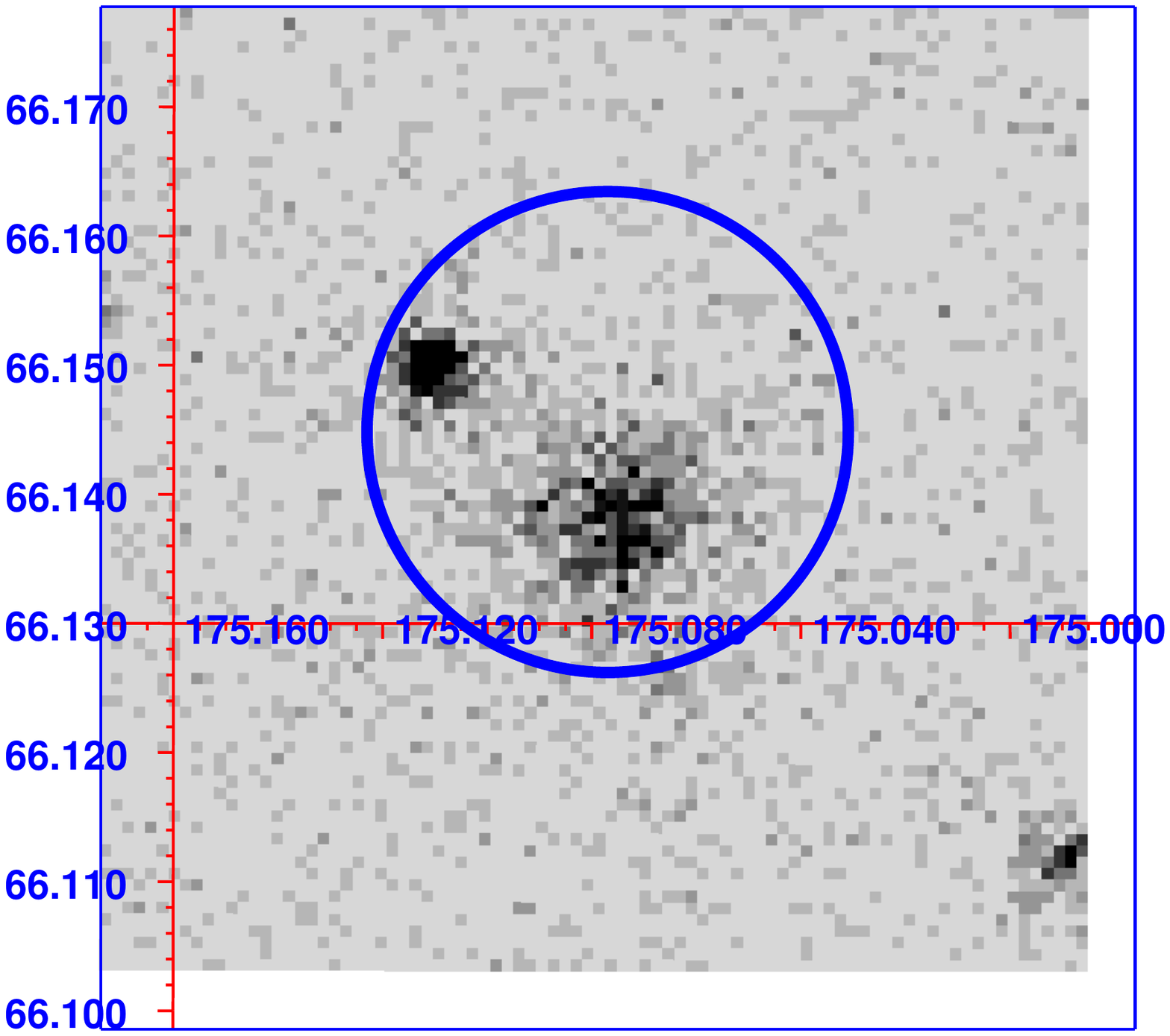}\includegraphics[width=2.7in,angle=0,bb=35 144 575 651,clip]{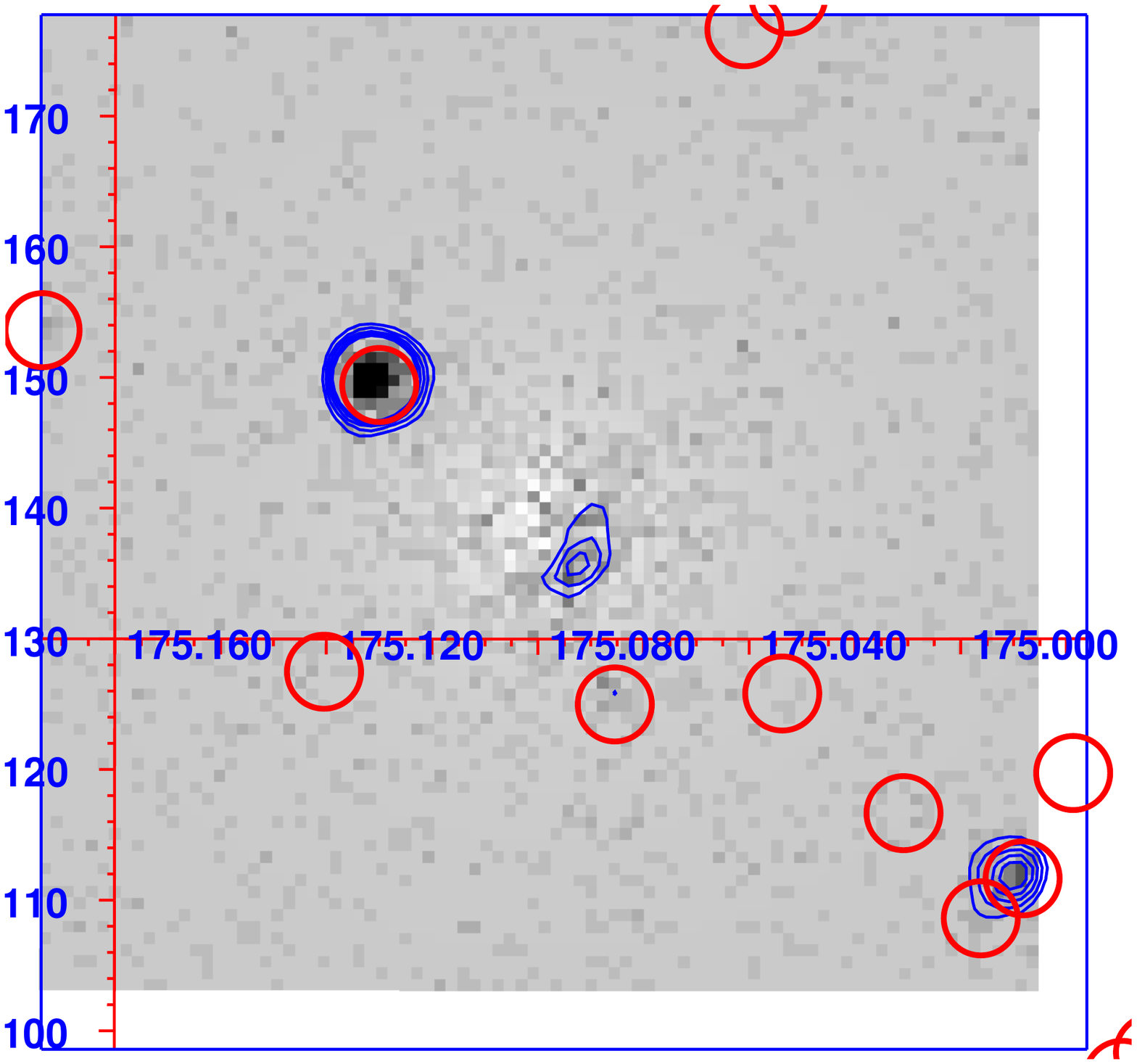}\\
    \includegraphics[width=2.7in,angle=0,bb=35 144 575 651,clip]{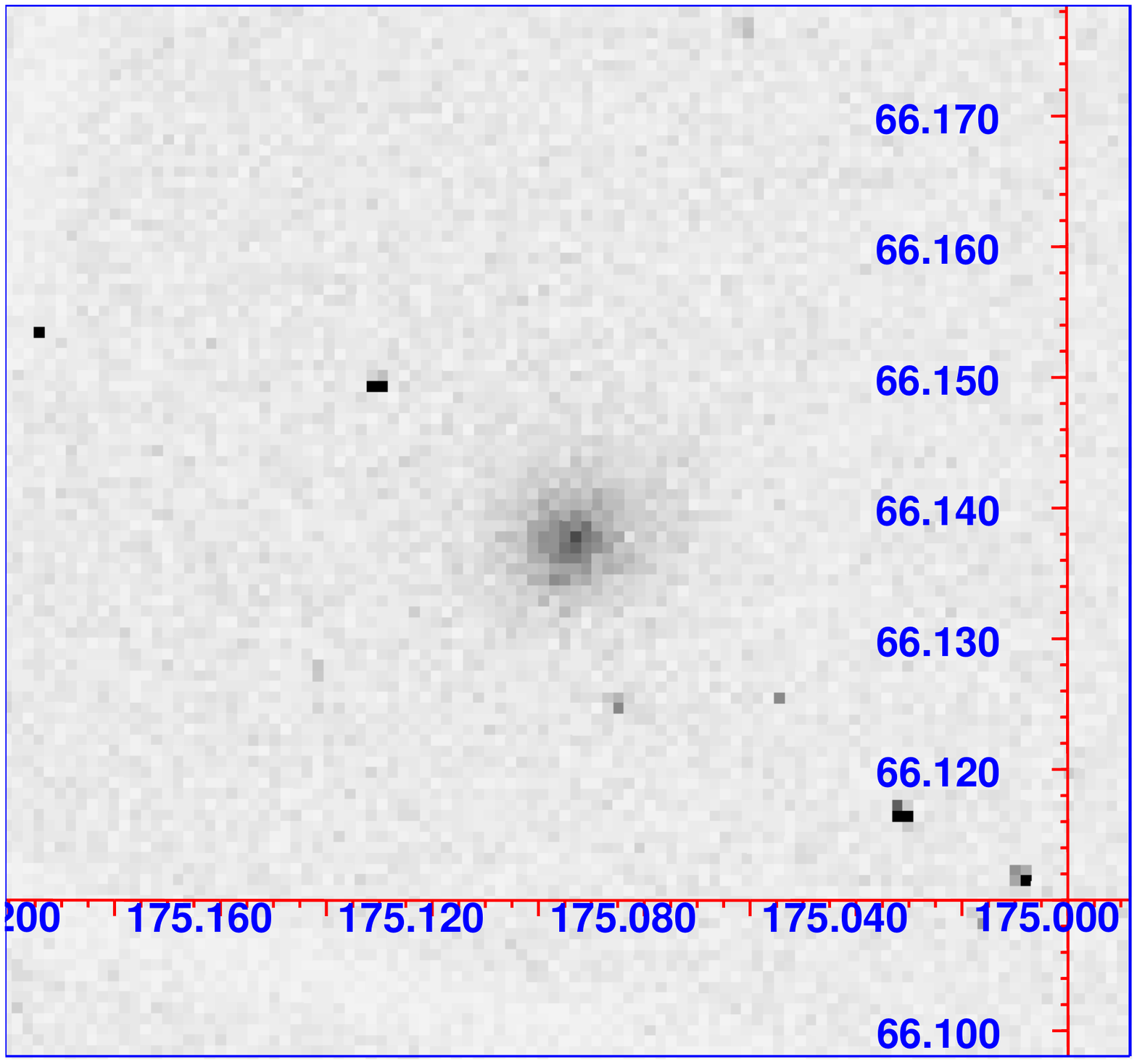}\includegraphics[width=2.7in,angle=0,bb=35 144 575 651,clip]{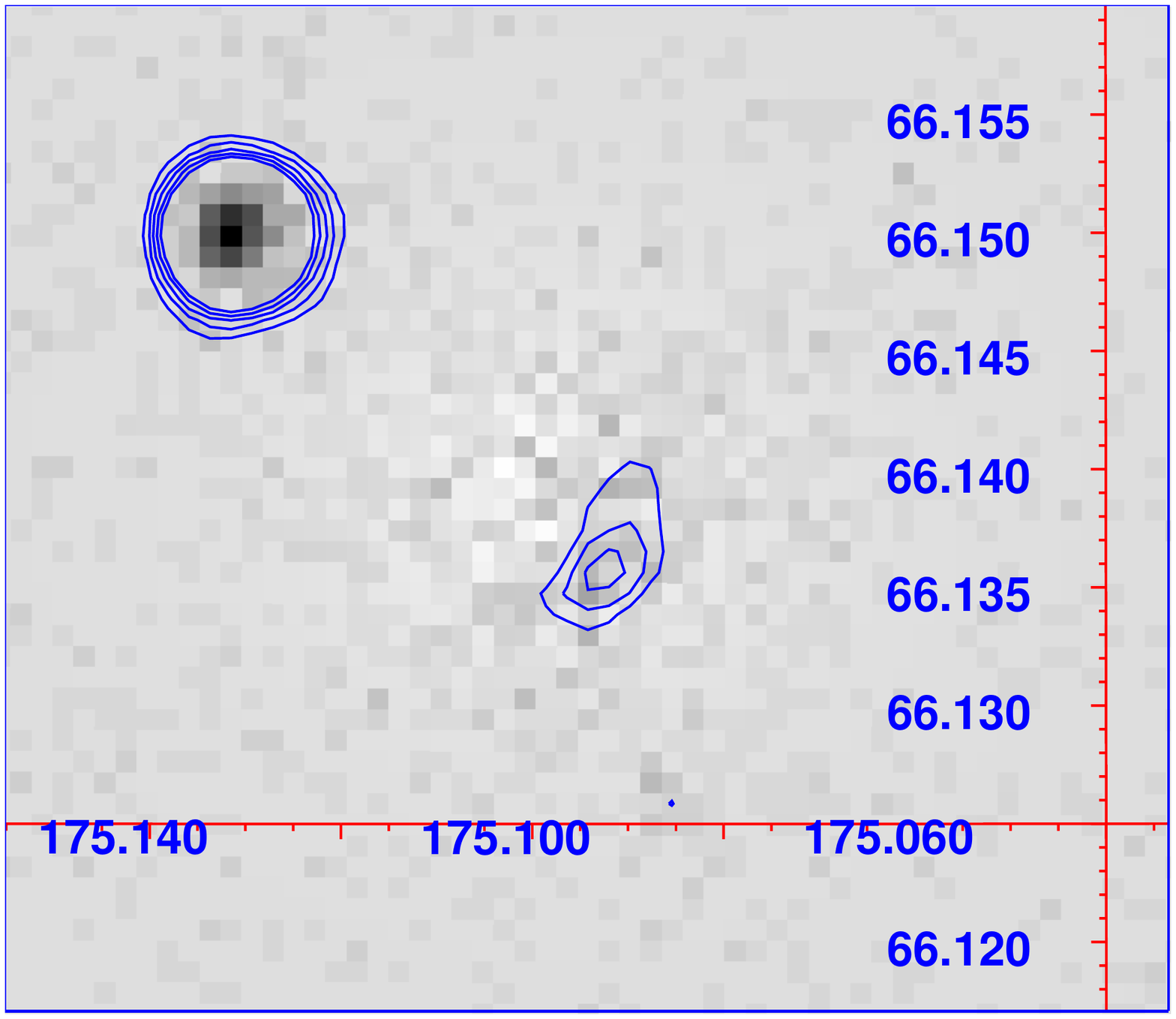}\\
    \includegraphics[width=2.7in,angle=0,bb=15 144 575 701,clip]{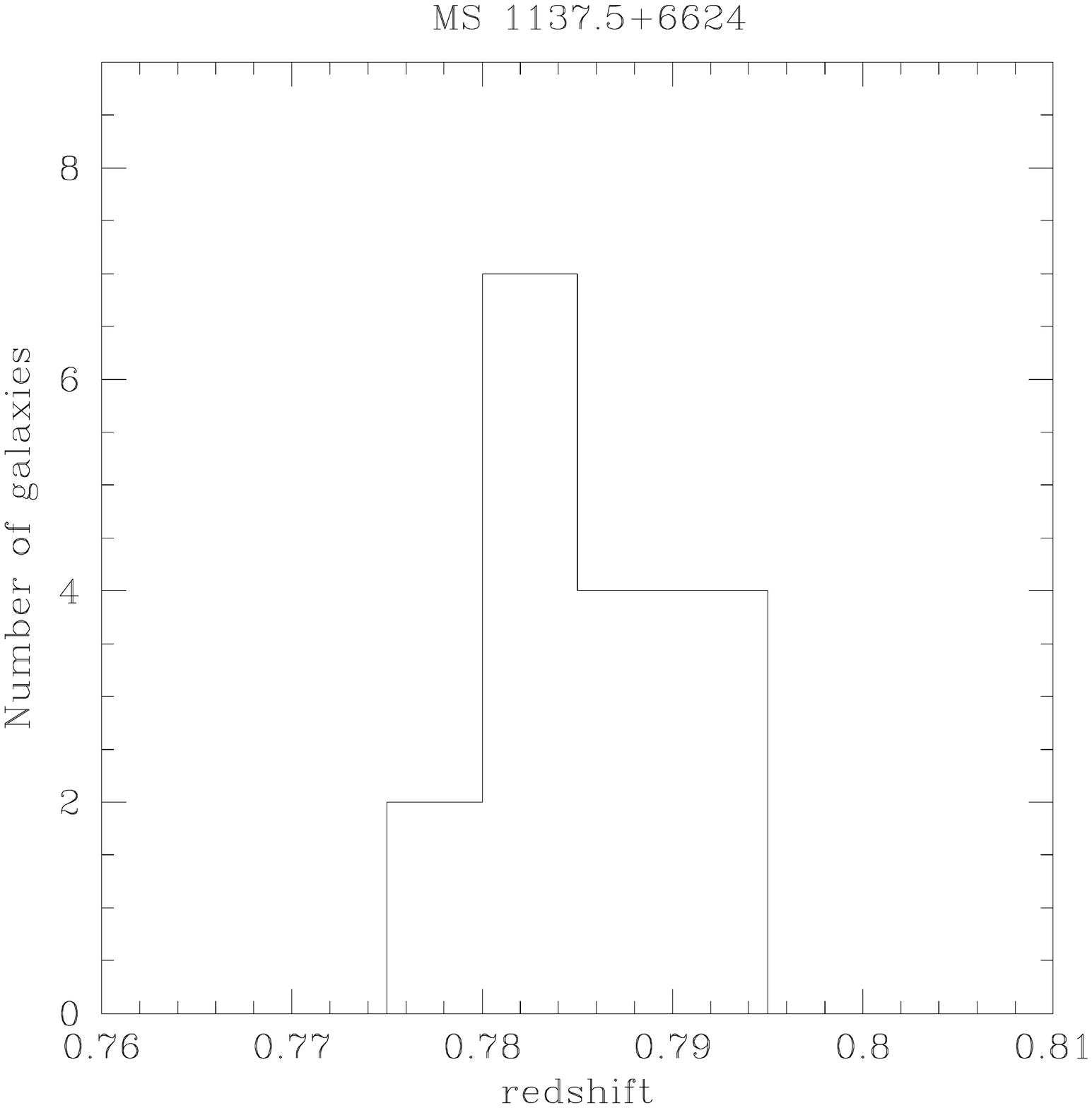}
    \caption{Same as Fig.~\ref{fig:cl0152_X} for MS 1137.5+6624.}
  \label{fig:MS1137_X}
  \end{center}
  \end{figure*}

  The residual X-ray image of MS 1137.5+6624 shows two compact sources
  to the north--east and the south--west (Fig.~\ref{fig:MS1137_X}), both
  identified with active objects on the line of sight (Gilmour et
  al. 2009).  Some excess diffuse emission is also left after
  subtracting the $\beta -$model, exactly at the cluster
  location. This is probably because a $\beta -$model is
  not able to reproduce the strong peak at the cluster centre visible
  in the Chandra image (Fig.~\ref{fig:MS1137_X}). We therefore
  consider that we are not dealing with a real X-ray substructure, as
  also suggested by the Chandra X-ray image of Maughan et al. (2008).

  On the optical side, the velocity histogram is asymmetric. There are
  17 redshifts in the cluster range (Fig.~\ref{fig:MS1137_X}), but the
  SG does not detect any substructure in the cluster. Since no
  substructures appear in the SG analysis or in X-rays, we are
  probably dealing with a fairly relaxed structure.

\subsection{CLG~J1205+4429 (181.46410$^o$, +44.4860$^o$, z=0.5915)} 

\begin{figure*}
  \begin{center}
    \includegraphics[width=2.7in,angle=0,bb=35 144 575 651,clip]{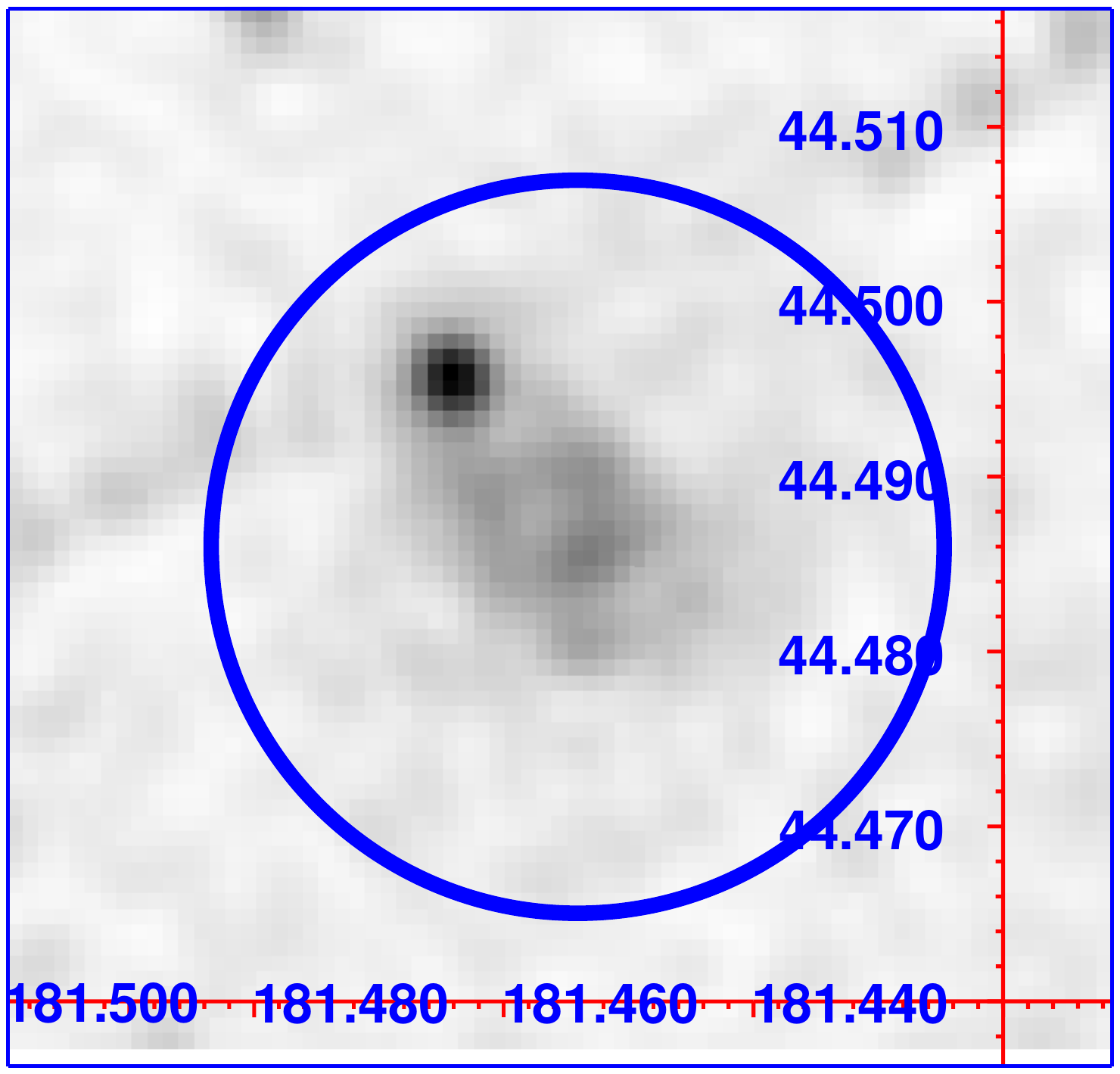}\includegraphics[width=2.7in,angle=0,bb=35 144 575 651,clip]{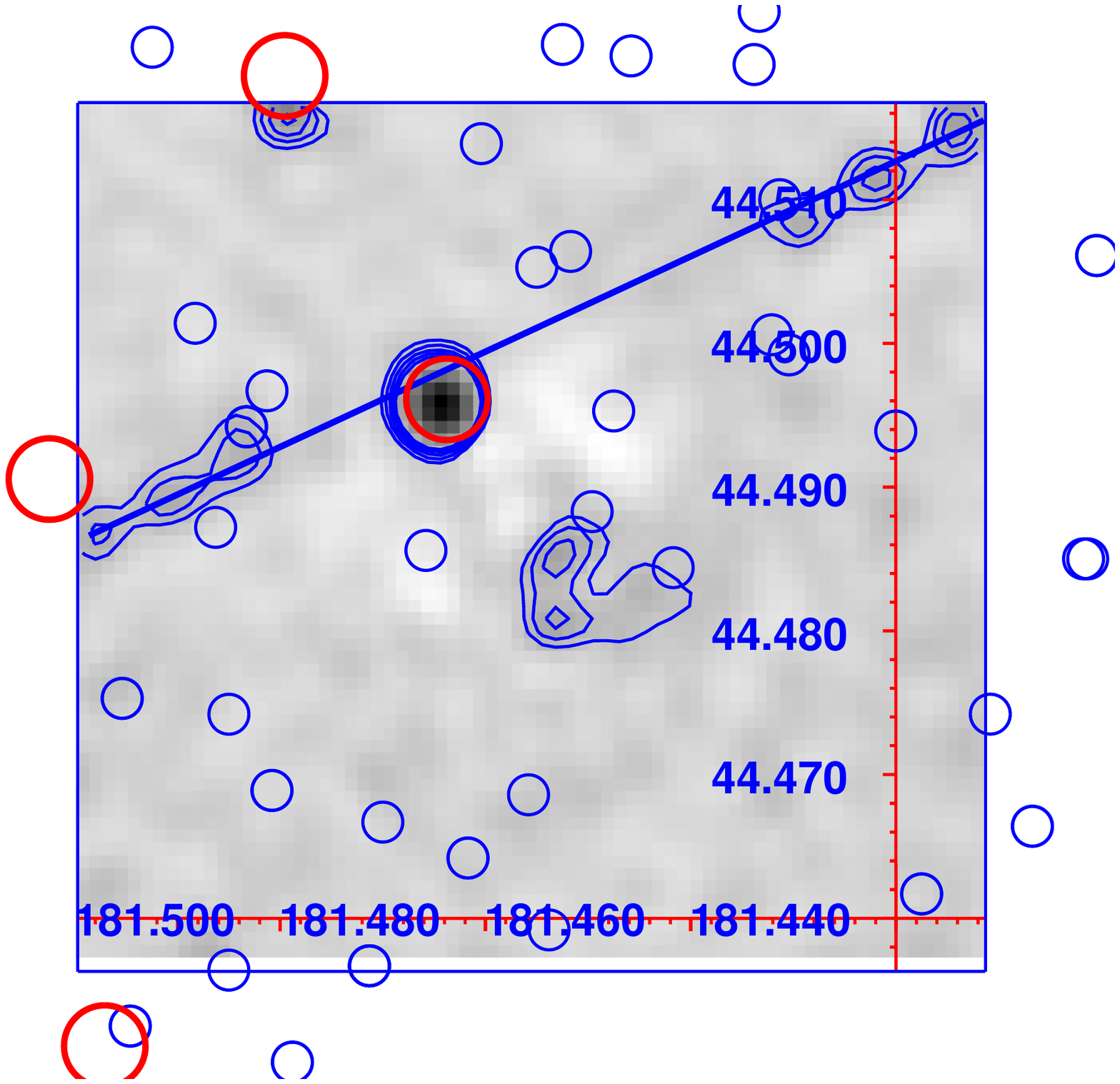}\\
    \includegraphics[width=2.7in,angle=0,bb=35 144 575 651,clip]{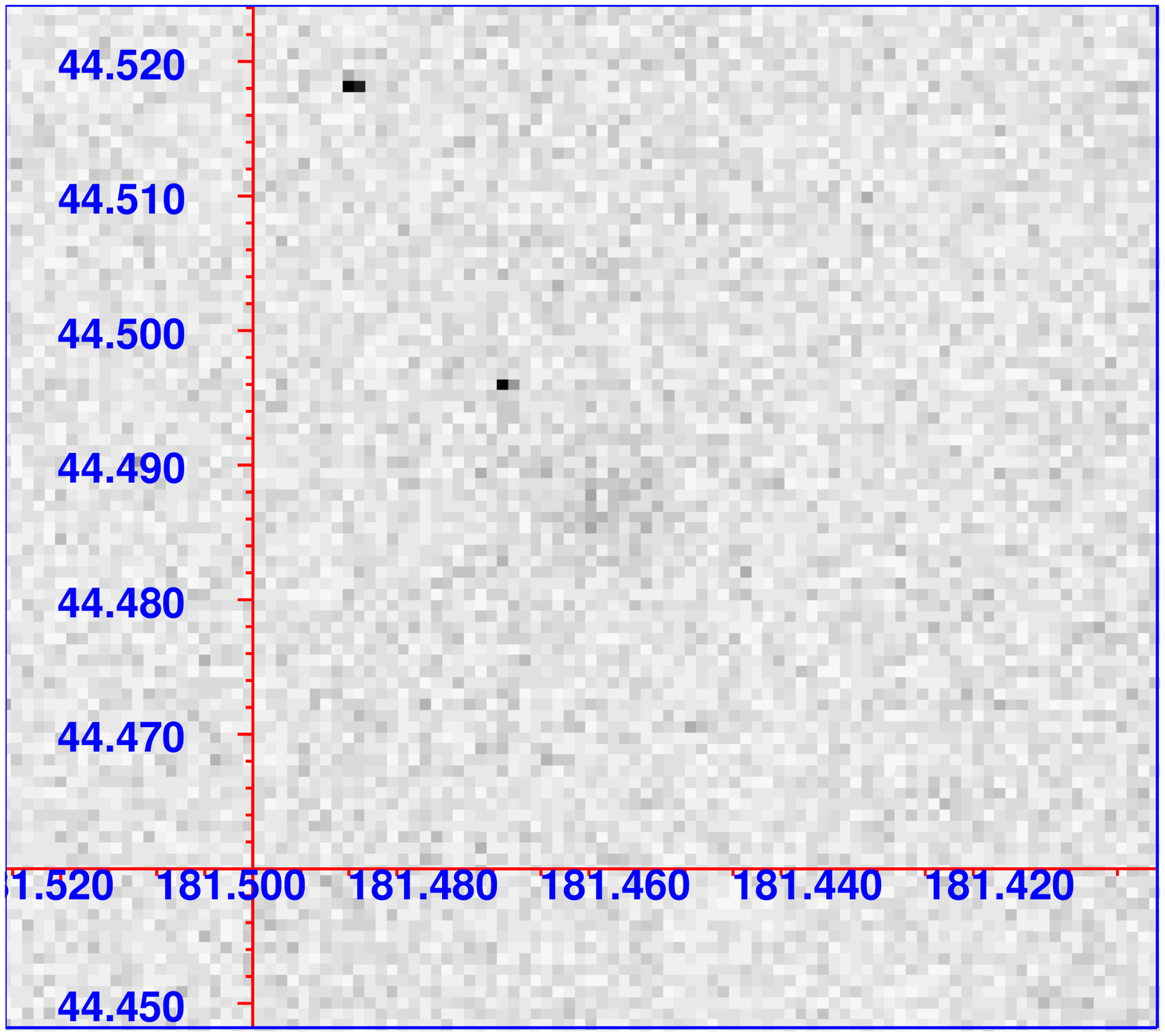}\includegraphics[width=2.7in,angle=0,bb=35 144 575 651,clip]{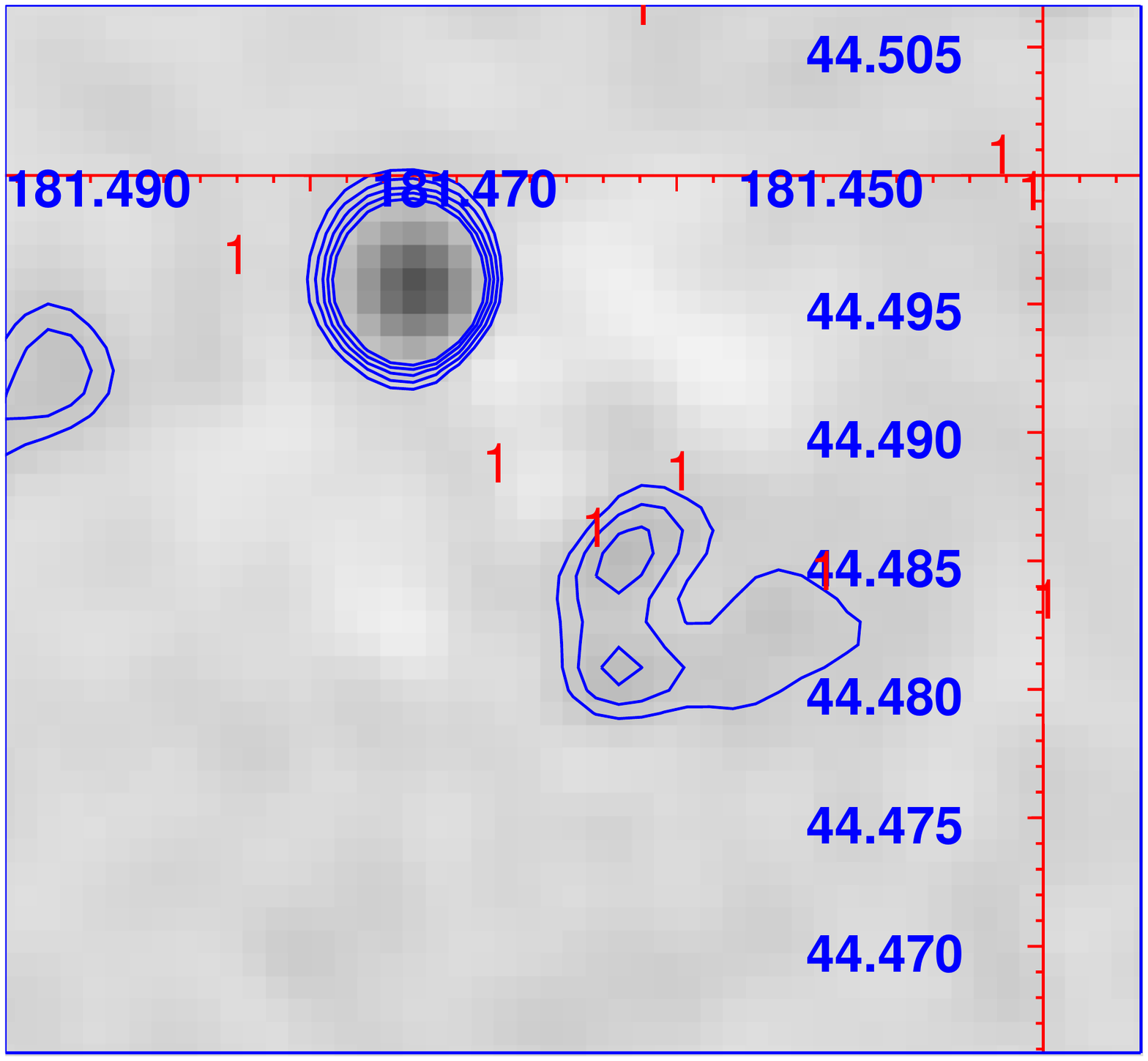}\\
    \includegraphics[width=2.7in,angle=0,bb=15 144 575 701,clip]{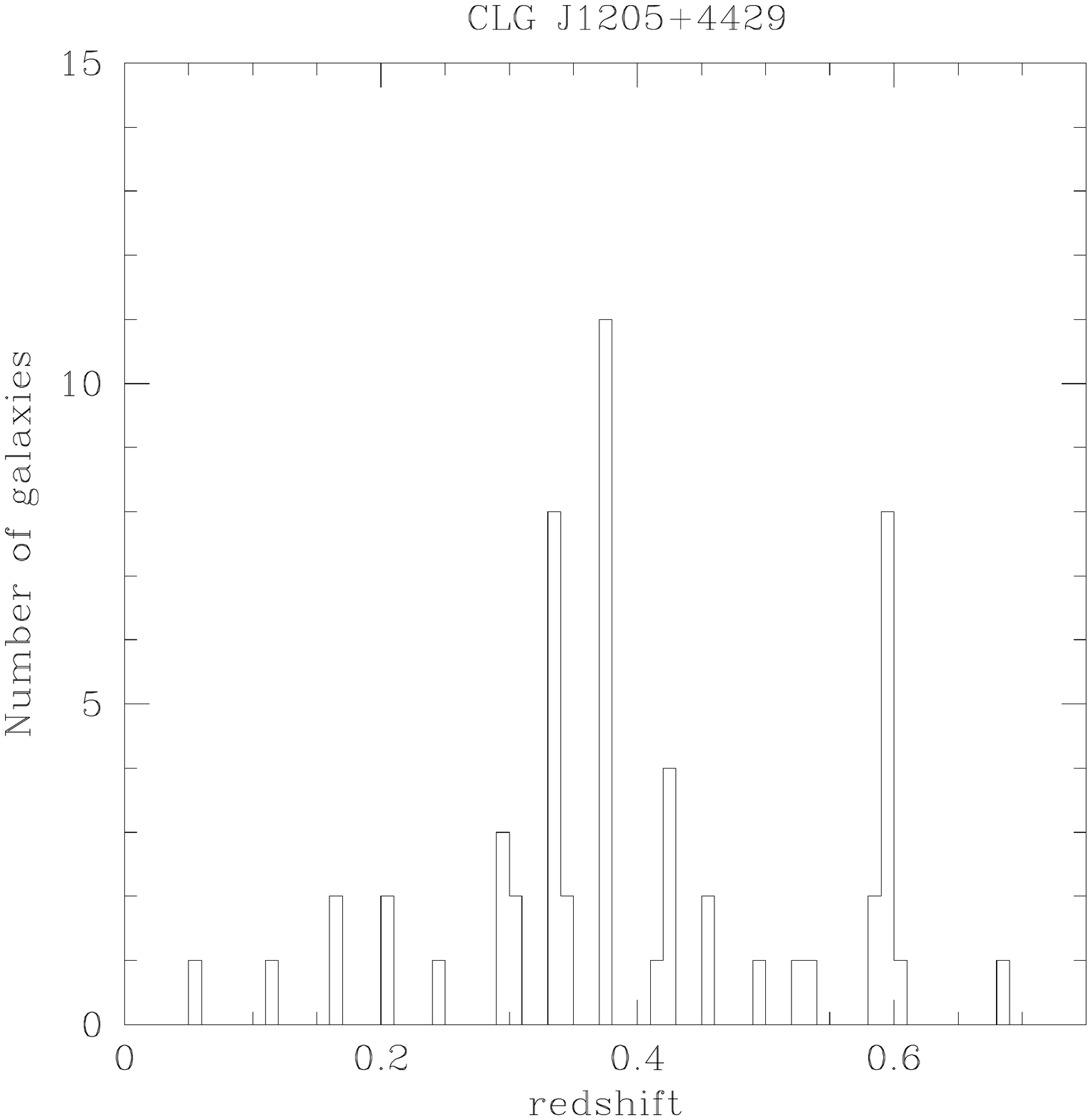}
    \caption{Same as Fig.~\ref{fig:cl0152_X} for CLG J1205+4429.}
  \label{fig:CL1205_X}
  \end{center}
  \end{figure*}

The model-subtracted XMM-Newton X-ray image of CLG~J1205+4429 
shows three main peaks (in addition to the detector interchip residuals
shown  in Fig.~\ref{fig:CL1205_X}).  Two of them are
compact sources  north--east of the cluster (identified as AGNs in
Gilmour et al. 2009), and the comparison with the Chandra image
clearly shows that these sources are point-like.  The third X-ray
source in the X-ray residual image seems extended and is roughly centred on
the cluster position.

On the optical side, the redshift histogram
shows that the line of sight to CLG J1205+4429 intercepts several
structures (Fig.~\ref{fig:CL1205_X}). 
A peak is apparent at z=0.5915 (the value given by NED) in the
redshift histogram, with 11 galaxies in the [0.582,0.600] range.  The
SG method interprets these galaxies as a single cluster without any
substructure. 

It is therefore likely that the X-ray residual corresponding to the
third X-ray source is not a classical substructure (a major infalling
group made of relatively bright galaxies), but it could possibly be
part of the central cluster X-ray emission incorrectly removed by the
subtraction of a $\beta -$model fit. However it is too extended to be
just a wrong $\beta -$model subtraction of the cluster central
emission and CLG~J1205+4429 is not a cool-core-like cluster (Ulmer et
al. 2005). This structure instead resembles what was observed in Coma
by Neumann et al. (2003), where relatively diffuse X-ray emission is
associated to both NGC~4874 and NGC~4889, in addition to the global
cluster $\beta -$model emission. This, together with the fact that
CLG~J1205+4429 is a structure with properties close to those of a
fossil group (Ulmer et al. 2005), speaks in favour of an old group with
its own original X-ray halo (seen as the central X-ray residual) which
acted as a primary seed to accrete more matter (without experiencing
major merging events) and is at the origin of the underlying X-ray
$\beta -$model.

\subsection{RXC~J1206.2-0848  (181.54991$^o$, --8.8000$^o$, z=0.4400) }  

\begin{figure*}
  \begin{center}
    \includegraphics[width=2.7in,angle=0,bb=35 144 575 651,clip]{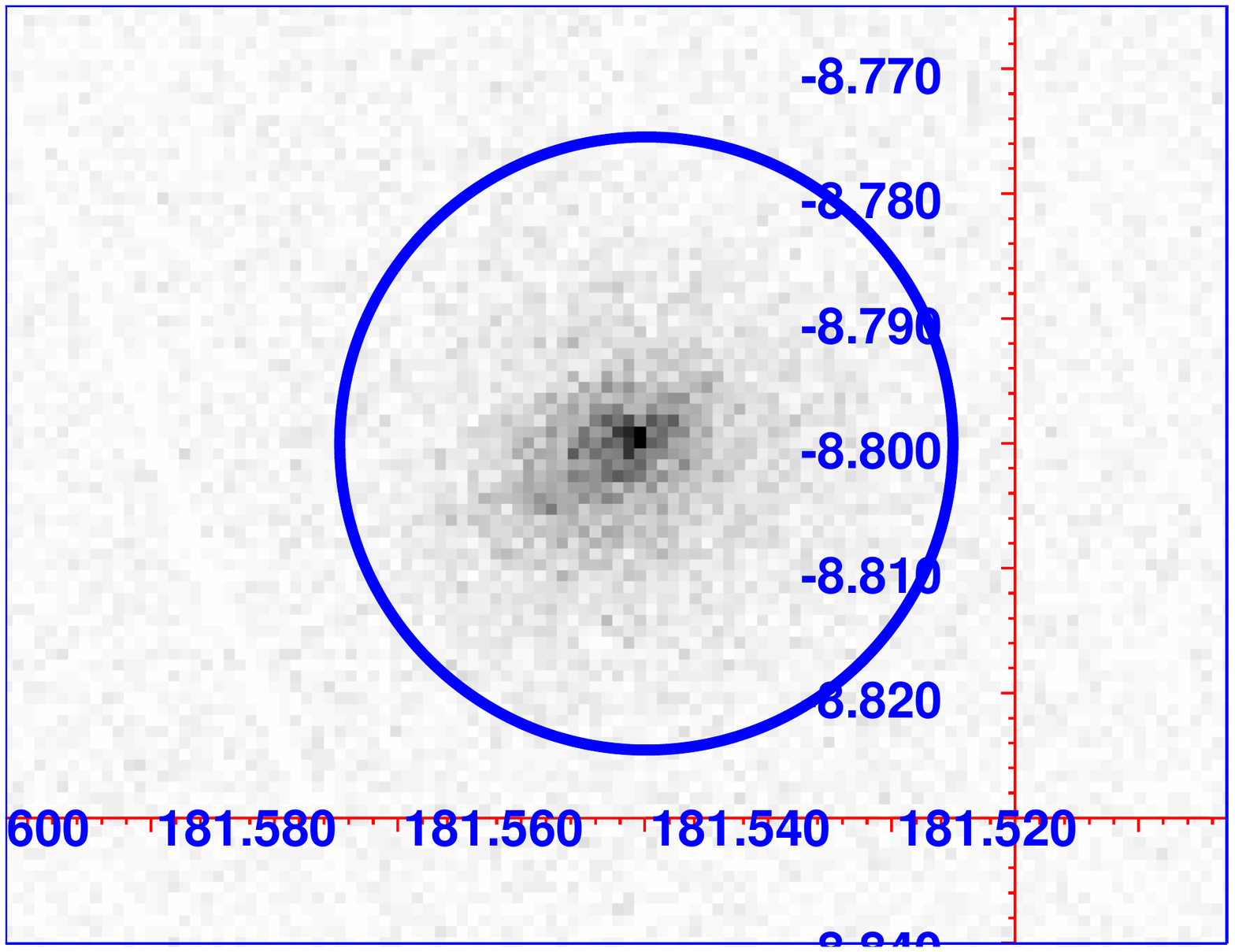}\includegraphics[width=2.7in,angle=0,bb=35 144 575 651,clip]{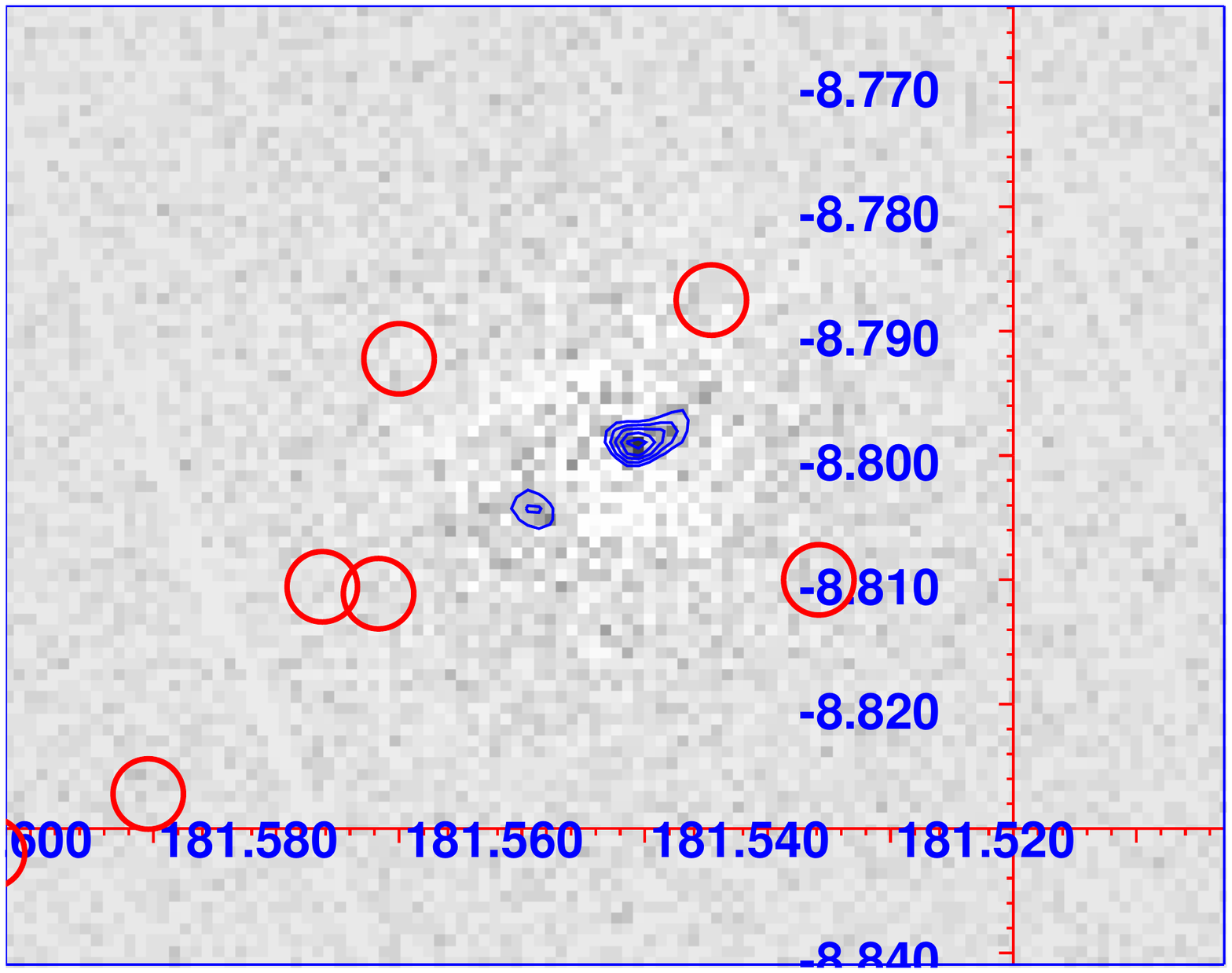}\\
    \includegraphics[width=2.7in,angle=0,bb=35 144 575 651,clip]{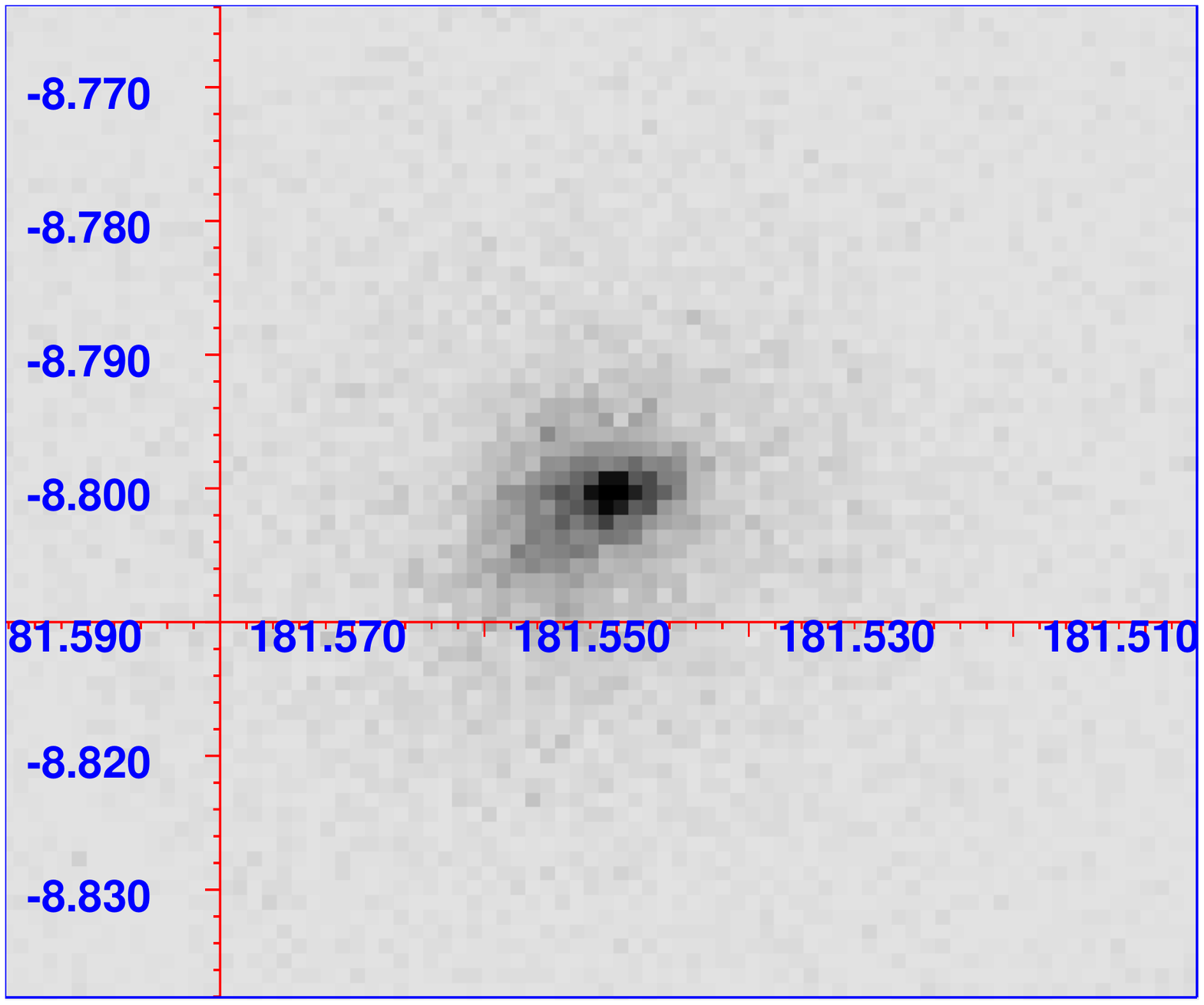}\includegraphics[width=2.7in,angle=0,bb=35 144 575 651,clip]{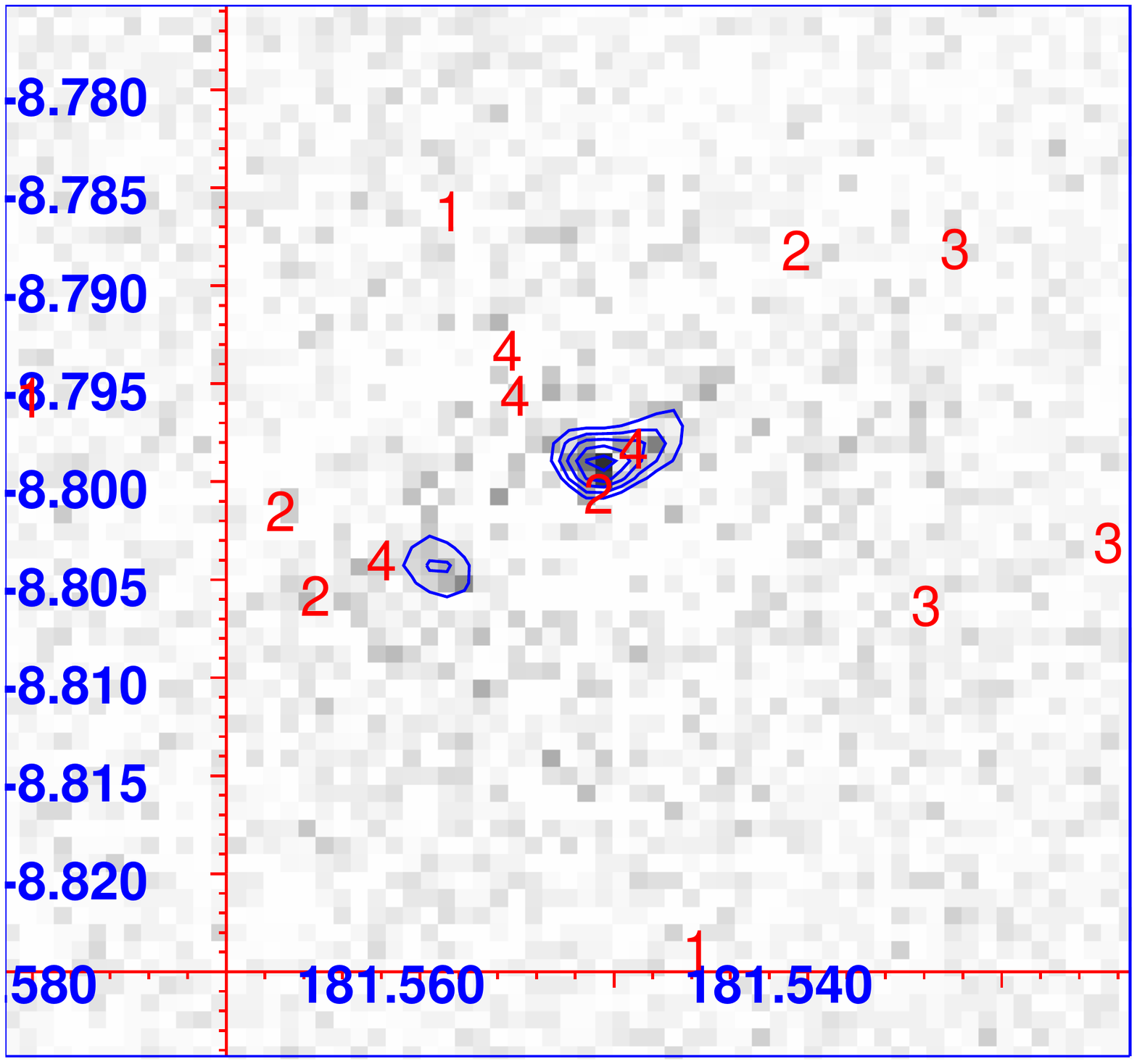}\\
    \includegraphics[width=2.7in,angle=0,bb=15 144 575 701,clip]{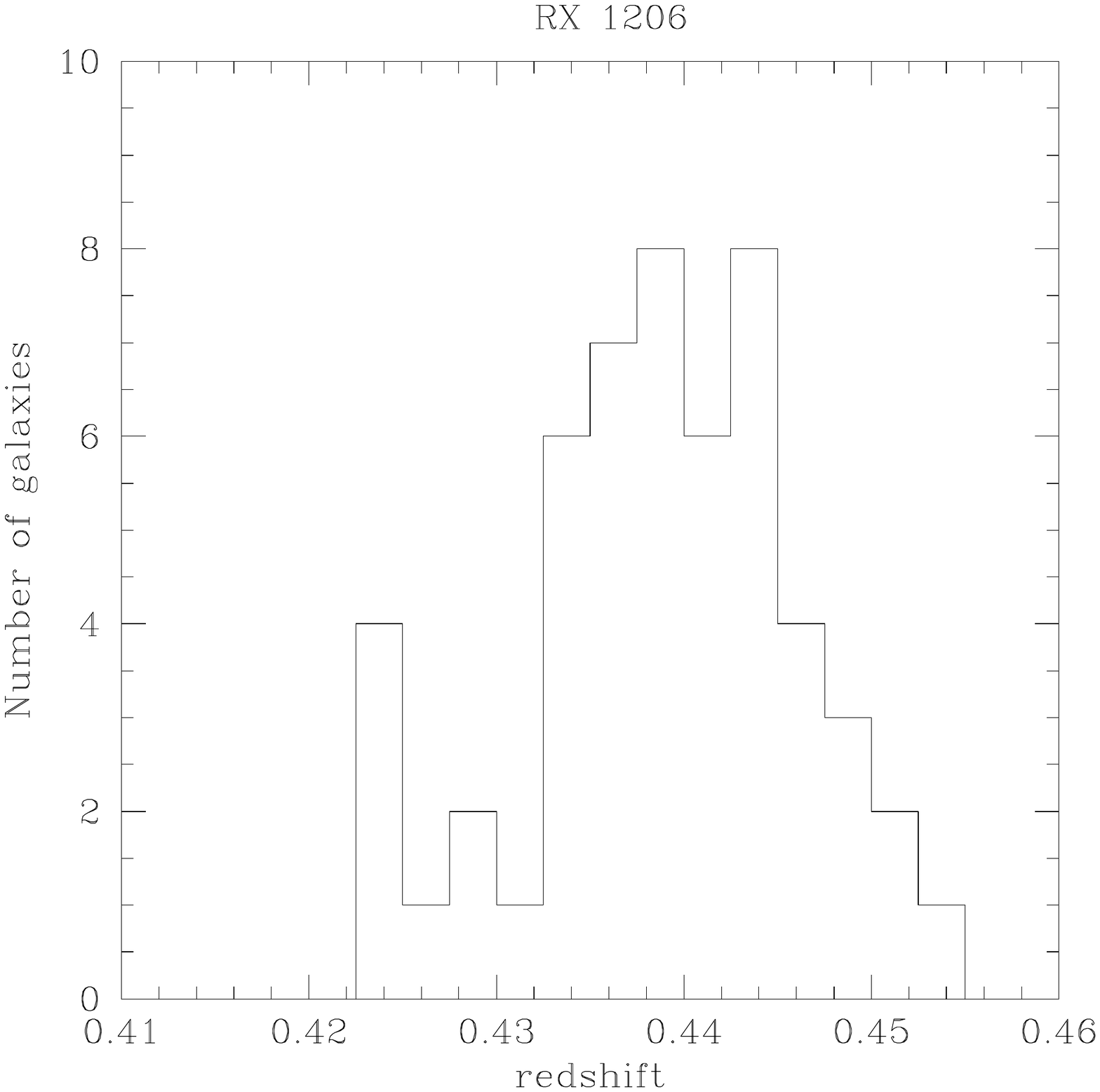}
    \caption{Same as Fig.~\ref{fig:cl0152_X} for RXC~J1206.2-0848.}
  \label{fig:RX1206_X}
  \end{center}
  \end{figure*}

  The X-ray emission is quite smooth. No point sources are detected in
  the Chandra image. The subtraction of a model shows the existence of
  two small excesses, one to the south--east and one at the cluster
  centre (Fig.~\ref{fig:RX1206_X}). The cluster--centred excess is
  probably due to the inability to correctly reproduce the cluster
  central surface brightness profile with a $\beta -$model, which is
  quite peaked. For this cluster, we have enough counts in the centre
  to measure a central temperature. Limiting our analysis to a 56 kpc
  radius, we obtain 6.00$\pm$0.75 keV, compared to 9.36$\pm$0.55 keV
  for the whole cluster.  This indicates the possible presence of a
  cool core that could explain the central X-ray excess.

  The redshift histogram of RXC J1206.2-0848 shows a broad double peak
  around z$\sim$0.44 (Fig.~\ref{fig:RX1206_X}), and looks quite
  asymmetric. There are 53 galaxies in the [0.42,0.46] range.  The SG
  analysis shows the presence of four rather small substructures with
  three to five galaxies (see Table~\ref{tab:SG}). SG4 could possibly
  be associated with the small south--east X-ray residual.

\subsection{LCDCS~0504 = Cl~1216.8-1201 (184.18845$^o$, --12.02147$^o$ (184.18792$^o$, --12.02139$^o$), z=0.7943)} 

\begin{figure*}
  \begin{center}
    \includegraphics[width=2.7in,angle=0,bb=35 144 575 651,clip]{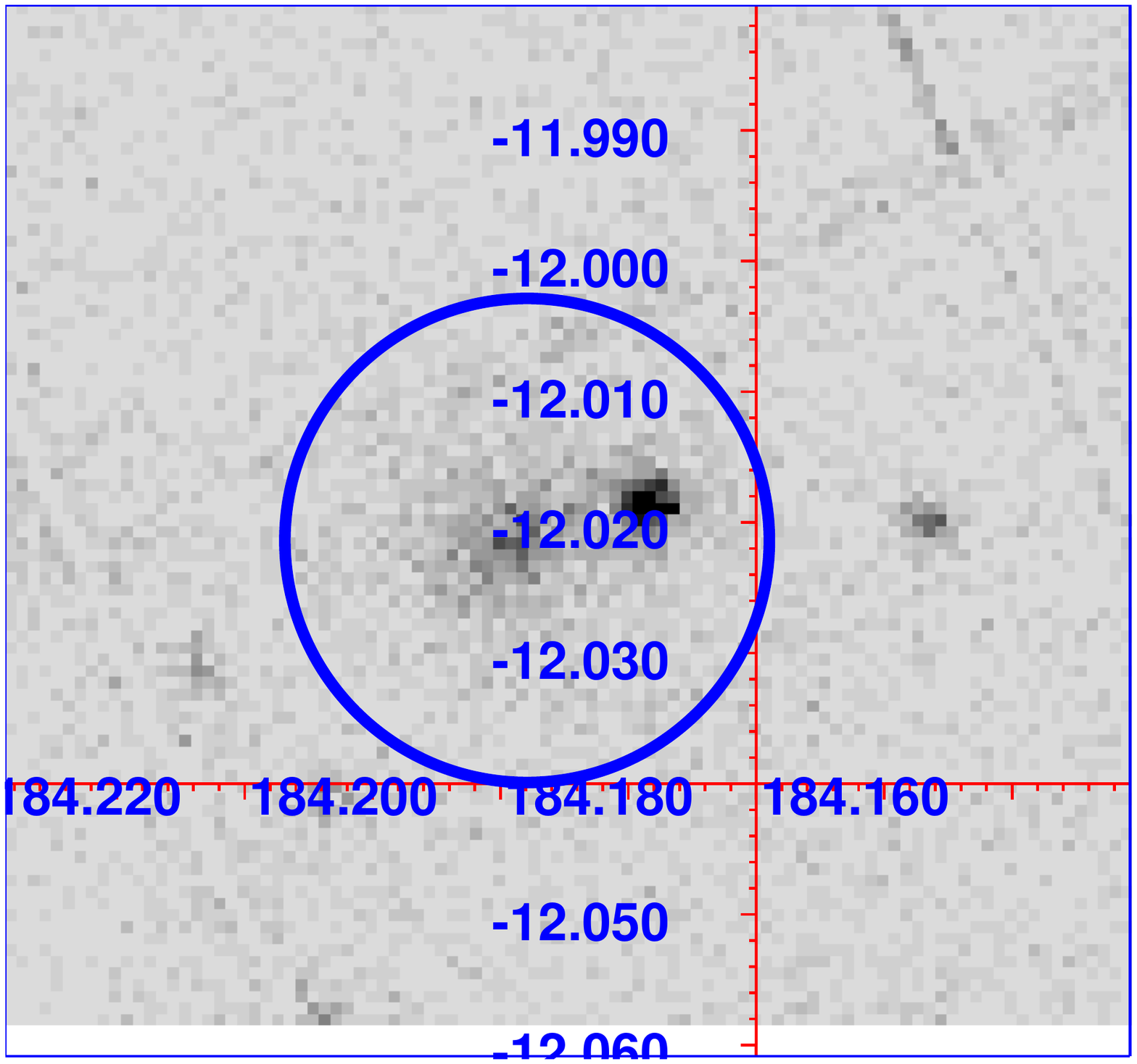}\includegraphics[width=2.7in,angle=0,bb=35 144 575 651,clip]{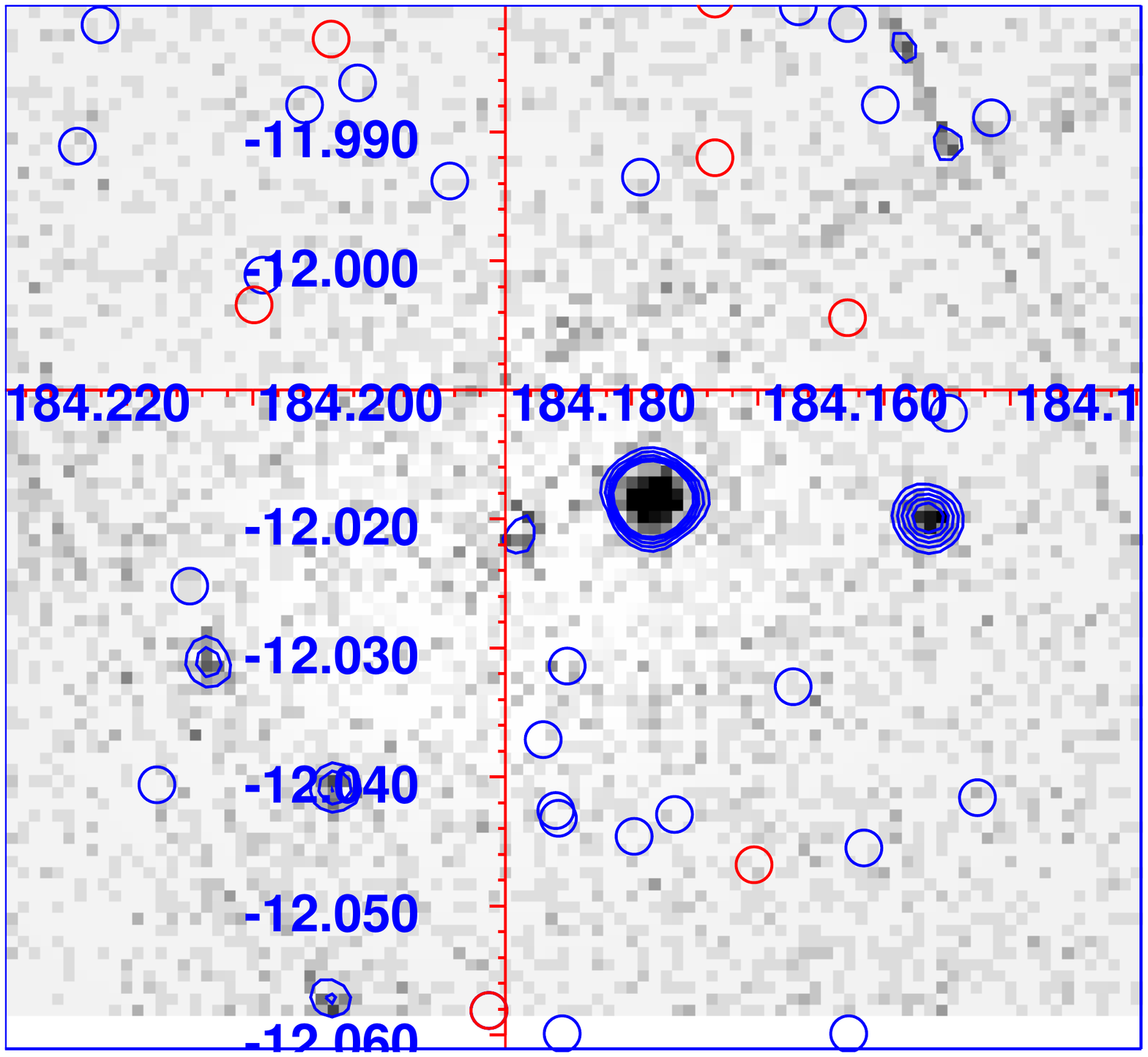}\\
    \includegraphics[width=2.7in,angle=0,bb=35 144 575 651,clip]{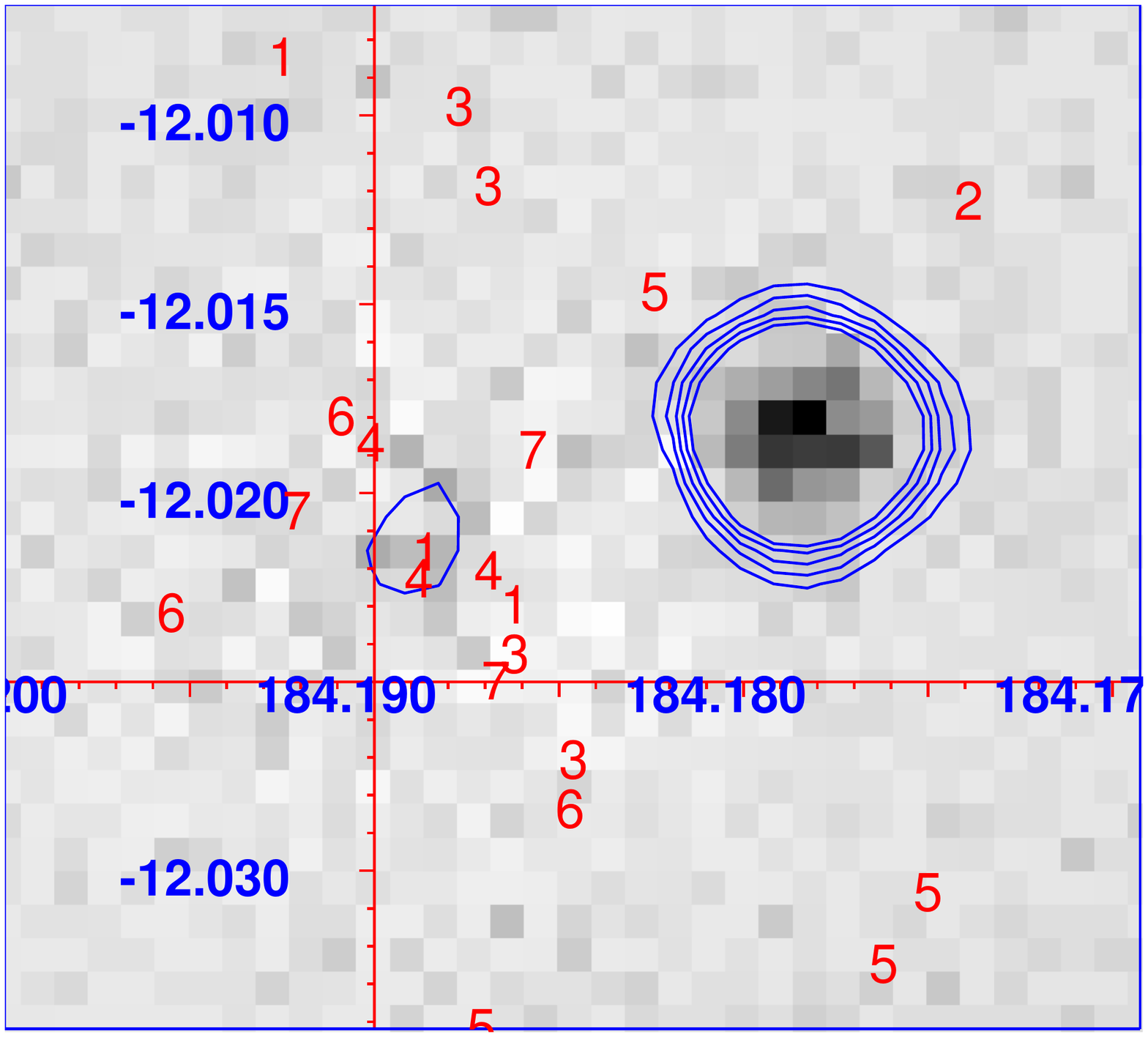}\includegraphics[width=2.7in,angle=0,bb=15 144 575 701,clip]{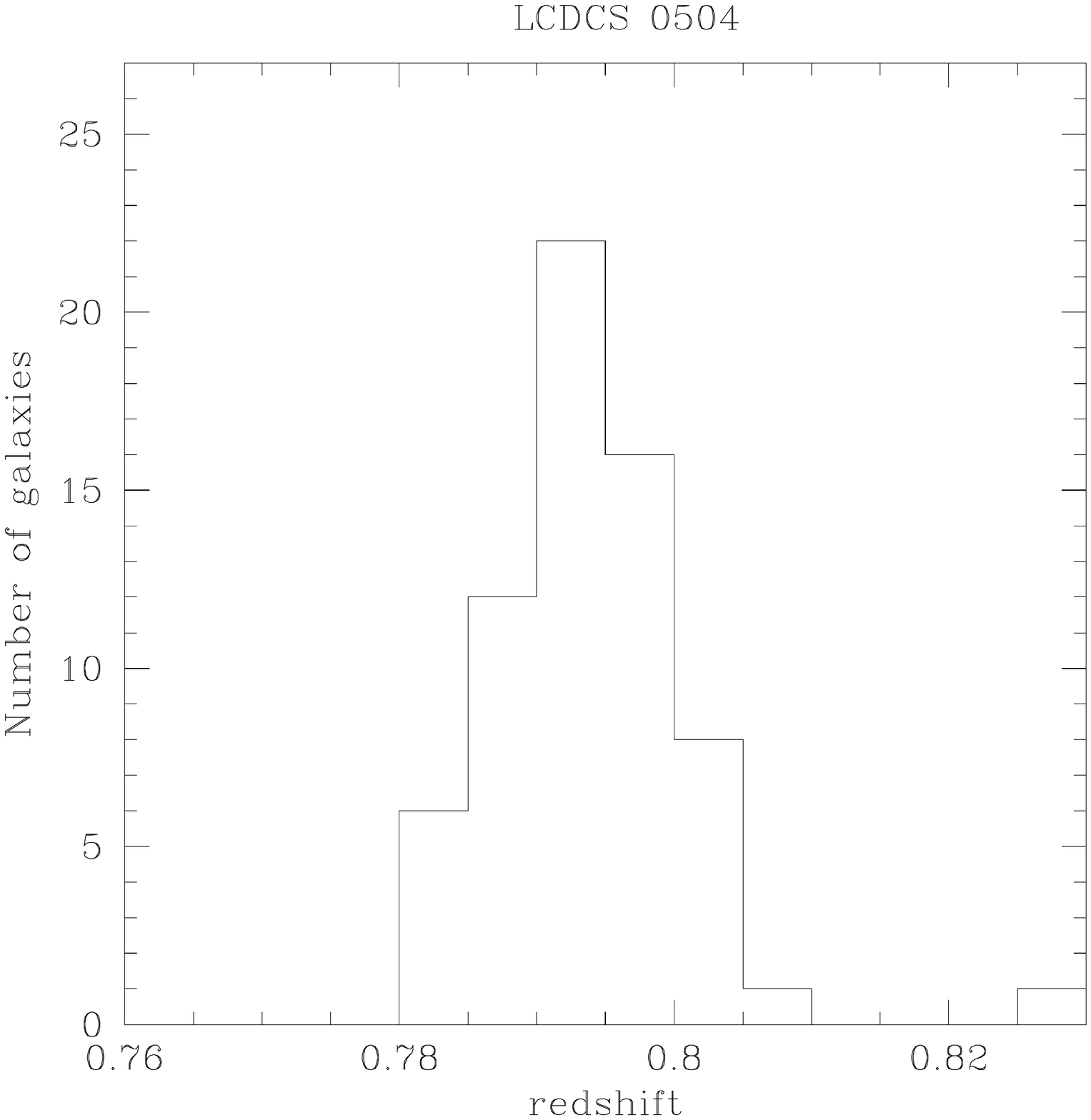}
    \caption{Same as Fig.~\ref{fig:cl0016_X} for LCDCS~0504.}
  \label{fig:LCDCS504_X}
  \end{center}
  \end{figure*}

The X-ray image of LCDCS~0504 shows a double structure
(Fig.~\ref{fig:LCDCS504_X}) after $\beta -$model subtraction. The
stronger source to the west
is likely to be a point-like source superimposed on the line of sight
(there is no Chandra image for this cluster, and no known AGN
along the line of sight).

We also have three other $\geq 3\sigma$ significant X-ray sources in
the residual map. None is associated with a galaxy structure detected
by the SG method on the line of sight. Since they are not close to the
cluster, they do not affect the substructure analysis.

On the optical side, there are 65 galaxies with redshifts in the
[0.78,0.81] range (Fig.~\ref{fig:LCDCS504_X}).  The characteristics of
the seven substructures found by the SG method are given in
Table~\ref{tab:SG}.  The eastern X-ray peak can be associated with SG1 or
SG4, but given the SG masses listed in Table~\ref{tab:SG}, we 
choose to associate the X-ray source with SG1 (because in view of the
relative masses of SG1 and SG4, SG4 would not be detected in
X-rays). This small group would therefore be located $\sim$5300 km/s
beyond the main cluster.

A more detailed analysis of this cluster will be presented elsewhere
(Guennou et al. 2013, in preparation).

\subsection{BMW-HRI~J122657.3+333253 (186.74167$^o$, 33.5484$^o$, z=0.8900)} 

\begin{figure*}
  \begin{center}
    \includegraphics[width=2.7in,angle=0,bb=35 144 575 651,clip]{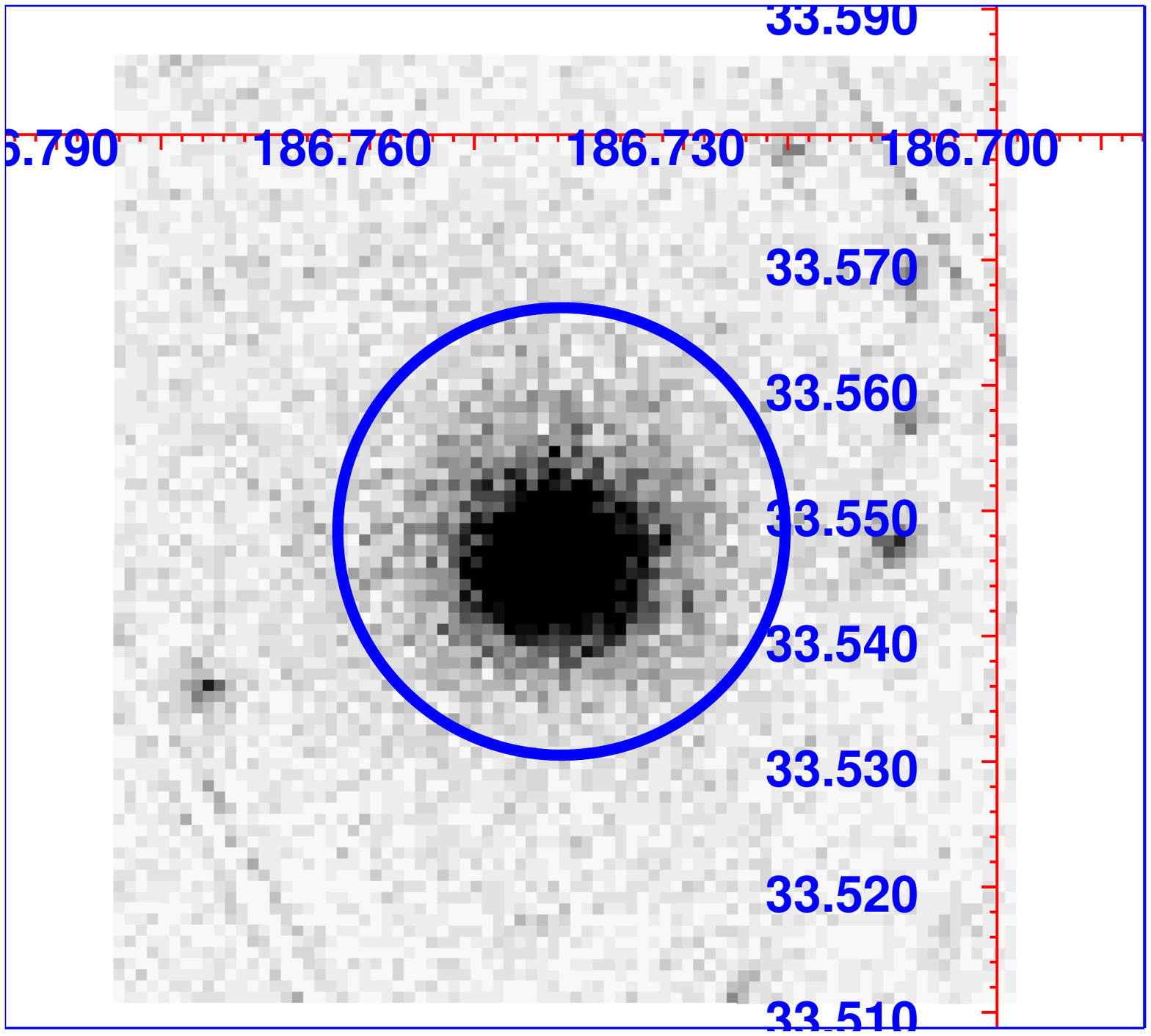}\includegraphics[width=2.7in,angle=0,bb=35 144 575 651,clip]{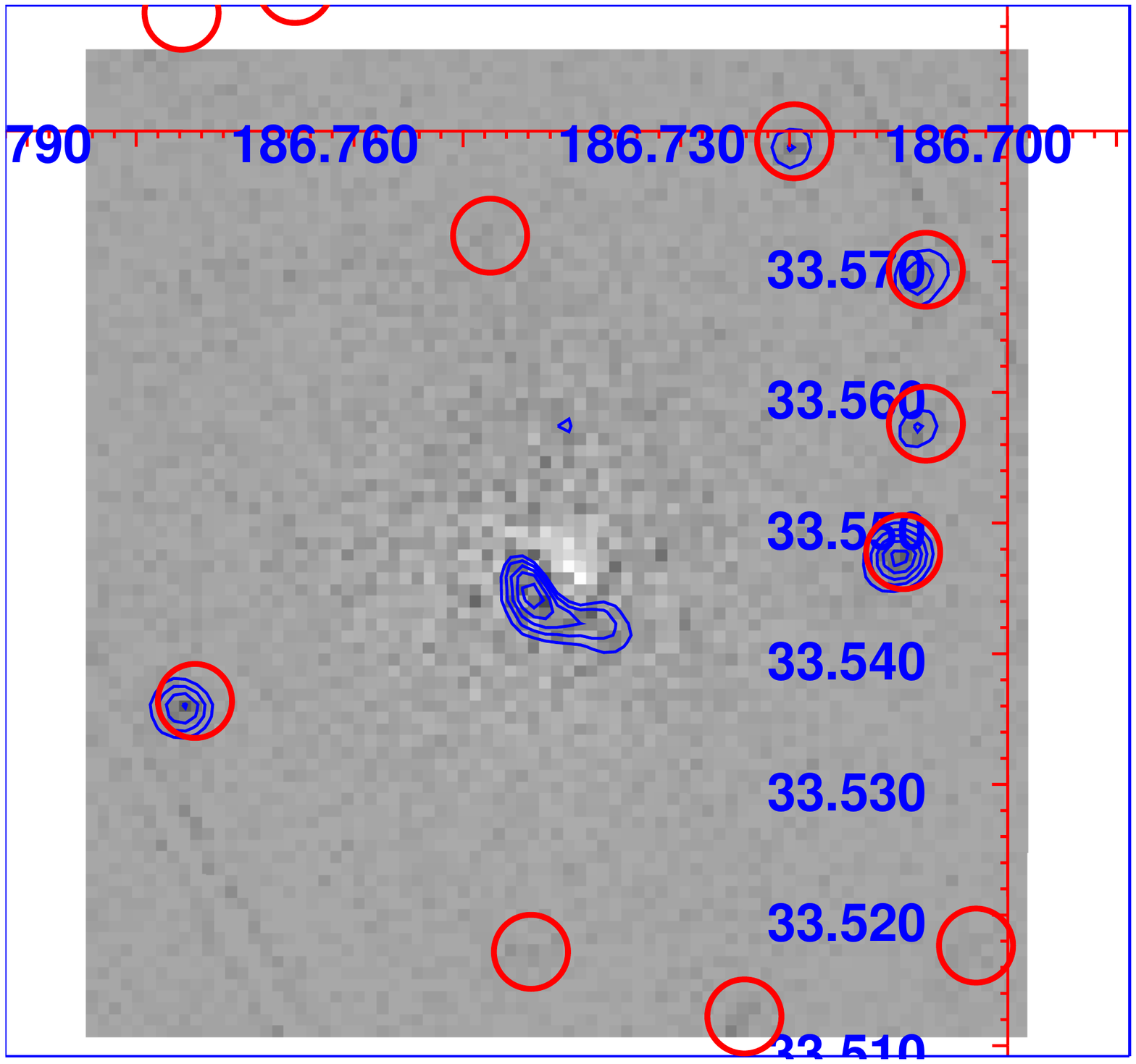}\\
    \includegraphics[width=2.7in,angle=0,bb=35 144 575 651,clip]{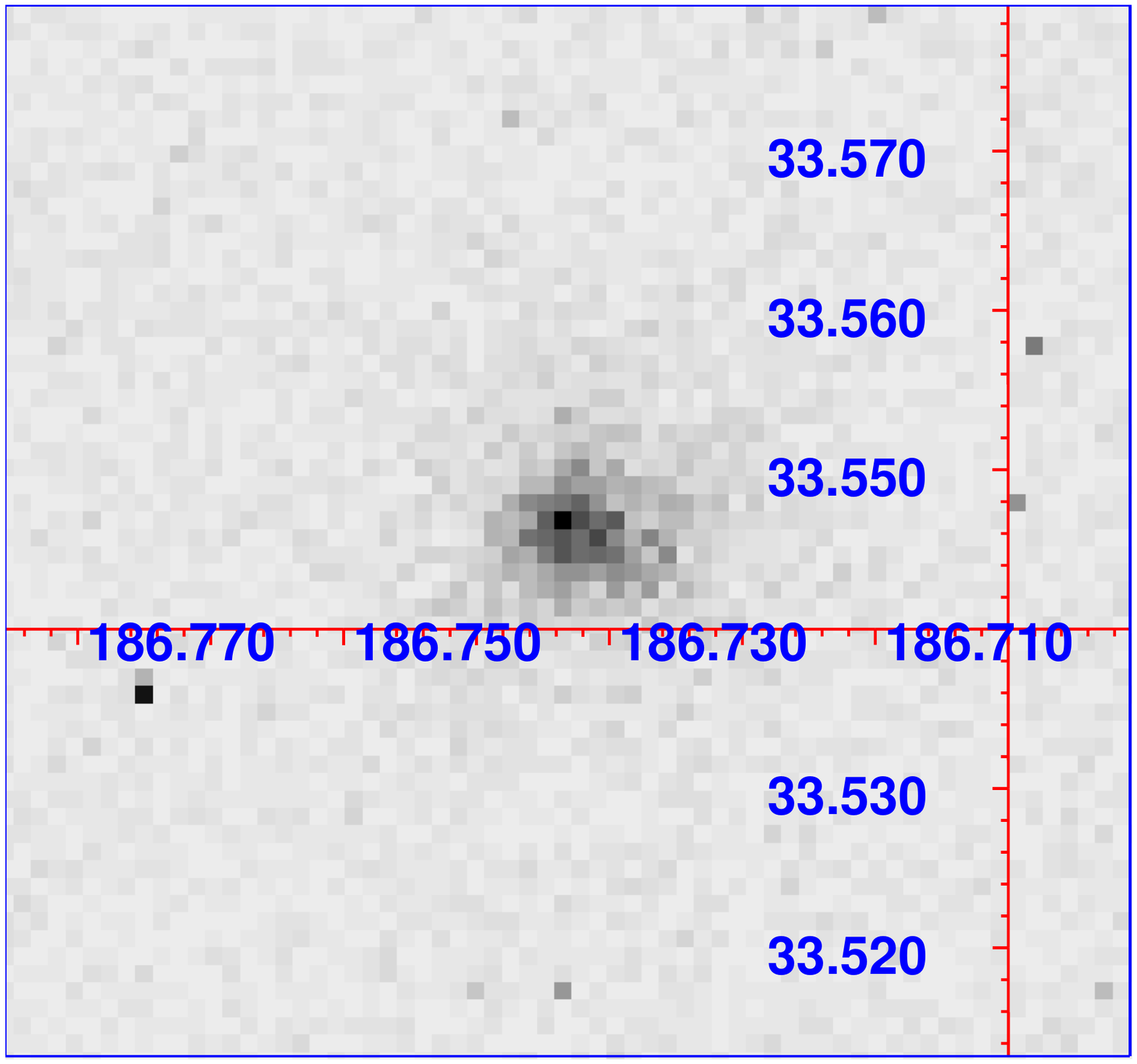}\includegraphics[width=2.7in,angle=0,bb=35 144 575 651,clip]{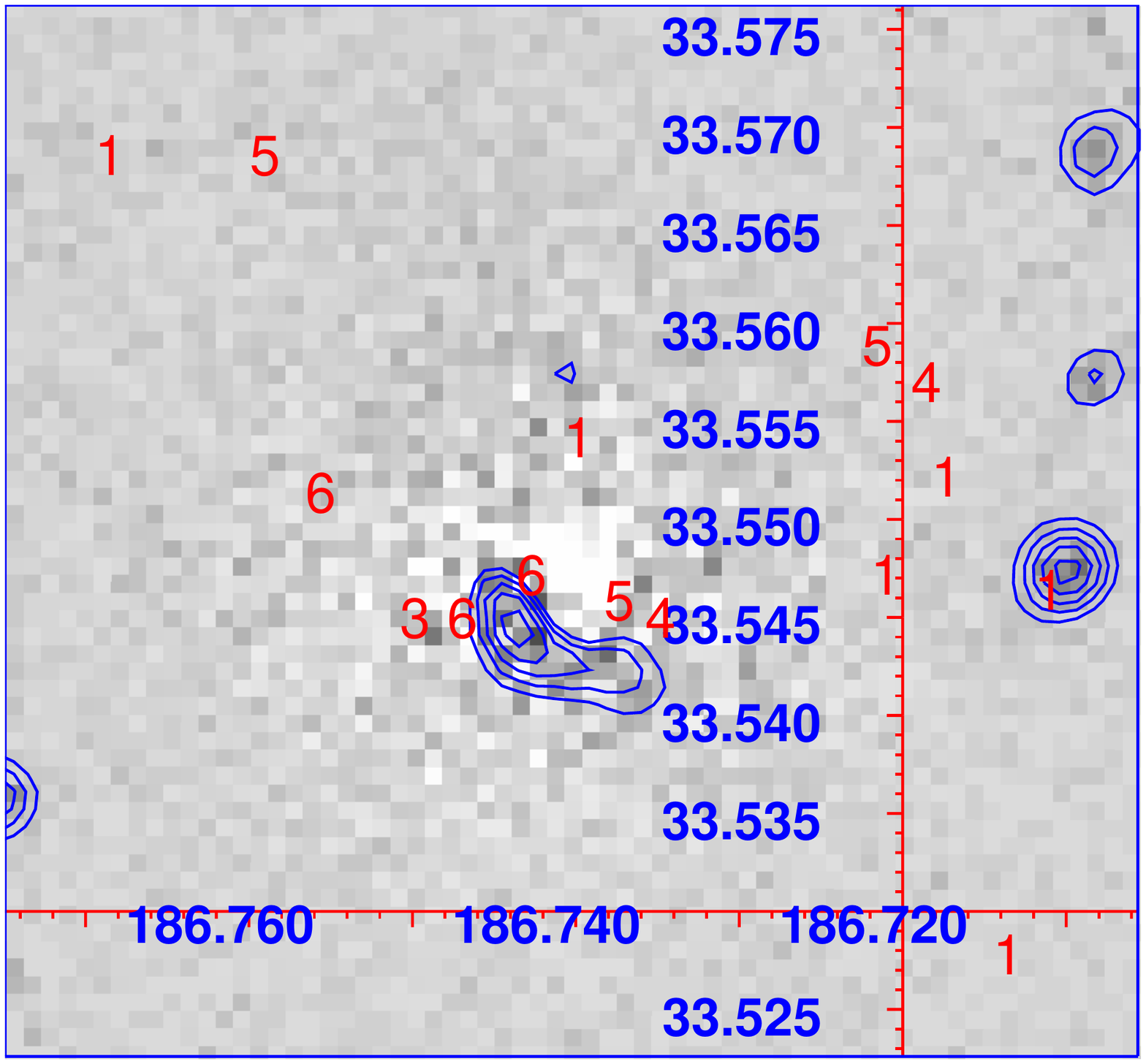}\\
    \includegraphics[width=2.7in,angle=0,bb=15 144 575 701,clip]{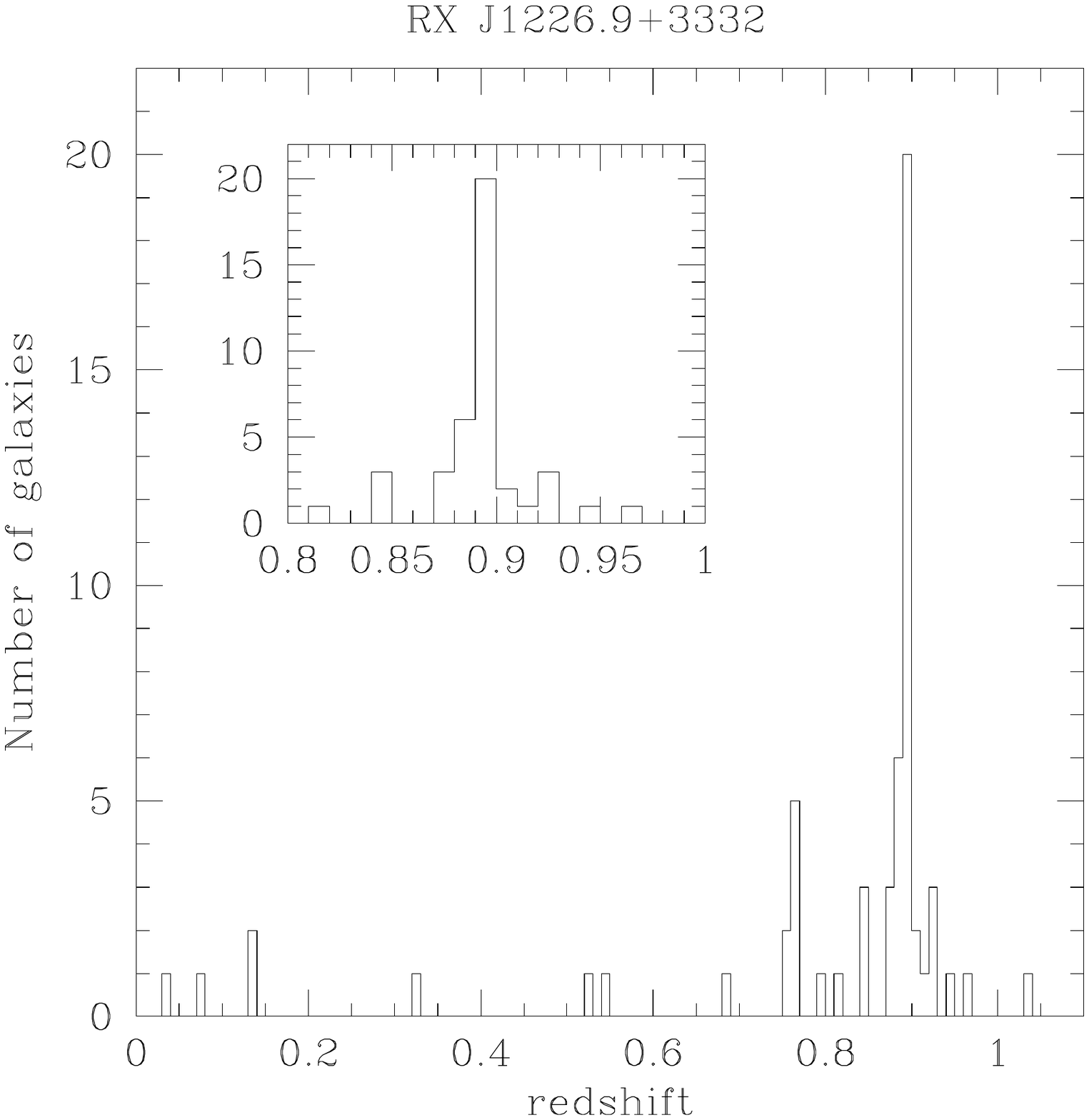}
    \caption{Same as Fig.~\ref{fig:cl0152_X} for BMW-HRI~J122657.3+333253.}
  \label{fig:RX1226_X}
  \end{center}
  \end{figure*}

  BMW-HRI~J122657.3+333253 was discovered in X-rays by Ebeling et
  al. (2001b). Its residual X-ray image shows several compact sources,
  all known as AGN (Gilmour et al. 2009), and most of them being
  visible as point sources in the Chandra image.  In addition, we
  detect a more extended source, directly south of the cluster itself,
  which is not correlated with any known structure along the line of
  sight.  Jee \& Tyson (2009) made a weak lensing mass reconstruction
  of this cluster and also found a subclump about 40~arcsec
  south--west of the cluster centre. This position corresponds to a
  temperature enhancement reported by Maughan et al. (2007). It is
  consistent with features detected in SZ with MUSTANG by Korngut et
  al. (2011), who interpret this cluster in a merger scenario where a
  small cluster has crossed a larger one on a trajectory oriented
  towards the south--west.

\clearpage

Even if it is not straightforward to assign this excess X-ray emission
to one of the SG detected substructures, it is likely that group SG6
(see Table~\ref{tab:SG}) is the optical manifestation of this
emission. However, this group is too sparsely sampled (only 3
redshifts) to provide a velocity dispersion. Its mean redshift puts it
$\sim$4600km/s beyond the cluster's main core (SG1).

\subsection{GHO 1322+3027 (201.20091$^o$, +30.1928$^o$, z=0.7550)}  

\begin{figure*}
  \begin{center}
    \includegraphics[width=2.7in,angle=0,bb=35 144 575 651,clip]{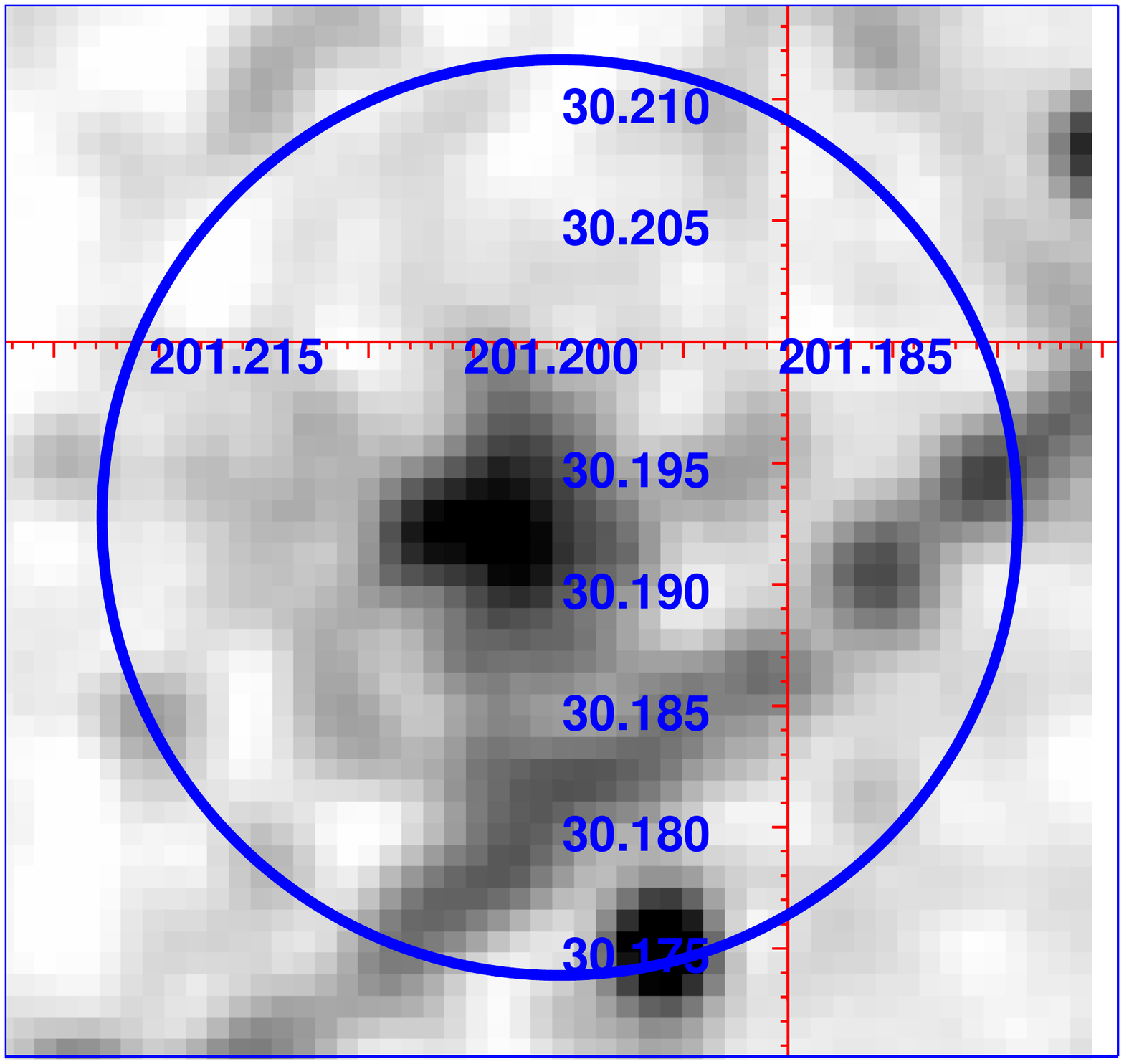}\includegraphics[width=2.7in,angle=0,bb=35 144 575 651,clip]{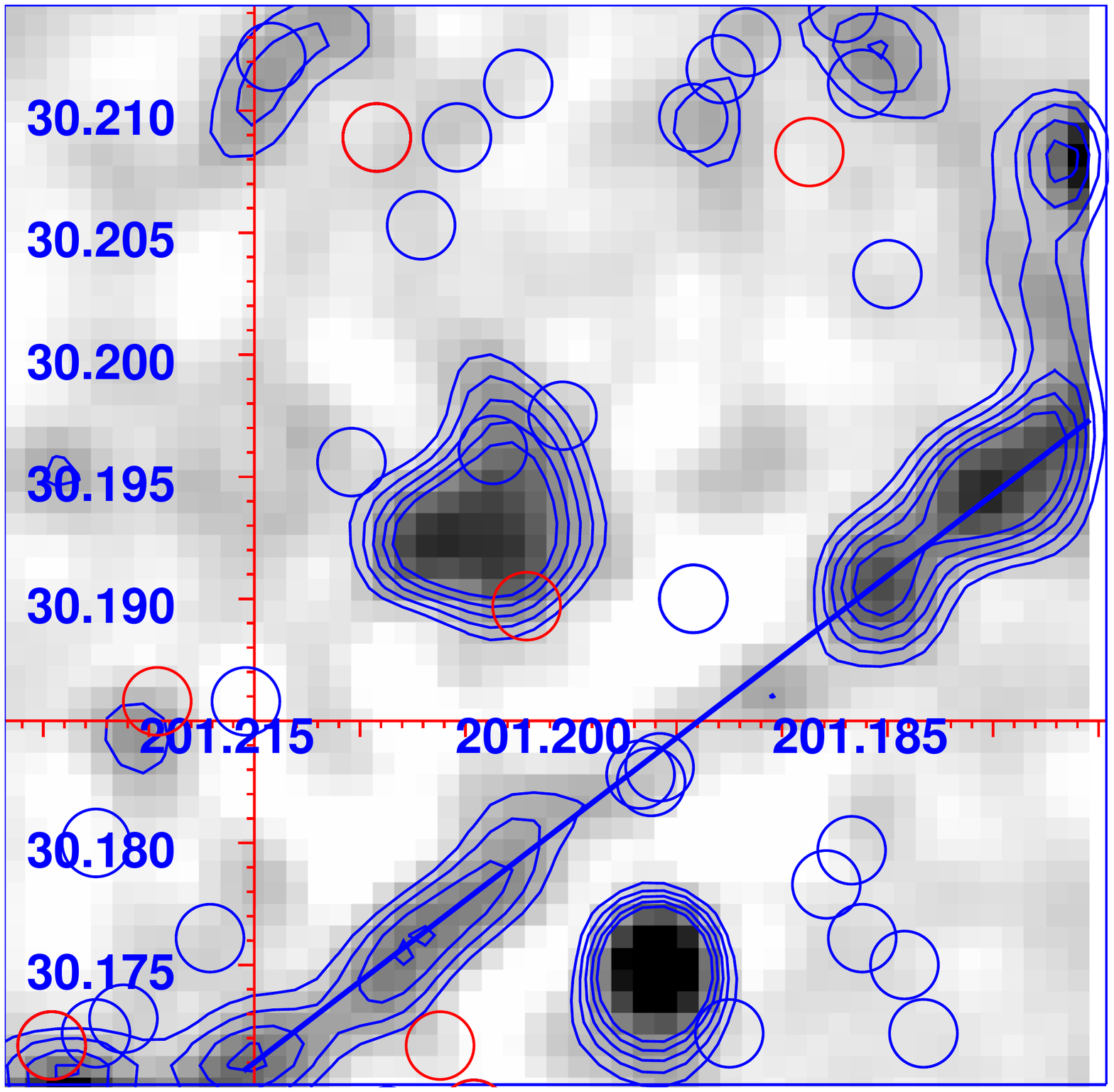}\\
    \includegraphics[width=2.7in,angle=0,bb=35 144 575 651,clip]{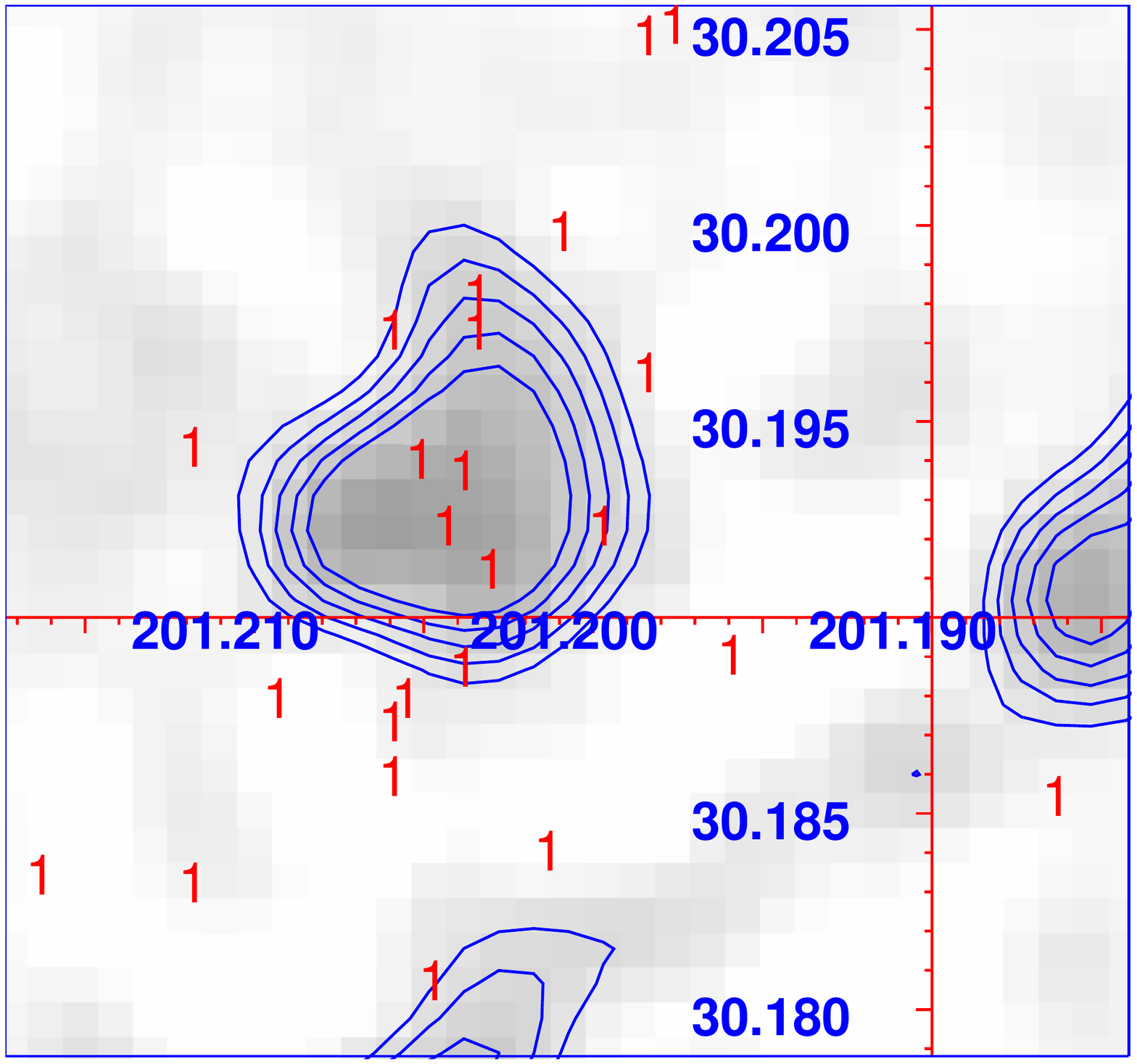}\includegraphics[width=2.7in,angle=0,bb=15 144 575 701,clip]{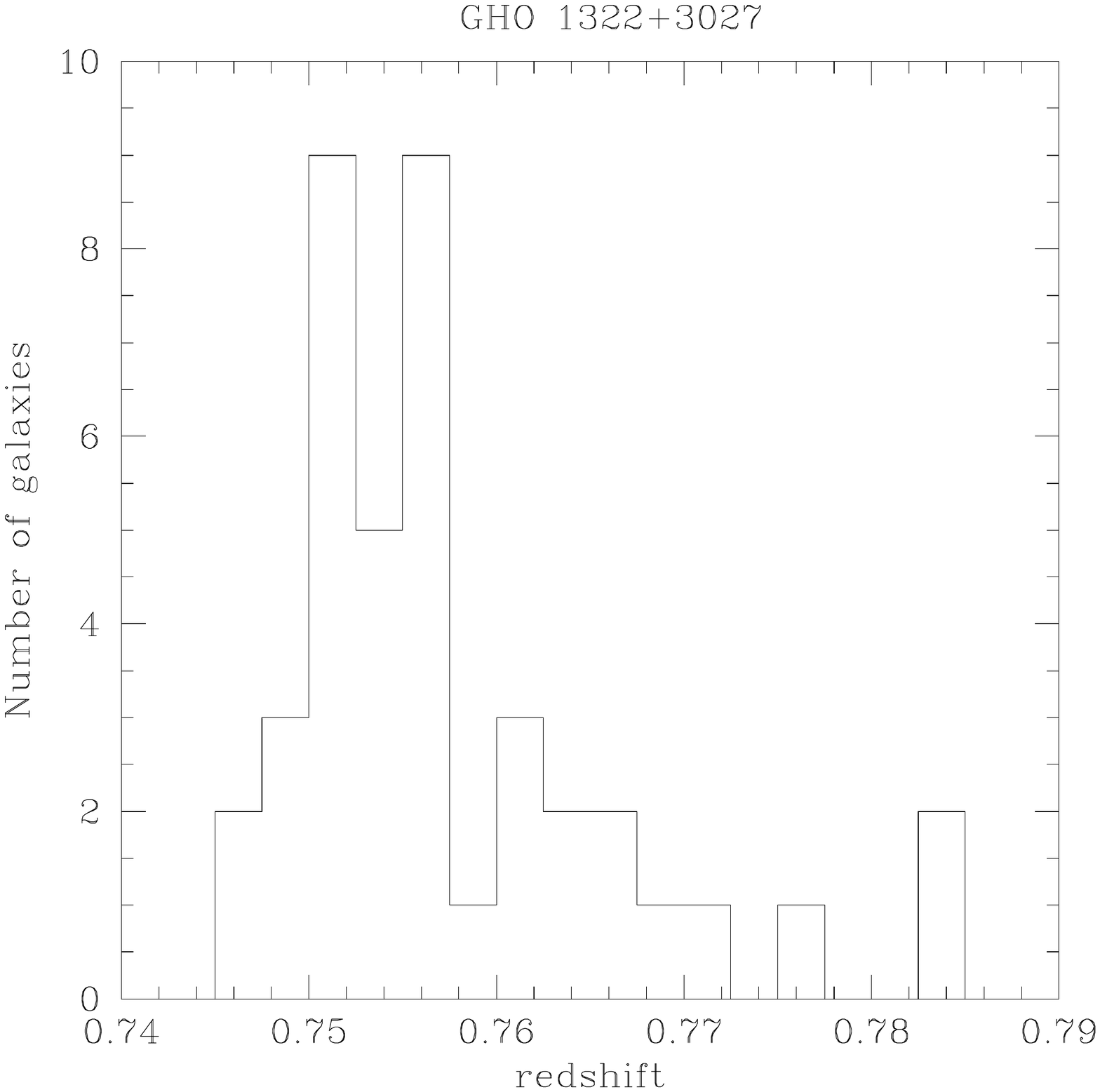}
    \caption{Same as Fig.~\ref{fig:cl0016_X} for GHO 1322+3027.}
  \label{fig:GHO1322_X}
  \end{center}
  \end{figure*}

  The residual X-ray image along the GHO~1322+3027 line of sight shows
  several X-ray sources (Fig.~\ref{fig:GHO1322_X}) that are more
  significant than the 2.5$\sigma$ level. Part of them are explained
  by the MOS1 interchip gap separation
  (Fig.~\ref{fig:GHO1322_X}). There is a point-like X-ray source
  directly south of the cluster. This source is not known as an AGN in
  NED or Vizier, and without Chandra images available, we cannot be
  100$\%$ sure that we are really dealing with an AGN.  We also detect
  three poorly significant X-ray sources north of the cluster that are
  all very well correlated with galaxies belonging to foreground
  galaxy structures on the line of sight.  Finally, we detect in the
  residual X-ray image a very significant X-ray emission superimposed
  on the cluster centre. It is very likely that we are dealing with an
  imperfect $\beta -$model subtraction, but without Chandra data, it
  is difficult to conclude whether it is an active galaxy in the
  middle of the cluster (the cD?)  or if this cluster has a cool-core.

The redshift histogram is rather asymmetric
(Fig.~\ref{fig:GHO1322_X}) with 38 galaxies in the [0.745,0.775]
redshift range.  However the SG analysis detects a single massive
structure: the cluster itself.
GHO~1322+3027 therefore does not exhibit significant substructures.

\subsection{ZwCl~1332.8+5043 (203.58333$^o$, +50.5151$^o$, z=0.6200)} 

\begin{figure*}
  \begin{center}
    \includegraphics[width=2.7in,angle=0,bb=35 144 575 651,clip]{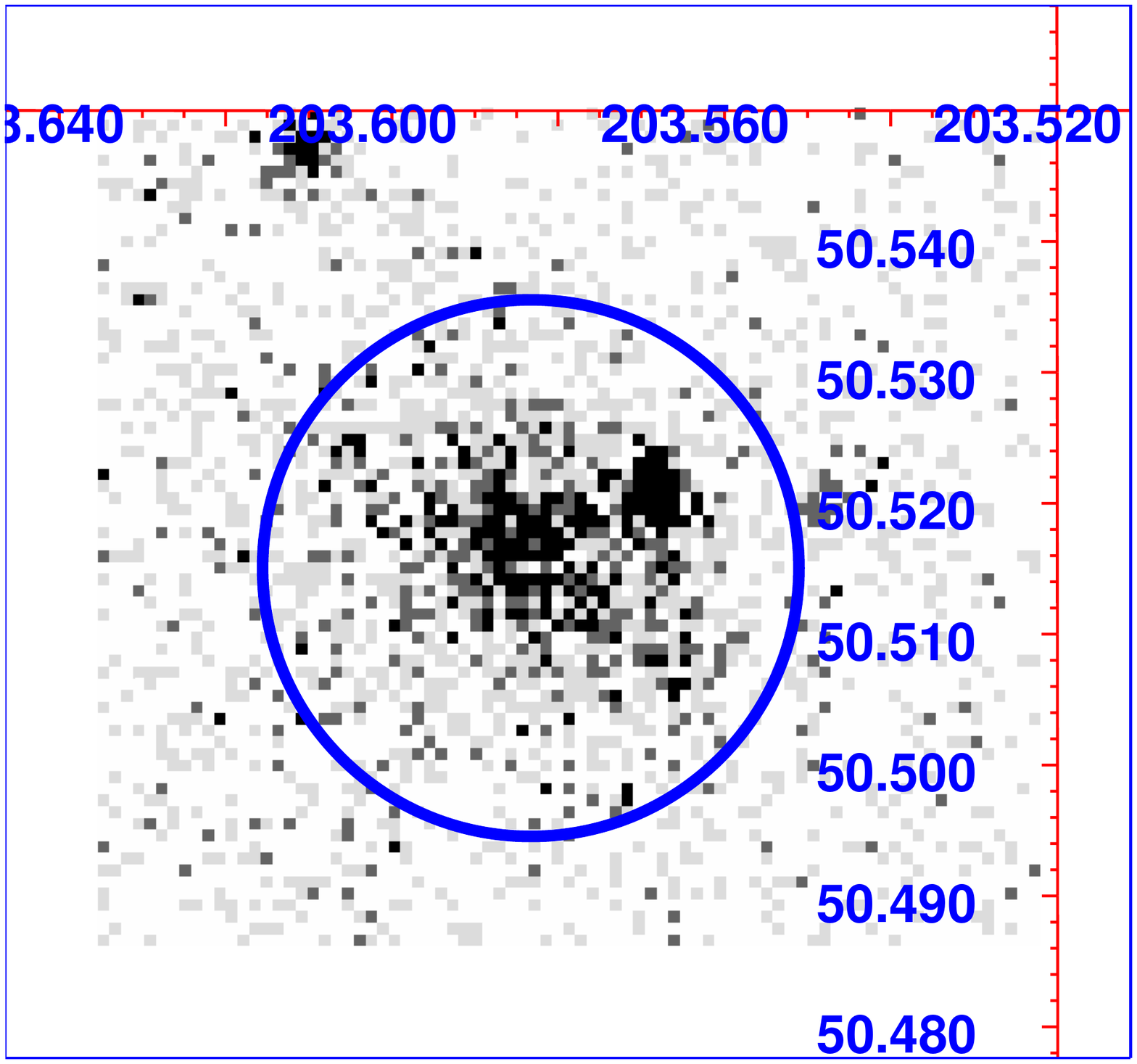}\includegraphics[width=2.7in,angle=0,bb=35 144 575 651,clip]{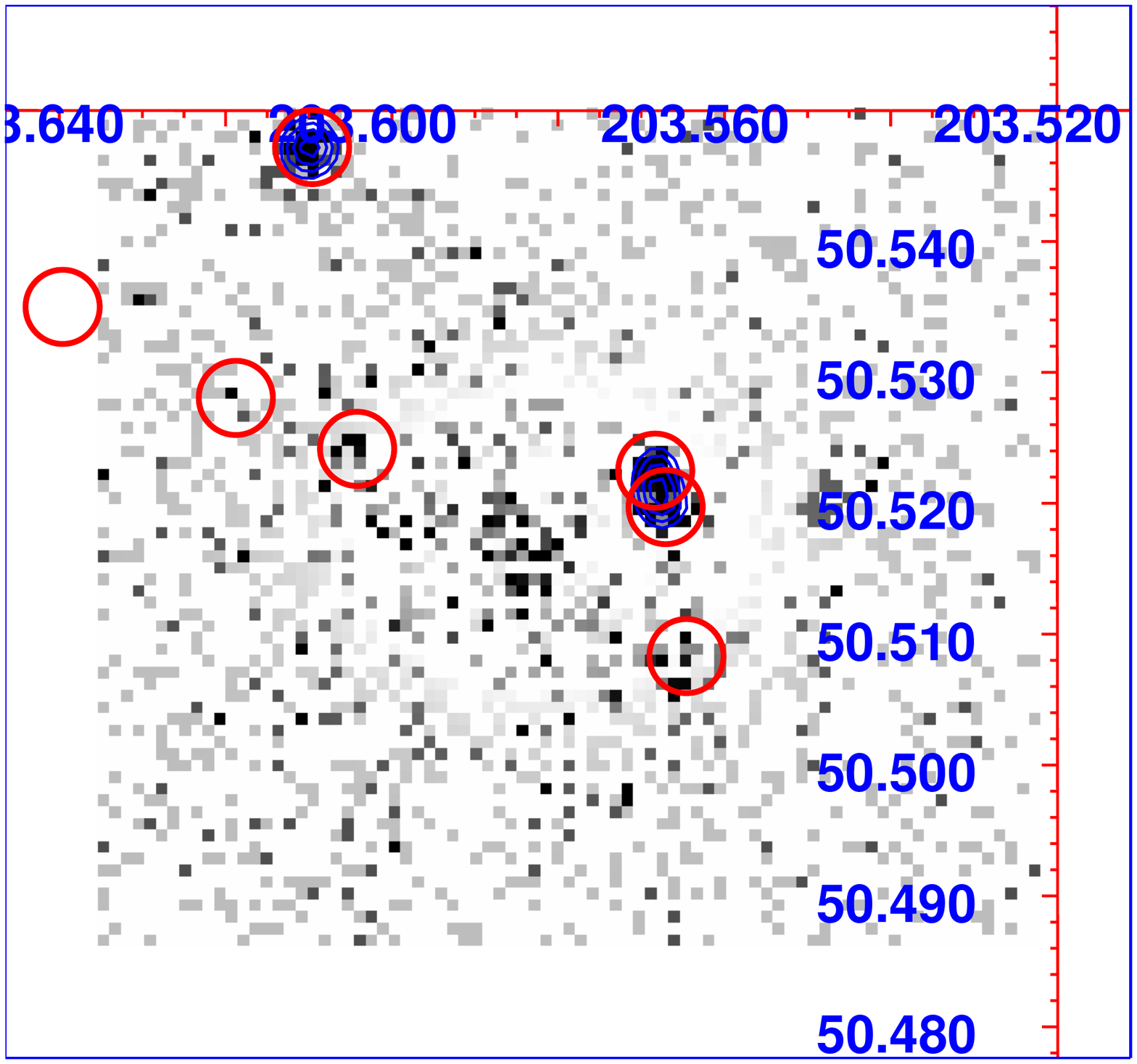}\\
    \includegraphics[width=2.7in,angle=0,bb=35 144 575 651,clip]{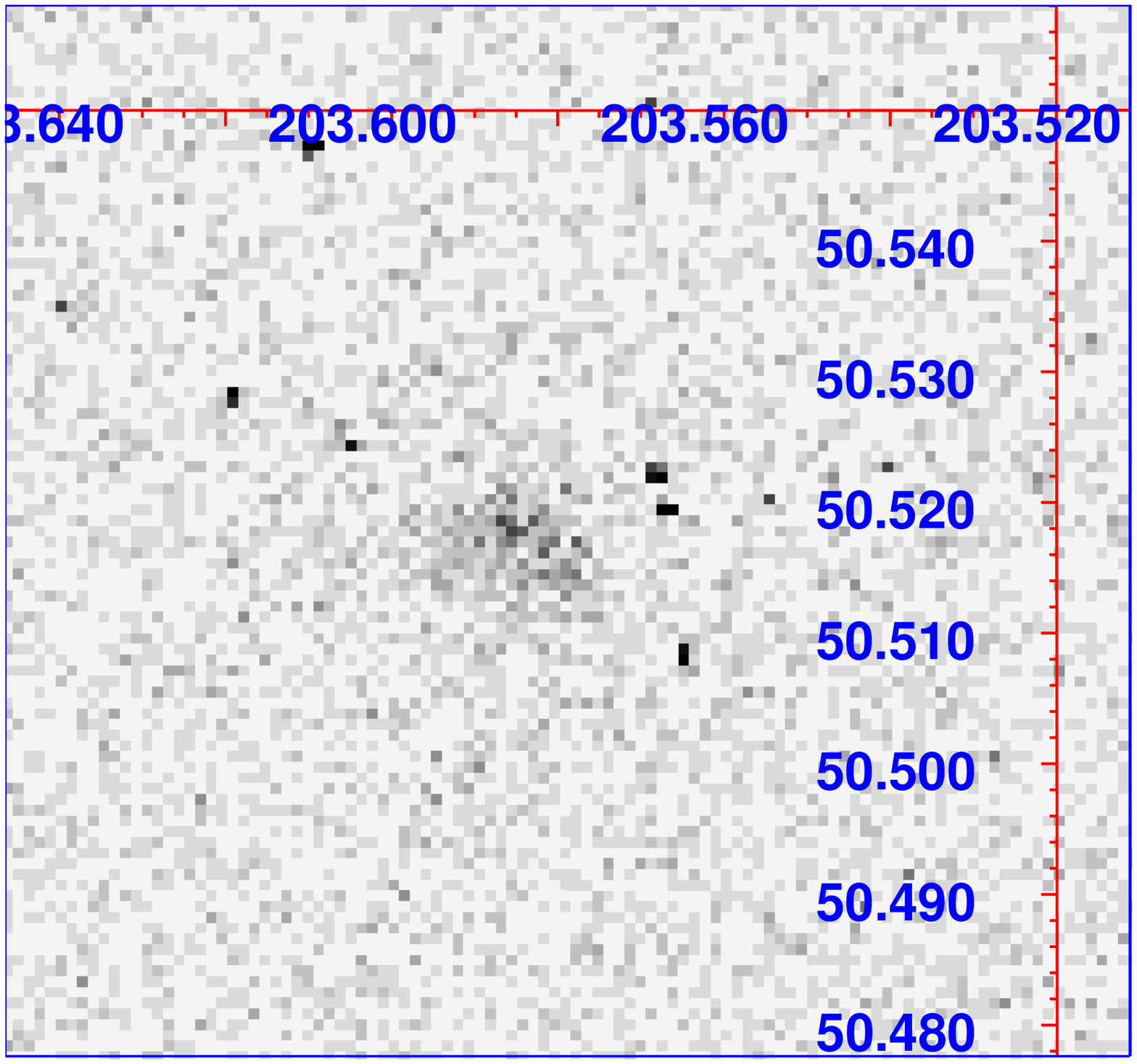}\includegraphics[width=2.7in,angle=0,bb=15 144 575 701,clip]{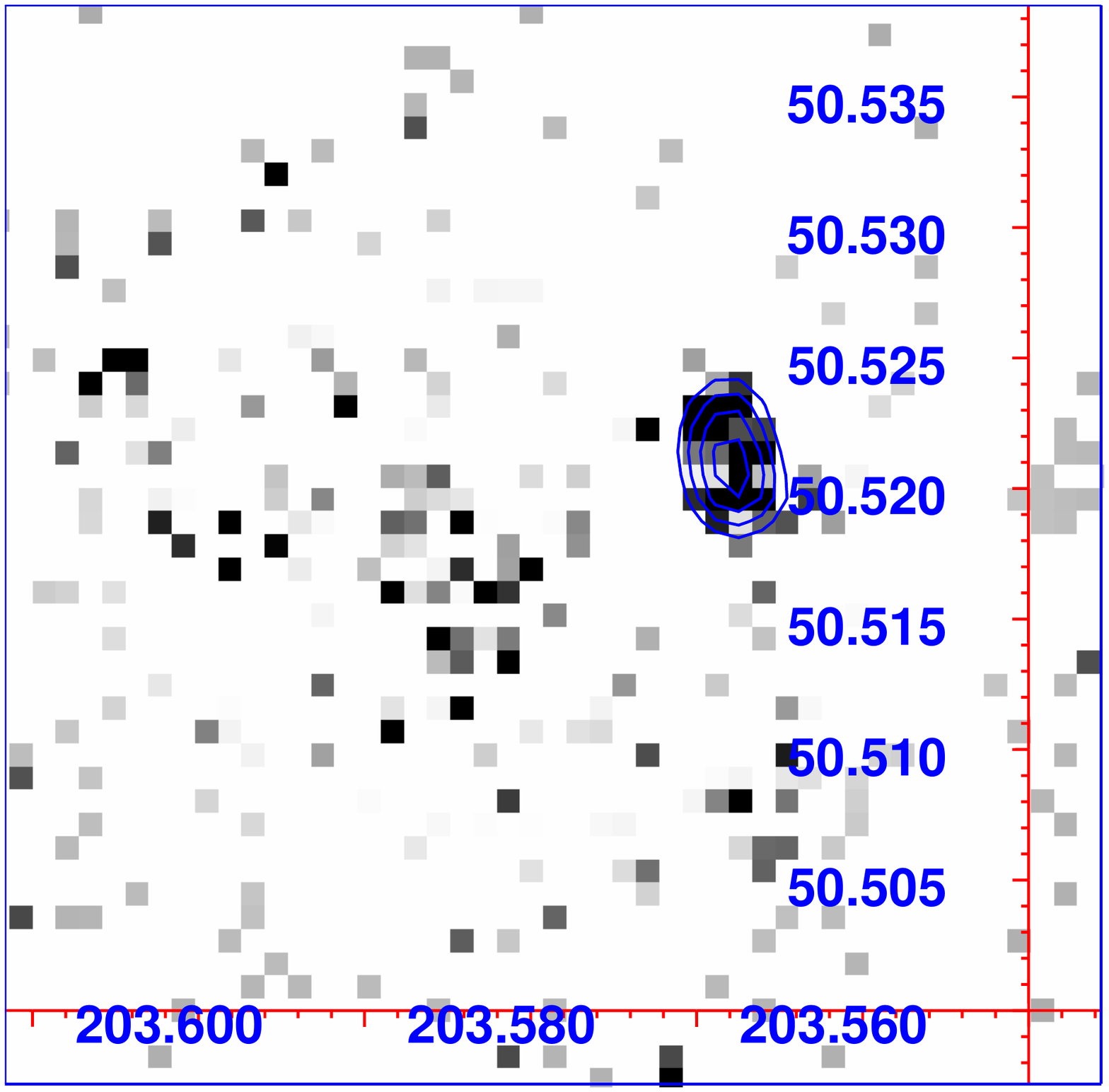}
    \caption{Same as Fig.~\ref{fig:bmw0522_X} for ZwCl~1332.8+5043.}
  \label{fig:Zw1332_X}
  \end{center}
  \end{figure*}

  The X-ray image of ZwCl~1332.8+5043 shows a diffuse source, and two
  compact sources to the north--west and north--east
  (Fig.~\ref{fig:Zw1332_X}). After model subtraction, only these two
  sources are left. The Chandra image confirms that we are dealing
  with a double point source to the north--west (known as two active
  galaxies in Gilmour et al.  2009) and a single point source to the
  north--east (also present in Gilmour et al.  2009). So we detect no
  substructures in this cluster.

NED provides a  single galaxy redshift, so the SG analysis is impossible.

\subsection{LCDCS~0829 = RXJ1347.5-1145 (206.88333, --11.7617$^o$, z=0.4510)} 

\begin{figure*}
  \begin{center}
    \includegraphics[width=2.7in,angle=0,bb=35 144 575 651,clip]{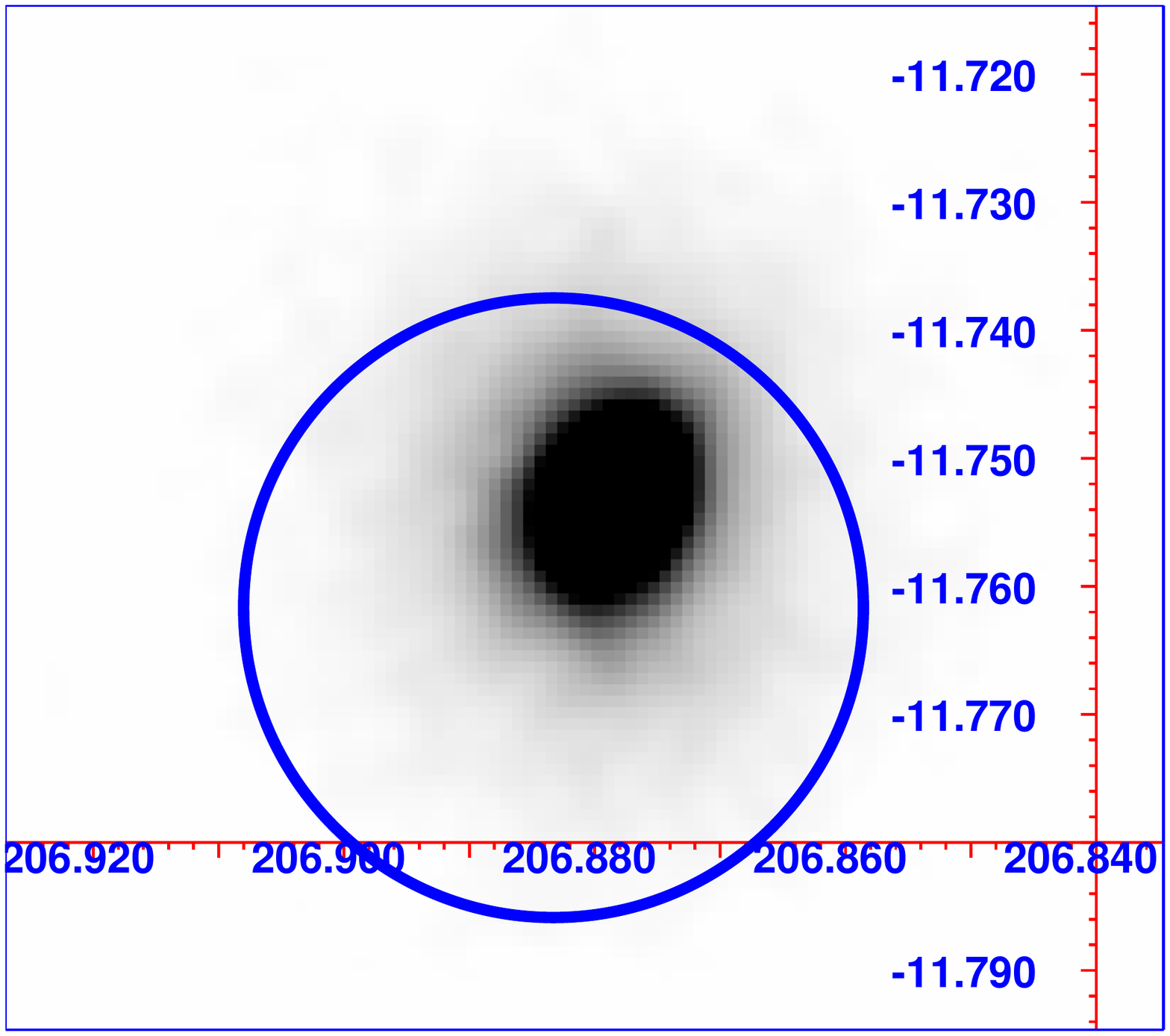}\includegraphics[width=2.7in,angle=0,bb=35 144 575 651,clip]{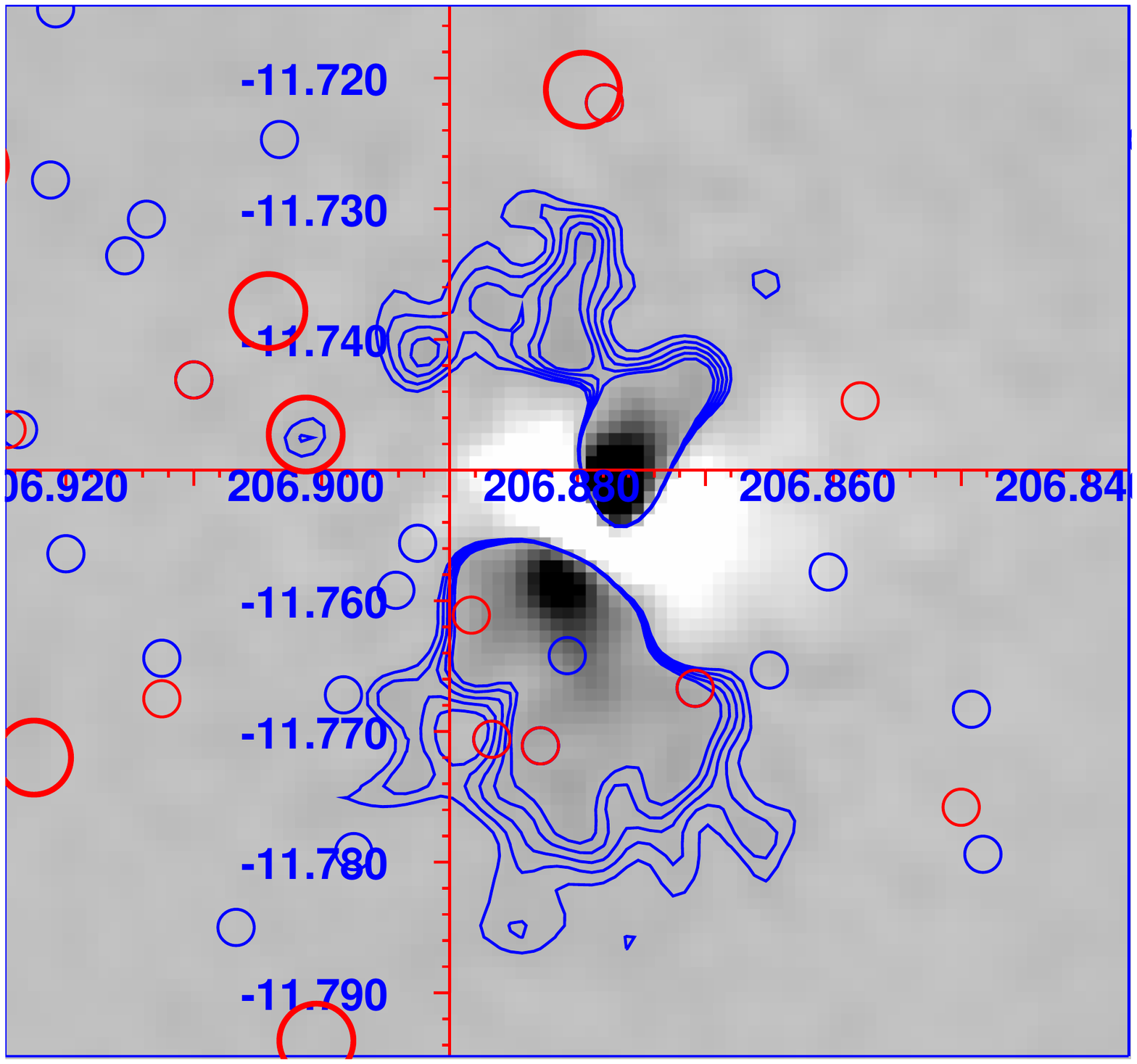}\\
    \includegraphics[width=2.7in,angle=0,bb=35 144 575 651,clip]{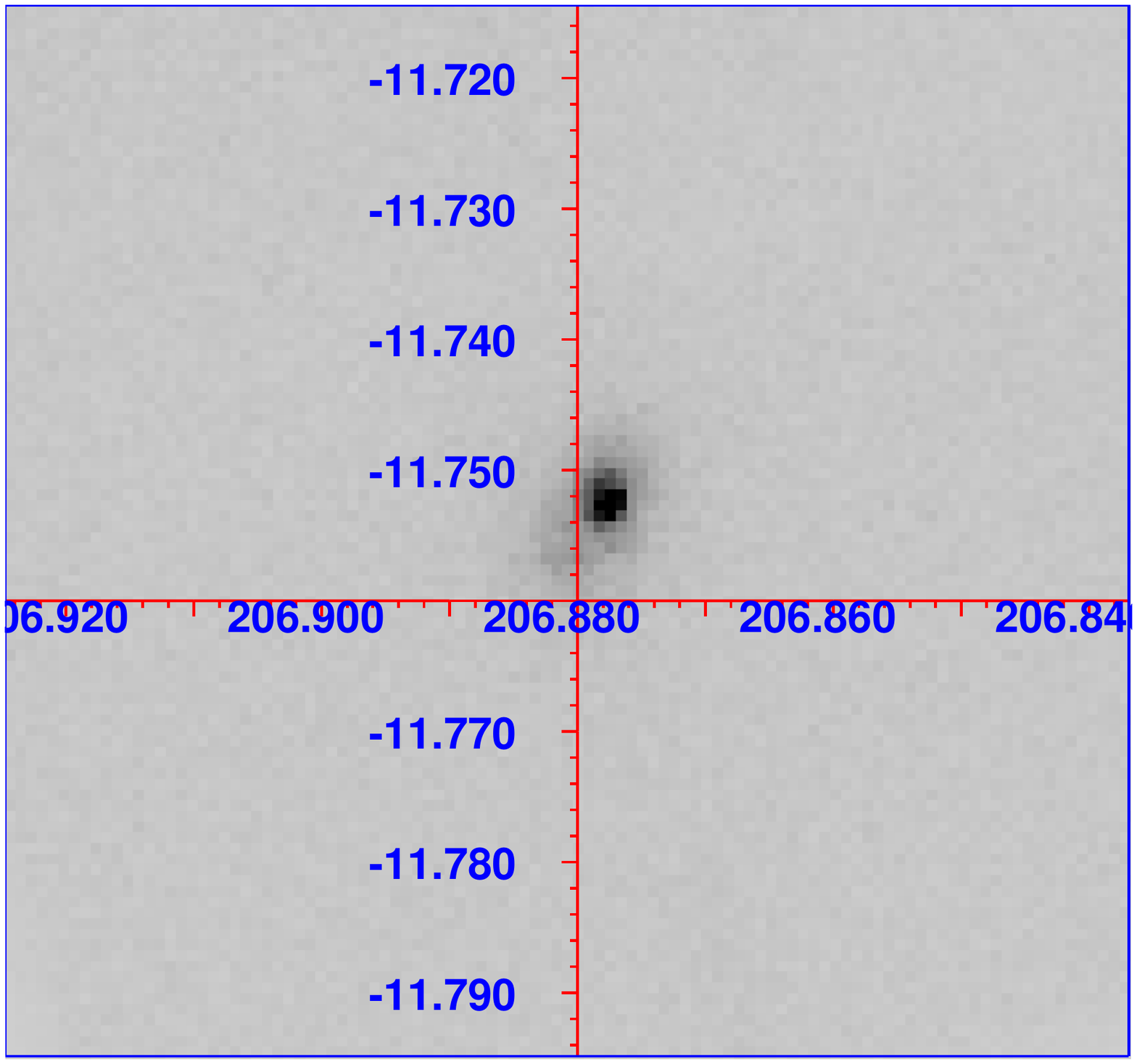}\includegraphics[width=2.7in,angle=0,bb=35 144 575 651,clip]{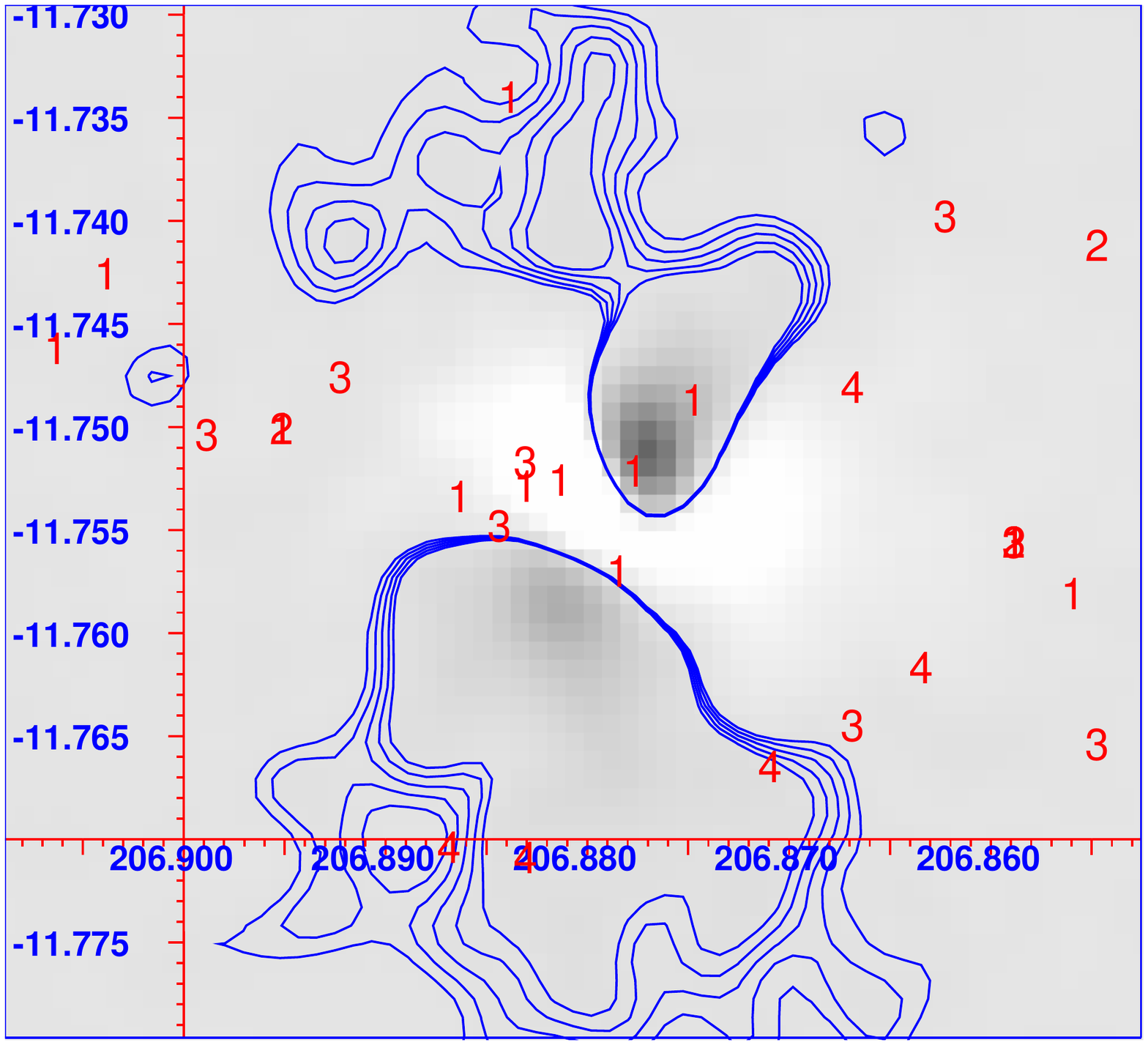}\\
    \includegraphics[width=2.7in,angle=0,bb=15 144 575 701,clip]{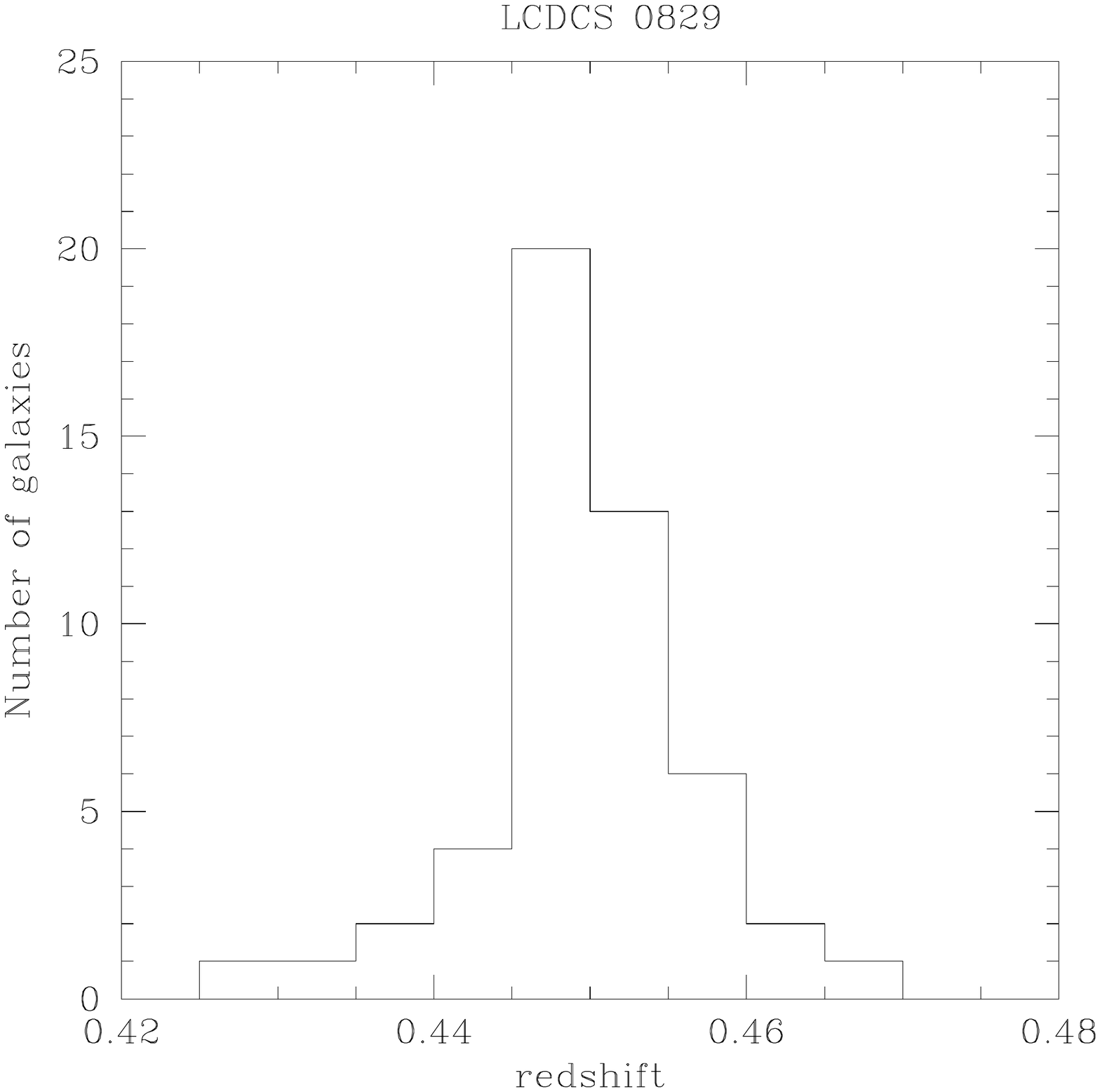}
    \caption{Same as Fig.~\ref{fig:cl0152_X} for LCDCS~0829.}
  \label{fig:LCDCS829_X}
  \end{center}
  \end{figure*}

  LCDCS~0829, also known as RXJ1347.5-1145, is a very hot and luminous
  X-ray cluster known to show several gravitational arcs (Schindler et
  al. 1995).  The XMM-Newton image of LCDCS~0829 is indeed bright
  (Fig.~\ref{fig:LCDCS829_X}), and the subtraction of a $\beta -$model
  shows excess emission south and north of the cluster. The northern
  part is probably the superposition of diffuse emission and of a
  point source (visible in the Chandra image). The southern excess
  roughly corresponds spatially to the emission detected in SZ by
  Komatsu et al. (2001), as later confirmed with better data by
  Korngut et al. (2011).  These two diffuse sources suggest there is a
  recent merger (Korngut et al. 2011 and references therein).  We also
  see a small deficit of emission just north of the cluster centre.

  On the optical side, the redshift histogram is somewhat asymmetric
  (Fig.~\ref{fig:LCDCS829_X}), with 50 galaxies with redshifts in the
  [0.425,0.47] range. The SG method first detects a large group of 50
  galaxies (not listed in Table~\ref{tab:SG}) at z$\sim$0.45 with an
  estimated mass of 3.83~10$^{14}$~M$_\odot$. This group is then split
  into four smaller substructures (see Table~\ref{tab:SG}). Among them,
  SG1 seems associated with the northern extended emission, while SG4
  seems associated with the southern extended emission.

\subsection{LCDCS~0853 (208.53958$^o$, --12.5164$^o$, z=0.7627)} 

\begin{figure*}
  \begin{center}
    \includegraphics[width=2.7in,angle=0,bb=35 144 575 651,clip]{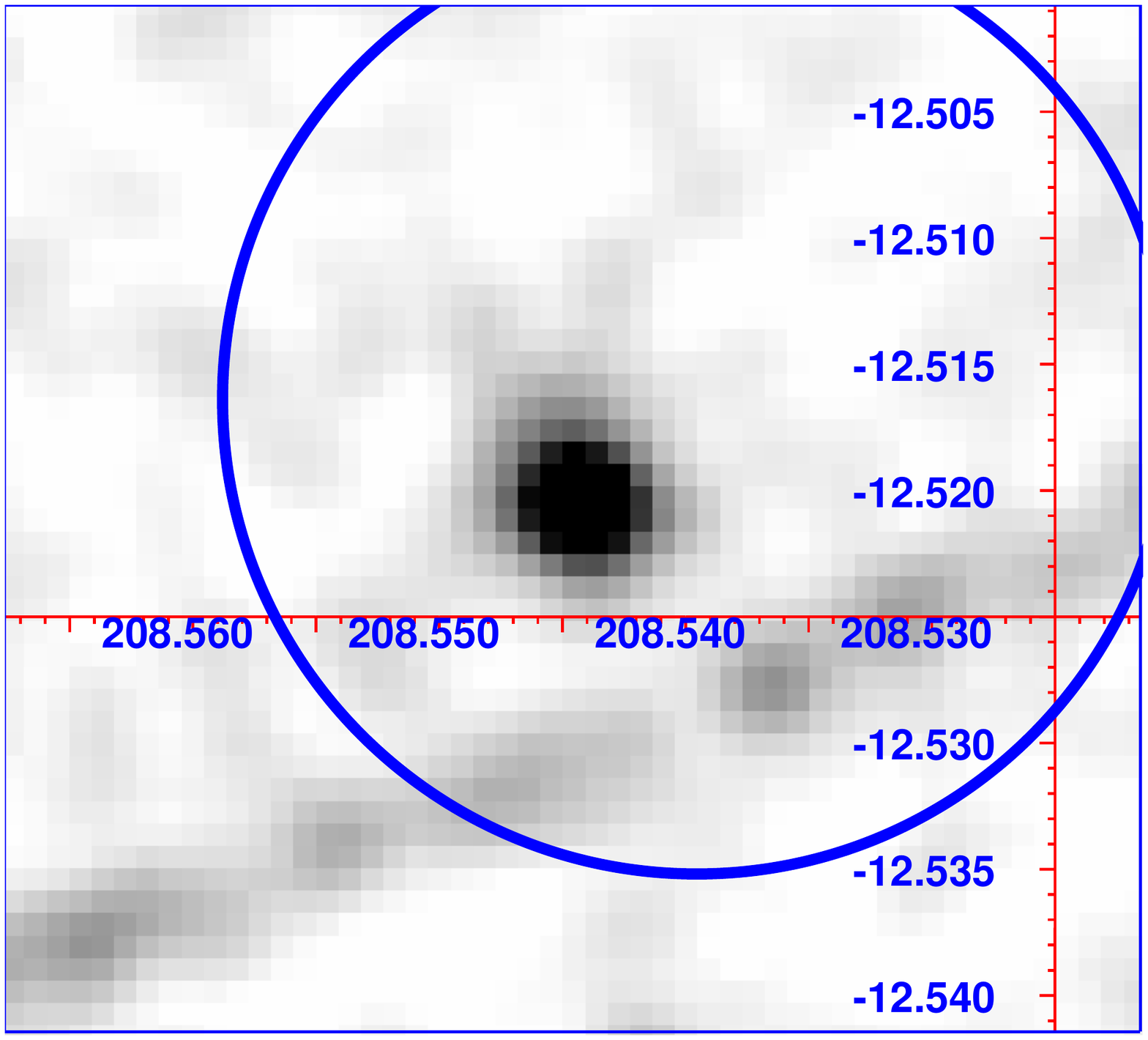}\includegraphics[width=2.7in,angle=0,bb=35 144 575 651,clip]{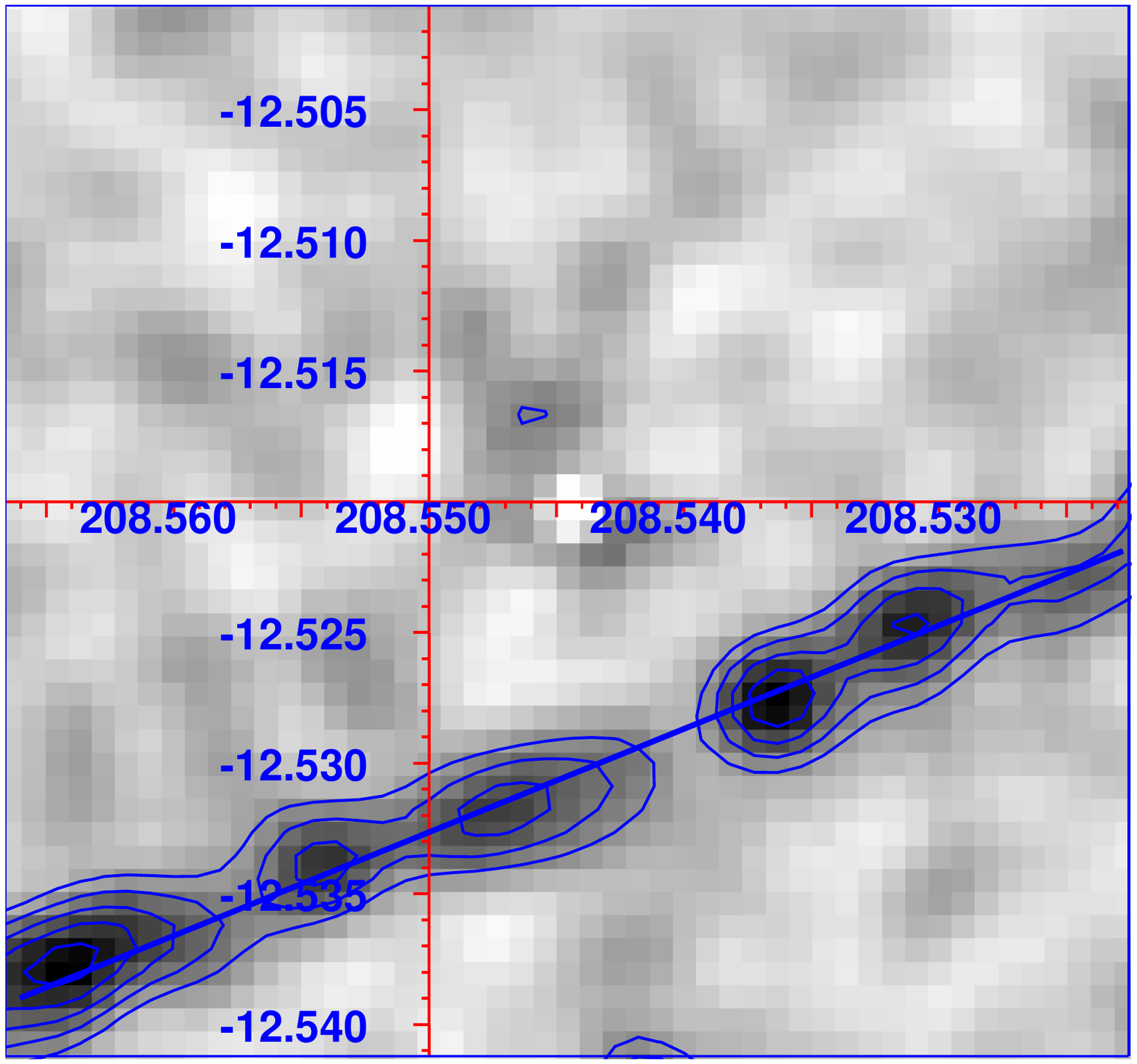}\\
    \includegraphics[width=2.7in,angle=0,bb=35 144 575 651,clip]{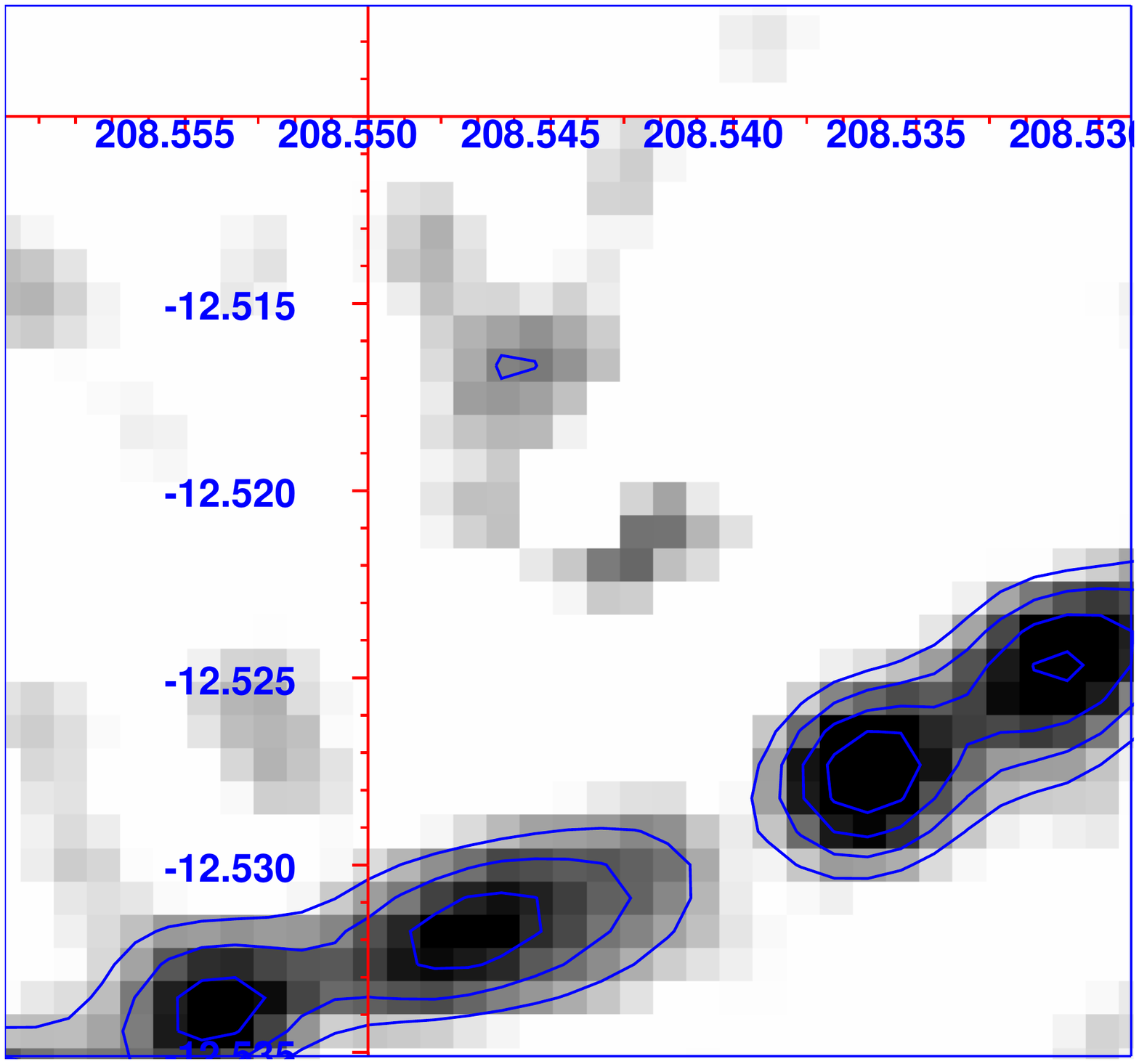}\includegraphics[width=2.7in,angle=0,bb=15 144 575 701,clip]{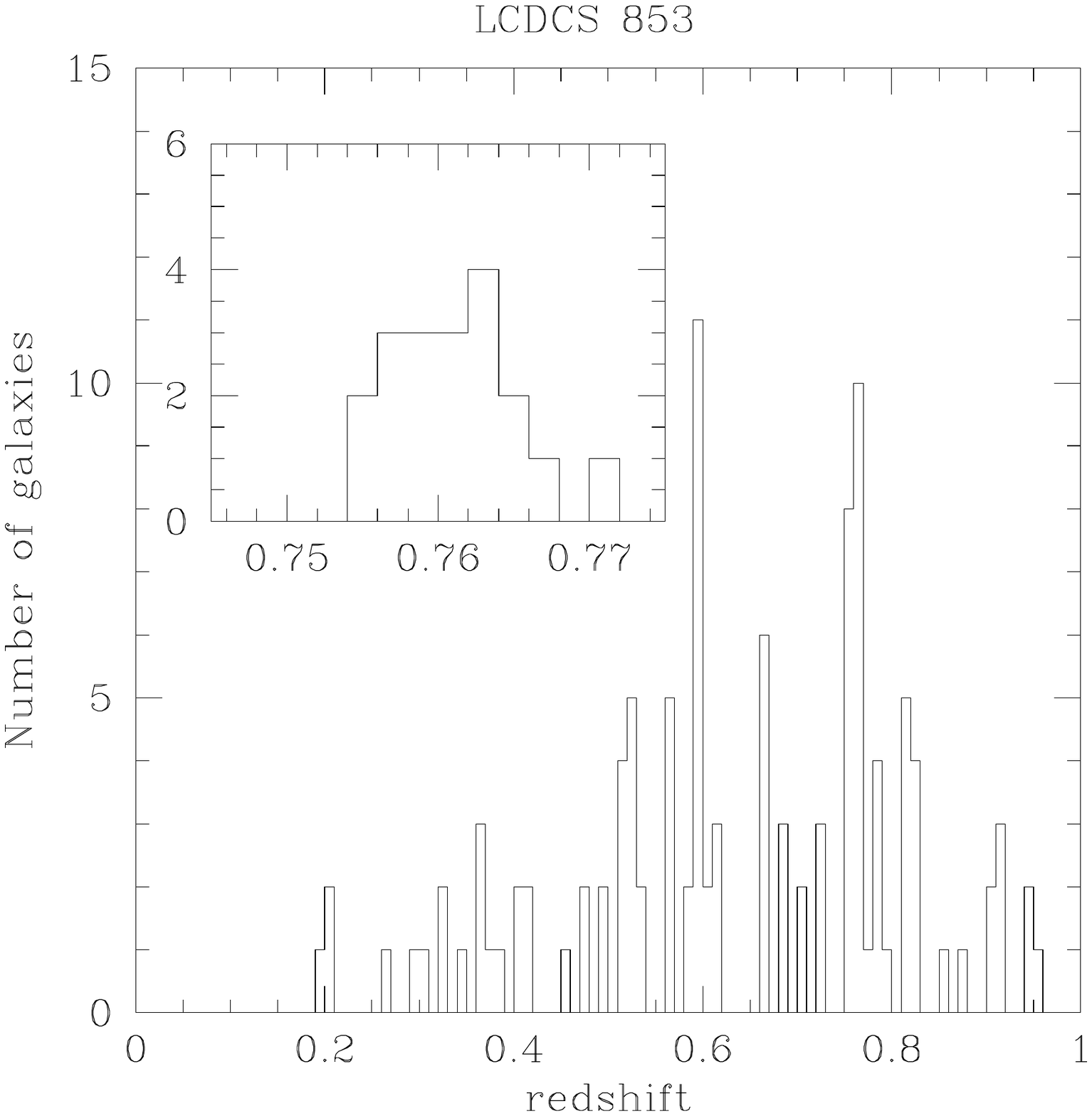}
    \caption{Same as Fig.~\ref{fig:cl0016_X} for LCDCS~0853.}
  \label{fig:LCDCS853_X}
  \end{center}
  \end{figure*}

  The X-ray image of LCDCS~0853 shows emission at the cluster position
  (Fig.~\ref{fig:LCDCS853_X}), which is fully accounted for by a
  simple model, besides a very minor residual close to the cluster
  centre (only significant at the 2.5$\sigma$ level and extended over
  less than 3 arcsec).  We also see the residuals of the PN interchip
  separation.

The full redshift histogram up to z=1 shows that the line of sight to
the cluster is intercepting several structures
(Fig.~\ref{fig:LCDCS853_X}). There are 18 galaxies with redshifts
in the [0.75,0.77] cluster range.  The SG method detects a main
structure (see Table~\ref{tab:SG}), but we must note that our
spectroscopic catalogue does not cover the full field of view.

\subsection{RX~J1354.2-0221 (208.57042$^o$, --2.3631$^o$, z=0.5460)} 

\begin{figure*}
  \begin{center}
    \includegraphics[width=2.7in,angle=0,bb=35 144 575 651,clip]{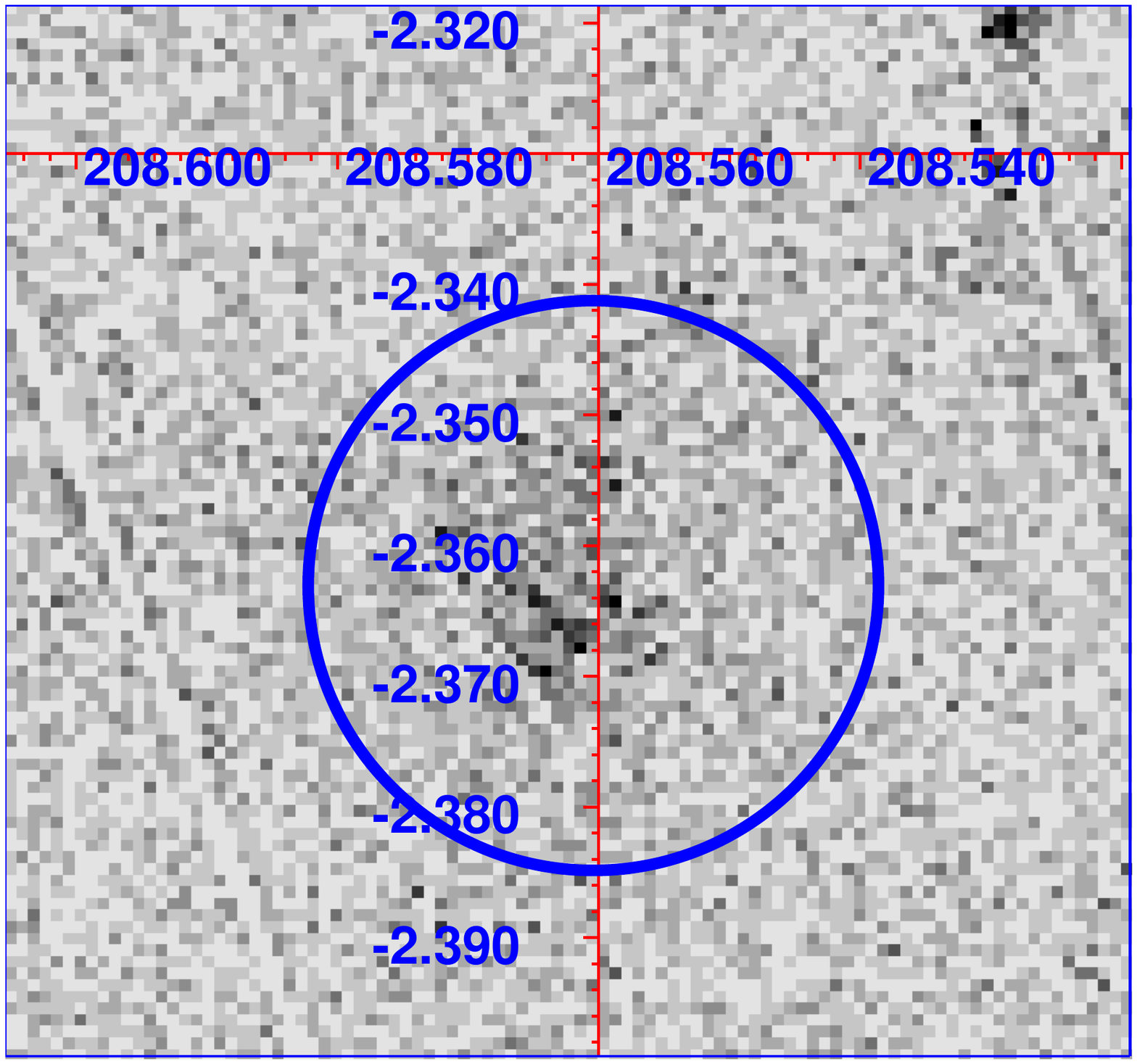}\includegraphics[width=2.7in,angle=0,bb=35 144 575 651,clip]{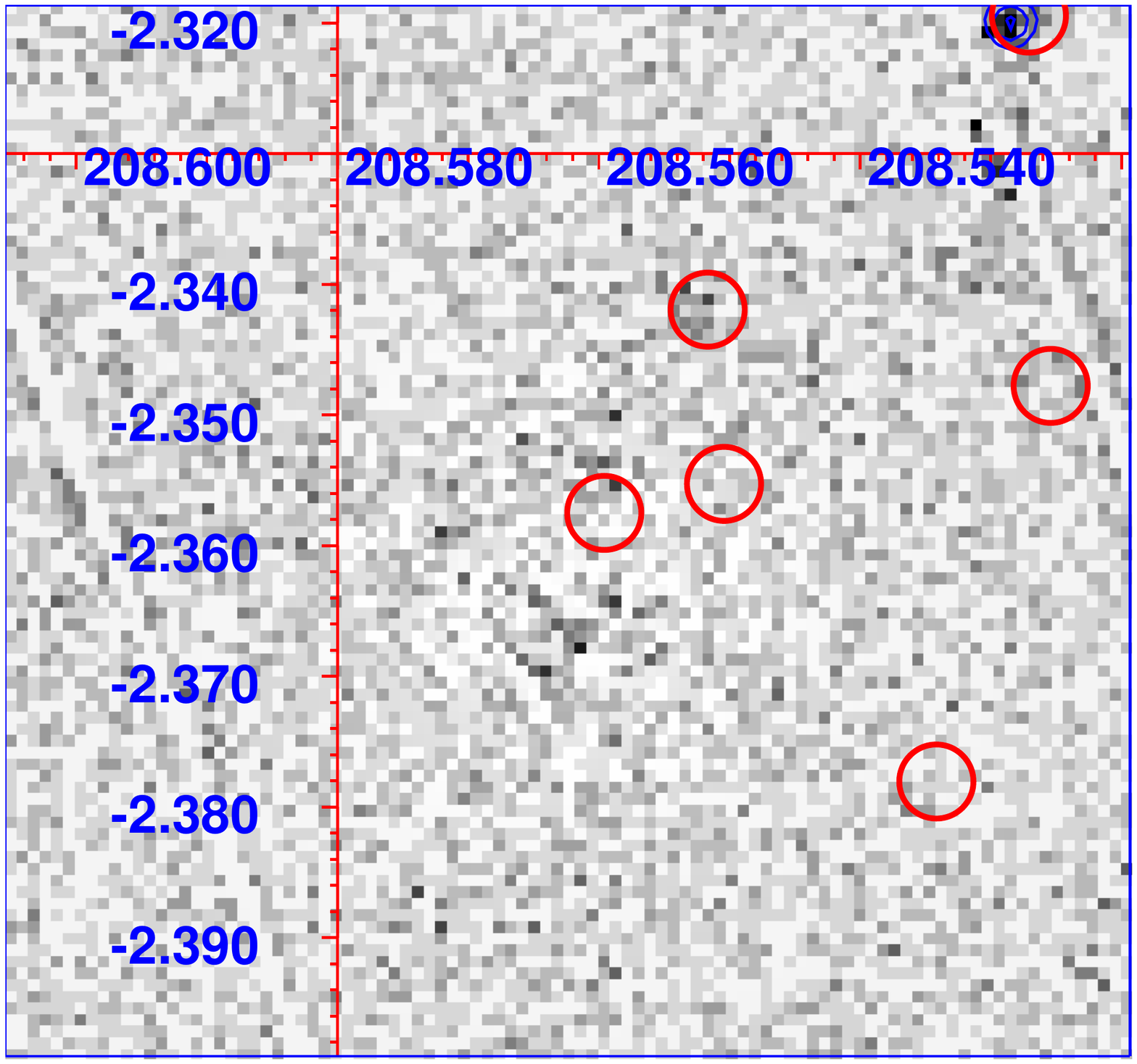}\\
    \includegraphics[width=2.7in,angle=0,bb=35 144 575 651,clip]{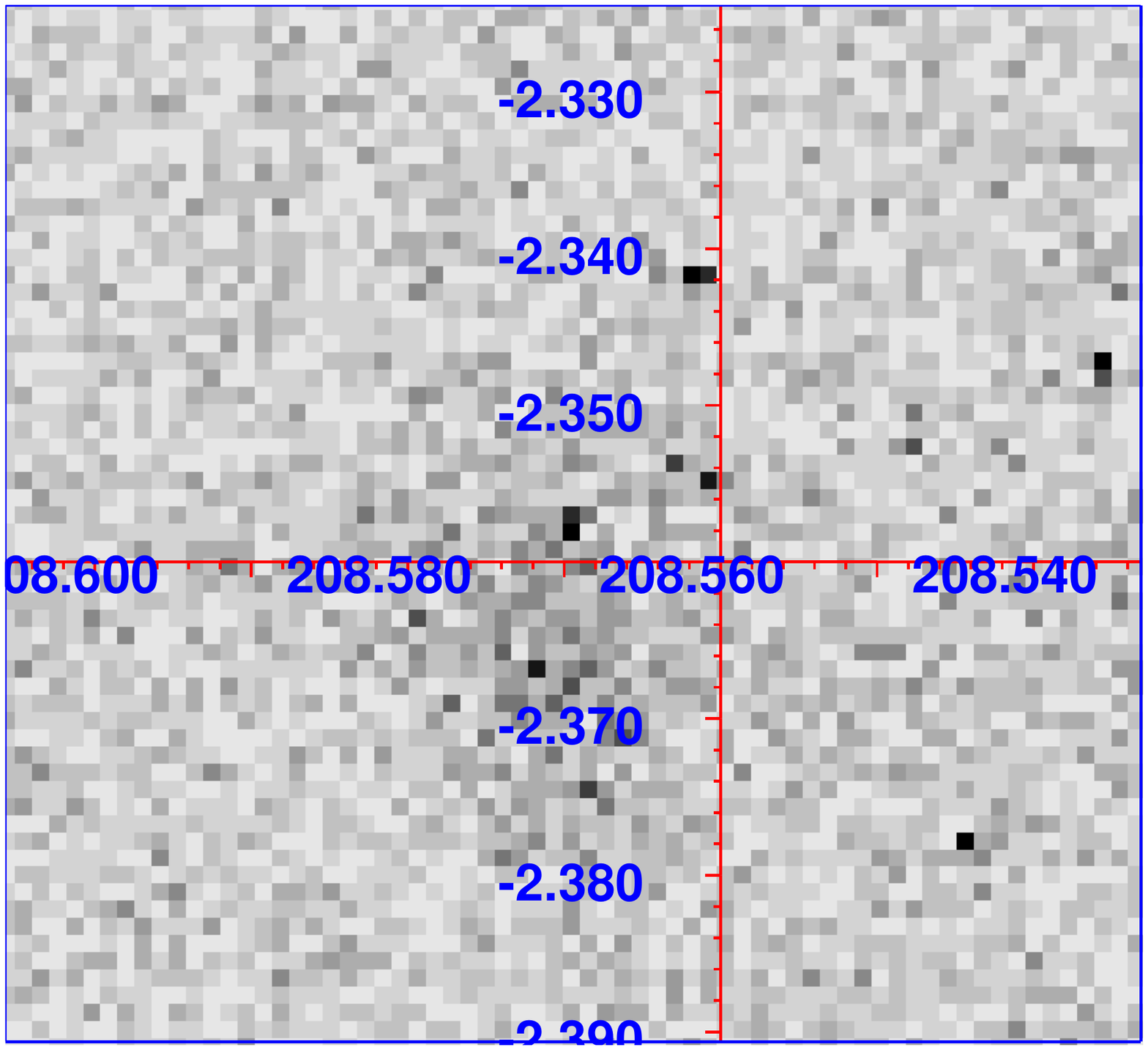}\includegraphics[width=2.7in,angle=0,bb=35 144 575 651,clip]{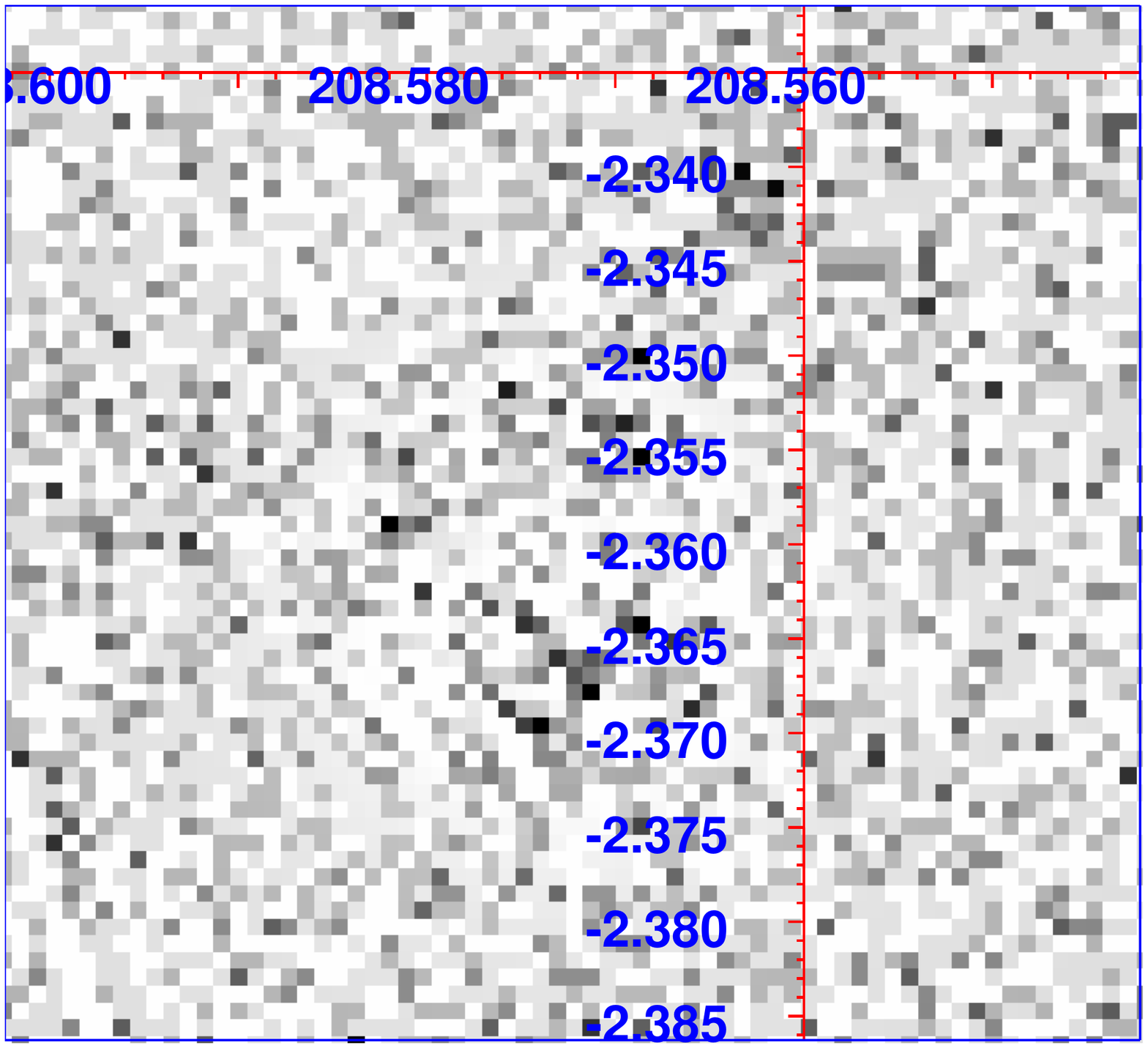}
    \caption{Same as Fig.~\ref{fig:bmw0522_X} for RX~J1354.2-0221.}
  \label{fig:rx1354_X}
  \end{center}
  \end{figure*}

  The X-ray image of RX~J1354.2-0221 (Vikhlinin et al. 1998) shows a
  very weak extended source (Fig.~\ref{fig:rx1354_X}), barely visible
  in the Chandra image, and no residuals appear after subtracting a
  model.  Several point sources appear in the Chandra image, most of
  them known in Gilmour et al. (2009). NED gives a redshift z=0.5460
  for the cluster, but only two redshifts are available in this range,
  preventing from any SG analysis.

\subsection{MACS~J1423.8+2404 (215.95125$^o$, +24.0797$^o$, z=0.5450)} 

\begin{figure*}
  \begin{center}
    \includegraphics[width=2.7in,angle=0,bb=35 144 575 651,clip]{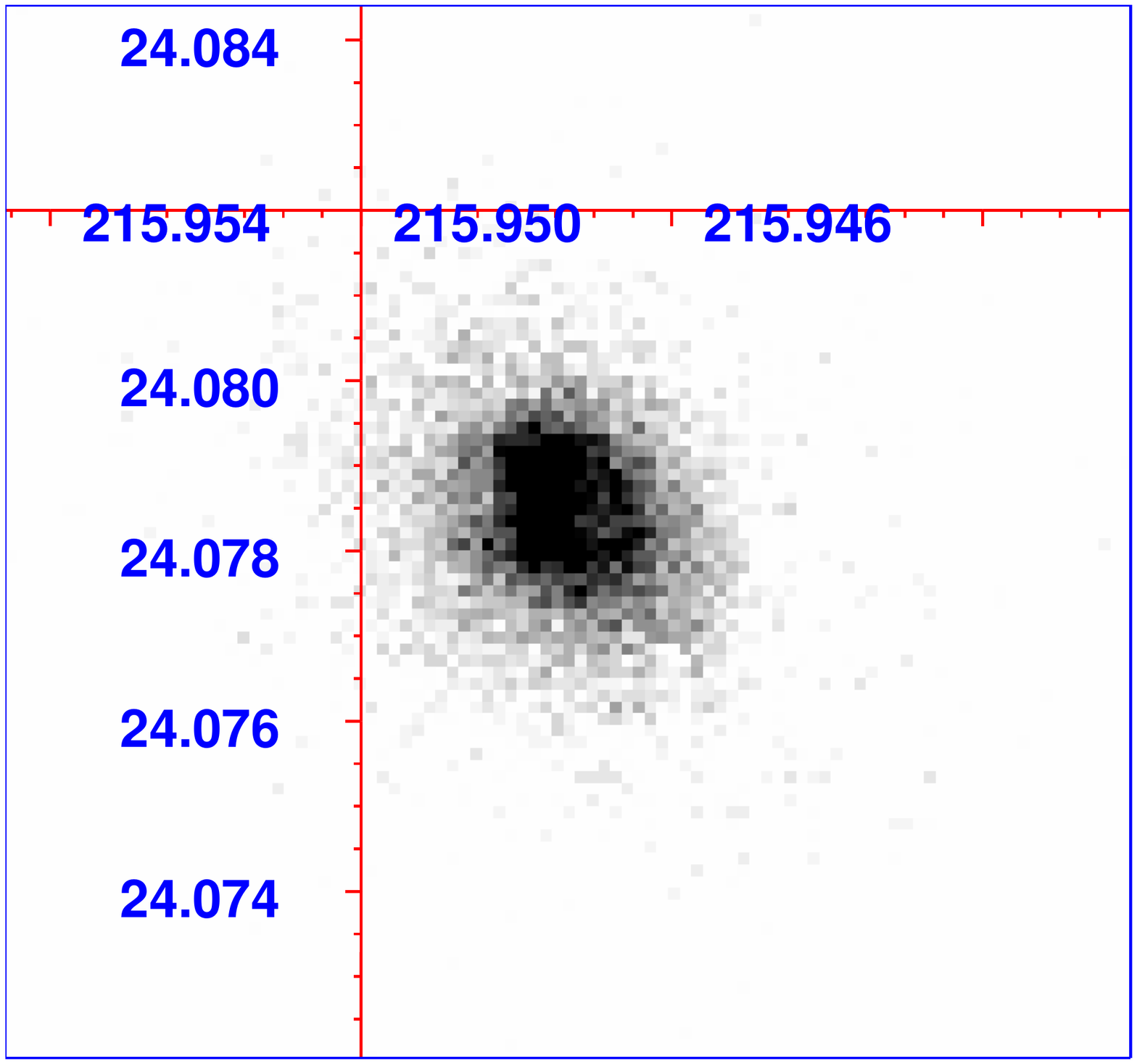}\includegraphics[width=2.7in,angle=0,bb=35 144 575 651,clip]{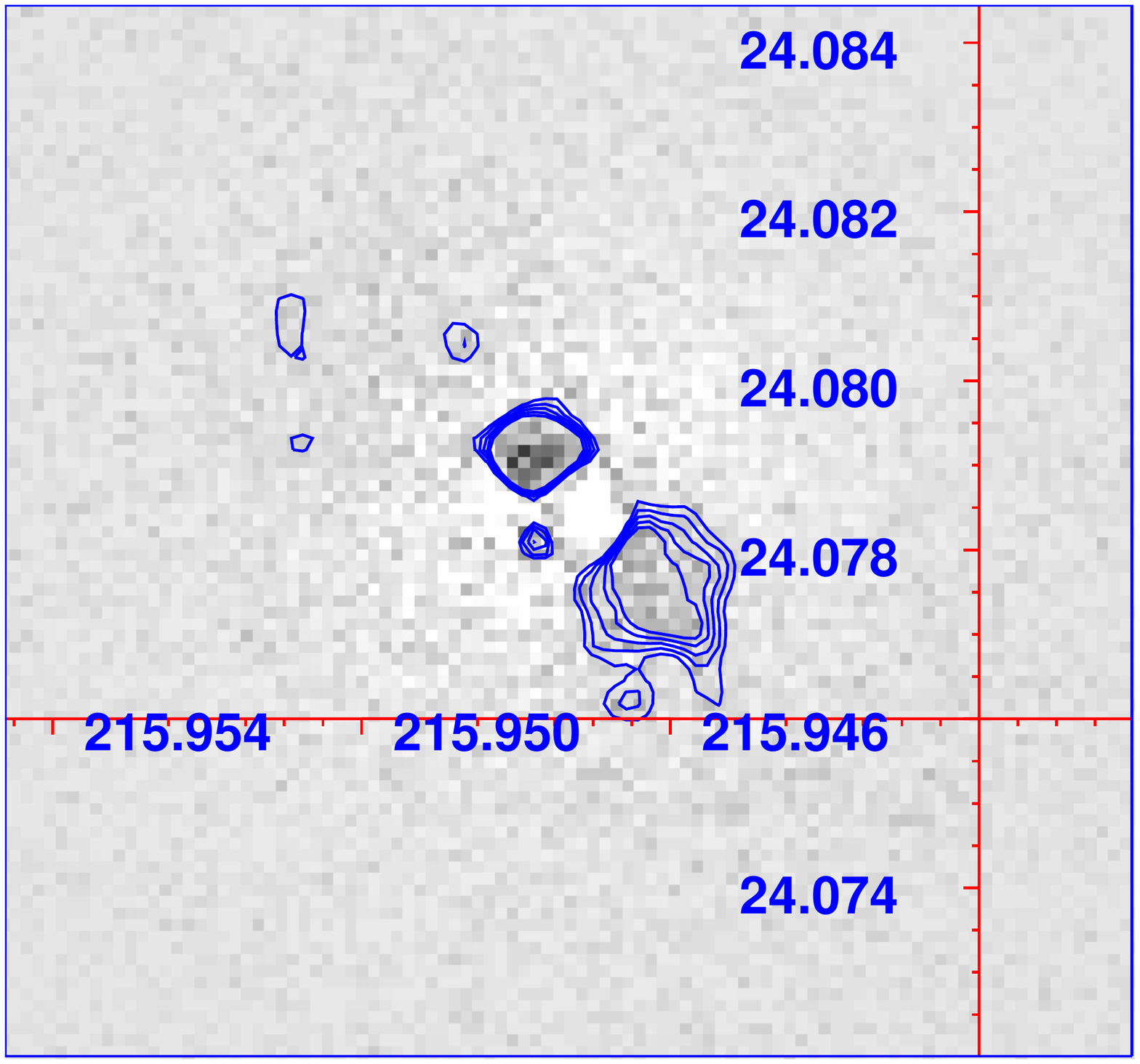}\\
    \includegraphics[width=2.7in,angle=0,bb=35 144 575 651,clip]{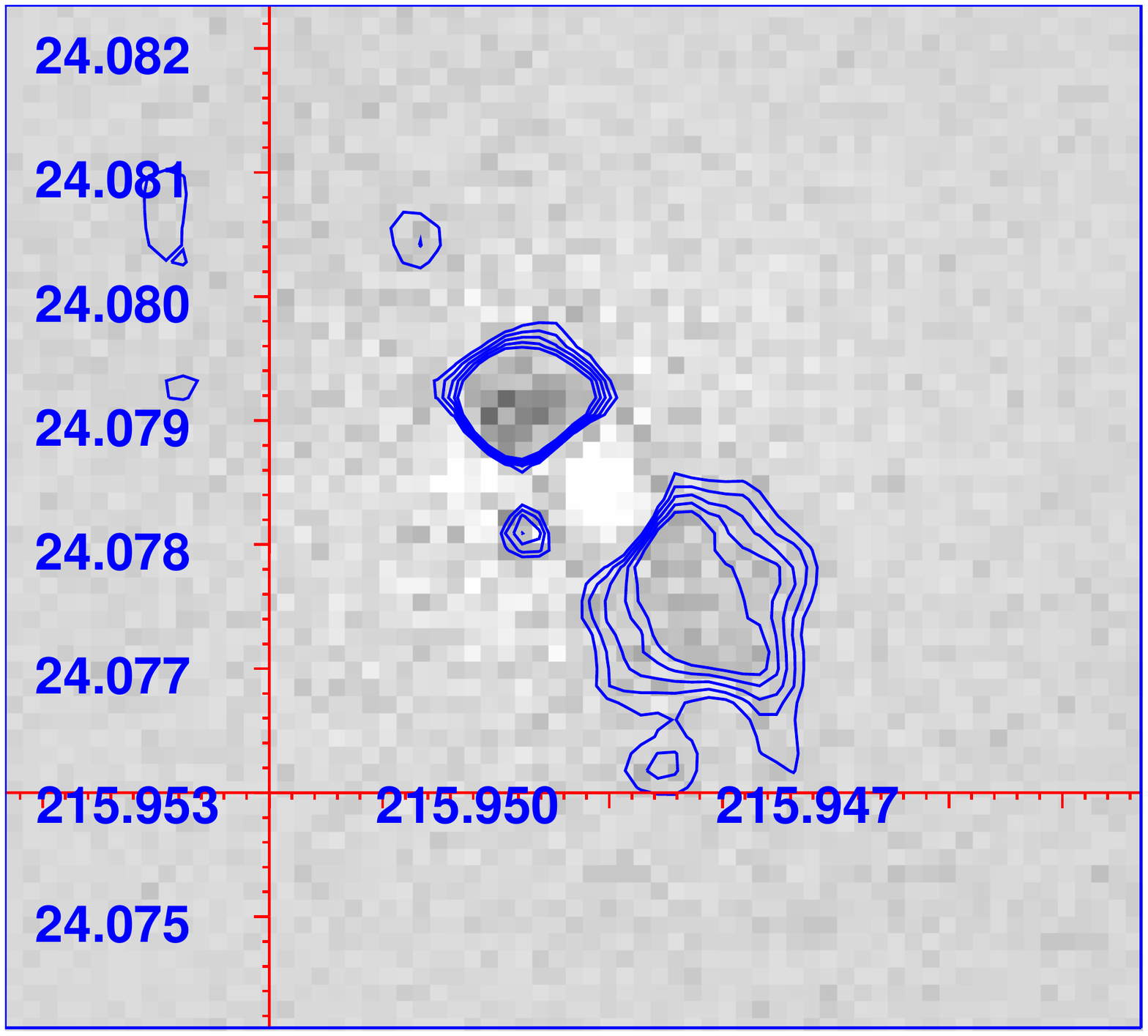}\includegraphics[width=2.7in,angle=0,bb=15 144 575 701,clip]{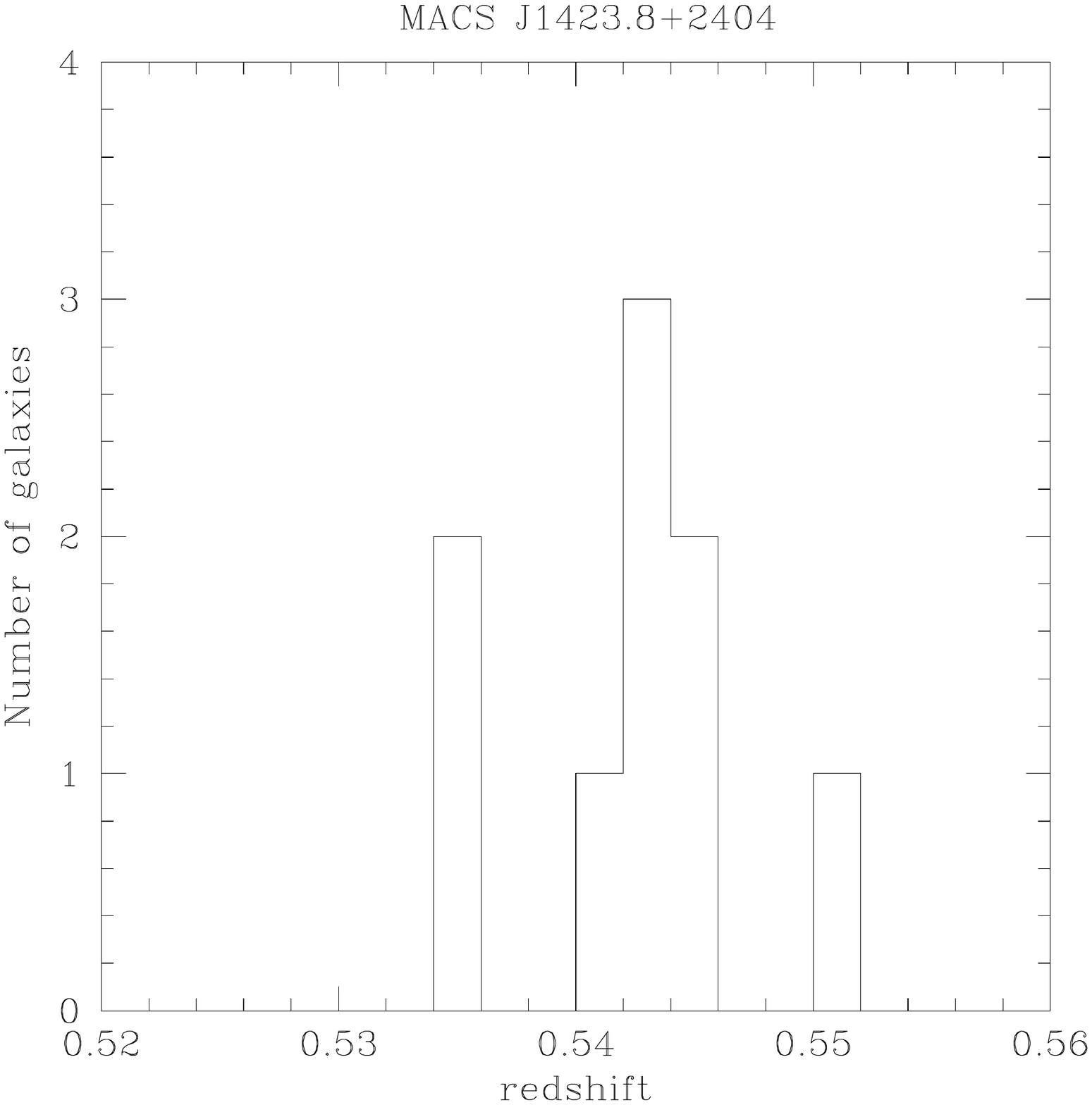}
    \caption{Same as Fig.~\ref{fig:cl0016_X} for MACS~J1423.8+2404. We only used Chandra X-ray data for this cluster.}
  \label{fig:figX1423}
  \end{center}
  \end{figure*}

  This is the only cluster for which there are no XMM-Newton data, but
  for which the Chandra image is deep enough to allow us to remove a
  $\beta -$model from the cluster image (Fig.~\ref{fig:figX1423}), so
  we include it in this section. The X-ray emission is strong and
  slightly elongated. After subtracting a $\beta -$model, we see
  several low significance X-ray residuals north east of the
  cluster. We also detect two brighter sources close to the cluster
  centre, respectively located close to the cluster centre and toward
  the south--west. The brighter one (close to the cluster centre) is
  probably due to the fact that the $\beta$-model subtraction is not
  perfect, while the other one is quite extended.

  We only have nine galaxy redshifts in the [0.536,0.552] range
  (Fig.~\ref{fig:figX1423}), and the SG method only detects the
  cluster as a three--galaxy structure, so we can only suggest that
  the extended X-ray source south--west of the cluster is a
  substructure of the MACS~J1423.8+2404 cluster.

\subsection{GHO~1602+4312 (241.10483$^o$, +43.0813$^o$, z=0.8950)} 

\begin{figure*}
  \begin{center}
    \includegraphics[width=2.7in,angle=0,bb=35 144 575 651,clip]{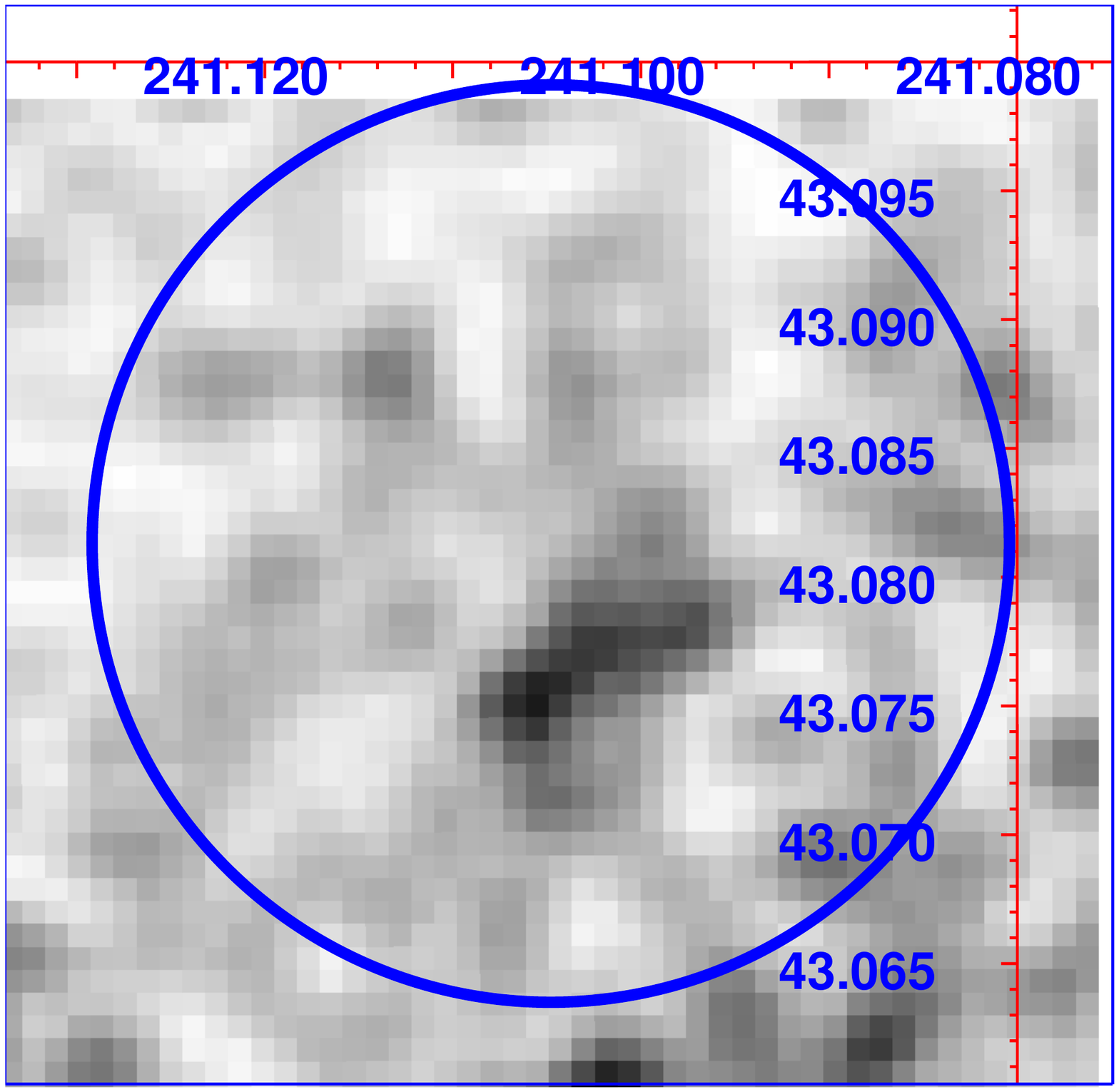}\includegraphics[width=2.7in,angle=0,bb=35 144 575 651,clip]{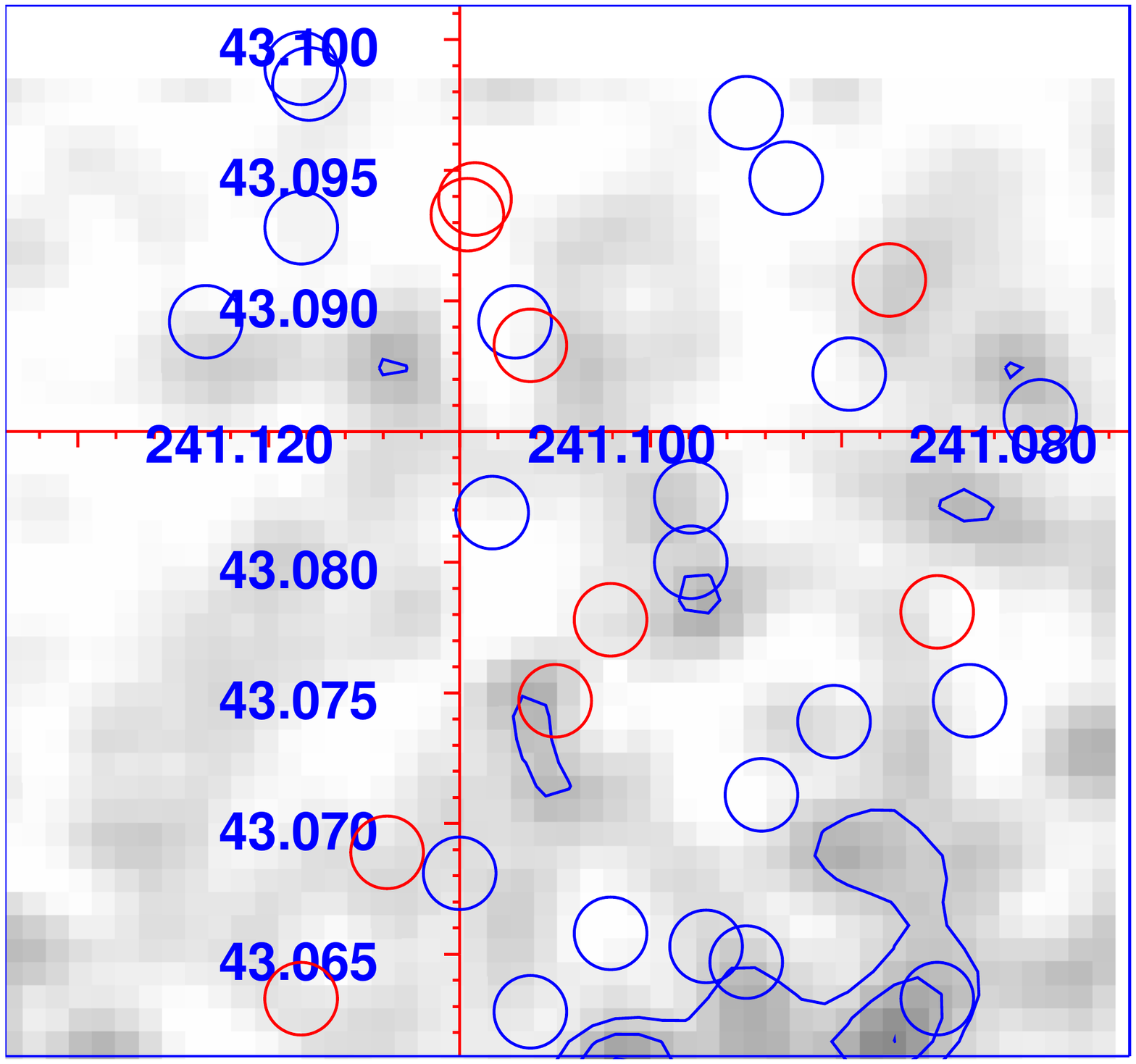}\\
    \includegraphics[width=2.7in,angle=0,bb=35 144 575 651,clip]{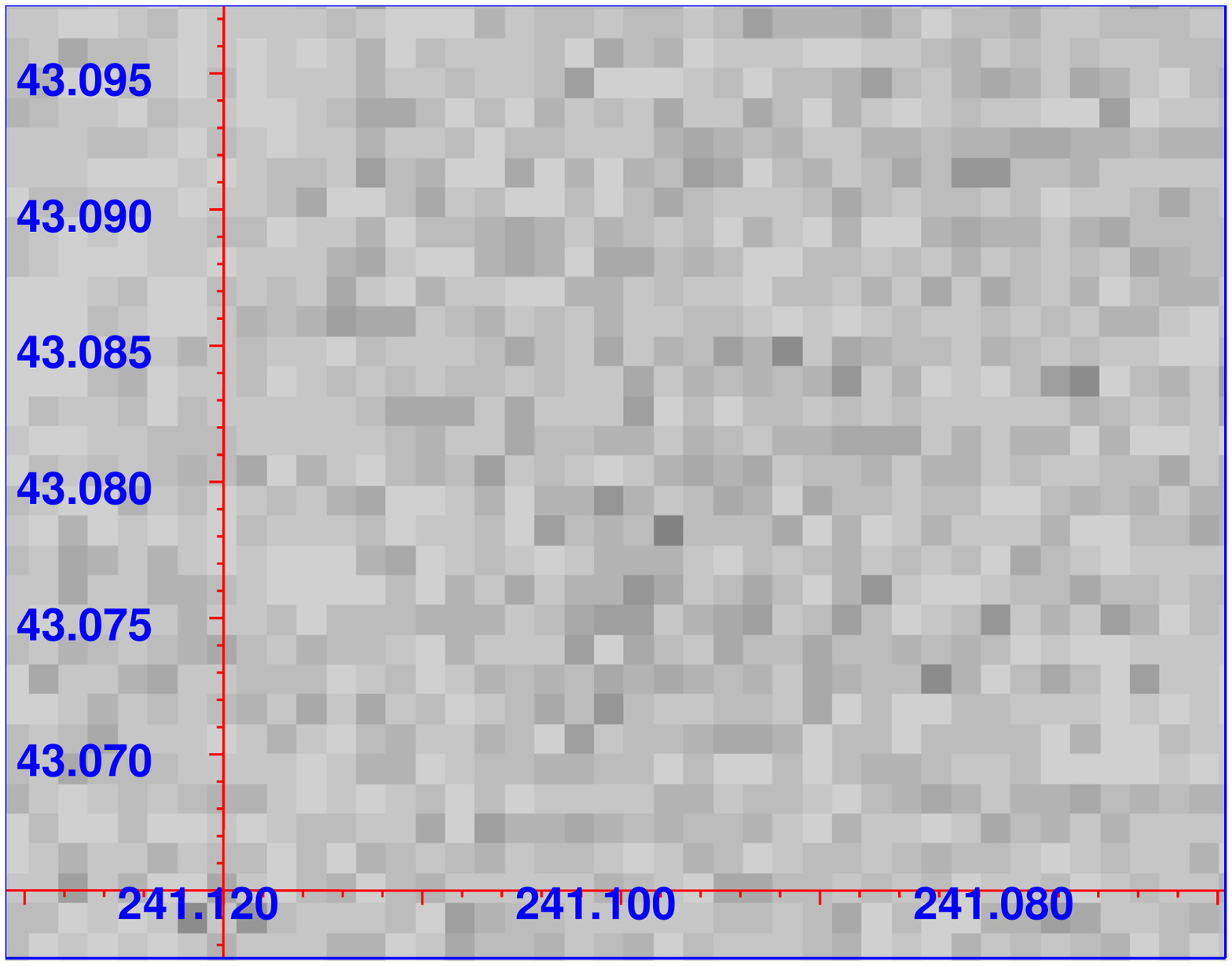}\includegraphics[width=2.7in,angle=0,bb=35 144 575 651,clip]{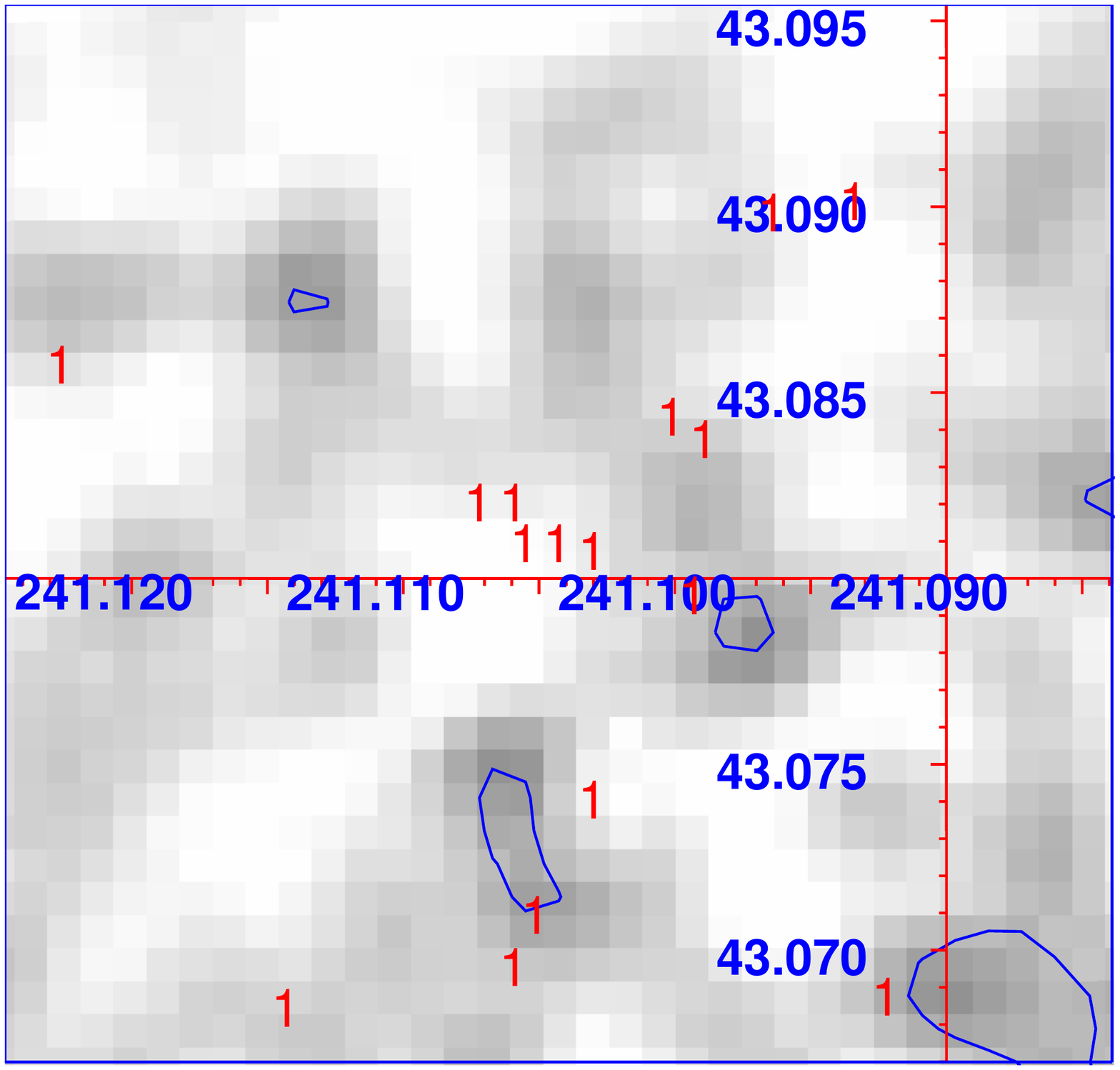}\\
    \includegraphics[width=2.7in,angle=0,bb=15 144 575 701,clip]{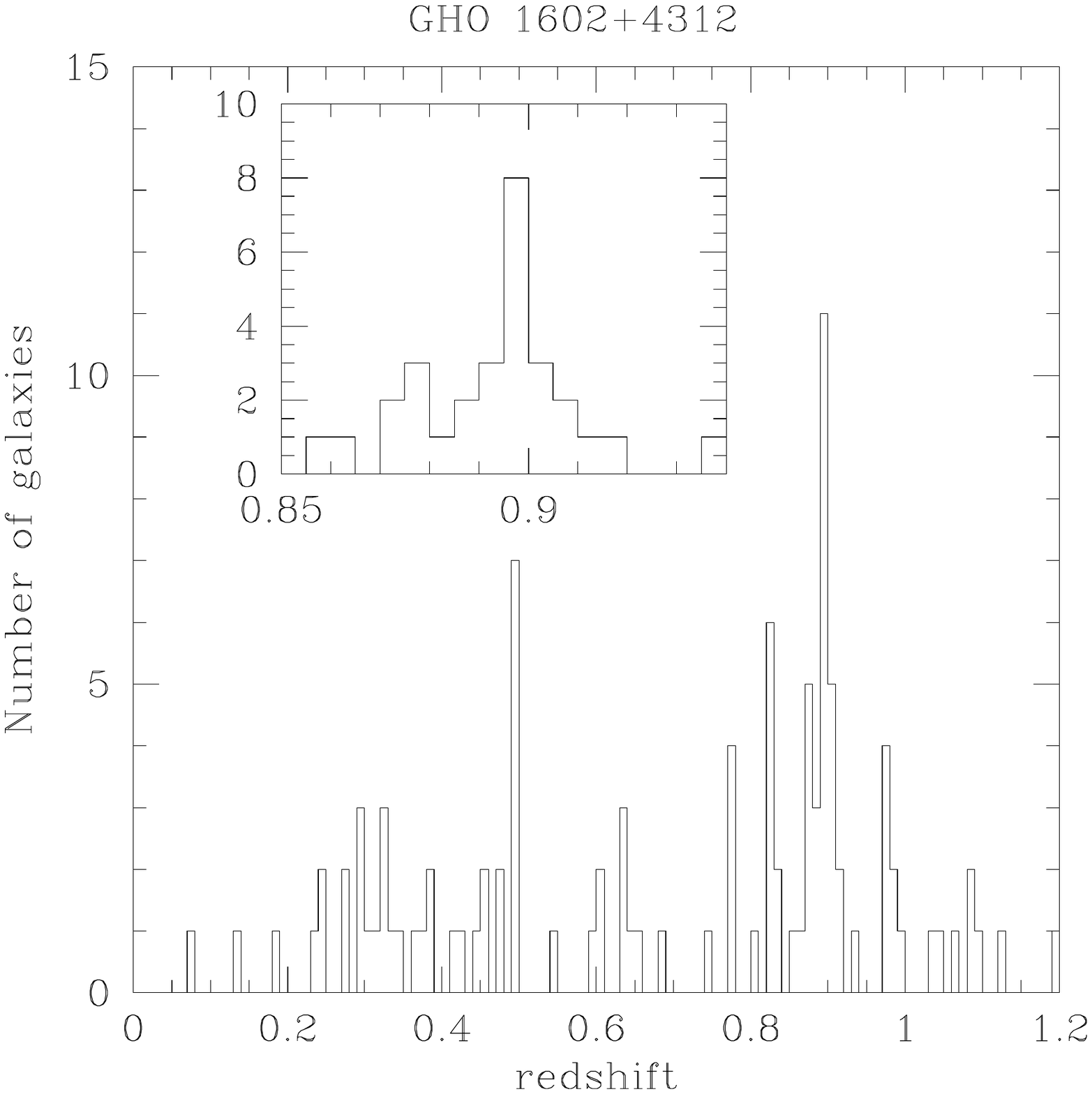}
    \caption{Same as Fig.~\ref{fig:cl0152_X} for GHO~1602+4312.}
  \label{fig:GHO1602_X}
  \end{center}
  \end{figure*}

  The X-ray image of GHO~1602+4312 shows a faint elongated
  structure. The residual image shows some emission due to the PN
  interchip separation (at the bottom of upper right of
  Fig.~\ref{fig:GHO1602_X}) and several barely significant
  (2.5$\sigma$ level) X-ray sources. Two of these sources are located
  at the edge of the elongated X-ray cluster emission. They may be due
  to structures on the line of sight (see upper right
  Fig.~\ref{fig:GHO1602_X}), as suggested by the redshift histogram
  showing the presence of several small structures on the line of
  sight (Fig.~\ref{fig:GHO1602_X}).  These two faint X-ray sources
  also could be minor infalling groups that are not detected by the SG
  analysis (see below). There are no X-ray point sources detected in
  the field of view, neither by Gilmour (2009) or in Chandra data.

  Only the cluster itself is detected by the SG analysis (group SG1 in
  Table~\ref{tab:SG}). The redshift histogram of the cluster shows a
  rather symmetric peak at z$\sim$0.89 with 26 galaxies in the
  [0.88,0.92] range, out of which five form the blue wing of the
  redshift histogram. This cluster is part of a supercluster,
  according to Gal et al. (2008) .

\subsection{MS 1621.5+2640 (245.89863$^o$, +26.5638$^o$, z=0.4260)} 

\begin{figure*}
  \begin{center}
    \includegraphics[width=2.7in,angle=0,bb=35 144 575 651,clip]{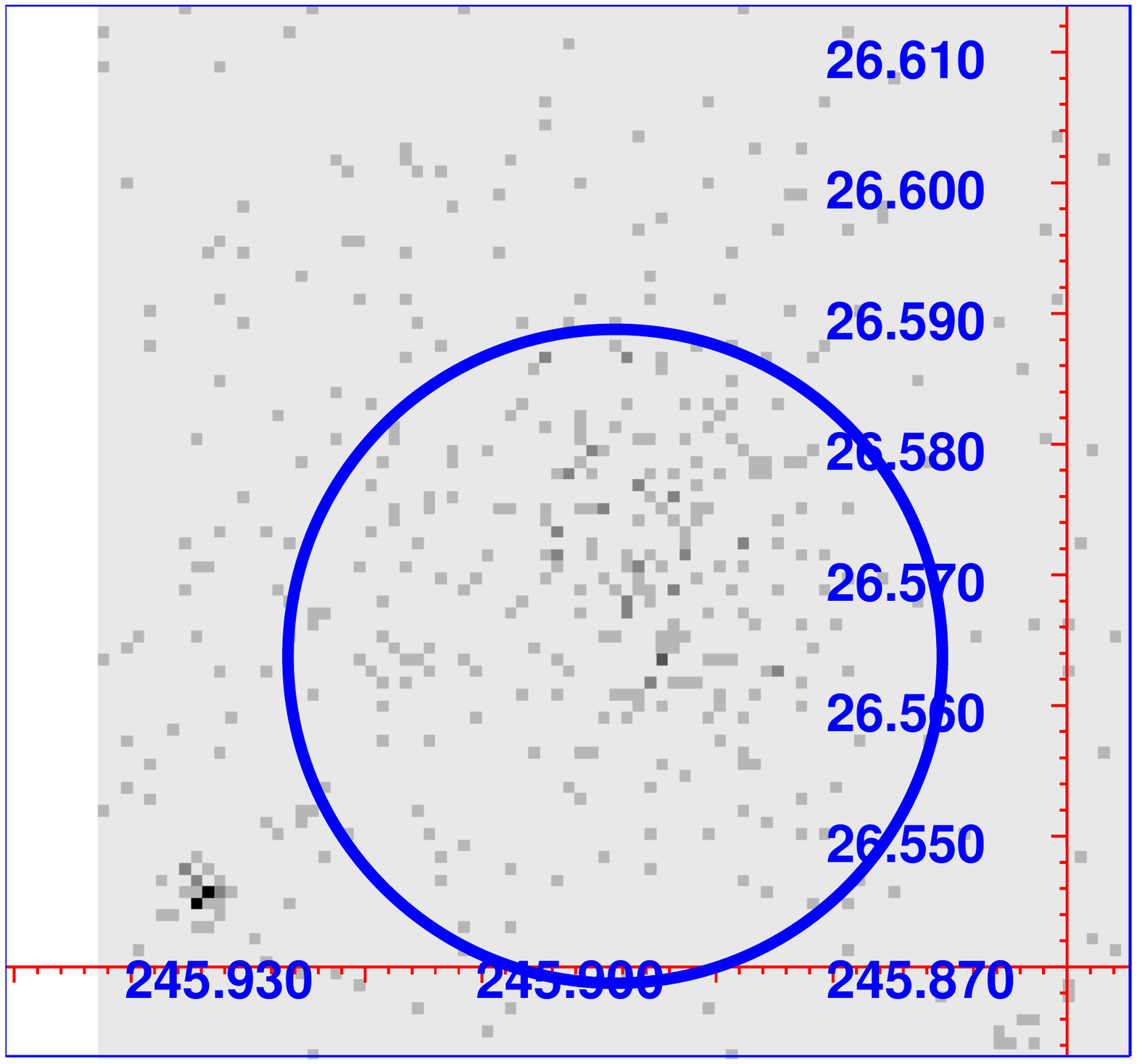}\includegraphics[width=2.7in,angle=0,bb=35 144 575 651,clip]{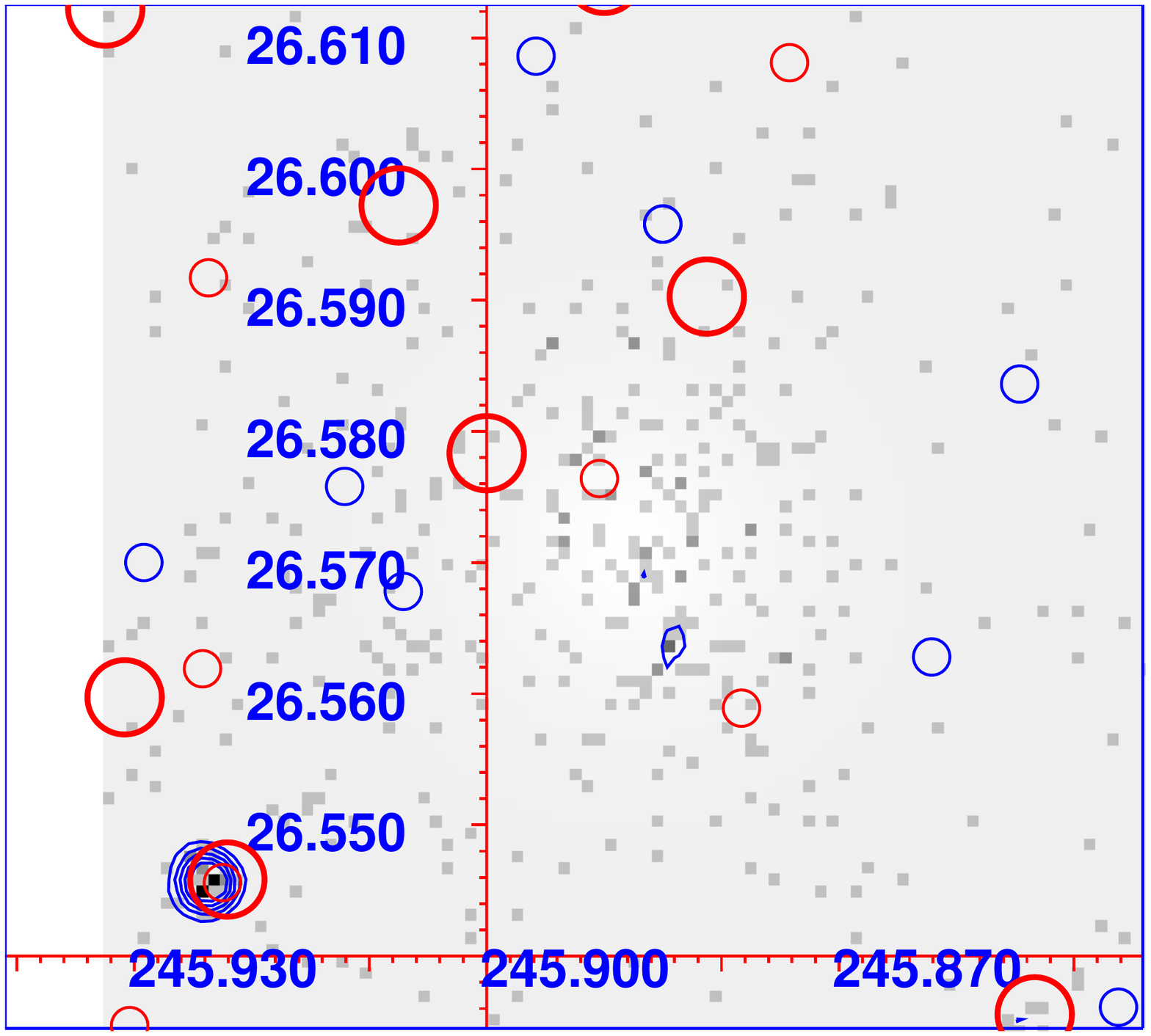}\\
    \includegraphics[width=2.7in,angle=0,bb=35 144 575 651,clip]{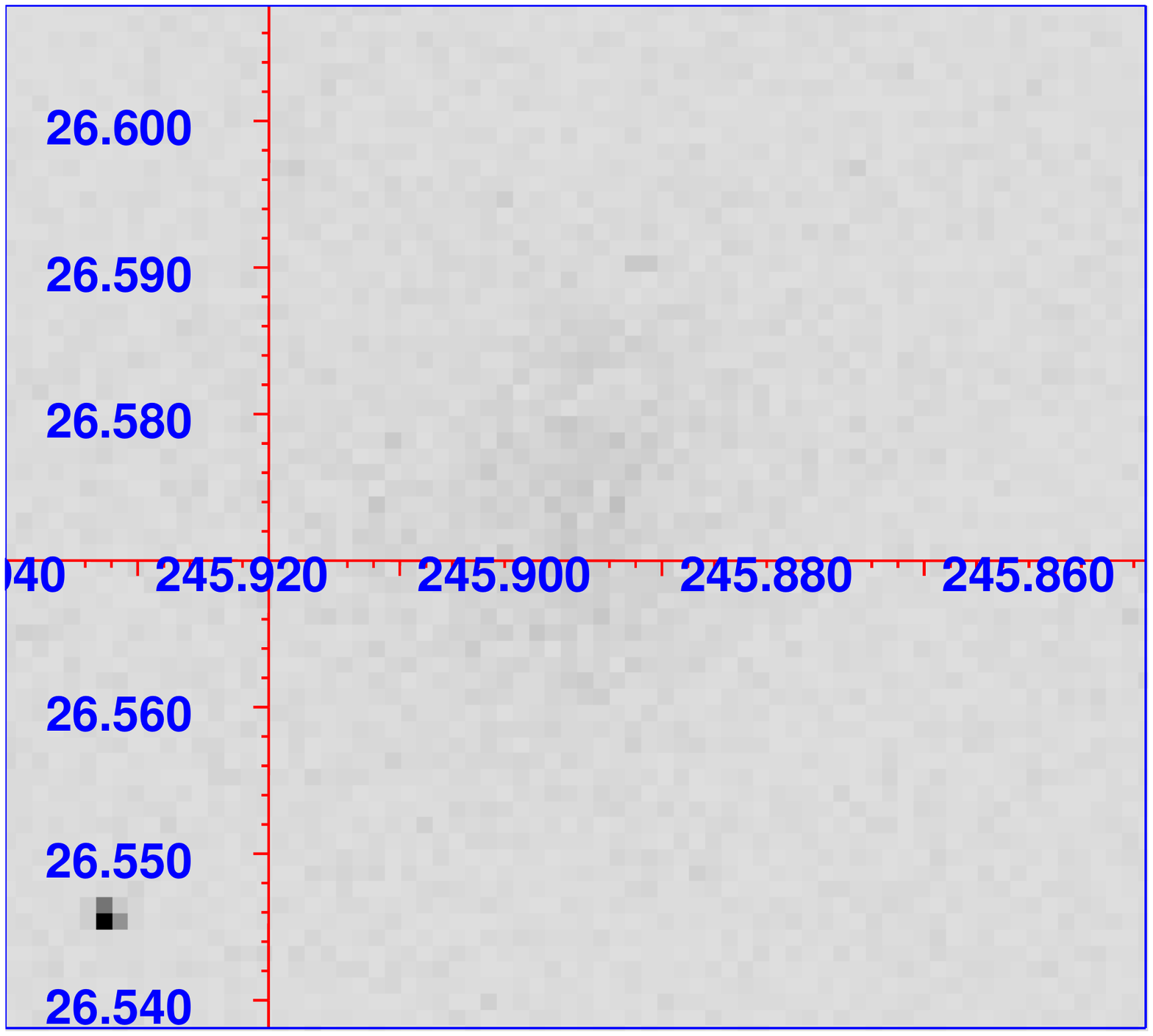}\includegraphics[width=2.7in,angle=0,bb=35 144 575 651,clip]{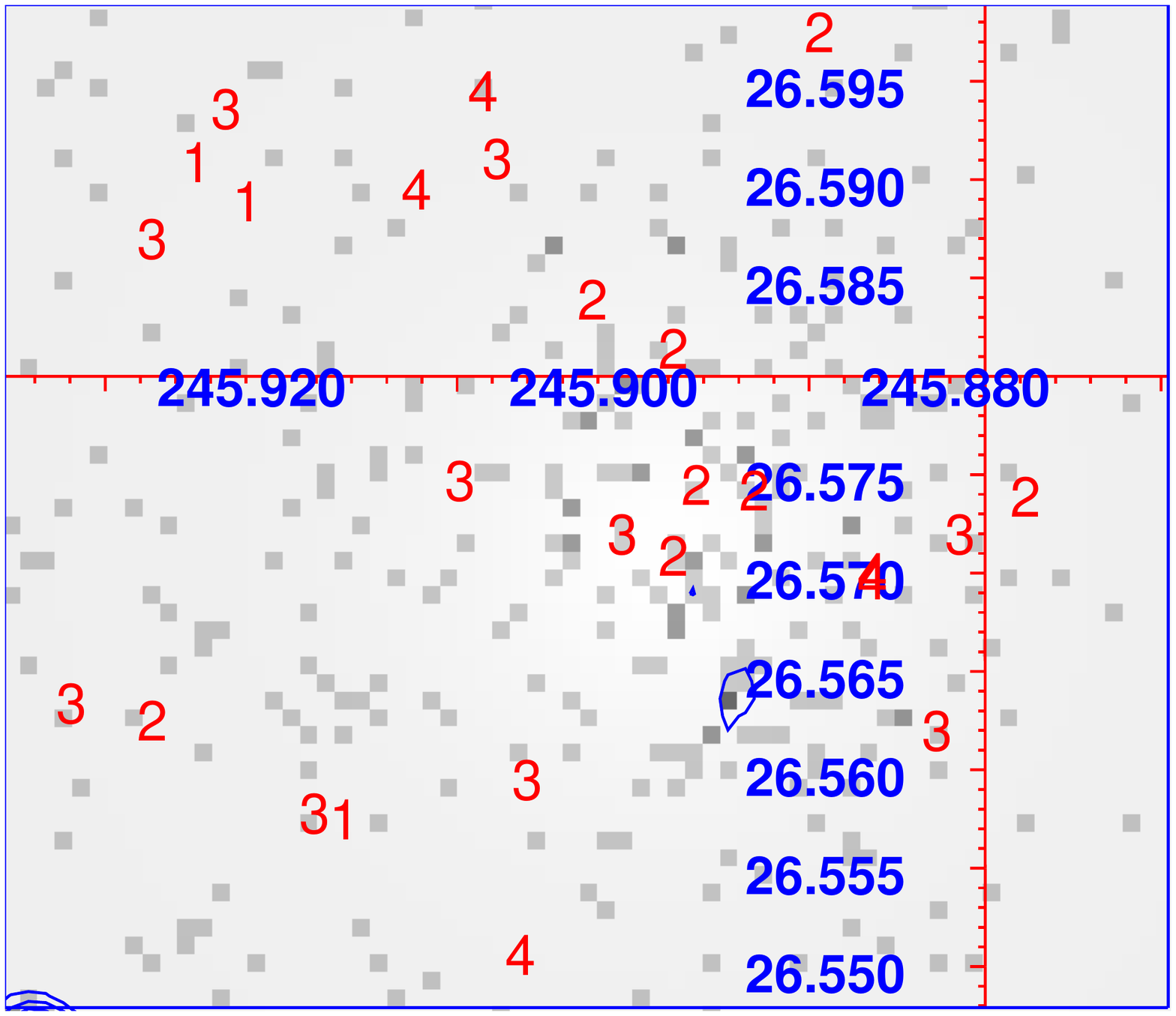}\\
    \includegraphics[width=2.7in,angle=0,bb=15 144 575 701,clip]{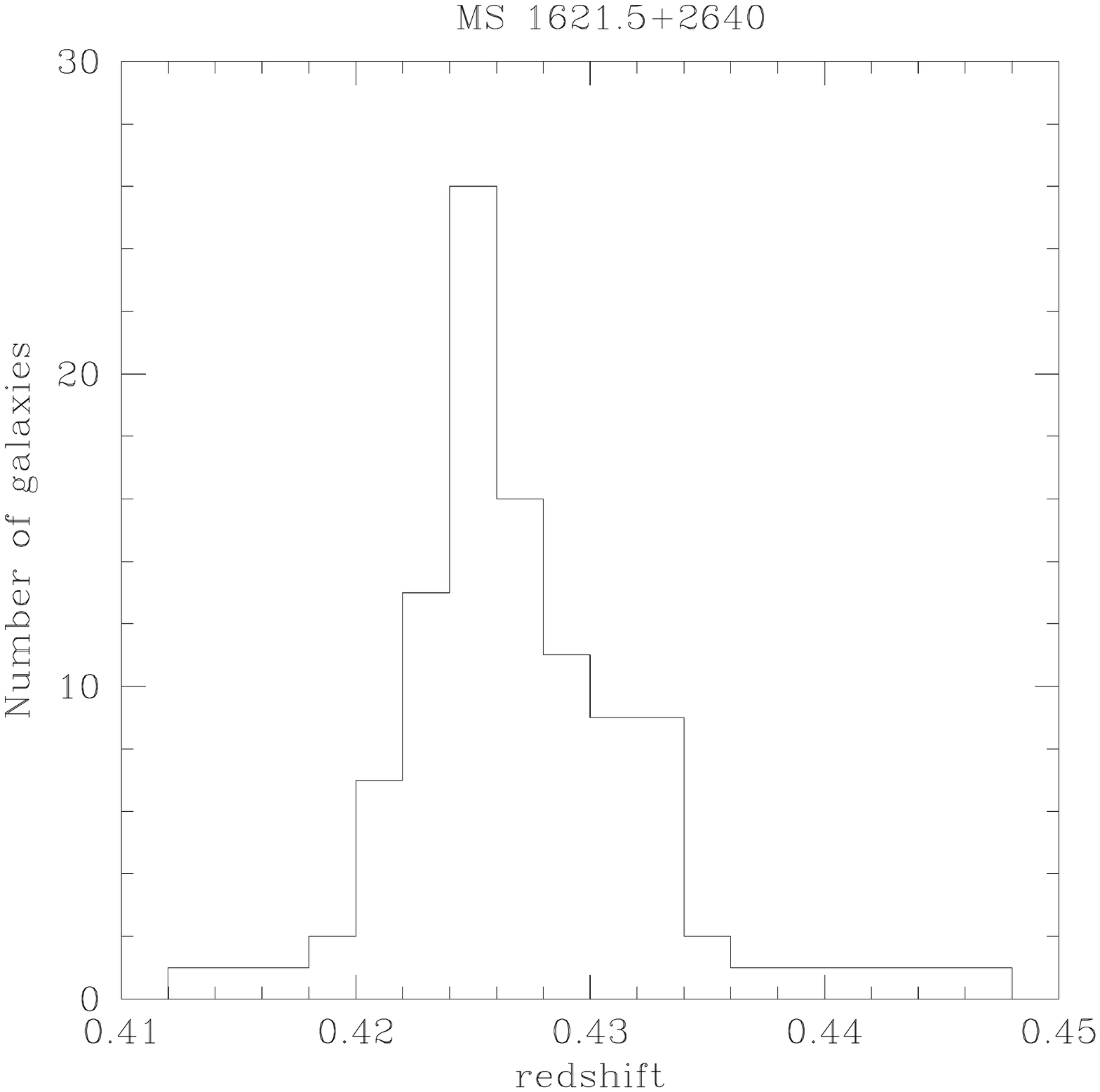}
    \caption{Same as Fig.~\ref{fig:cl0152_X} for MS 1621.5+2640.}
  \label{fig:MS1621_X}
  \end{center}
  \end{figure*}

  The XMM-Newton exposure is short (2.2 ksec), but though the image is
  quite faint, it shows the cluster core (Fig.~\ref{fig:MS1621_X}). We
  detect in the residual map a compact X-ray source south--east of MS
  1621.5+2640 identified as an active object by Gilmour et al. (2009)
  and also visible in the Chandra image.

The redshift histogram of MS~1621.5+2640 shows a strong and somewhat
asymmetric peak at z$\sim$0.427 (Fig.~\ref{fig:MS1621_X}), with 104
galaxies in the [0.412,0.448] range.  The SG method detects a main
structure and three smaller substructures (see
Table~\ref{tab:SG2}). Given the very short XMM-Newton exposure time,
it is impossible to detect them in X-rays, even if some of them seem
massive from the SG analysis.

\subsection{MS~2053.7-0449 (314.09321$^o$,, --4.62873$^o$, z=0.5830)   and CXOU J205617.1-044155 = CXOSEXSI J205617.1-044155
  (314.07150$^o$,, --4.69864$^o$, z=0.6002) } 

\begin{figure*}
  \begin{center}
    \includegraphics[width=2.7in,angle=0,bb=35 144 575 651,clip]{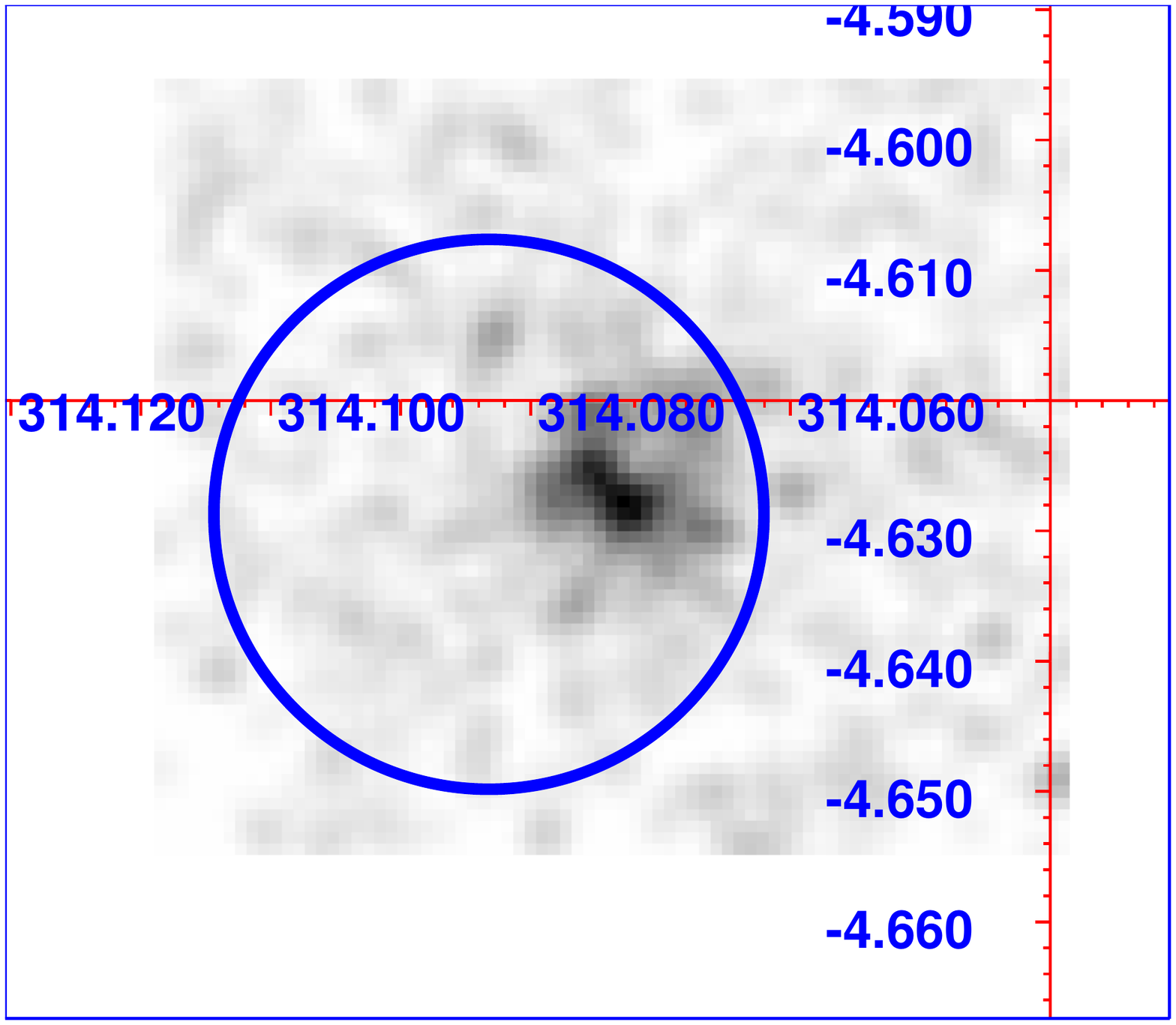}\includegraphics[width=2.7in,angle=0,bb=35 144 575 651,clip]{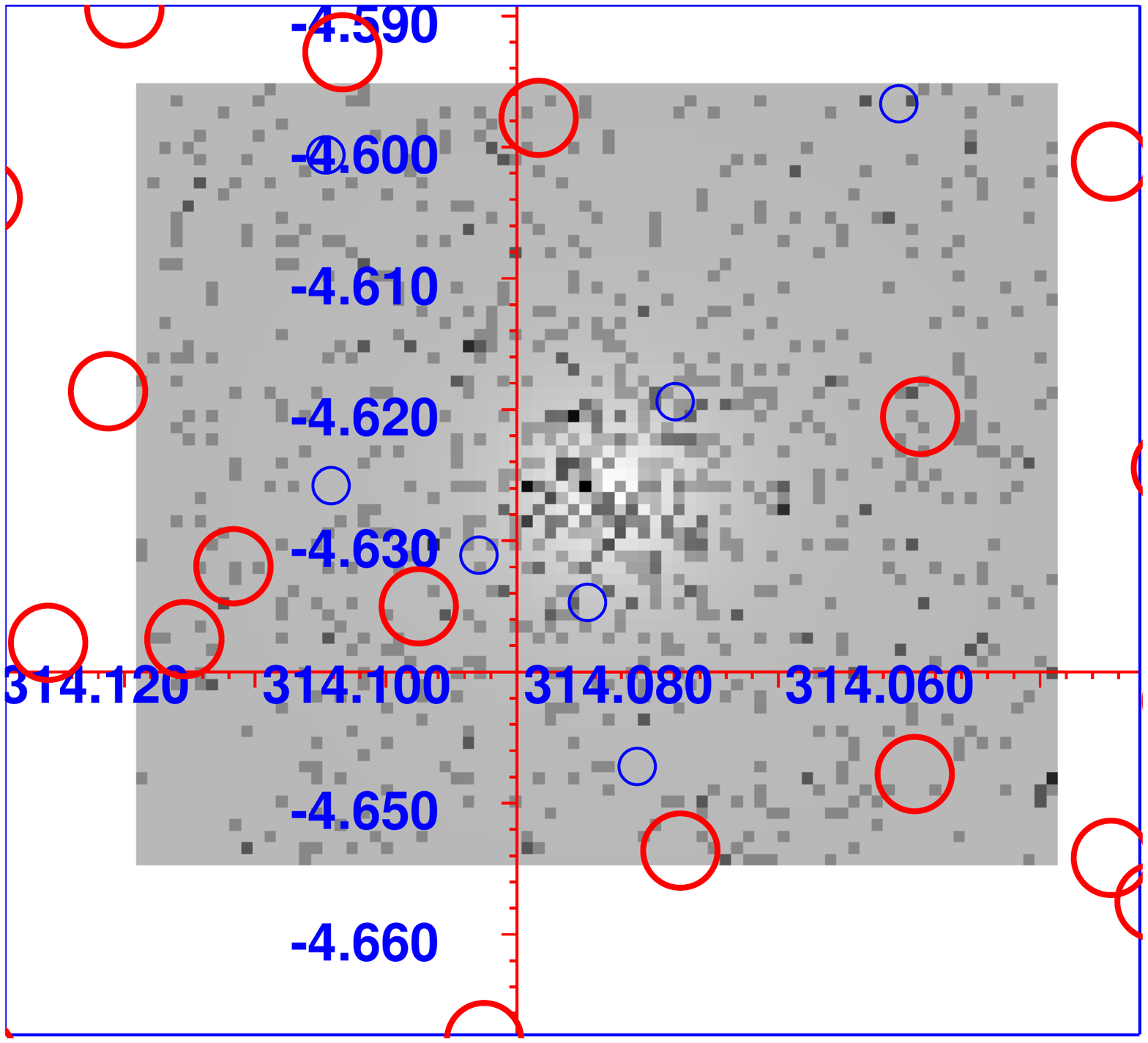}\\
    \includegraphics[width=2.7in,angle=0,bb=35 144 575 651,clip]{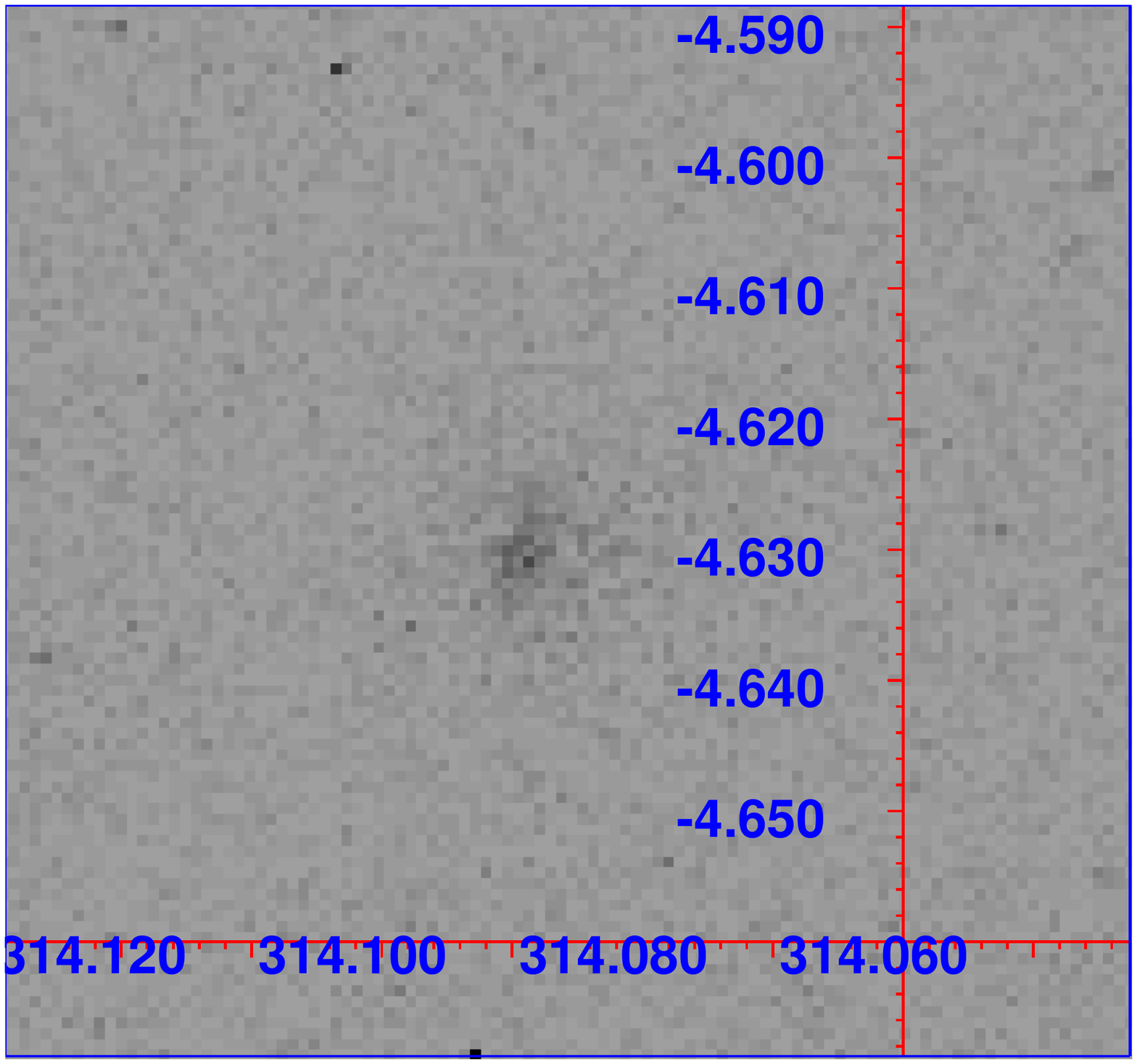}\includegraphics[width=2.7in,angle=0,bb=35 144 575 651,clip]{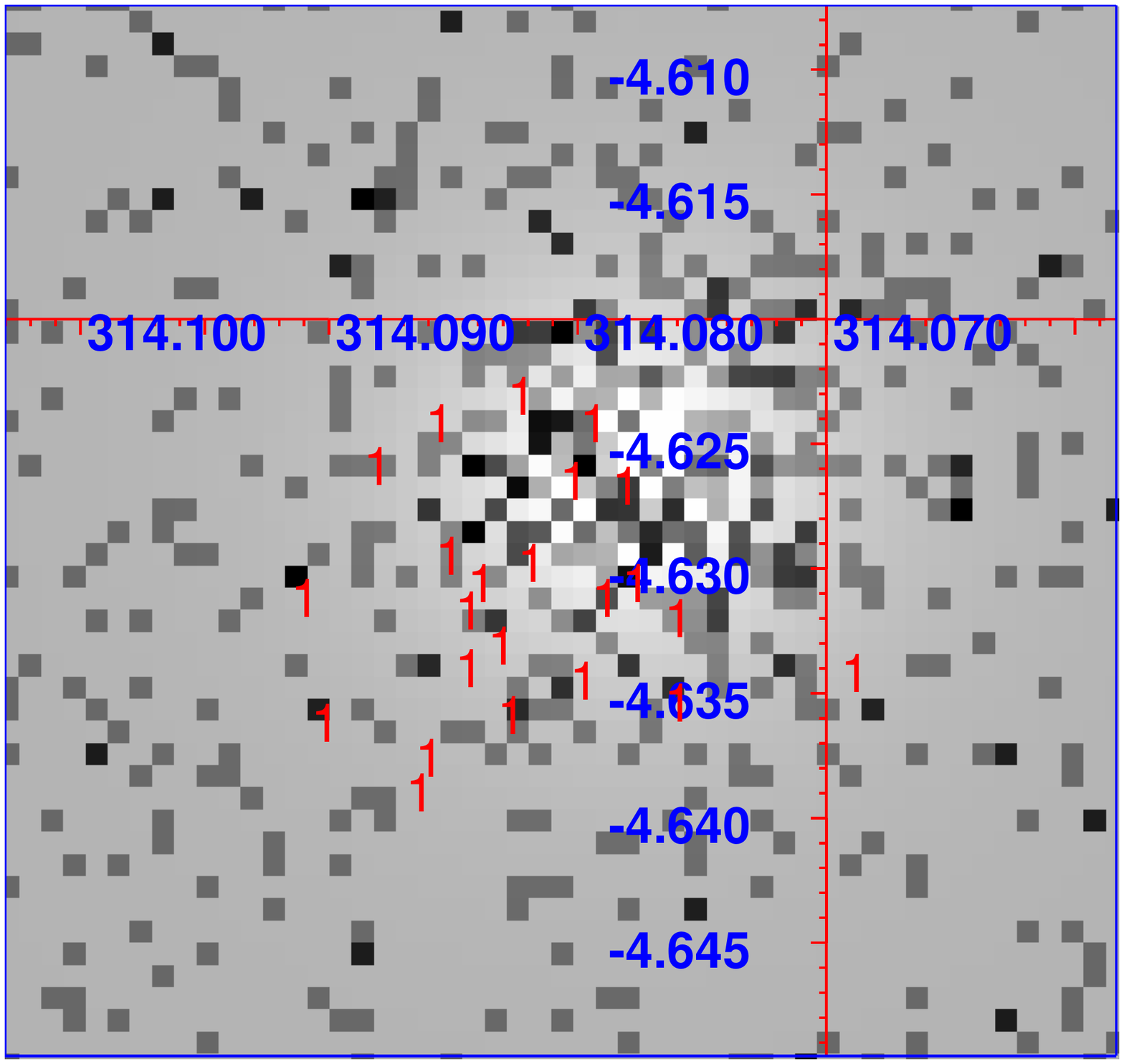}\\
    \includegraphics[width=2.7in,angle=0,bb=15 144 575 701,clip]{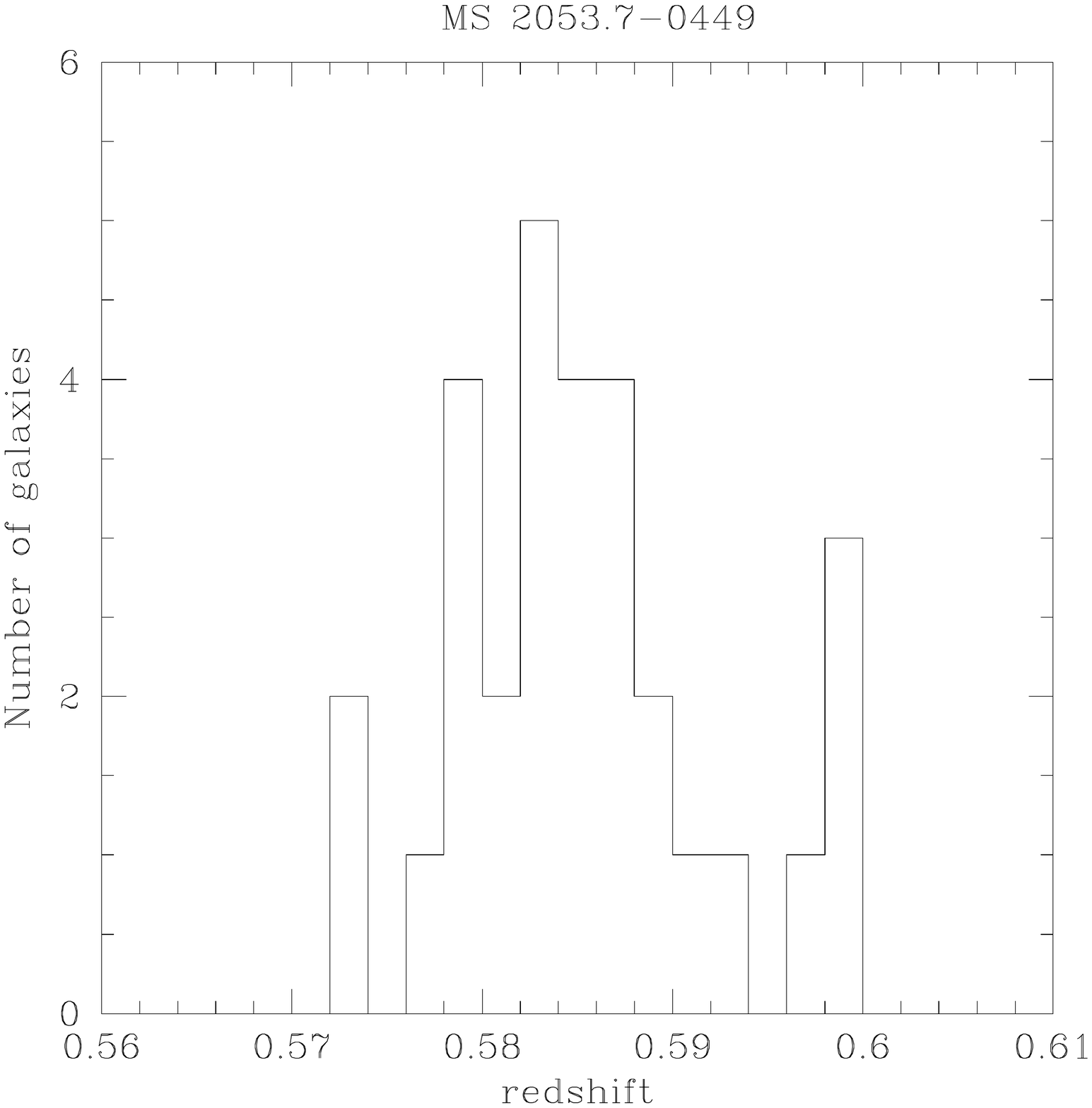}
    \caption{Same as Fig.~\ref{fig:cl0152_X} for MS~2053.7-0449.}
  \label{fig:MS2053_X}
  \end{center}
  \end{figure*}

\begin{figure*}
  \begin{center}
    \includegraphics[width=2.7in,angle=0,bb=35 144 575 651,clip]{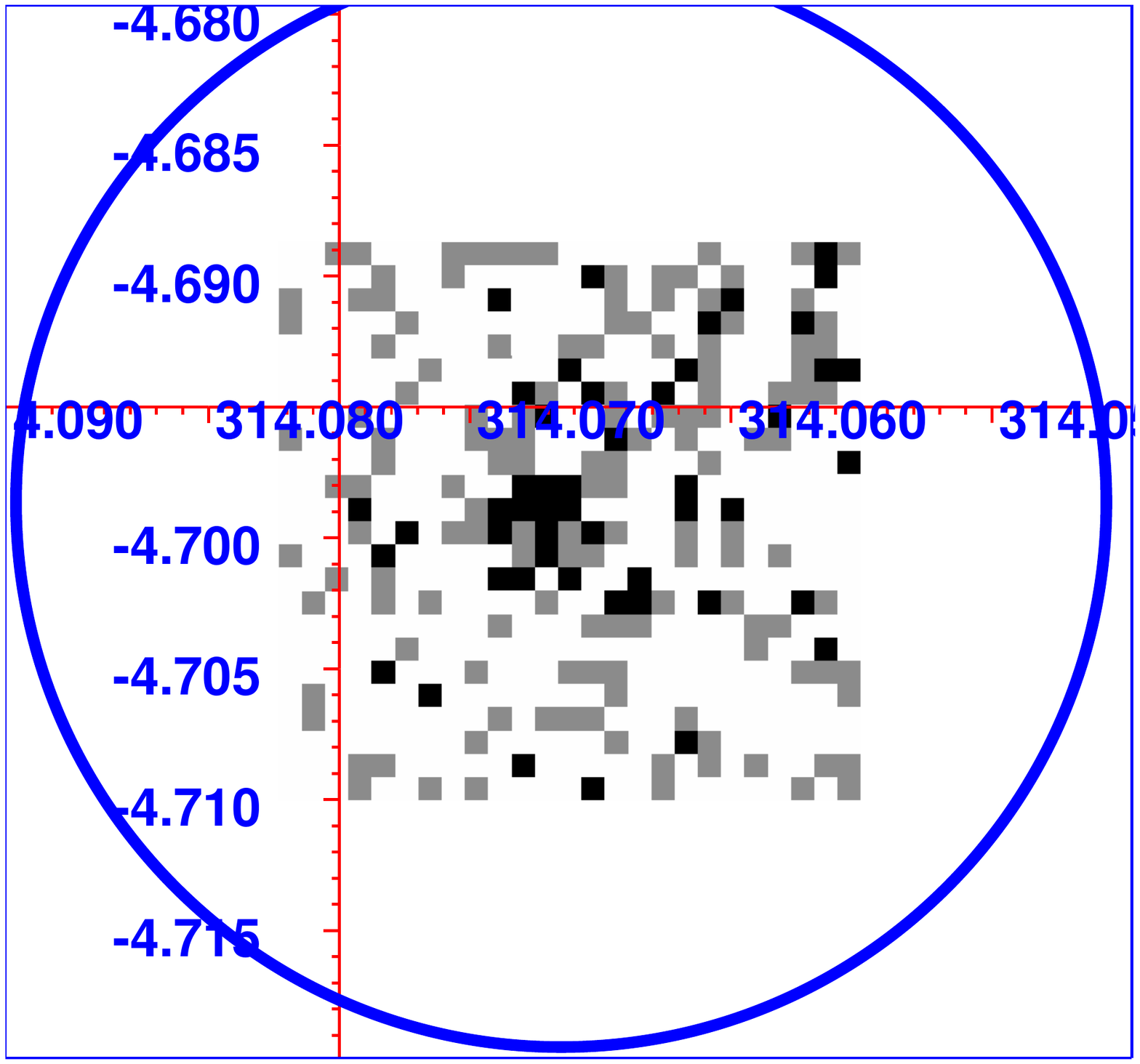}\includegraphics[width=2.7in,angle=0,bb=35 144 575 651,clip]{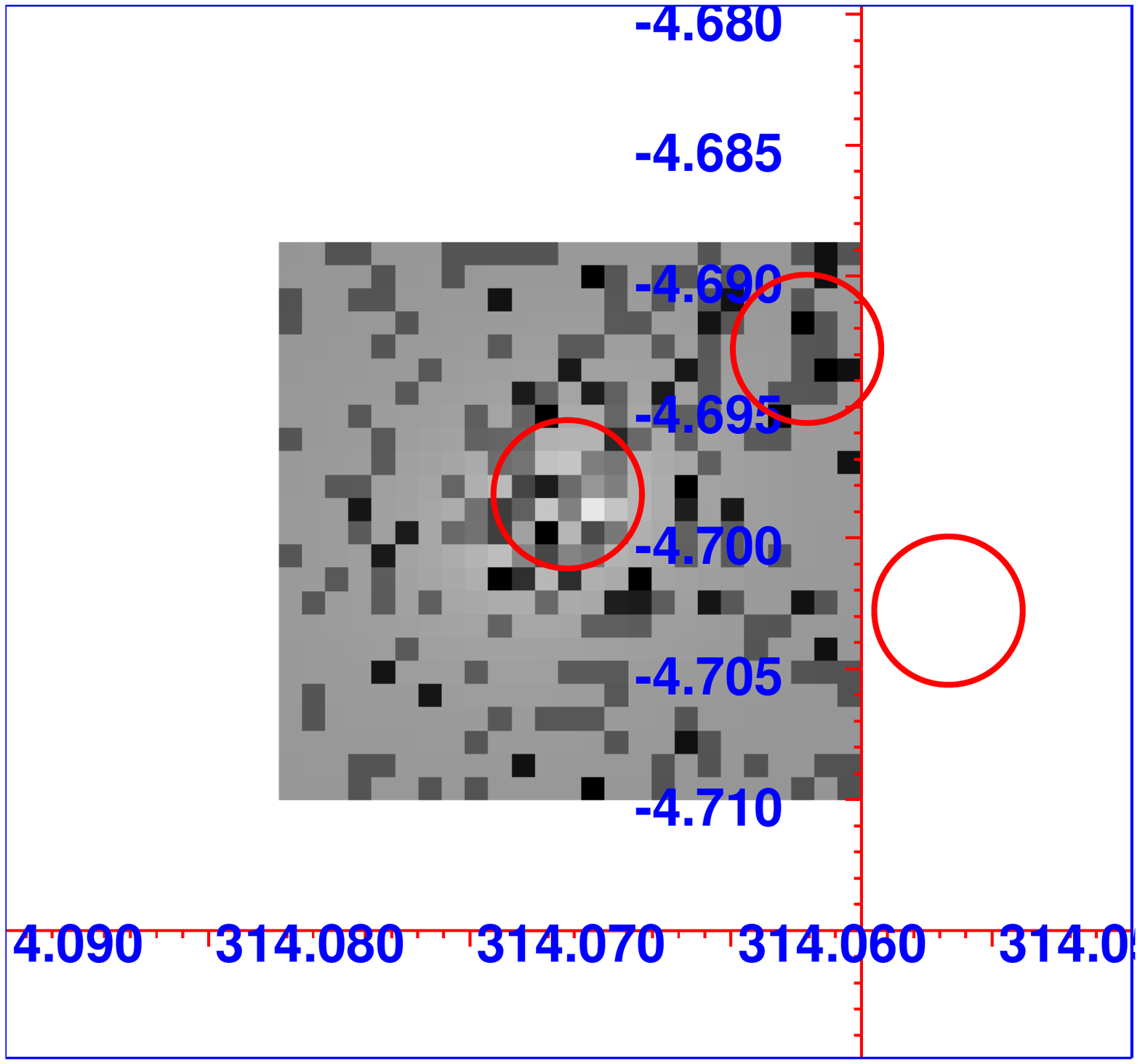}\\
    \includegraphics[width=2.7in,angle=0,bb=35 144 575 651,clip]{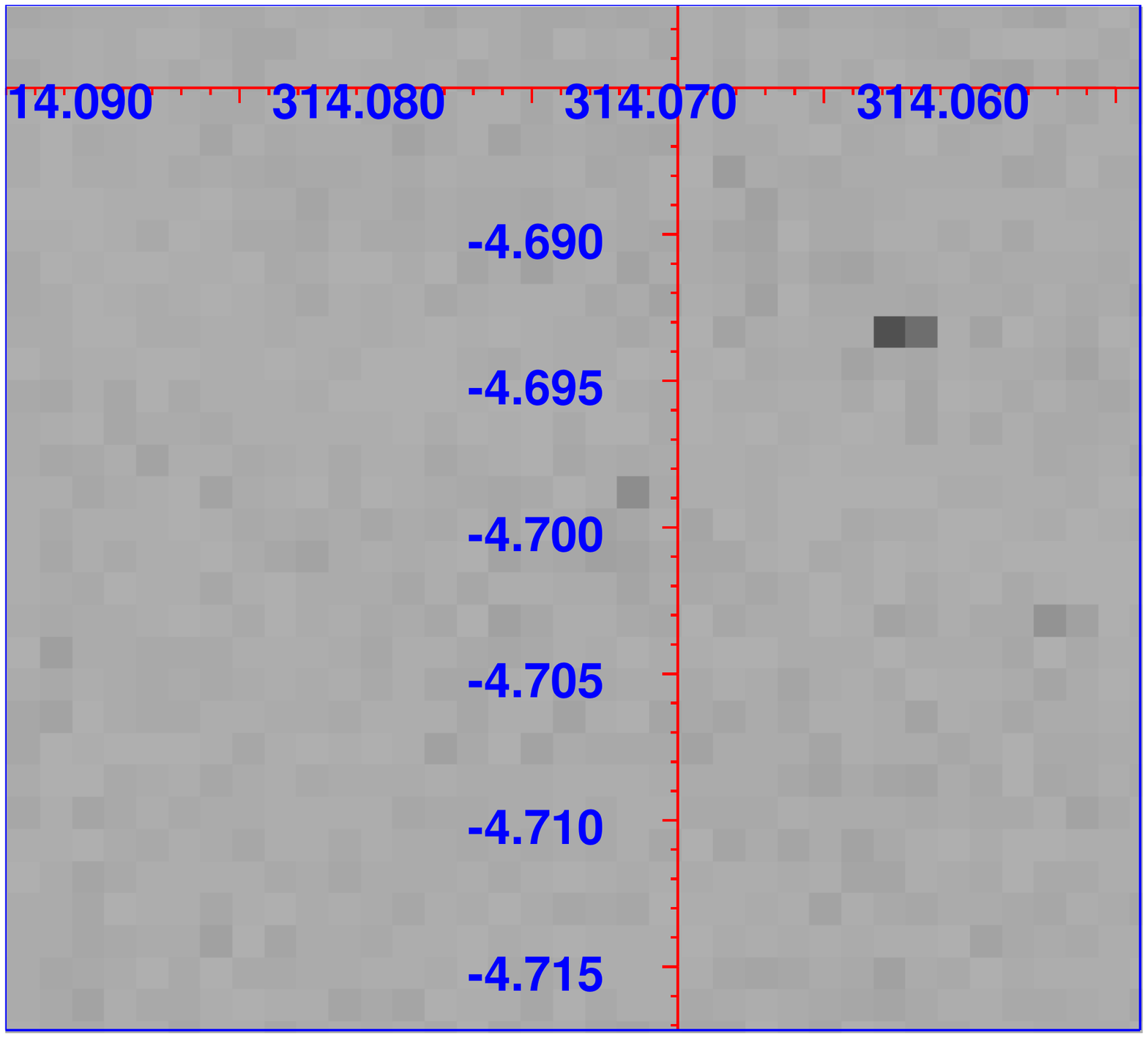}\includegraphics[width=2.7in,angle=0,bb=35 144 575 651,clip]{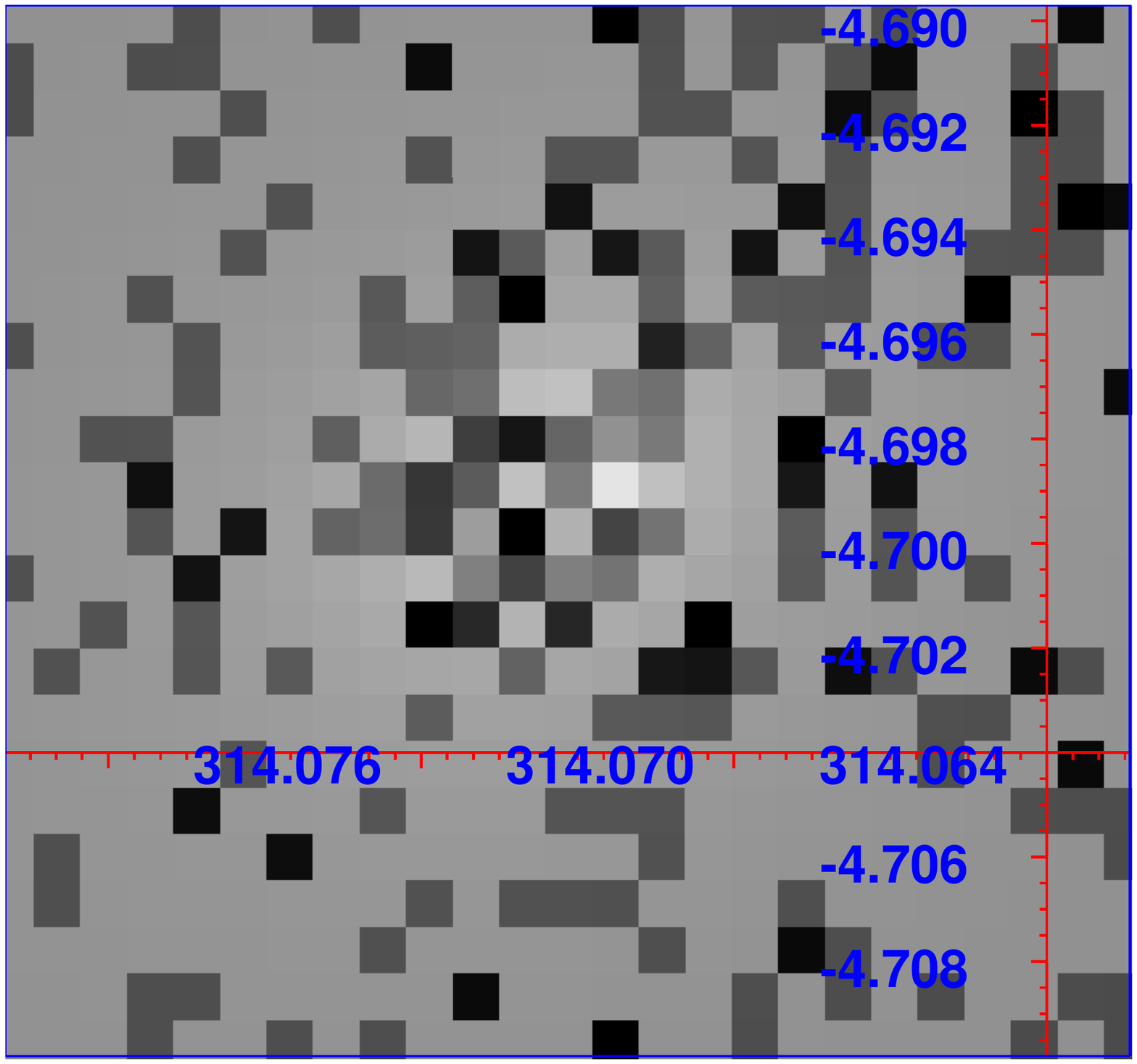}
    \caption{Same as Fig.~Fig.~\ref{fig:bmw0522_X} for CXOSEXSI J205617.1-044155.}
  \label{fig:CXOSEXSI}
  \end{center}
  \end{figure*}

  MS~2053.7-0449 (Harrison et al. 2003) is a fairly bright cluster,
  probably physically associated with the smaller cluster CXOSEXSI
  J205617.1-044155. The X-ray residual image does not show any
  substructures or point sources more significant than the 2.5$\sigma$
  level (Fig.~\ref{fig:MS2053_X}). Gilmour et al. (2009) detected
  several AGN, but they are probably too faint to appear as
  significant sources. This statement is in good agreement with the
  Chandra image not showing any bright point source.

On the optical side, the redshift histogram in an area of 2.2~arcmin
radius around MS~2053.7-0449 is shown in
Fig.~\ref{fig:MS2053_X}. There are 30 galaxies in the [0.57,0.60]
redshift range, and the SG method finds a single massive structure (see
Table~\ref{tab:SG}) associated with cluster MS~2053.7-0449 itself.
MS~2053.7-0449 is therefore a cluster without any detectable
substructures, whether in X-rays or in the optical. It may
merge with CXOSEXSI J205617.1-044155 in the future.

As seen in Fig.~\ref{fig:CXOSEXSI}, CXOSEXSI J205617.1-044155
(Harrison et al. 2003) is rather faint in X-rays (X-ray luminosity of
4.37~10$^{43}$~erg/s) and not visible in the Chandra image.  The X-ray
image may show an X-ray point source north--west of the cluster also
listed in Gilmour et al. (2009). This point source, as well as two
other ones known in Gilmour et al. (2009), is visible in the Chandra
image.  The subtraction of a $\beta -$model on the XMM-Newton image
does not allow detecting any significant X-ray residual. In a zone of
2.2~arcmin around CXOSEXSI J205617.1-044155, there is only one redshift
(z=0.6002) corresponding to the cluster dominant galaxy.

\subsection{GHO~2143+0408  (326.52000$^o$, +4.3886$^o$, z=0.5310) } 

\begin{figure*}
  \begin{center}
    \includegraphics[width=2.7in,angle=0,bb=35 144 575 651,clip]{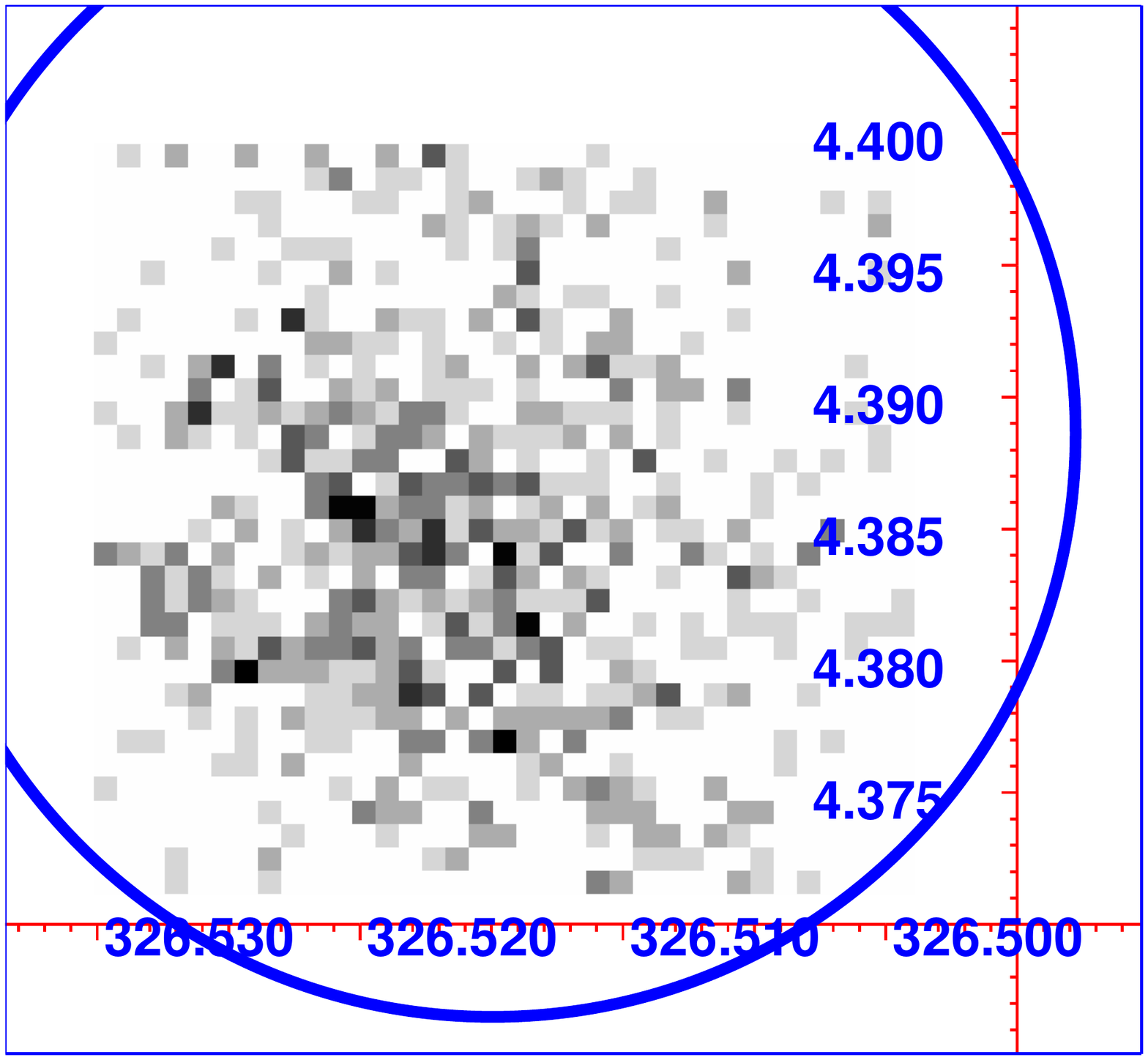}\includegraphics[width=2.7in,angle=0,bb=35 144 575 651,clip]{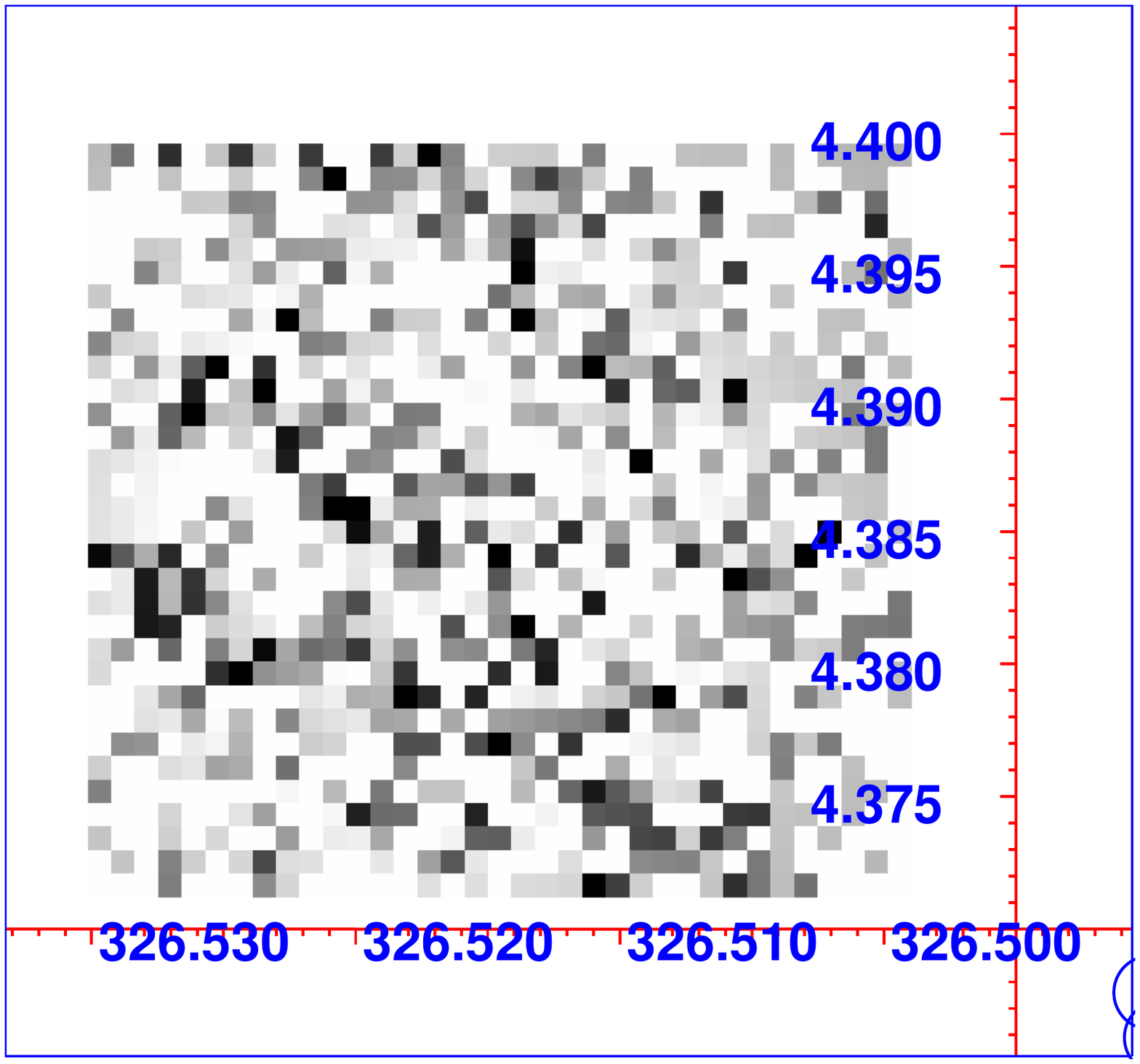}\\
    \includegraphics[width=2.7in,angle=0,bb=35 144 575 651,clip]{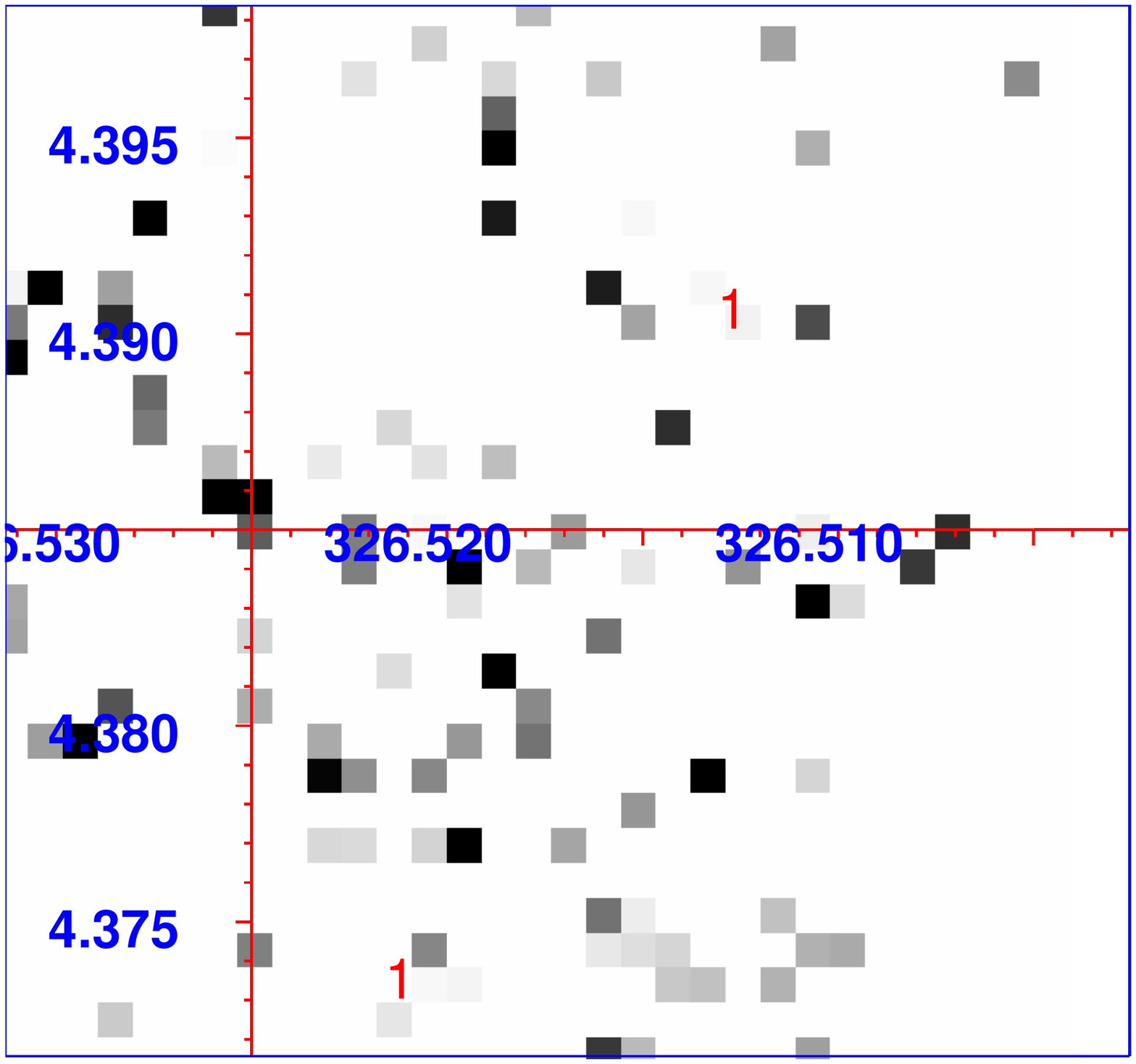}\includegraphics[width=3.5in,angle=0,bb=15 200 575 600,clip]{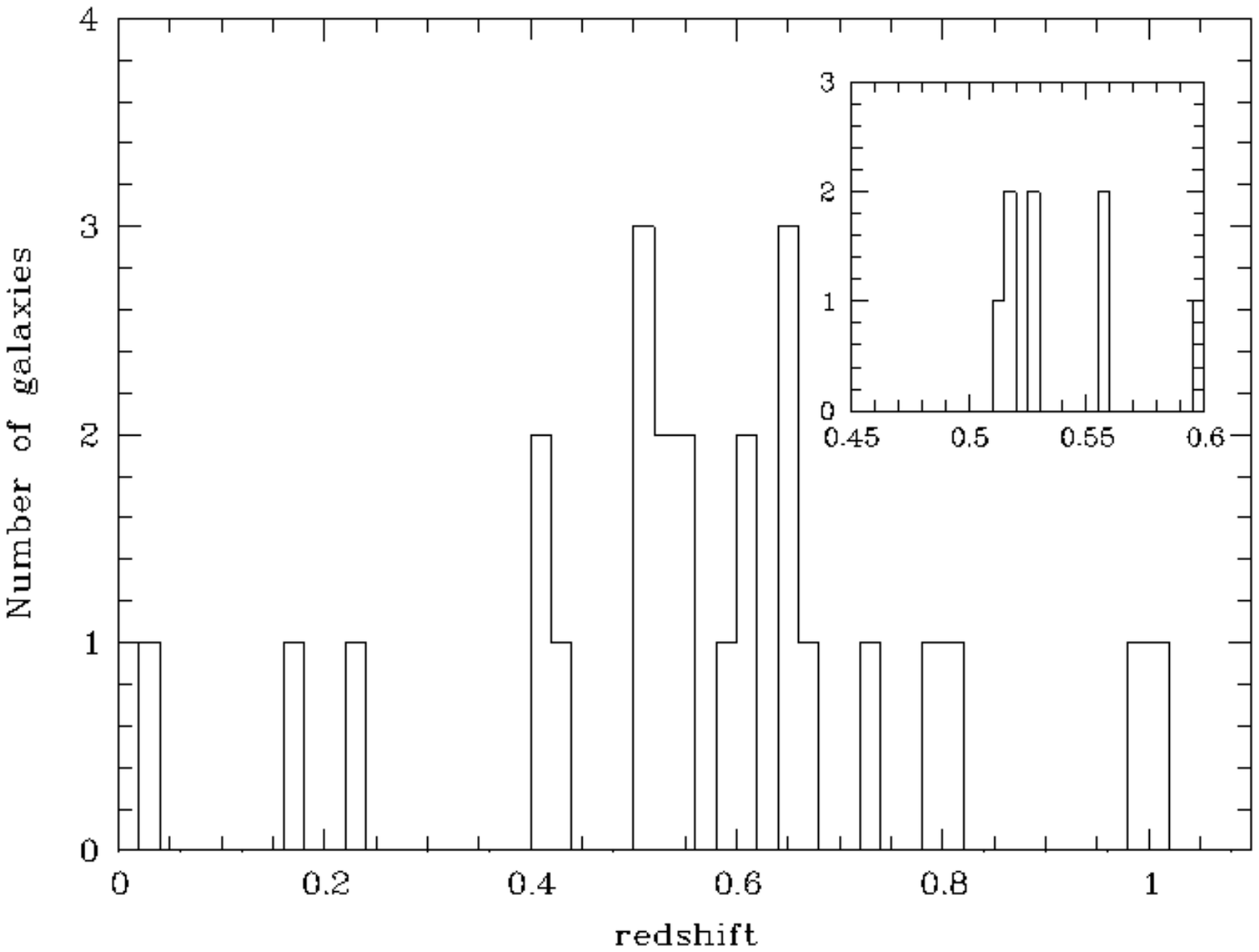}
    \caption{Same as Fig.~\ref{fig:cl0016_X} for GHO~2143+0408.}
  \label{fig:GHO2143_X}
  \end{center}
  \end{figure*}

This is a massive X-ray cluster discovered by Vikhlinin et al. (1998)
and member of the WARPS survey (WARP~J2146.0+0423, Perlman et
al. 2002).  The X-ray residual image does not show any X-ray source
more significant than the 2.5$\sigma$ level, and there is no known
active object along the line of sight (Fig.~\ref{fig:GHO2143_X}).

We have obtained 28 redshifts with the ESO VLT and FORS2 (see
Table~\ref{tabzspecus} at the end of Appendix~C).  The SG method
detects a few structures along the line of sight, but none in the
cluster region. The cluster itself is detected by the SG method
without any additional substructure at the same redshift.

\subsection{RX~J2202.7-1902  (330.68708$^o$, --19.0361$^o$, z=0.4380) } 

\begin{figure*}
  \begin{center}
    \includegraphics[width=2.7in,angle=0,bb=35 144 575 651,clip]{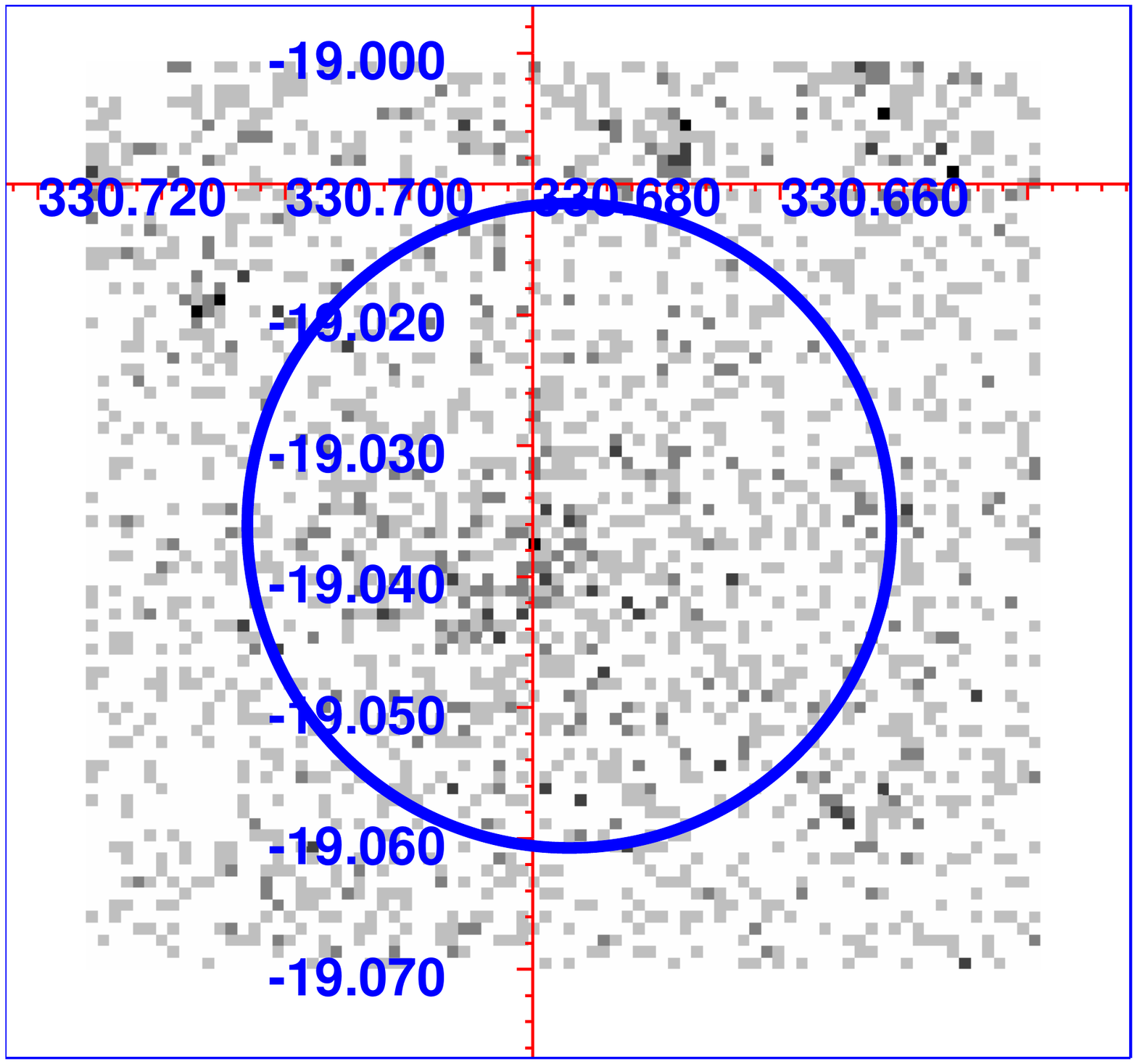}\includegraphics[width=2.7in,angle=0,bb=35 144 575 651,clip]{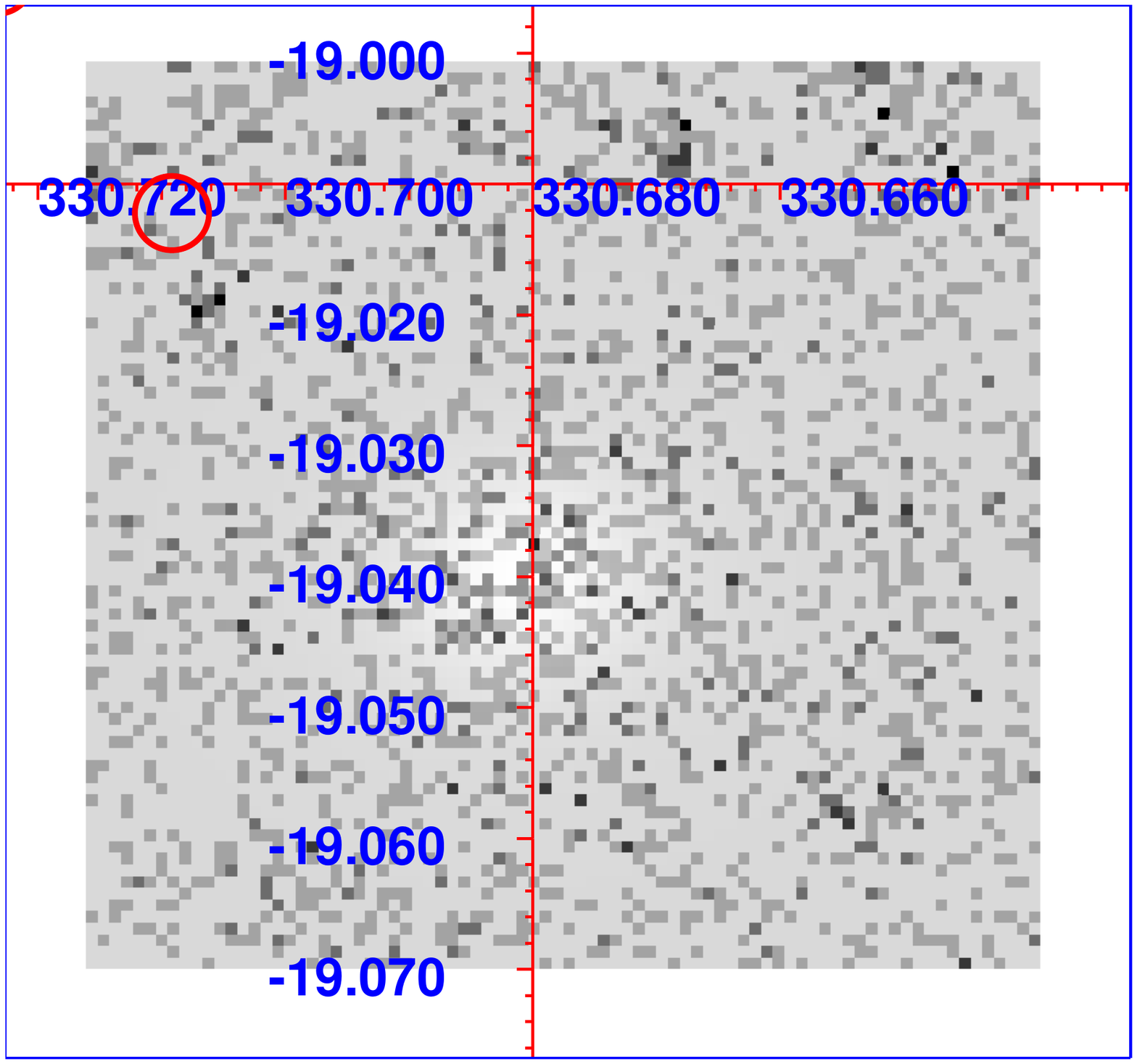}\\
    \includegraphics[width=2.7in,angle=0,bb=35 144 575 651,clip]{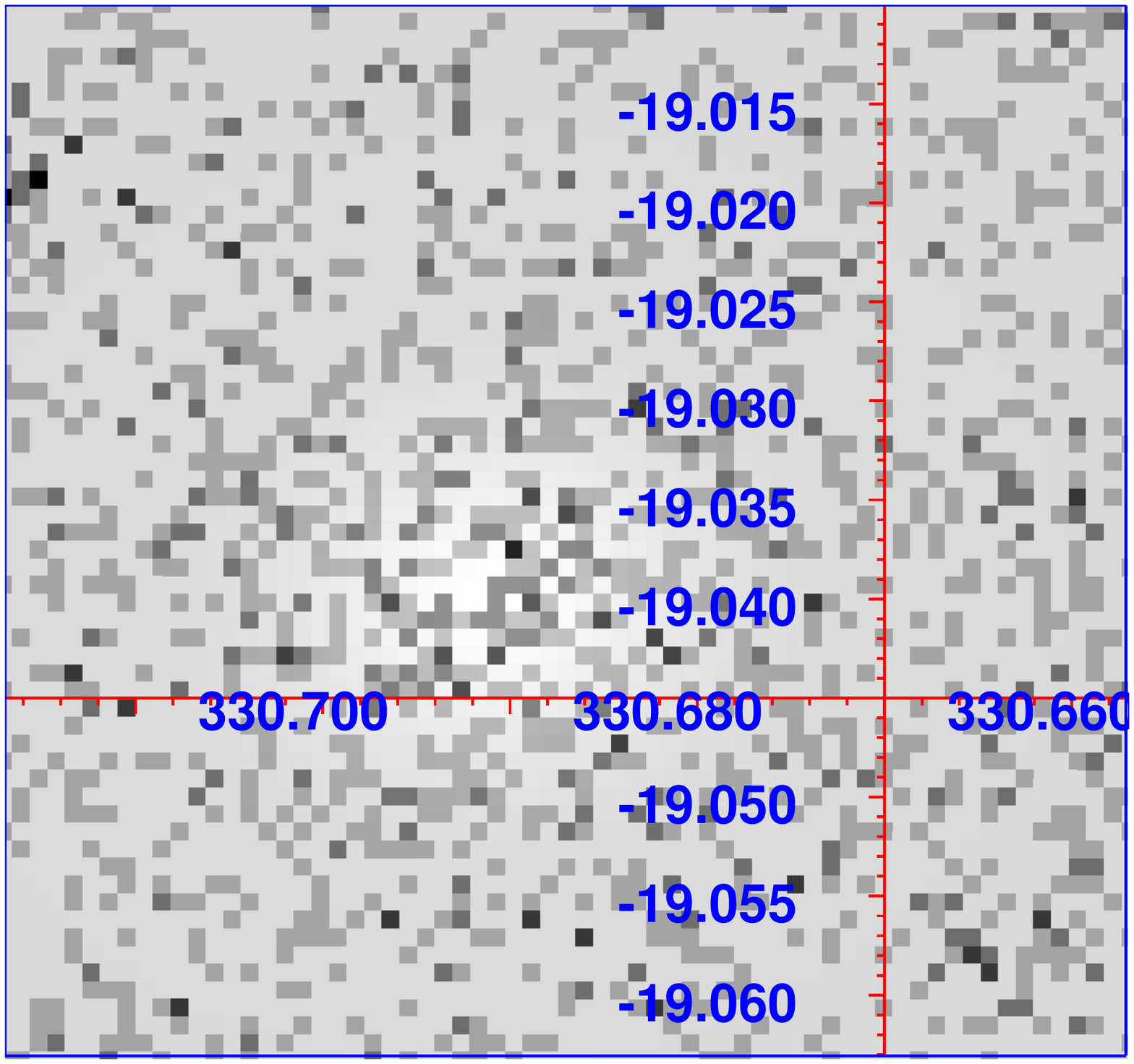}
    \caption{Same as Fig.~\ref{fig:ms0302X} for RX~J2202.7-1902.}
  \label{fig:RX2202_X}
  \end{center}
  \end{figure*}

  The X-ray image of RX~J2202.7-1902 (Vikhlinin et al. 1998) shows a
  very weak extended source (Fig.~\ref{fig:RX2202_X}). No significant
  residuals appear after subtracting a model. NED gives only a few
  redshifts along the cluster line of sight, preventing any SG
  analysis.

\subsection{RX~J2328.8+1453 (352.20792$^o$, +14.8867$^o$, z=0.4970)} 

\begin{figure*}
  \begin{center}
    \includegraphics[width=2.7in,angle=0,bb=35 144 575 651,clip]{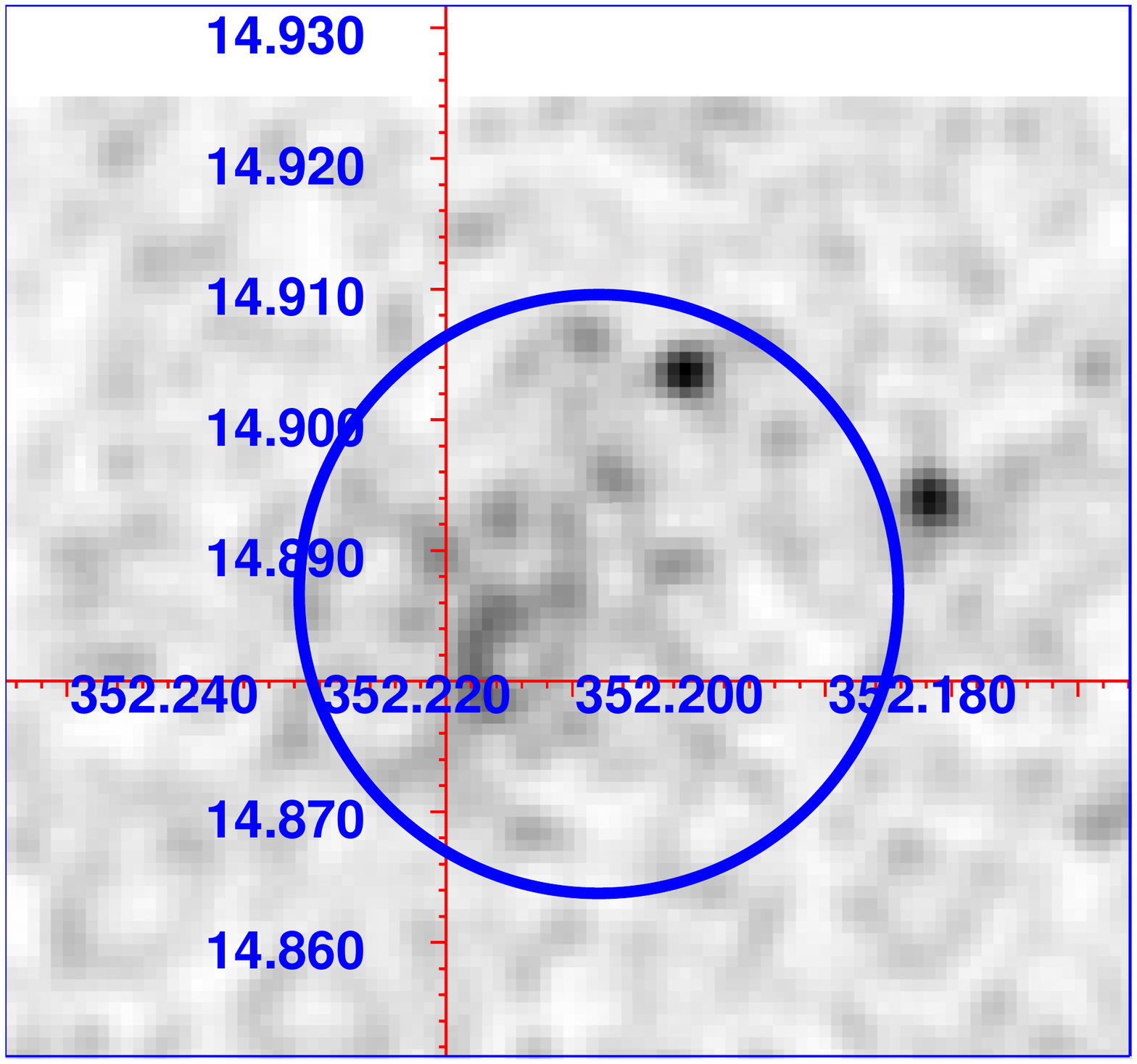}\includegraphics[width=2.7in,angle=0,bb=35 144 575 651,clip]{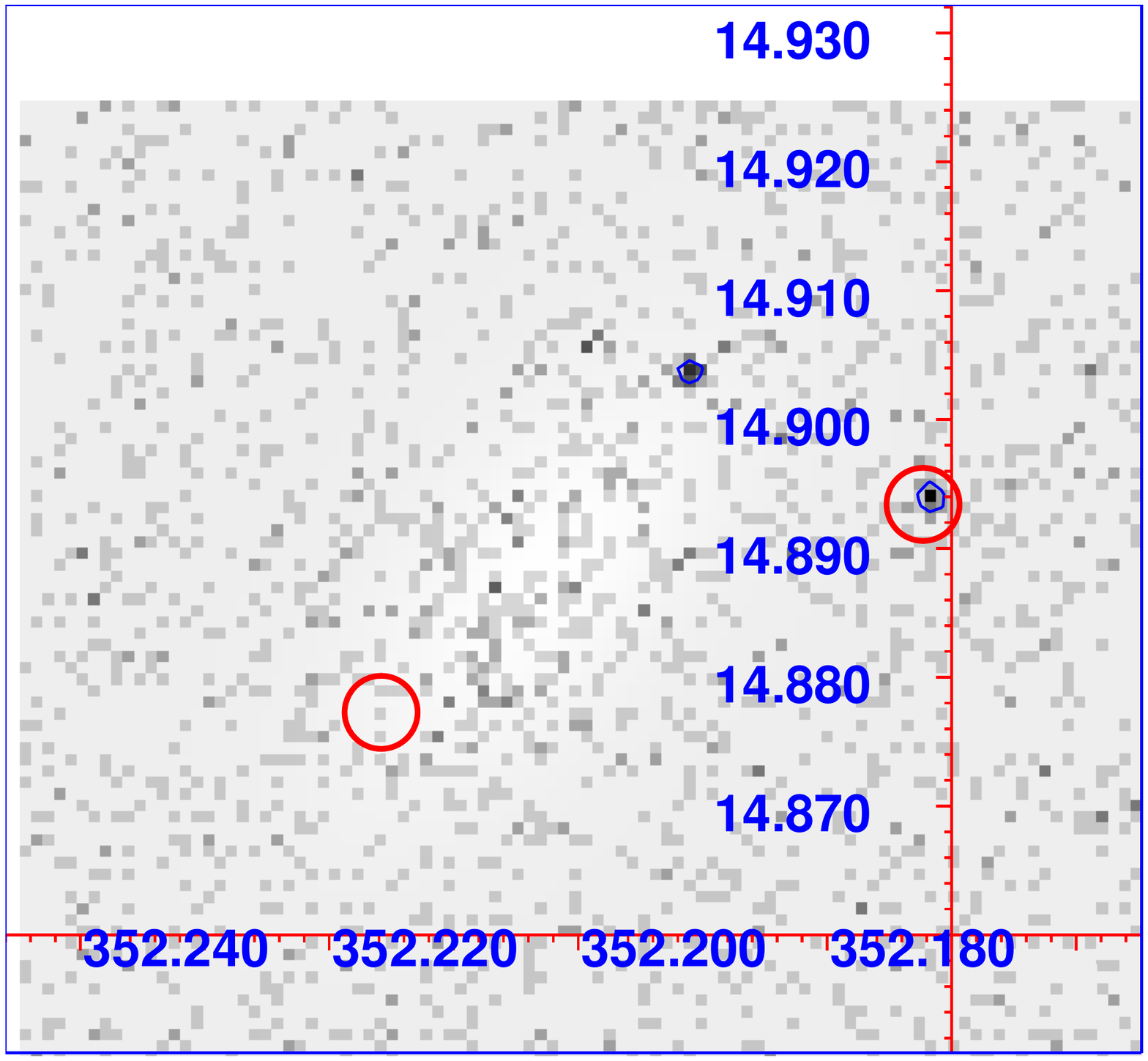}\\
    \includegraphics[width=2.7in,angle=0,bb=35 144 575 651,clip]{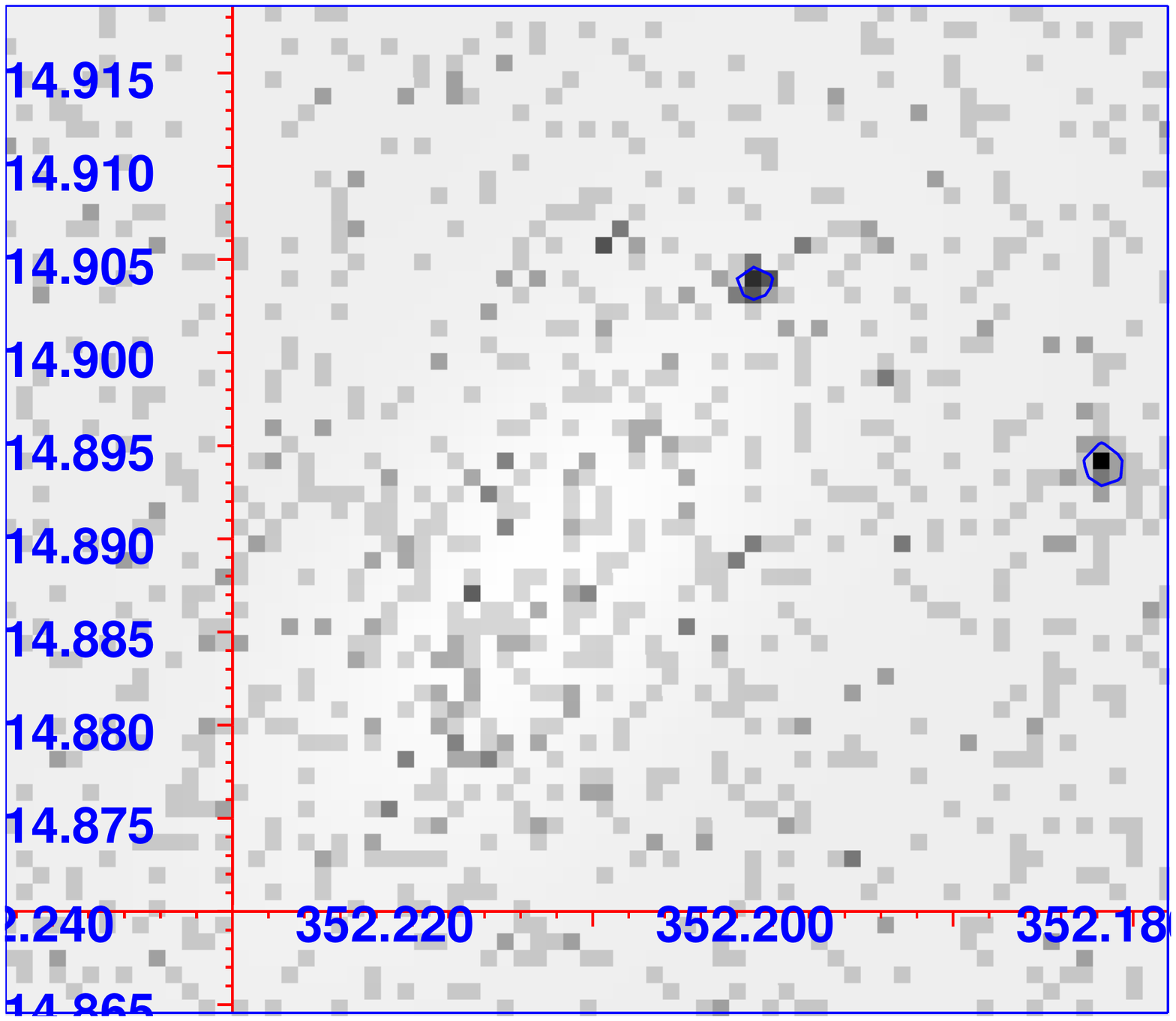}
    \caption{Same as Fig.~\ref{fig:ms0302X} for RX~J2328.8+1453.}
  \label{fig:RX2328_X}
  \end{center}
  \end{figure*}

The X-ray image of RX~J2328.8+1453 (Vikhlinin et al. 1998) appears
quite diffuse (Fig.~\ref{fig:RX2328_X}), and the subtraction of a model
shows two small 2.5$\sigma$ level sources (too faint to provide a
reliable X-ray luminosity).  One of them is identified with an active
object in Vizier.

NED gives for this cluster a redshift z=0.4970. There are only 5
other redshifts available in an area of 5~arcmin radius around this
cluster, and no galaxy is at a similar redshift, so an SG analysis is
impossible.


\clearpage

\section{SG analysis for clusters with no usable XMM-Newton data}
\label{sec:onlyzSG1}

We present in Section~\ref{sec:onlyzSG1} the results of the SG analysis
for the clusters with no usable XMM-Newton data, but with more than 15
redshifts in the cluster range.

\subsection{Cl J0023+0423B (5.96587$^o$, +4.3780$^o$, z=0.8453)} 

\begin{figure}
\begin{center}
\includegraphics[width=6cm]{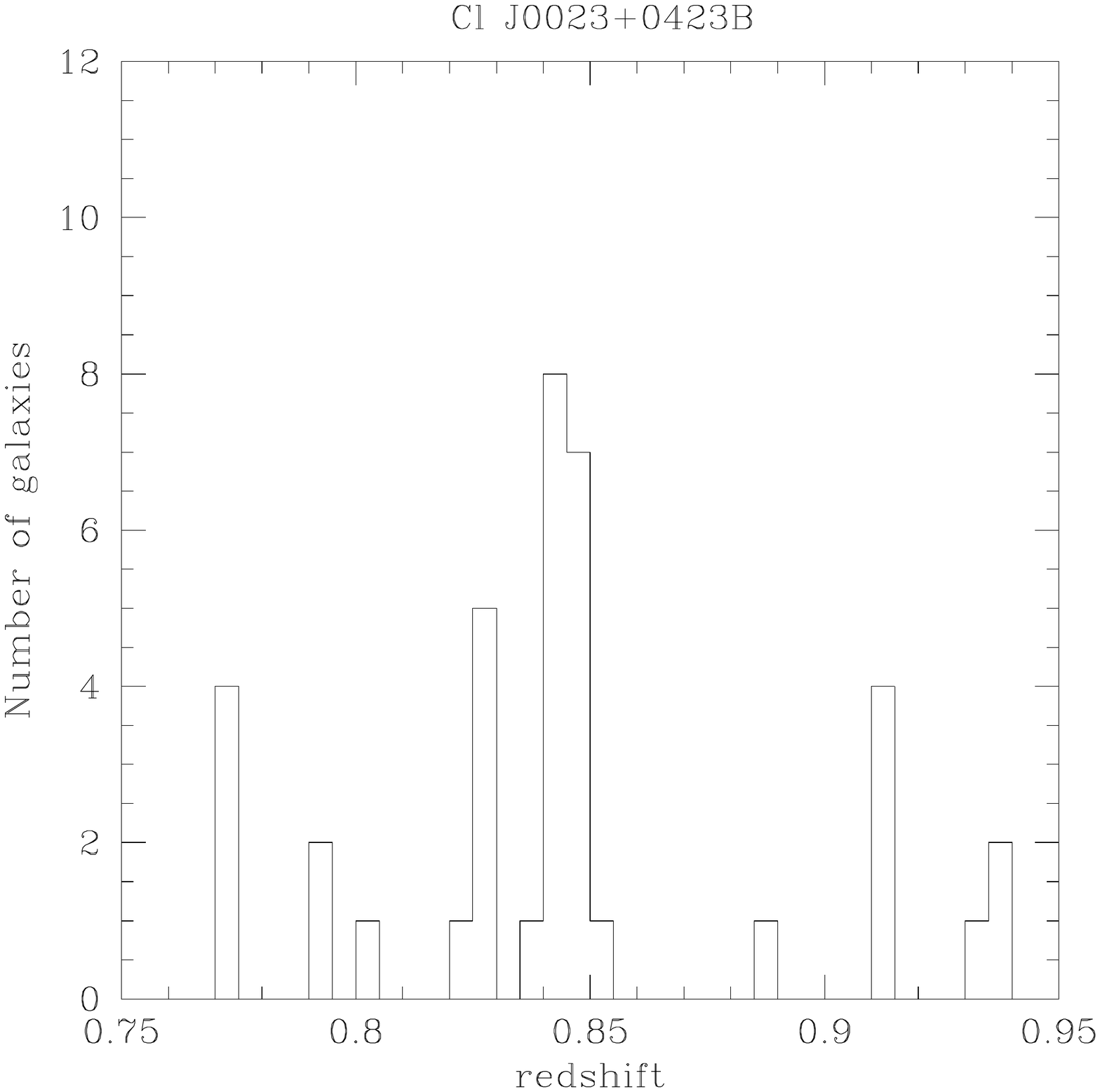} 
\caption{Redshift histogram for Cl J0023+0423B. }
\label{fig:cl0023_histoz}
\end{center}
\end{figure}

There are 23 galaxies in the [0.82,0.855] redshift range
(Fig.~\ref{fig:cl0023_histoz}), and the SG method detects a single
structure. We note the presence of several other structures on the line
of sight at redshifts that are not very different from that of the cluster,
suggesting we may be observing a filament seen close to face-on.

\subsection{CXOMP J002650.2+171935 (6.70917$^o$, +17.3266$^o$, z=0.4907)} 

\begin{figure}
\begin{center}
\includegraphics[width=6cm]{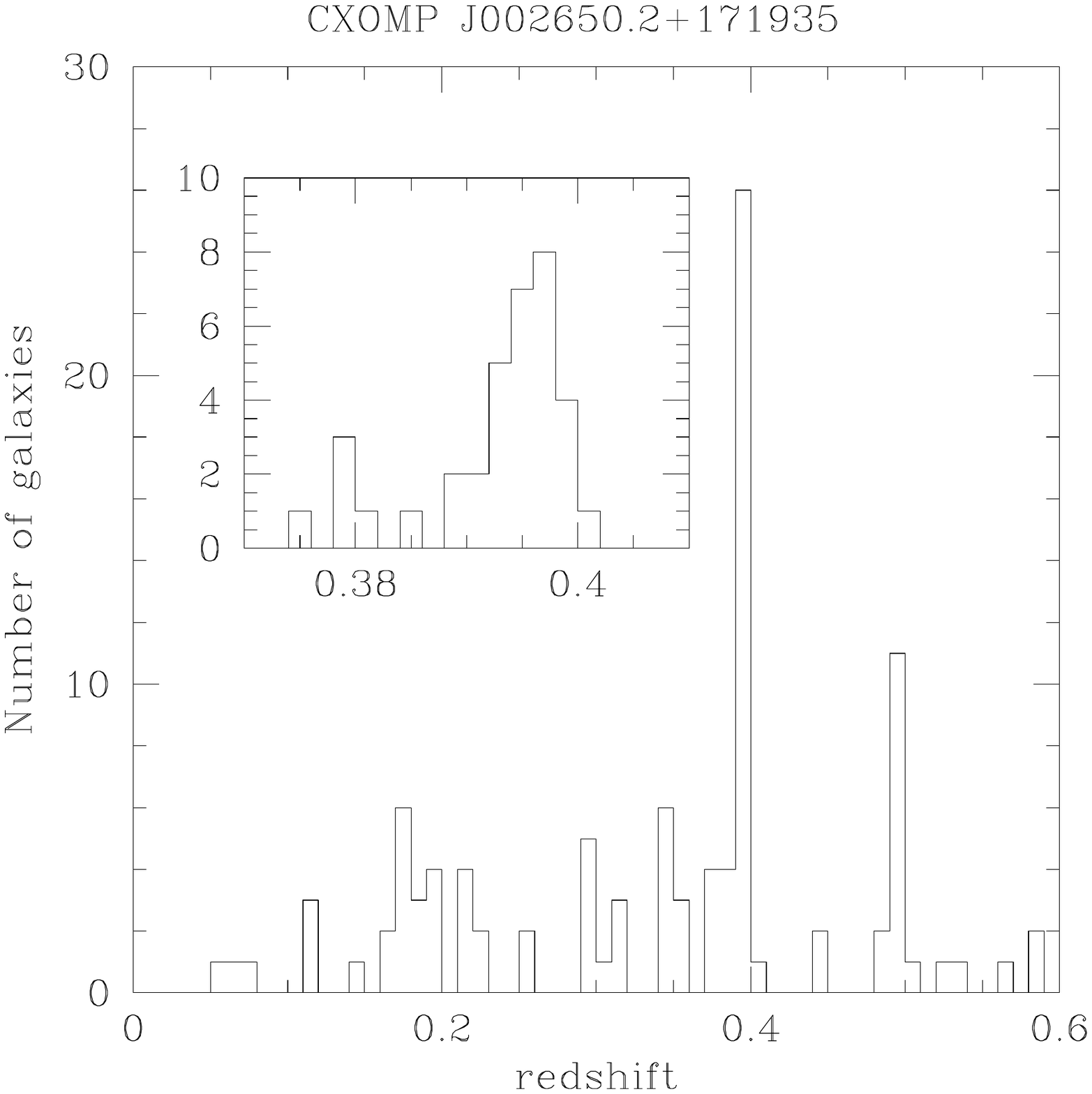} 
\caption{Redshift histogram for CXOMP J002650.2+171935. The insert shows a zoom around
  the cluster redshift. }
\label{fig:cx0026_histoz}
\end{center}
\end{figure}

CXOMP J002650.2+171935 was discovered by Barkhouse et al. (2006) in
the CHaMP survey. Its redshift histogram shows several small
foreground structures and a background structure at z$\sim 0.5$. The
main structure, corresponding to the cluster, has 29 galaxies in the
[0.388,0.402] range (Fig.~\ref{fig:cx0026_histoz}), and its redshift
distribution is roughly symmetric.

The SG analysis detects a single structure.

\subsection{CXOMP~J091126.6+055012 (137.86083$^o$, +5.8368$^o$, z=0.7682)} 

\begin{figure}
\begin{center}
\includegraphics[width=6cm]{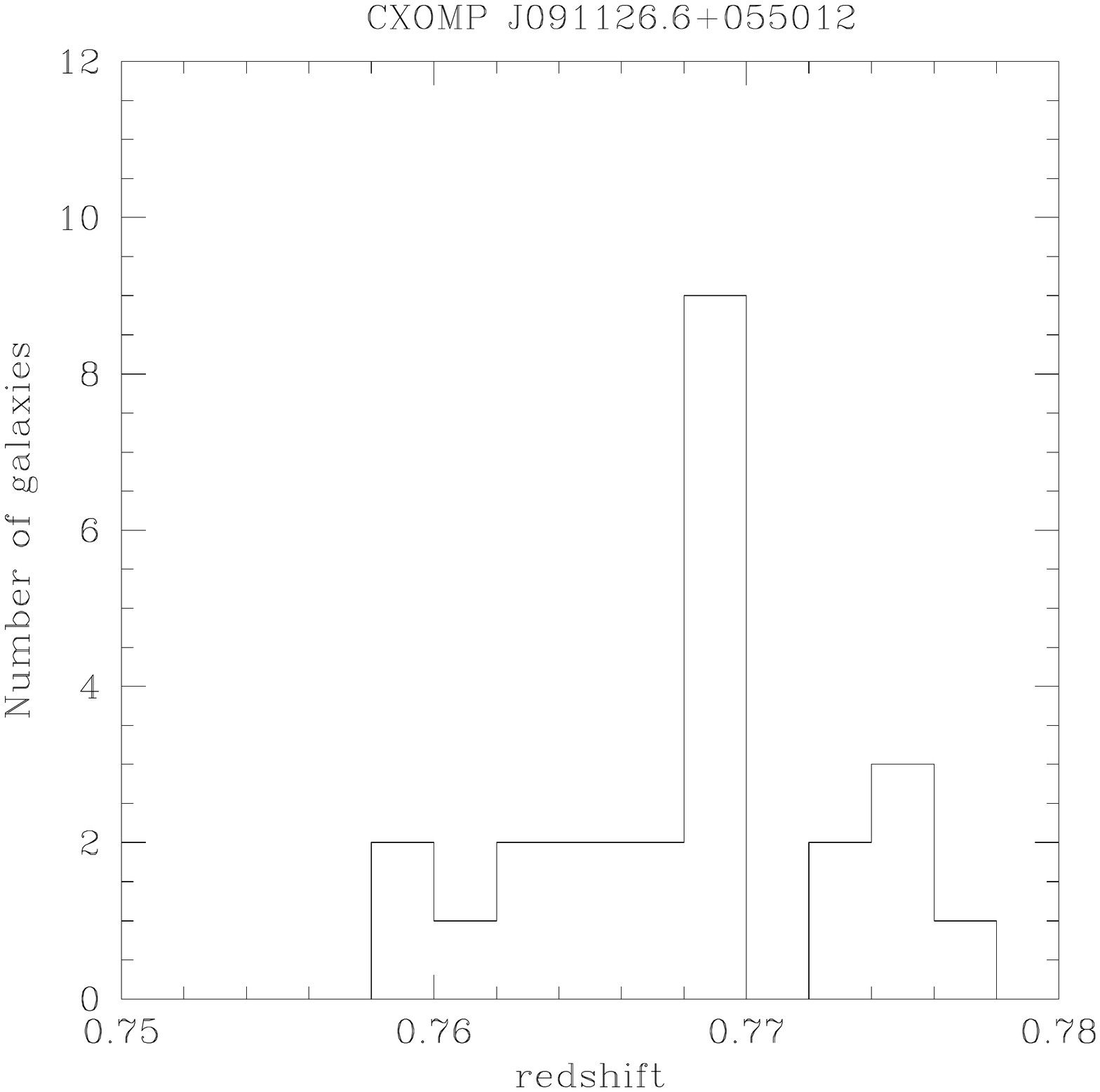} 
\includegraphics[width=6cm,angle=270]{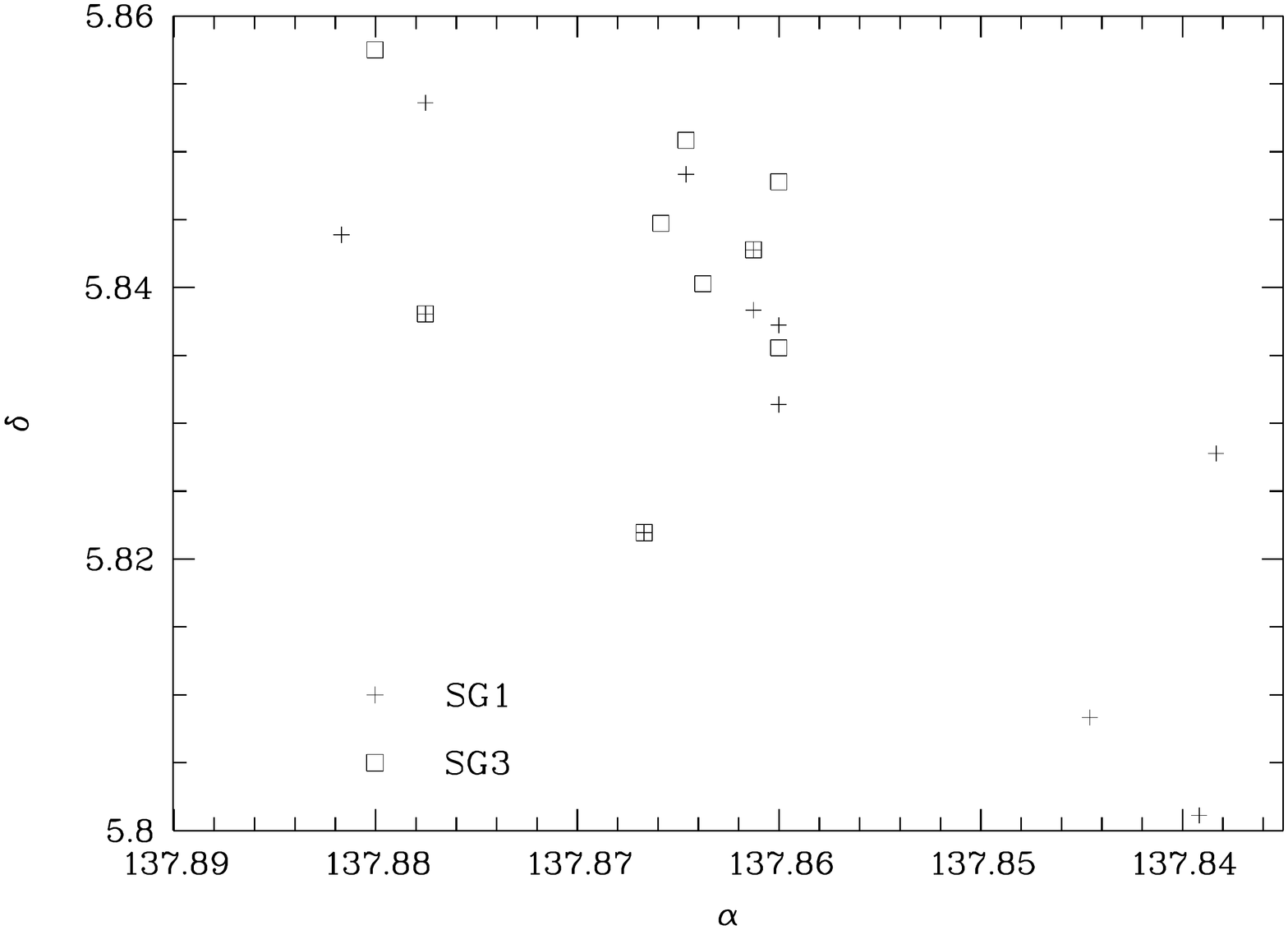} 
\caption{Upper figure: redshift histogram for CXOMP~J091126.6+055012. Lower
figure: SG1 and SG3 galaxy spatial distribution.}
\label{fig:CX0911_histoz}
\end{center}
\end{figure}

CXOMP~J091126.6+055012 was also discovered by Barkhouse et al. (2006)
in the CHaMP survey.  The XMM-Newton image of CXOMP~J091126.6+055012
shows a very faint source at the position given by NED, too faint to
be properly analysed, so we are including it in this section.

There is a peak in the redshift histogram at z$\sim$0.769 (see
Fig.~\ref{fig:CX0911_histoz}), with 18 galaxies with redshifts in the
[0.758,0.778] range. The SG analysis separates two structures within
the main structure of the velocity histogram, and a third structure at
redshift z$\sim$0.775 (see Table~\ref{tab:SG2}). SG3 tends to be
located north-east of the cluster centre.

\subsection{LCDCS 0110 (159.46917$^o$, --12.7289$^o$, z=0.5789)} 

\begin{figure}
\begin{center}
\includegraphics[width=6cm]{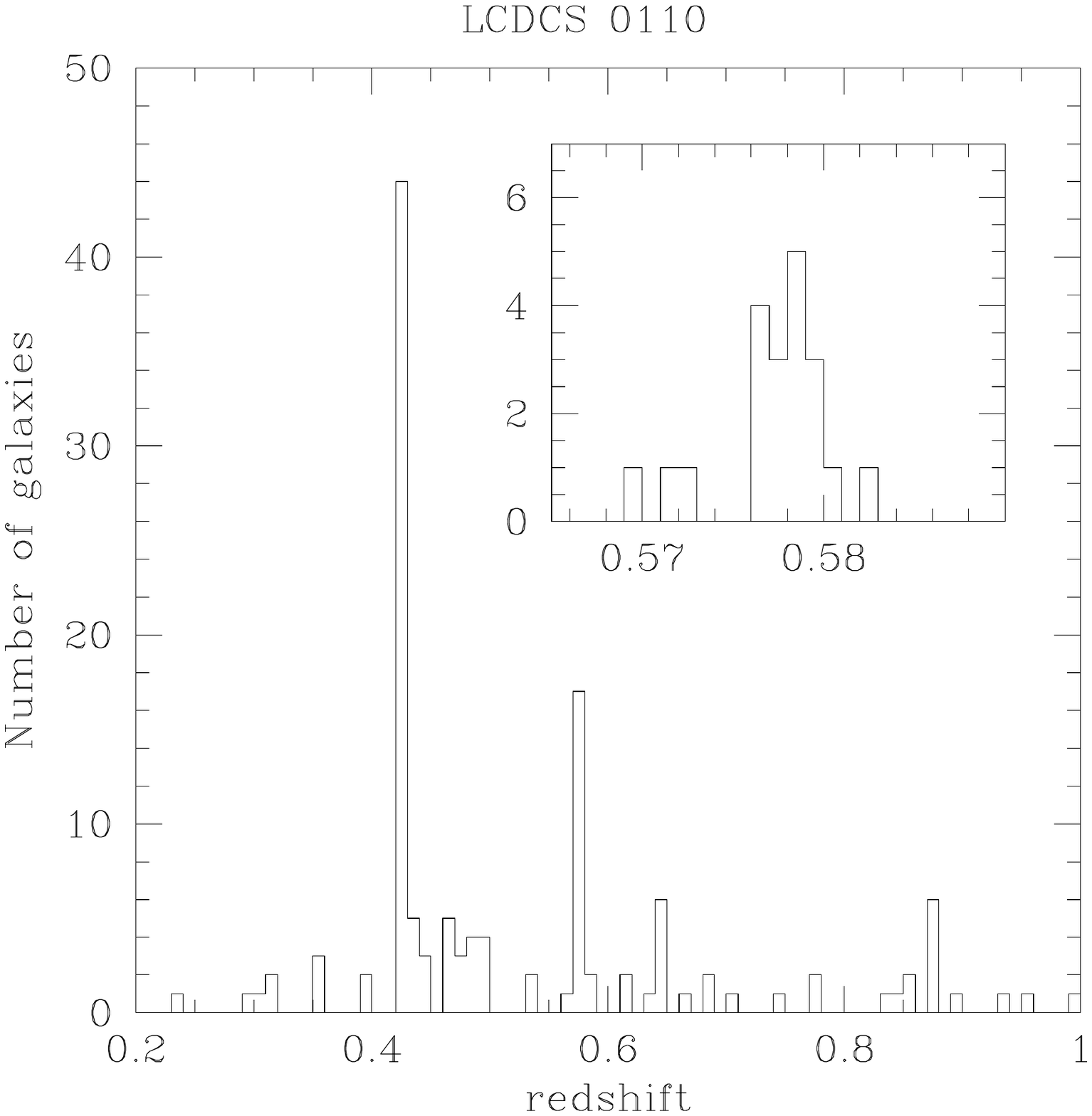} 
\includegraphics[width=6cm,angle=270]{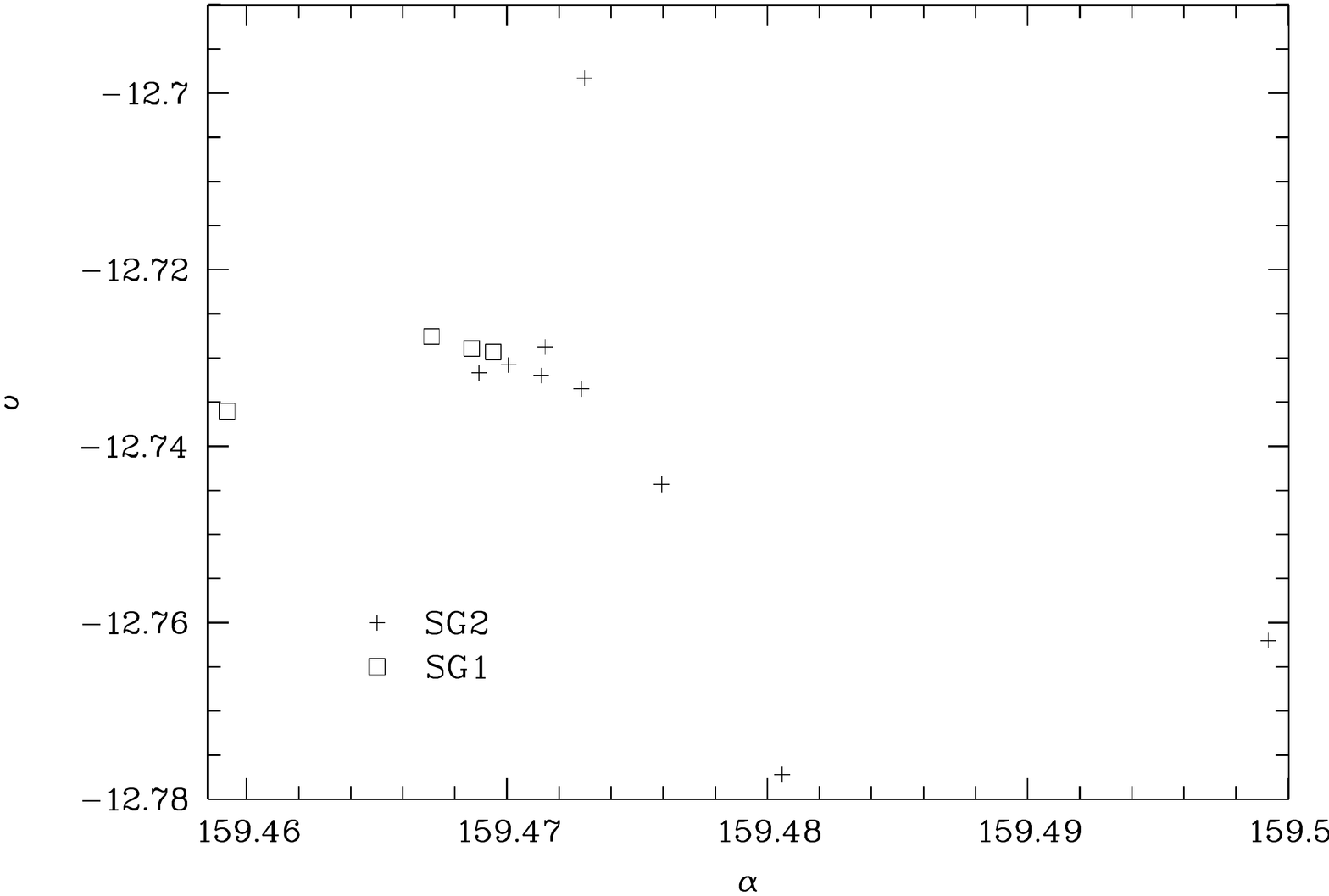} 
\caption{Upper figure: redshift histogram for LCDCS 0110. The insert shows 
a zoom around the cluster redshift. Lower figure: SG1 and SG2 galaxy spatial 
distribution.}
\label{fig:lcdcs0110_histoz}
\end{center}
\end{figure}

The full redshift histogram shows a peak at z=0.5789 (the redshift
given by NED for this cluster), and a larger peak at z$\sim$0.42
(Fig.~\ref{fig:lcdcs0110_histoz}). A weak lensing mass reconstruction
shows the presence of two clusters close to each other on the line of
sight (Clowe et al. 2006).

There are 18 galaxies in the [0.575,0.585] redshift range. The SG
method detects two main structures (see Table~\ref{tab:SG2}), with
redshifts roughly corresponding to the two peaks of the zoomed
redshift histogram. SG1 is east of the cluster centre and SG2 is west of the 
cluster centre.

\subsection{LCDCS 0130 (160.17333$^o$, --11.9308$^o$, z=0.7043)} 

\begin{figure}
\begin{center}
\includegraphics[width=6cm]{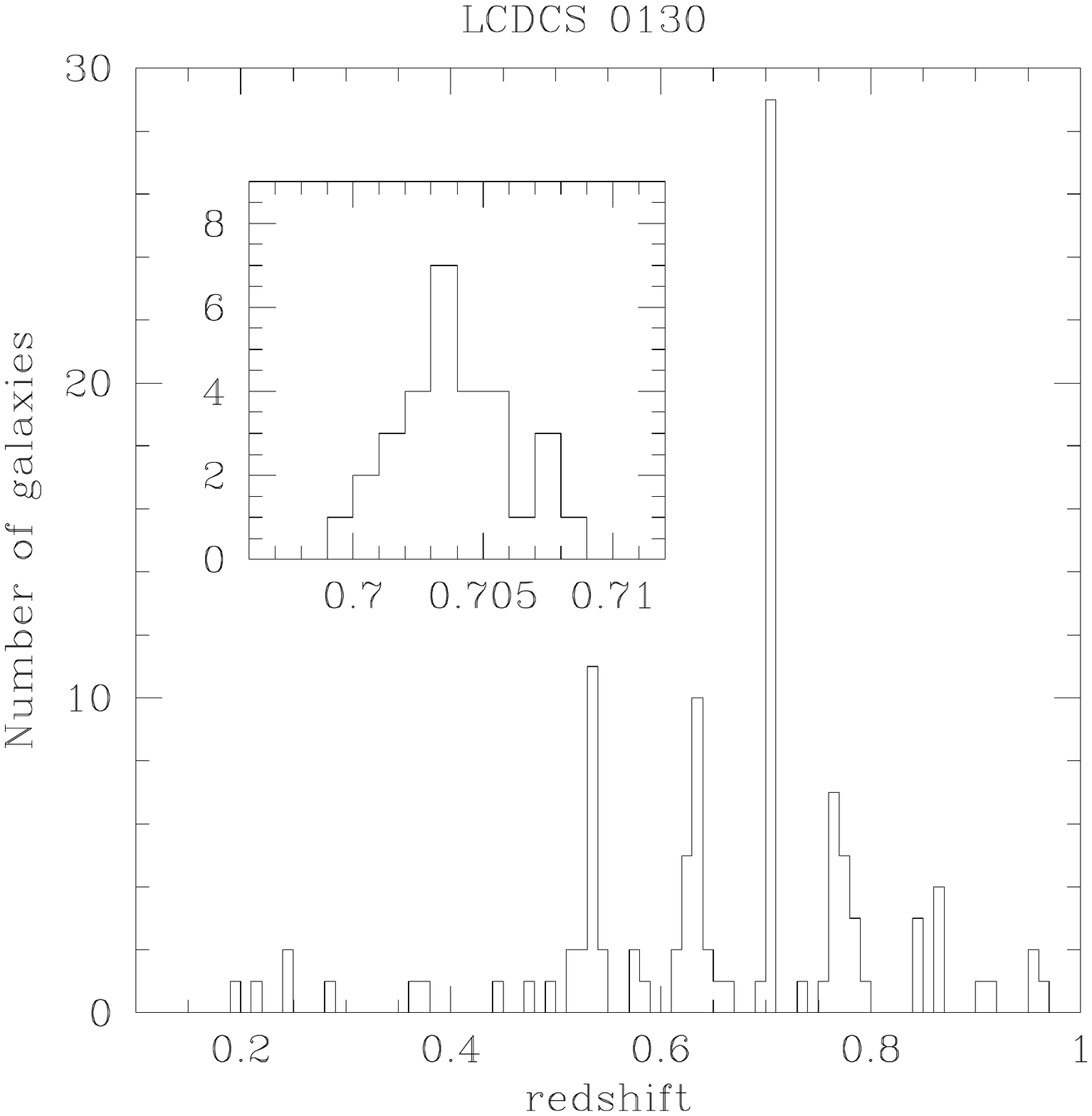} 
\includegraphics[width=6cm,angle=270]{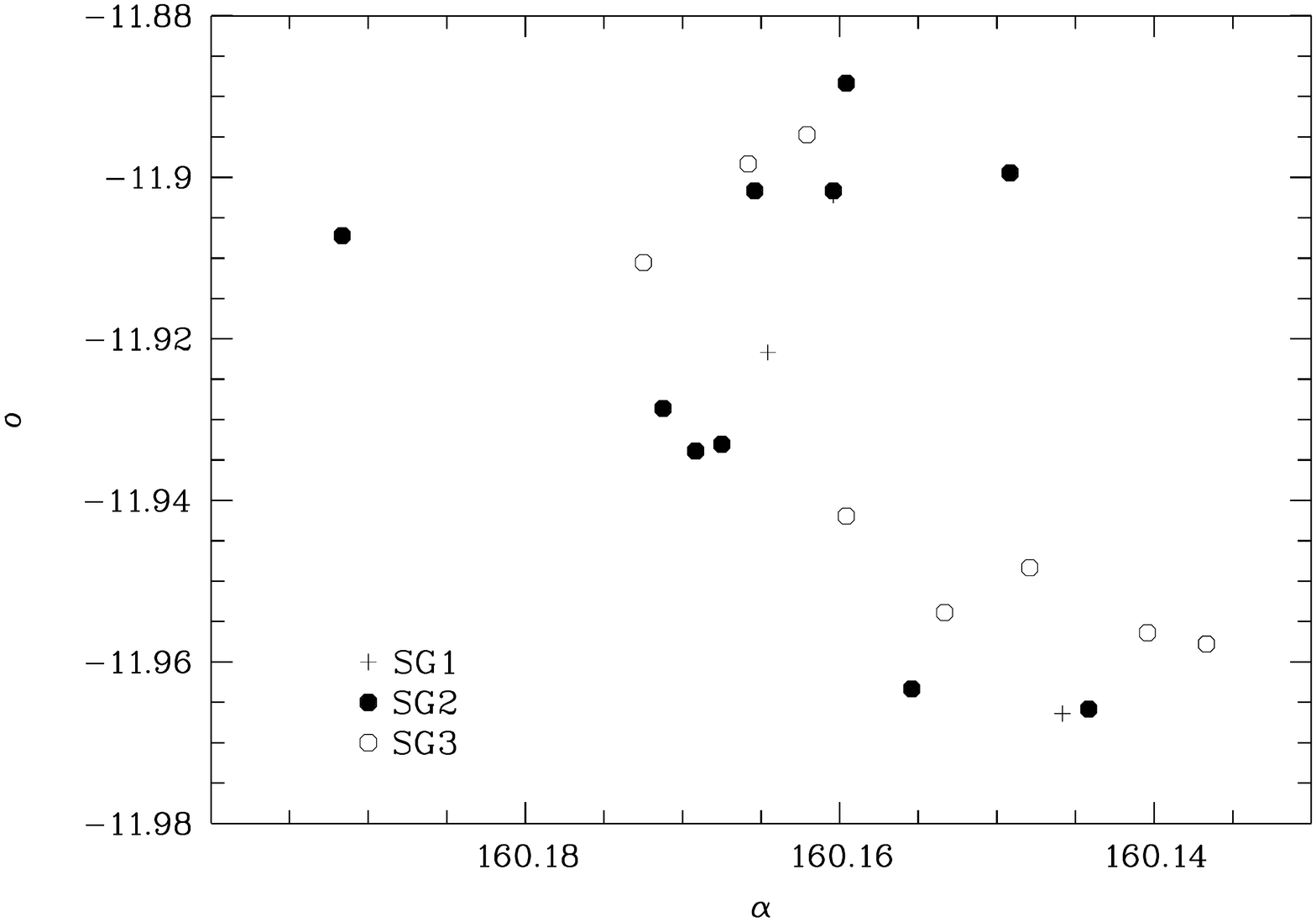} 
\caption{Upper figure: redshift histogram for LCDCS 0130. The insert shows a 
zoom around the cluster redshift. Lower figure: SG1, SG2, and SG3 galaxy 
spatial distribution.}
\label{fig:lcdcs0130_histoz}
\end{center}
\end{figure}

The full redshift histogram shows several peaks, in particular two in
front of and one behind the cluster, implying that this line of sight
is intercepting various galaxy structures
(Fig.~\ref{fig:lcdcs0130_histoz}).  There are 30 galaxies in the
[0.699,0.71] redshift range, where the redshift histogram is clearly
asymmetric. The SG method detects a main structure and two less
massive ones (see Table~\ref{tab:SG2}), in agreement with possible
substructuring found by Halliday et al. (2004).

\subsection{LCDCS 0172 ( (163.60083$^o$, --11.7717$^o$, z=0.6972)} 

\begin{figure}
\begin{center}
\includegraphics[width=6cm]{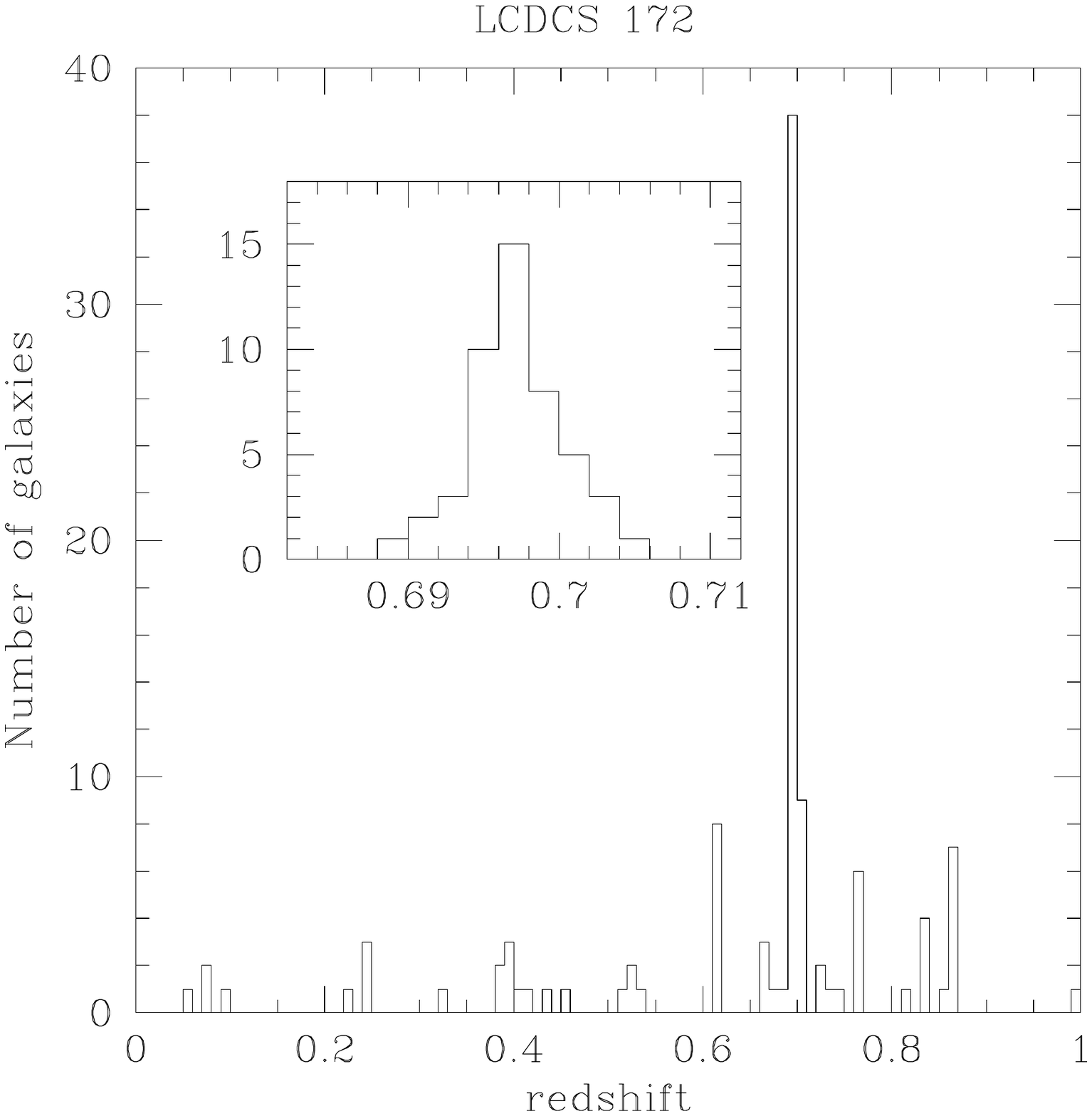} 
\includegraphics[width=6cm,angle=270]{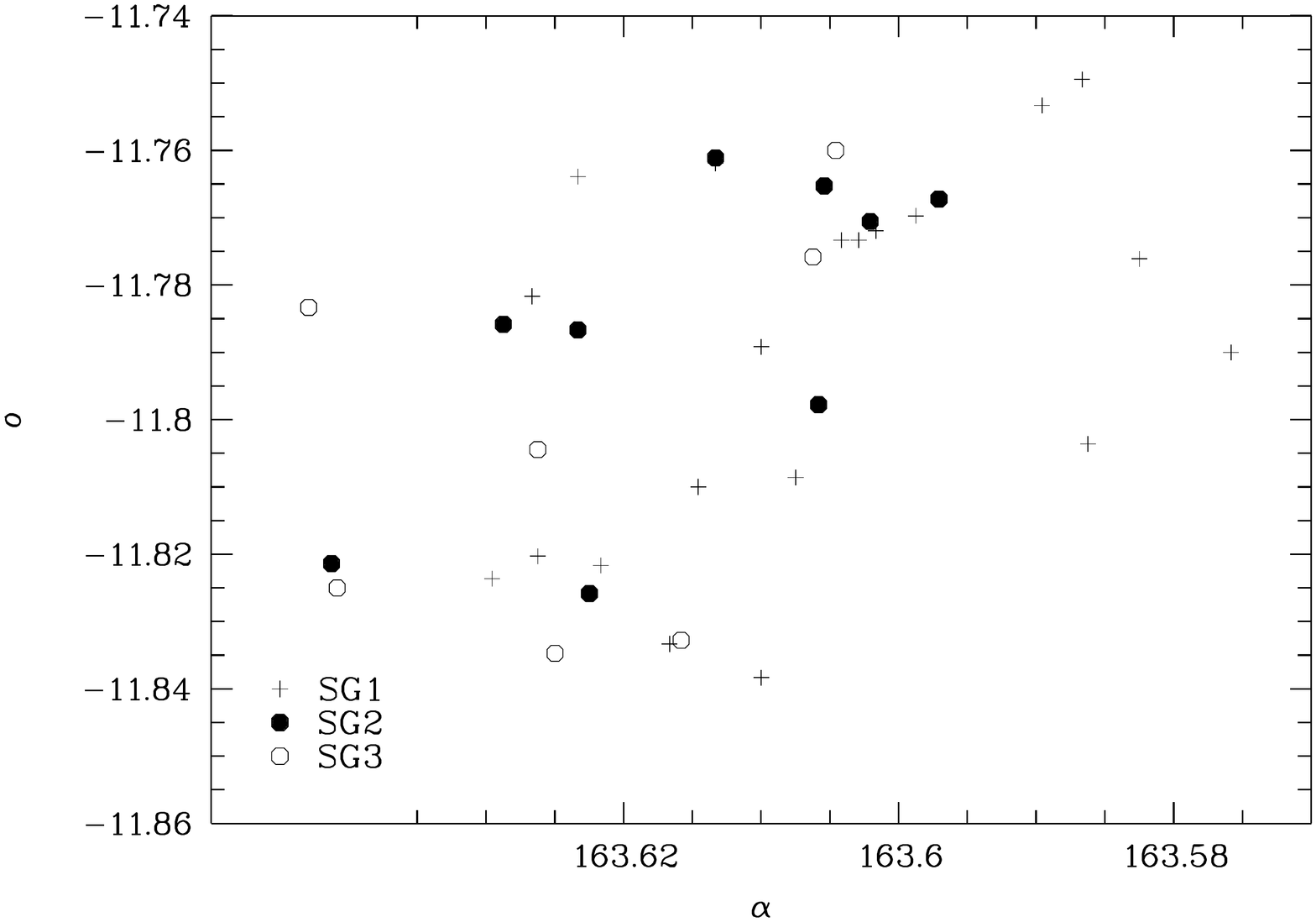} 
\caption{Upper figure: redshift histogram for LCDCS 0172. The insert shows a 
zoom around the cluster redshift. Lower figure: SG1, SG2, and SG3 galaxy 
spatial distribution.}
\label{fig:lcdcs172_histoz}
\end{center}
\end{figure}

The redshift histogram of LCDCS 0172 shows a prominent peak at z$\sim$0.697
(Fig.~\ref{fig:lcdcs172_histoz}), with 48 galaxies in the
[0.688,0.705] range. There are also several smaller peaks, one in the
foreground and two or three in the background of the cluster.
The SG method finds three structures in the cluster (see
Table~\ref{tab:SG2}).

\subsection{LCDCS 0173 (163.68125$^o$, --12.76389$^o$, z=0.7498)} 

\begin{figure}
\begin{center}
\includegraphics[width=6cm]{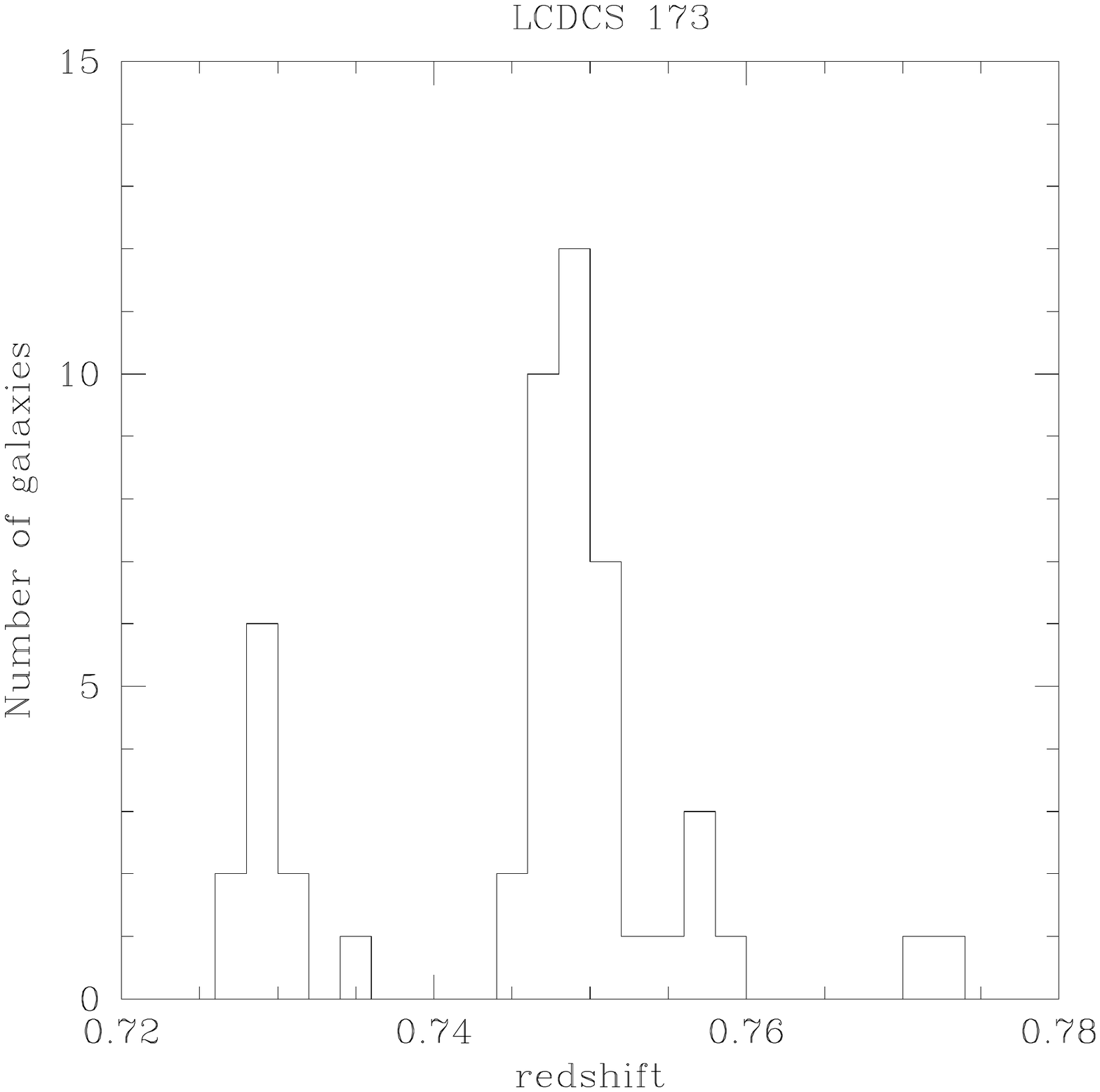} 
\includegraphics[width=6cm,angle=270]{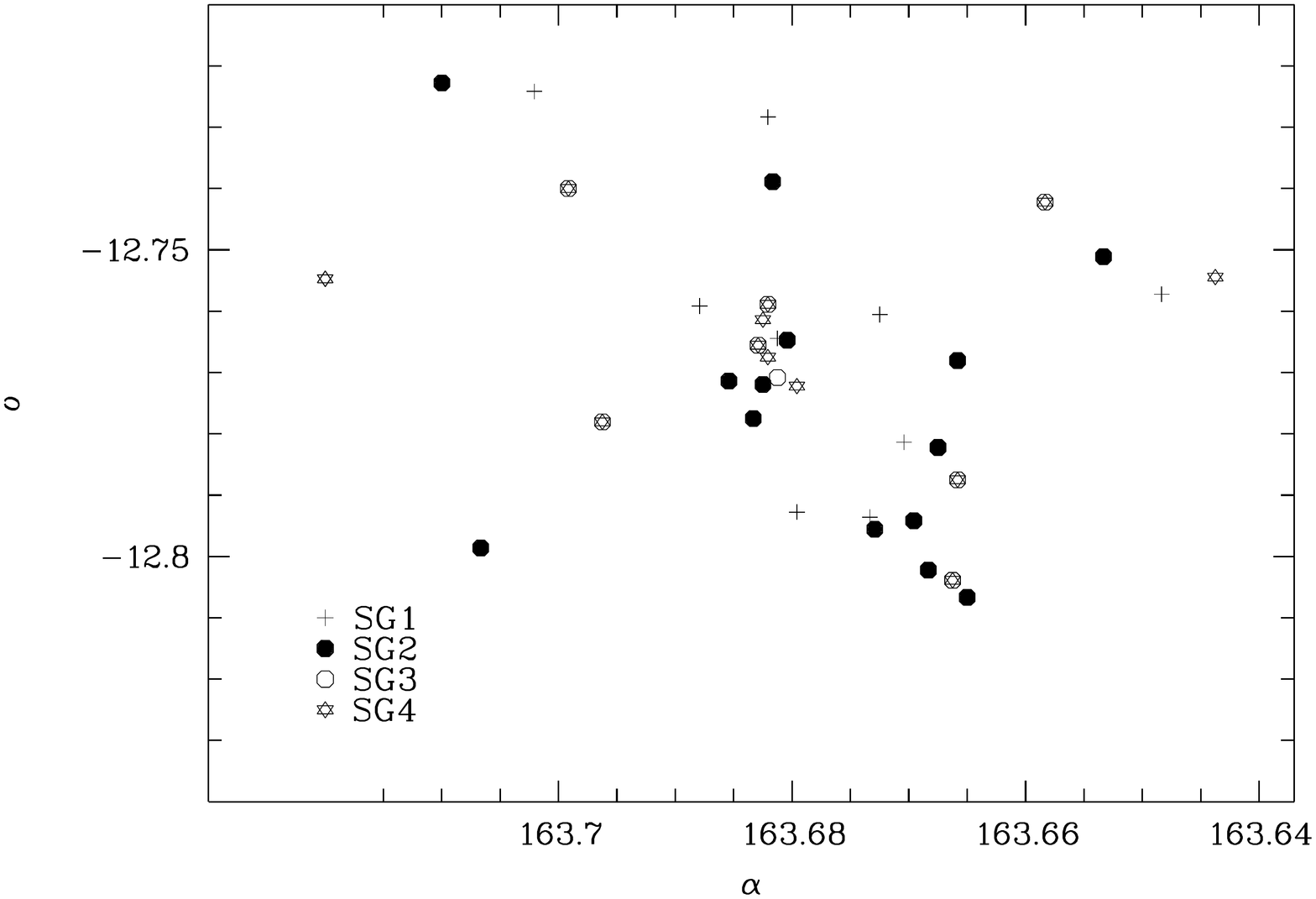} 
\caption{Upper figure: redshift histogram for LCDCS 0173. Lower figure: SG1, 
SG2, SG3, and SG4 galaxy spatial distribution.}
\label{fig:lcdcs173_histoz}
\end{center}
\end{figure}

The redshift histogram LCDCS 0173 shows a prominent peak at
z$\sim$0.744 (Fig.~\ref{fig:lcdcs173_histoz}), with 37 galaxies in the
[0.741,0.760] range.  The SG method finds four structures with
comparable masses (Table~\ref{tab:SG2}).  A smaller peak is detected
at z$\sim$0.73, as also noted by Halliday et al. (2004).

\subsection{CL~J1103.7-1245a (165.89542$^o$, --12.7794$^o$, z=0.6300)} 

\begin{figure}
\begin{center}
\includegraphics[width=6cm]{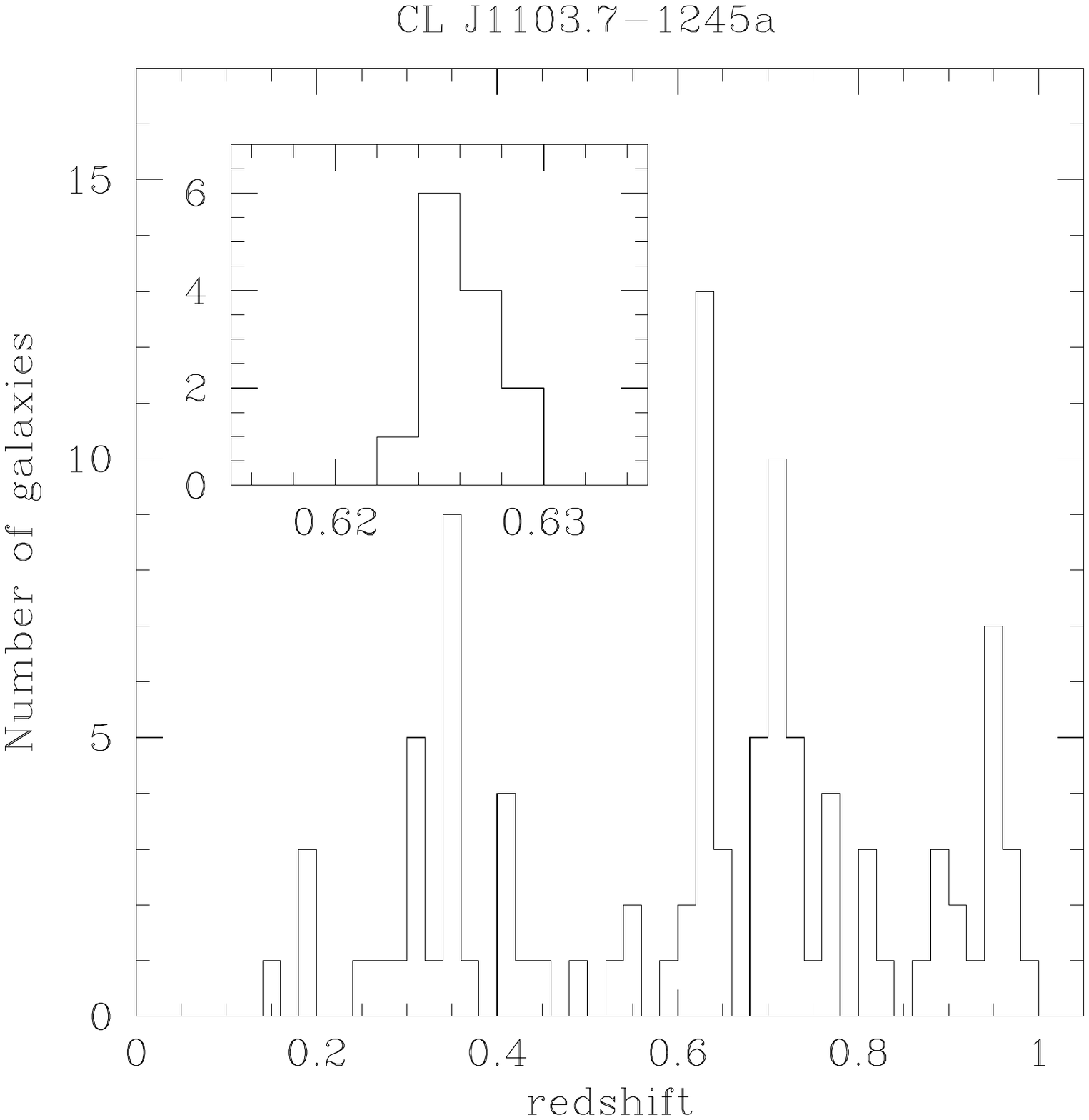} 
\caption{Redshift histogram for CL~J1103.7-1245a. The insert 
shows a zoom around the cluster redshift. }
\label{fig:cl1103_histoz}
\end{center}
\end{figure}

Though available, the XMM-Newton image of CL~J1103.7-1245a is too
faint to be usable.

The full redshift histogram shows a peak at redshift 0.6300 (the
cluster redshift given by NED) but another peak is also observed at
z$\sim$0.72, and this line of sight seems to be intersecting several
galaxy structures (Fig.~\ref{fig:cl1103_histoz}), as already seen by
Milvang-Jensen (2008).
There are 19 galaxies with measured redshifts in the [0.58,0.63]
range, and the SG analysis finds a single massive structure.

\subsection{CXOMP~J111726.1+074335 (169.358875$^o$, +7.7264$^o$, z=0.4770)} 

\begin{figure}
\begin{center}
\includegraphics[width=6cm]{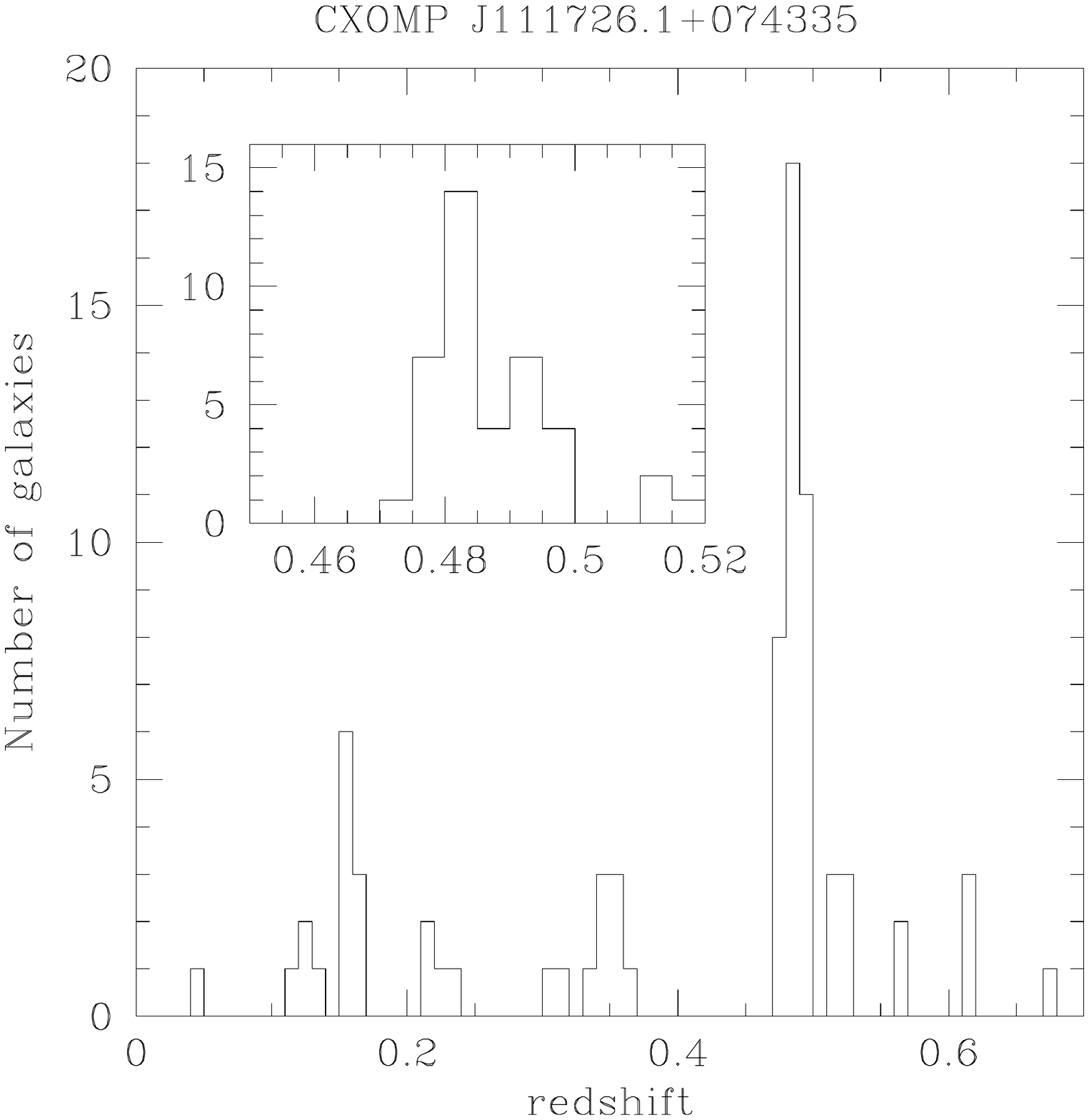} 
\includegraphics[width=6cm,angle=270]{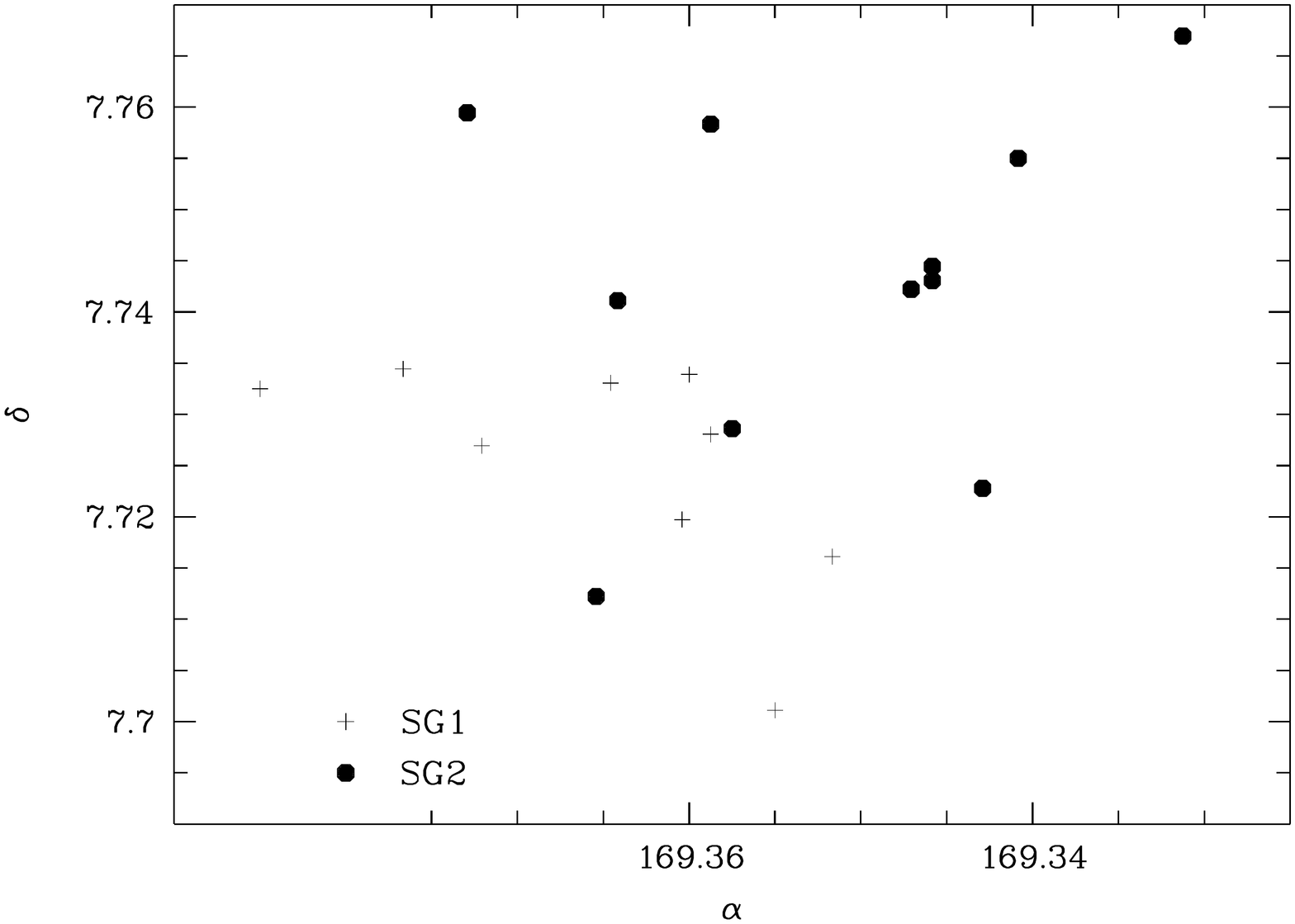} 
\caption{Upper figure: redshift histogram for CXOMP~J111726.1+074335. Lower 
figure: SG1 and SG2 galaxy spatial distribution.}
\label{fig:cx1117_histoz}
\end{center}
\end{figure}

The redshift histogram of CXOMP~J111726.1+074335 (discovered by
Barkhouse et al. 2006) shows a strong peak at the value given by NED:
z$\sim$0.477 (Fig.~\ref{fig:cx1117_histoz}). There are 36 galaxies in
the [0.476,0.500] range, with a clearly asymmetric distribution.  The
SG method finds two substructures of comparable masses
(Table~\ref{tab:SG2}).

\subsection{LCDCS 0340 (174.54292$^o$, --11.5664$^o$, z=0.4798)} 

\begin{figure}
\begin{center}
\includegraphics[width=6cm]{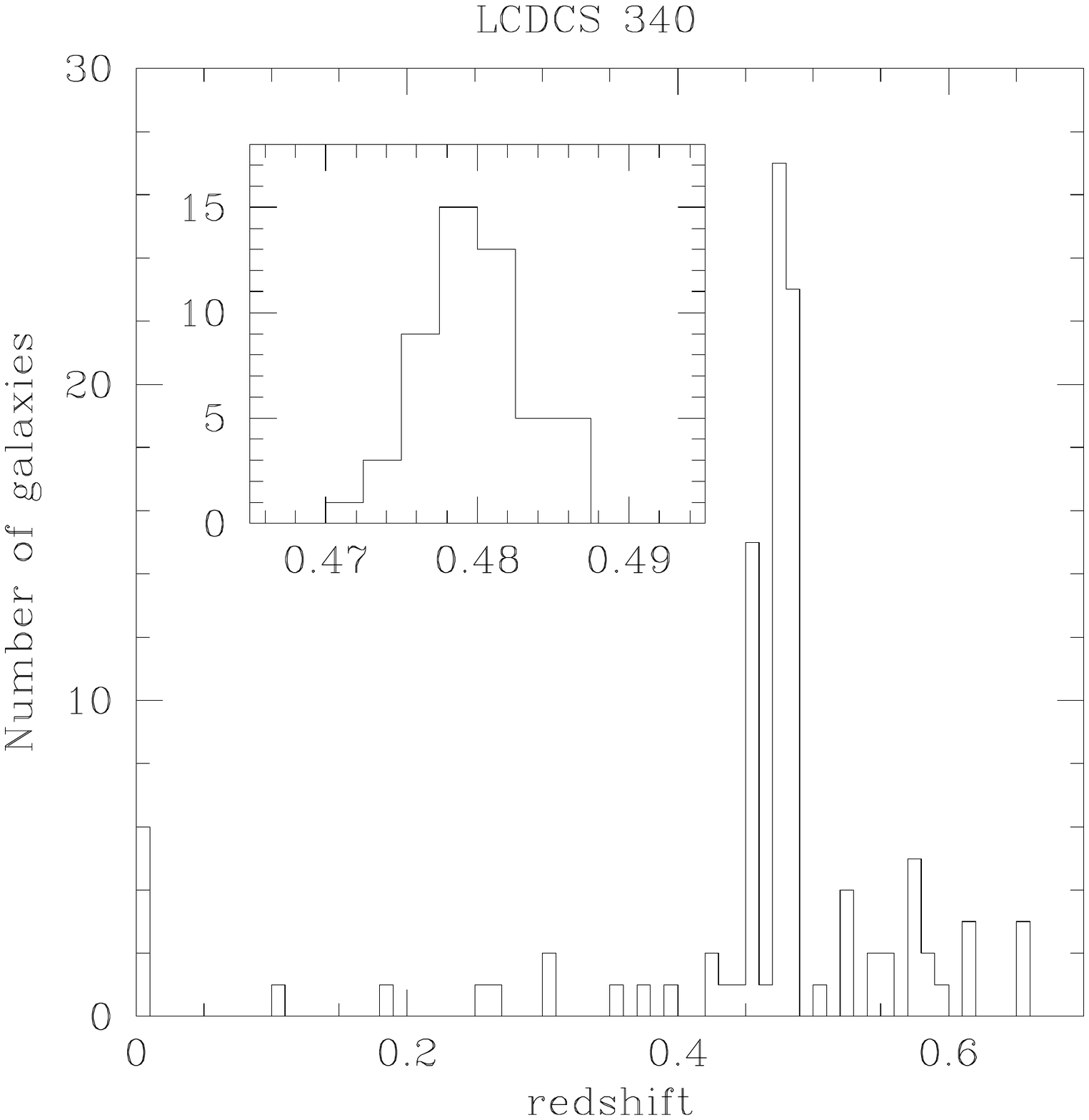} 
\includegraphics[width=6cm,angle=270]{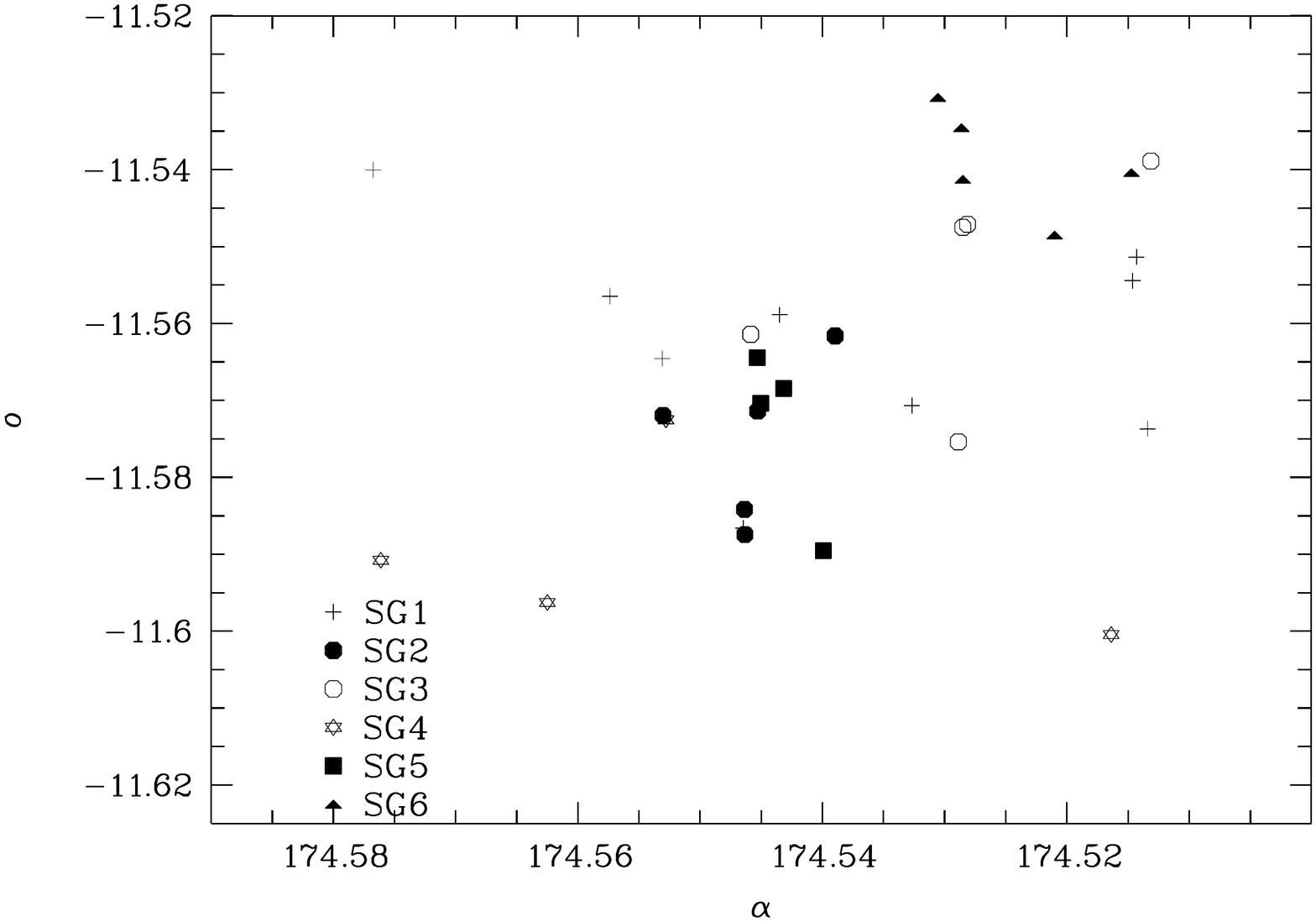} 
\caption{Upper figure: redshift histogram for LCDCS 0340. The insert shows a 
zoom around the cluster redshift. Lower figure: SG1, SG2, SG3, SG4, SG5, and 
SG6 galaxy spatial distribution.}
\label{fig:lcdcs340_histoz}
\end{center}
\end{figure}

The redshift histogram of LCDCS 0340 shows a strong peak at
z$\sim$0.48 (Fig.~\ref{fig:lcdcs340_histoz}), with 51 galaxies in the
[0.47,0.49] range.  The SG method finds six substructures with quite
comparable masses (Table~\ref{tab:SG2}). Note that a foreground
structure of 16 galaxies is detected in the [0.4500,0.4585] redshift
range, suggesting the presence of a second smaller cluster on the line
of sight, as already noted by Milvang-Jensen et al. (2008).

\subsection{LCDCS~0531 (186.97458$^o$, --11.6389$^o$, z=0.6355)} 

\begin{figure}
\begin{center}
\includegraphics[width=6cm]{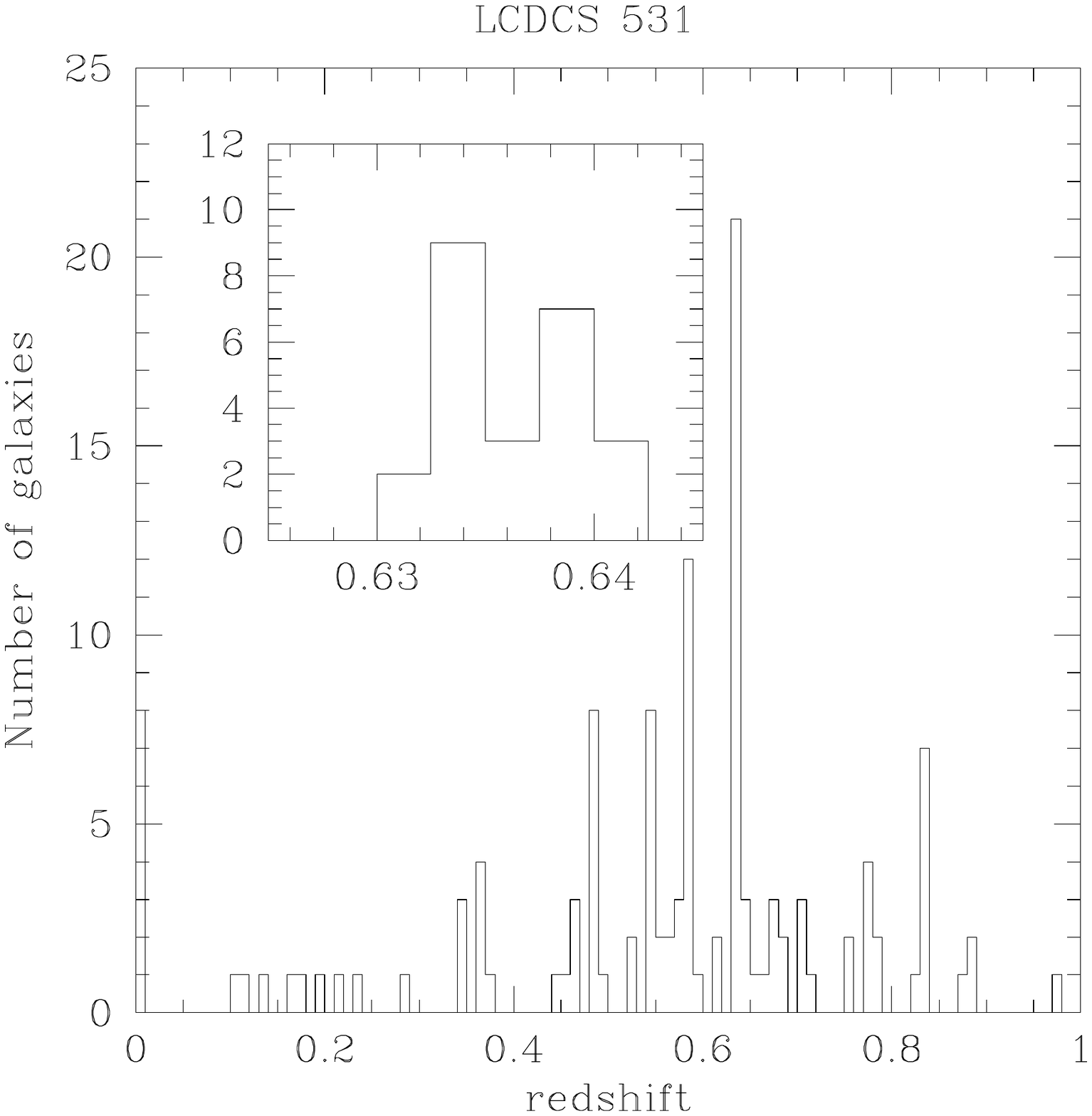} 
\caption{Redshift histogram for LCDCS~0531. The insert shows a zoom around
  the cluster redshift.}
\label{fig:lcdcs531_histoz}
\end{center}
\end{figure}

The XMM-Newton image is  too faint to be usable. 

There are 24 galaxies in the [0.63,0.65] redshift range, but though
the redshift histogram appears double-peaked
(Fig.~\ref{fig:lcdcs531_histoz}), the SG method detects a single
massive structure, in agreement with the lack of substructuring found
by Clowe et al. (2006). Several smaller structures are detected on the
line of sight.

\subsection{LCDCS 0541 (188.12708$^o$, --12.8425$^o$, z=0.5414)} 

\begin{figure}
\begin{center}
\includegraphics[width=6cm]{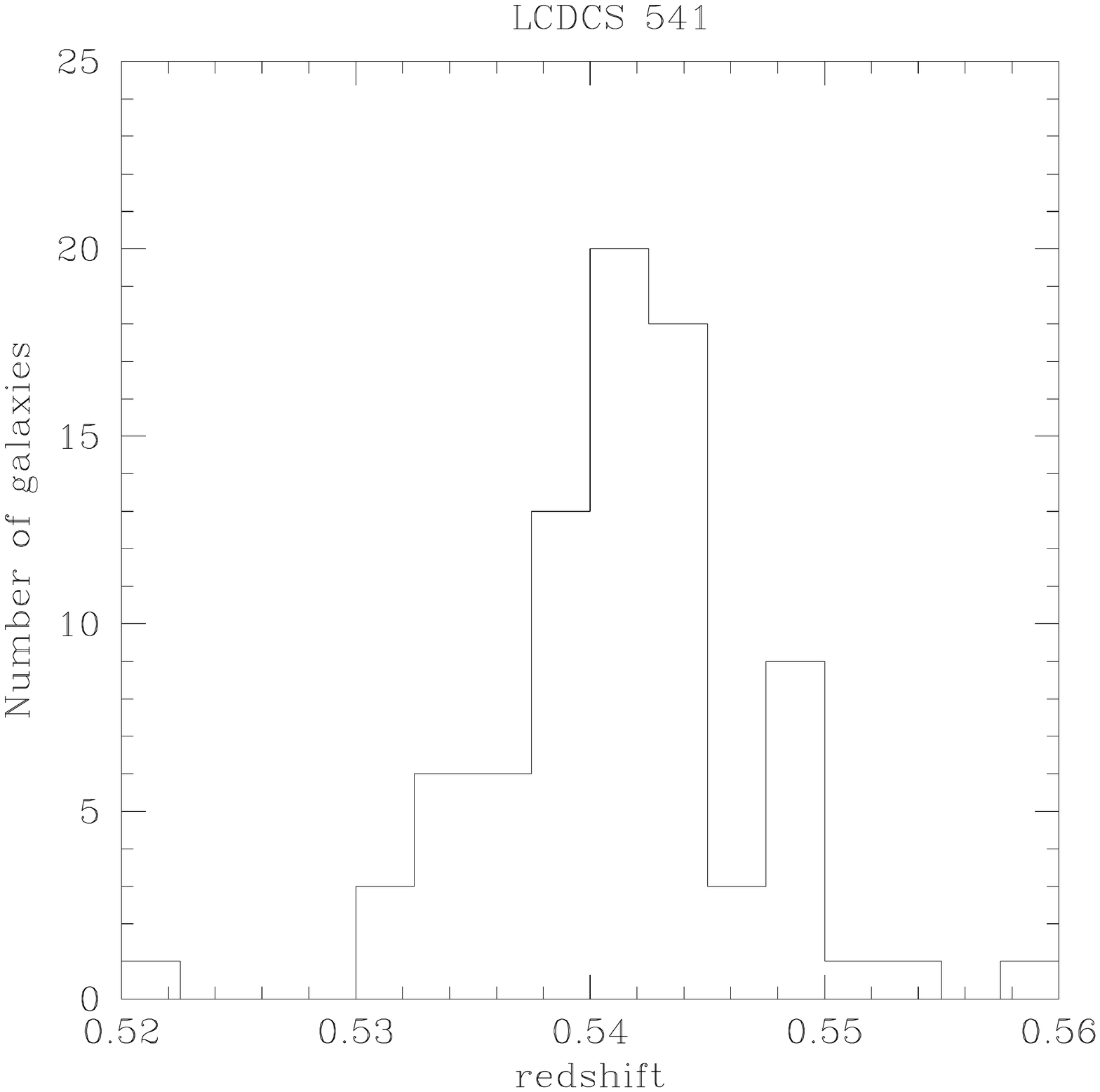} 
\includegraphics[width=6cm,angle=270]{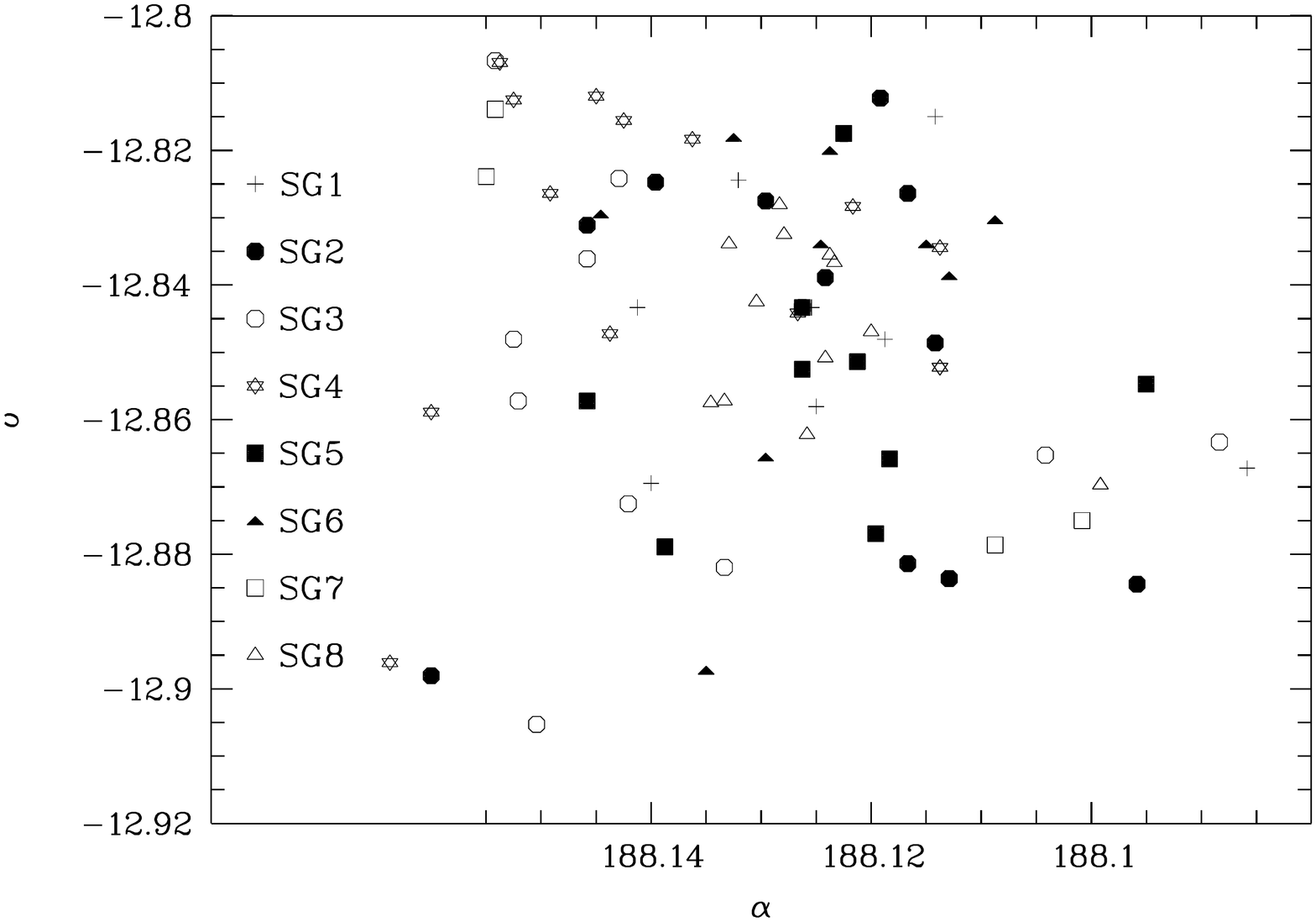} 
\caption{Upper figure: redshift histogram for LCDCS 0541. Lower figure: SG1, 
SG2, SG3, SG4, SG5, SG6, SG7, and SG8 galaxy spatial distribution.}
\label{fig:lcdcs541_histoz}
\end{center}
\end{figure}

The redshift histogram of LCDCS 0541 shows a strong peak at
z$\sim$0.54, with 80 galaxies in the [0.53,0.555] range
(Fig.~\ref{fig:lcdcs541_histoz}).

The SG analysis shows the existence of eight substructures of
comparable mass, suggesting that this cluster is far from relaxed
(Table~\ref{tab:SG2}). Based on a weak lensing analysis, Clowe et
al. (2006) detected at least two mass peaks.  X-ray data on this
cluster would be very interesting for relating the optical
substructures with the distribution of the X-ray gas.

\subsection{ClG J1236+6215 (189.16500$^o$, +62.2650$^o$, z=0.8500)} 

\begin{figure}
\begin{center}
\includegraphics[width=6cm]{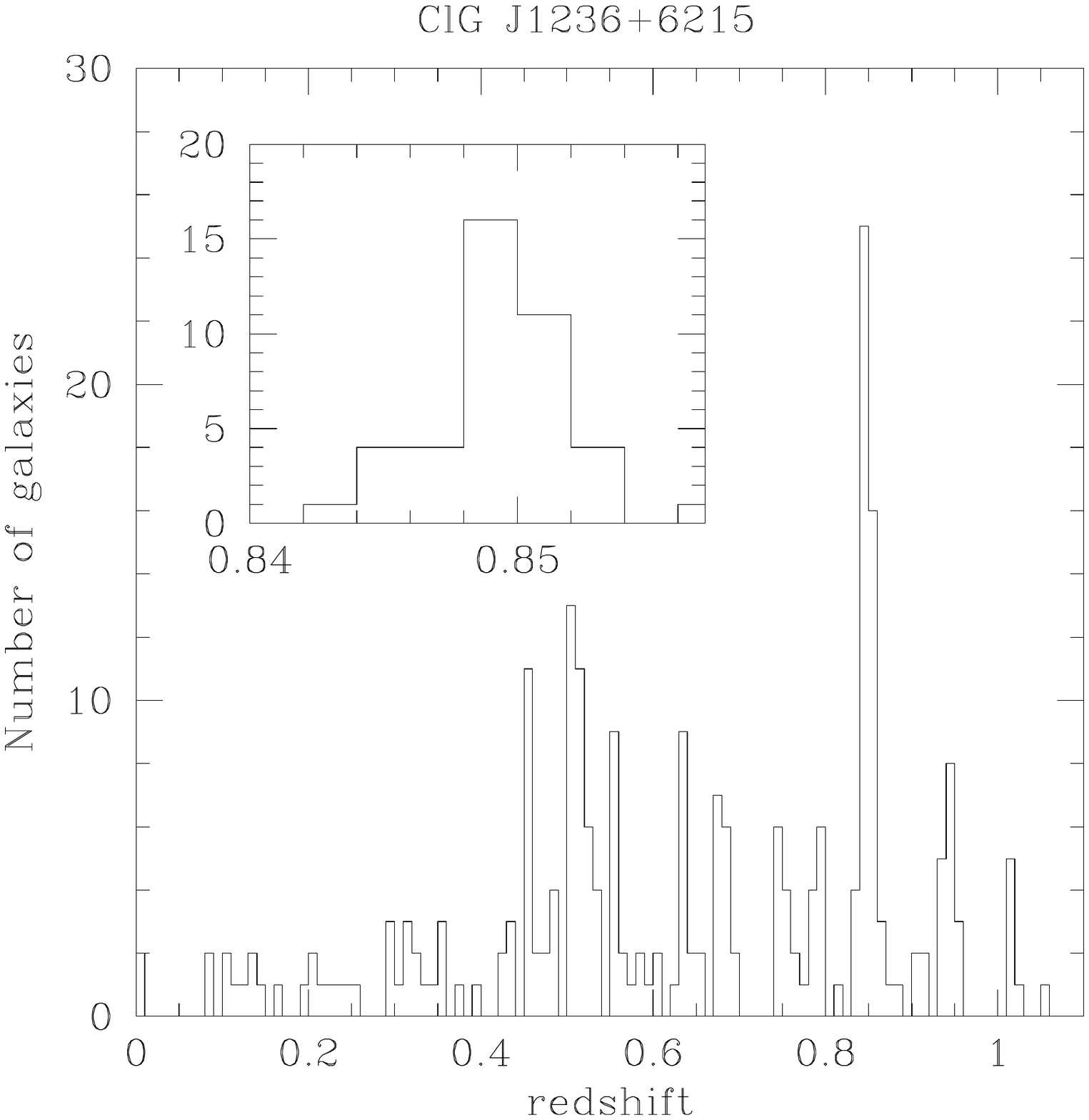} 
\includegraphics[width=6cm,angle=270]{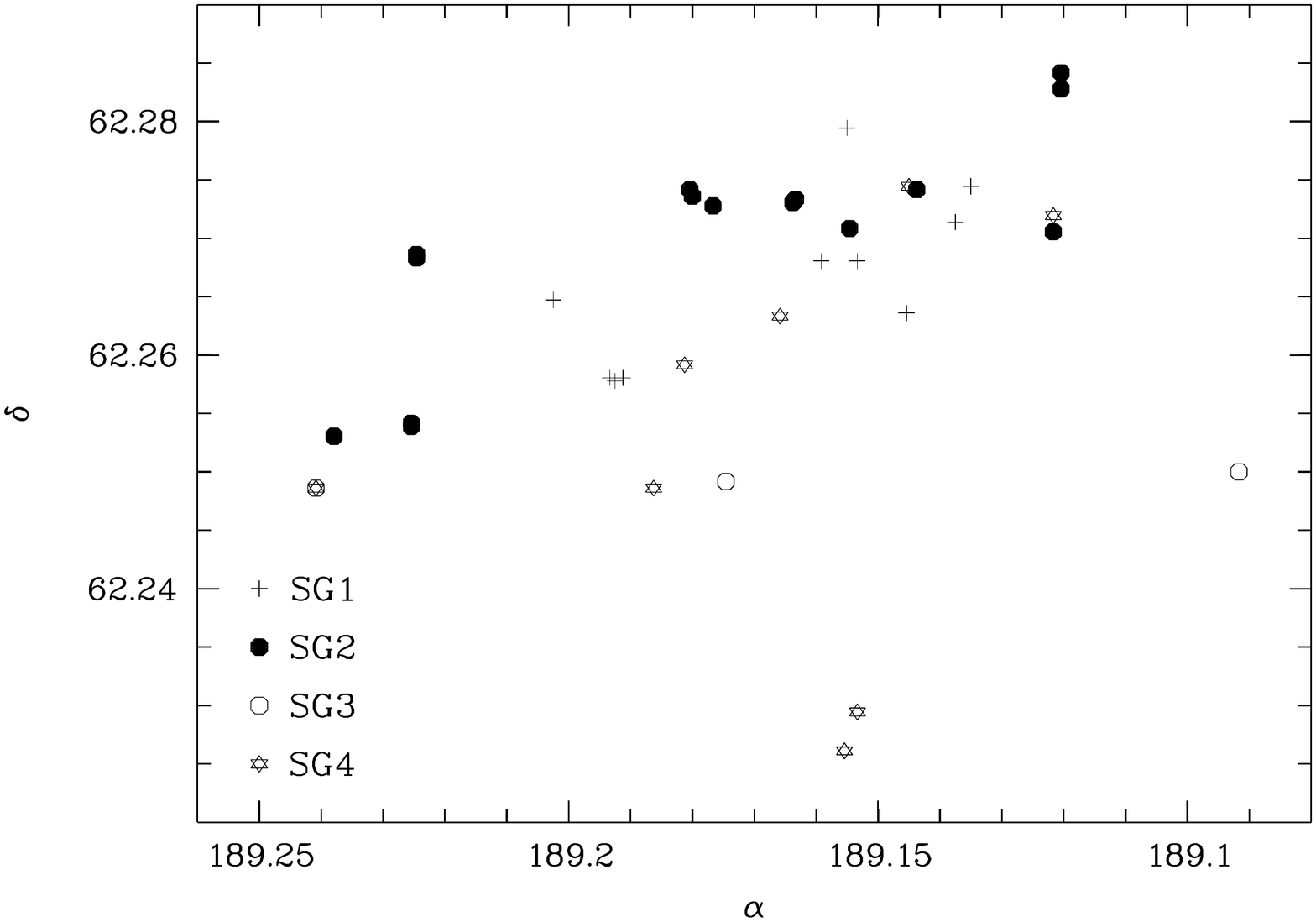} 
\caption{Upper figure: redshift histogram for ClG J1236+6215. The insert 
shows a zoom around the cluster redshift. Lower figure: SG1, 
SG2, SG3, and SG4 galaxy spatial distribution.}
\label{fig:cl1236_histoz}
\end{center}
\end{figure}

ClG J1236+6215 was serendipitously detected by Dawson et al. (2001) in
the Hubble Deep Field North.  The redshift histogram in the direction
of this cluster shows a strong peak at z$\sim$0.85, as well as many
smaller peaks along the line of sight, including a foreground peak at
z$\sim$0.53 (Fig.~\ref{fig:cl1236_histoz}). There are 40 galaxies in
the [0.842,0.855] cluster range.  The SG method finds four main
substructures.

\subsection{XDCS mf J131001.9+322110 (197.50792$^o$, +32.3528$^o$, z=0.4370)} 

\begin{figure}
\begin{center}
\includegraphics[width=6cm]{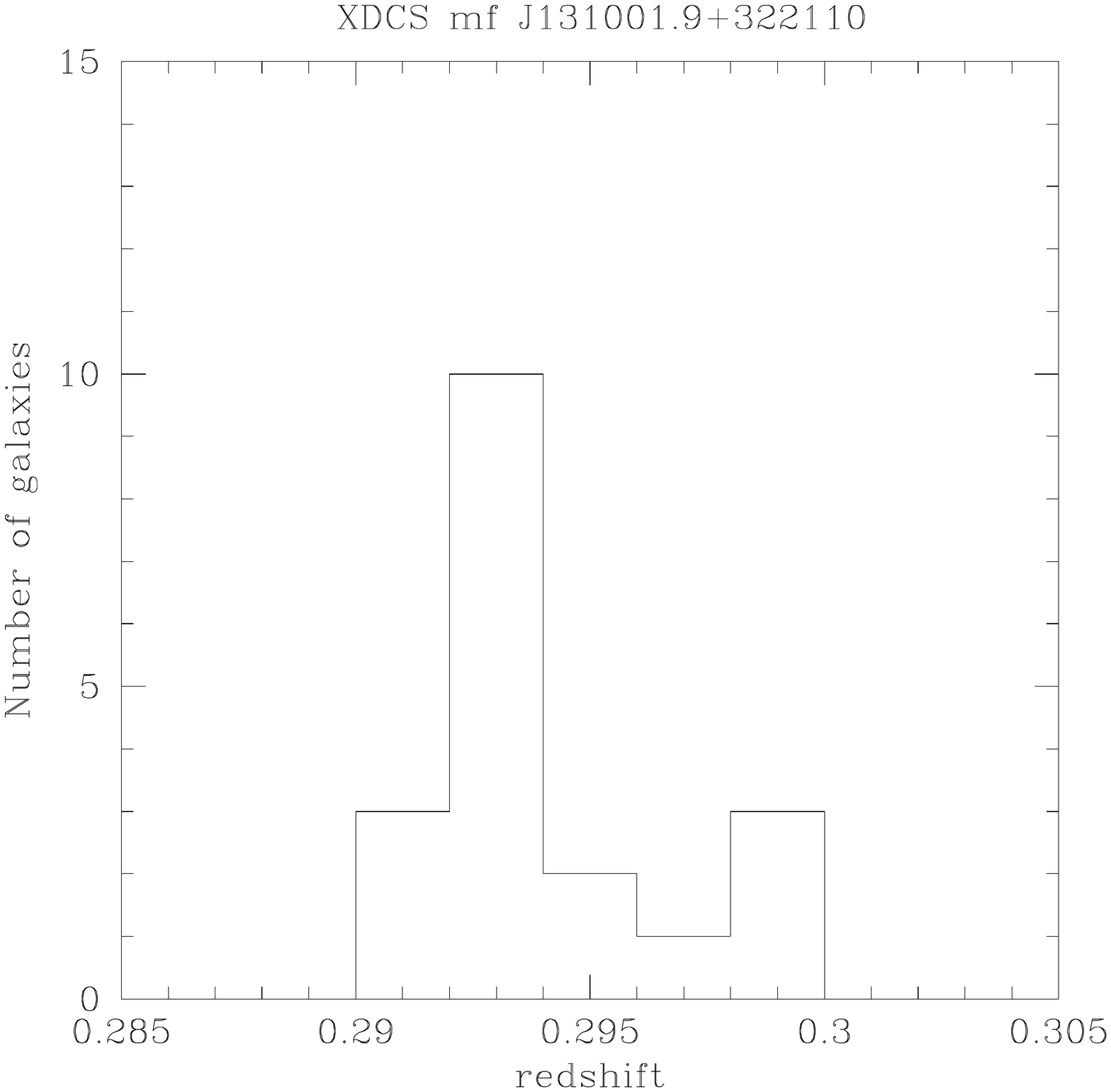} 
\caption{Redshift histogram for XDCS~mf~J131001.9+322110. }
\label{fig:xdcs1310_histoz}
\end{center}
\end{figure}

XDCS~mf~J131001.9+322110 was discovered by Vikhlinin et al. (1998) but
its XMM-Newton image is too faint to be usable.

The redshift histogram of XDCS mf J131001.9+322110 shows a strong,
somewhat asymmetric, peak at z$\sim$0.293. There are 19 galaxies in the
[0.287,0.300] cluster range (Fig.~\ref{fig:xdcs1310_histoz}), and the
SG analysis finds a single structure.

\subsection{NSCS J132336+302223 (200.91500$^o$, +30.3760$^o$, z=0.5080)} 

\begin{figure}
\begin{center}
\includegraphics[width=6cm]{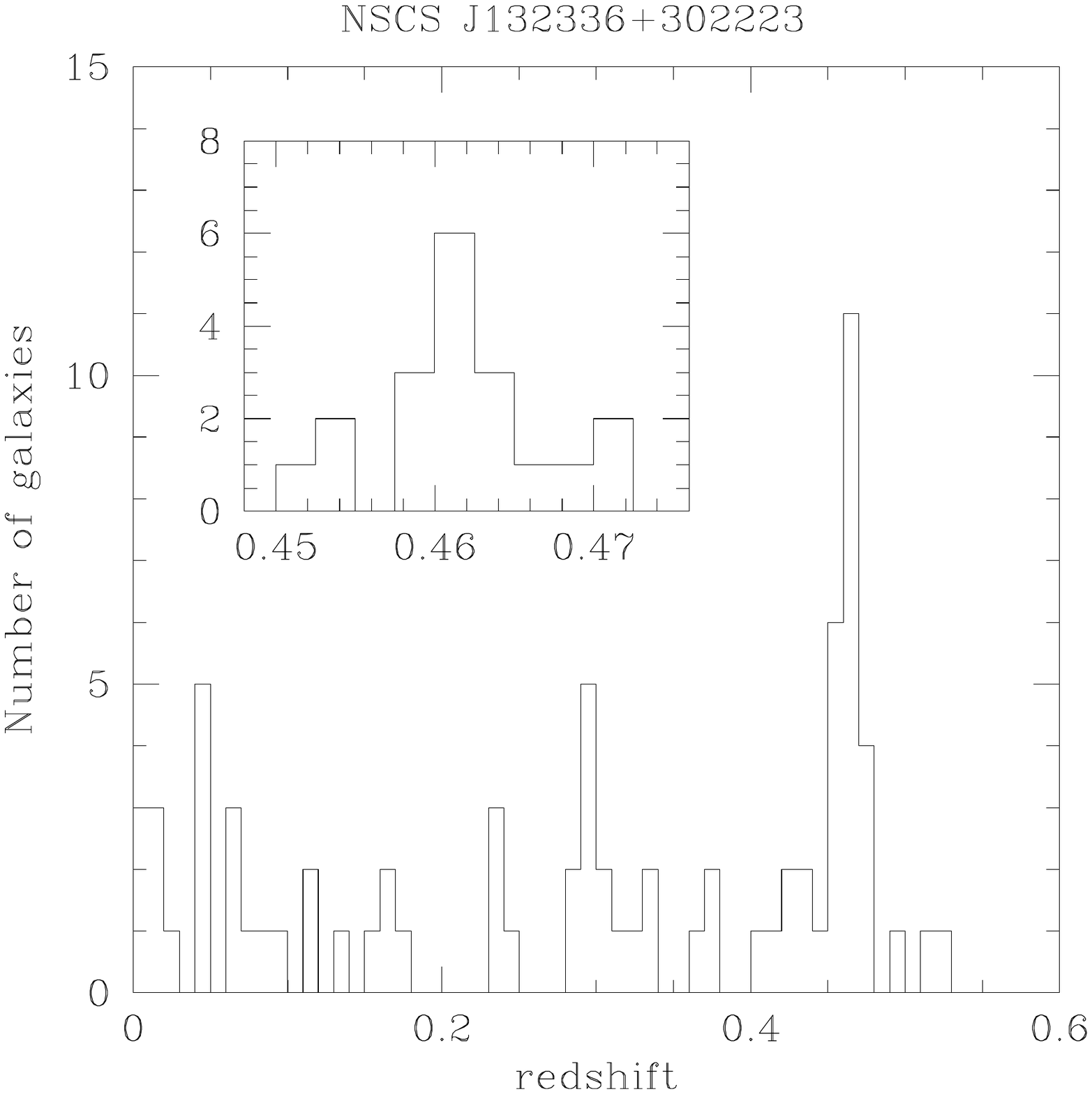} 
\caption{Redshift histogram for NSCS~J132336+302223. }
\label{fig:nscs1323_histoz}
\end{center}
\end{figure}

NSCS J132336+302223 was discovered by Gladders \& Yee (2005). Its
redshift histogram shows a peak at z$\sim$0.46, with 19 galaxies in
the [0.450,0.473] range (Fig.~\ref{fig:nscs1323_histoz}). Since the
number of galaxies is not very large and the redshift peak is broad,
it is not straightforward to define  the cluster redshift
range exactly.  We note, however, that the mean redshift (z$\sim$0.461) differs
from the value z=0.5080 given by NED, at which we detect no peak in
the redshift histogram. The SG analysis finds a single structure.

\subsection{[MJM98]~034 (203.80742$^o$, +37.8156$^o$, z=0.5950)} 

\begin{figure}
\begin{center}
\includegraphics[width=6cm]{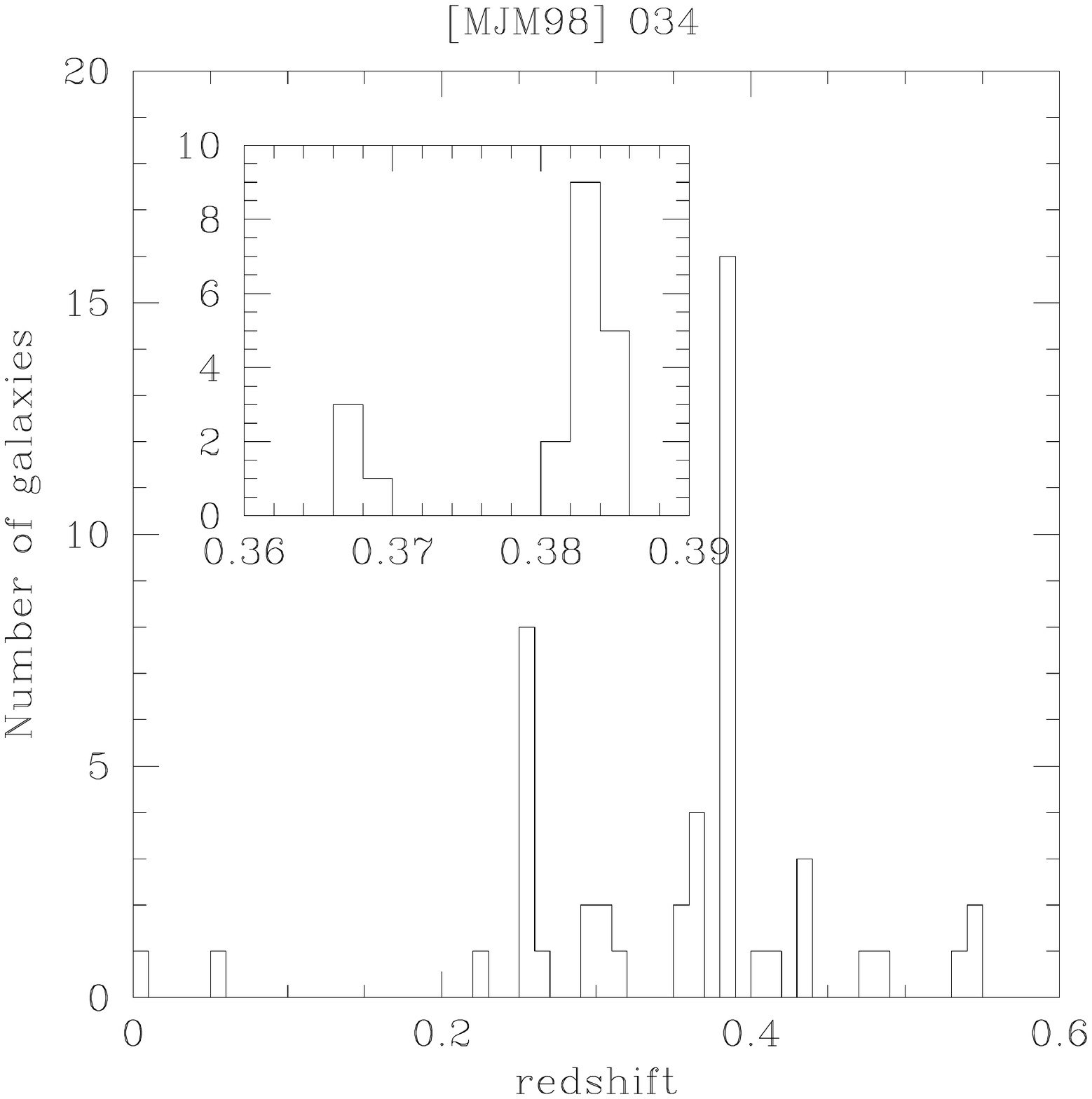} 
\caption{Redshift histogram for [MJM98]~034. }
\label{fig:mjm98_034_histoz}
\end{center}
\end{figure}

[MJM98]~034 was first identified in a ROSAT field by McHardy et
al. (1998).  Its redshift histogram shows a peak at z$\sim$0.383, with
16 galaxies in the [0.380,0.386] range
(Fig.~\ref{fig:mjm98_034_histoz}). Here again, the number of galaxies
is not very large, and there is a small foreground structure of four
galaxies at z$\sim$0.367, which may (or may not) be related to the main
cluster, so it is not straightforward to define  the cluster
redshift range exactly.  Here also, the mean redshift (z$\sim$0.383)
differs notably from the one given by NED (0.5950), where we detect no
peak in the redshift histogram. The SG analysis finds a single
structure.

\subsection{3C 295 Cluster (212.83396$^o$, +52.2025$^o$, z=0.4600)} 

\begin{figure}
\begin{center}
\includegraphics[width=6cm]{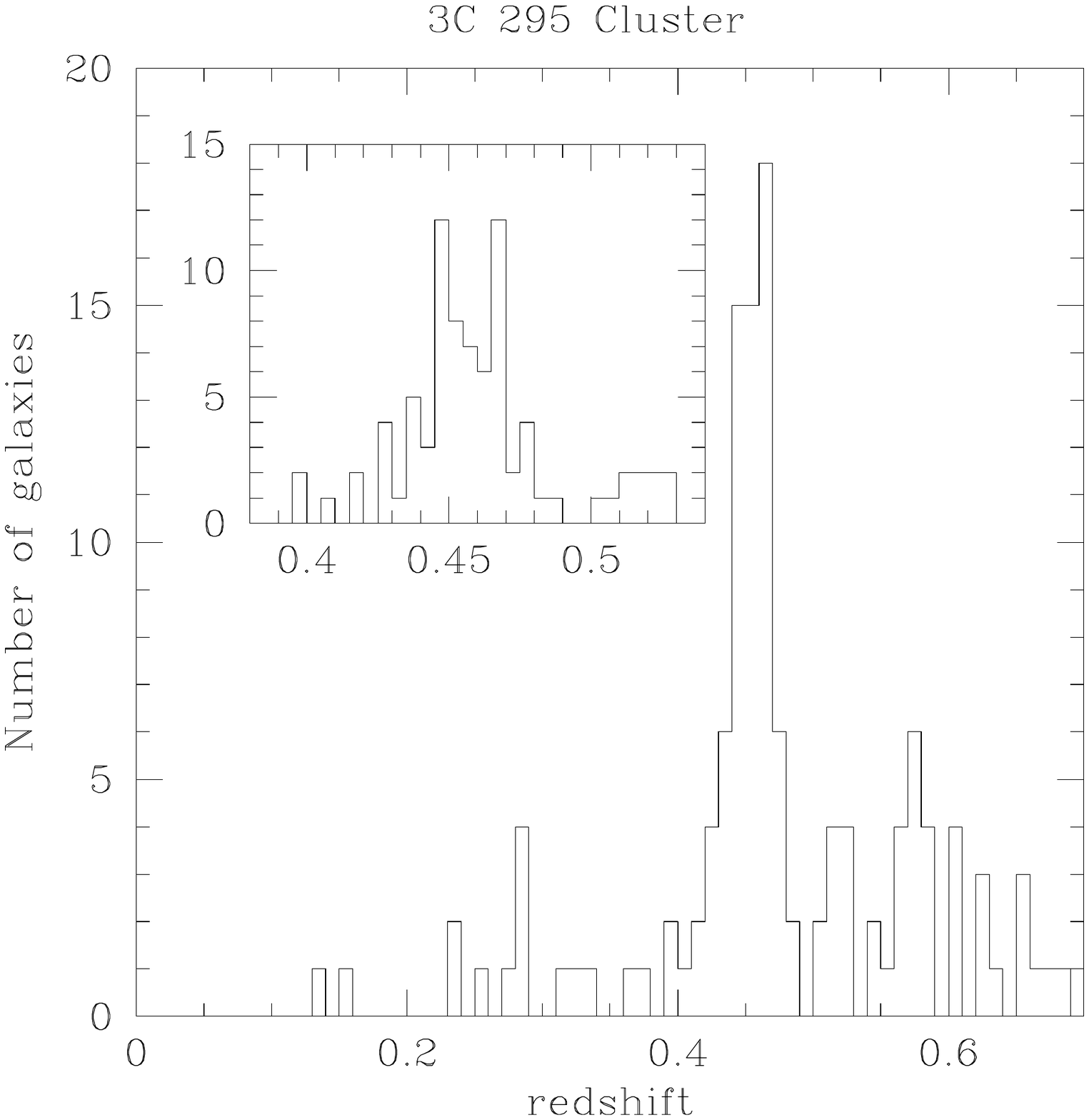} 
\includegraphics[width=6cm,angle=270]{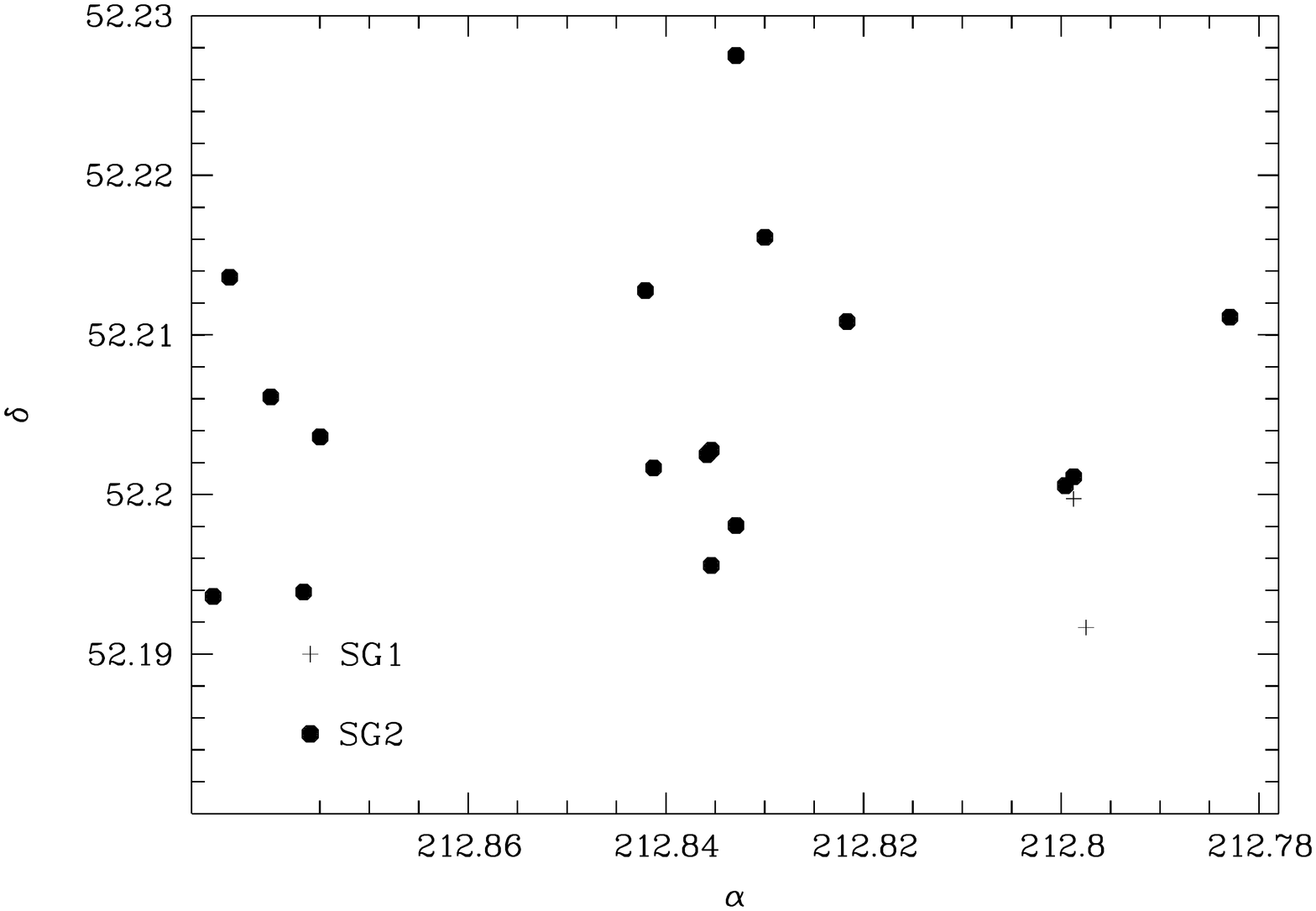} 
\caption{Upper figure: redshift histogram for 3C~295~Cluster. Lower figure: 
SG1 and SG2 galaxy spatial distribution.}
\label{fig:3c295_histoz}
\end{center}
\end{figure}

The redshift histogram of the 3C 295 cluster shows a strong and rather
broad peak at z$\sim$0.46 (Fig.~\ref{fig:3c295_histoz}), with 66
galaxies in the [0.43,0.49] range. The cluster redshift range is
probably narrower but in view of the redshift histogram alone, it is
difficult to estimate it robustly. We can note, however, that the
redshift histogram shows two peaks, and this is confirmed by the SG
method, which finds two substructures of comparable masses (see
Table~\ref{tab:SG2}), with velocities differing by only about
1000~km~s$^{-1}$.

We note that NED gives a redshift z=0.2317 for the 3C~295 cluster and
z=0.4641 for the 3C~295 radio galaxy, which appears rather confusing.
In view of the large number of galaxies at redshift $\sim 0.46$,
the cluster lies most probably at this redshift and not at z=0.2317,
unless we are intercepting a filament along the line of sight.

\subsection{GHO~1601+4253 (240.80762$^o$, +42.7601$^o$, z=0.5391)} 

\begin{figure}
\begin{center}
\includegraphics[width=6cm]{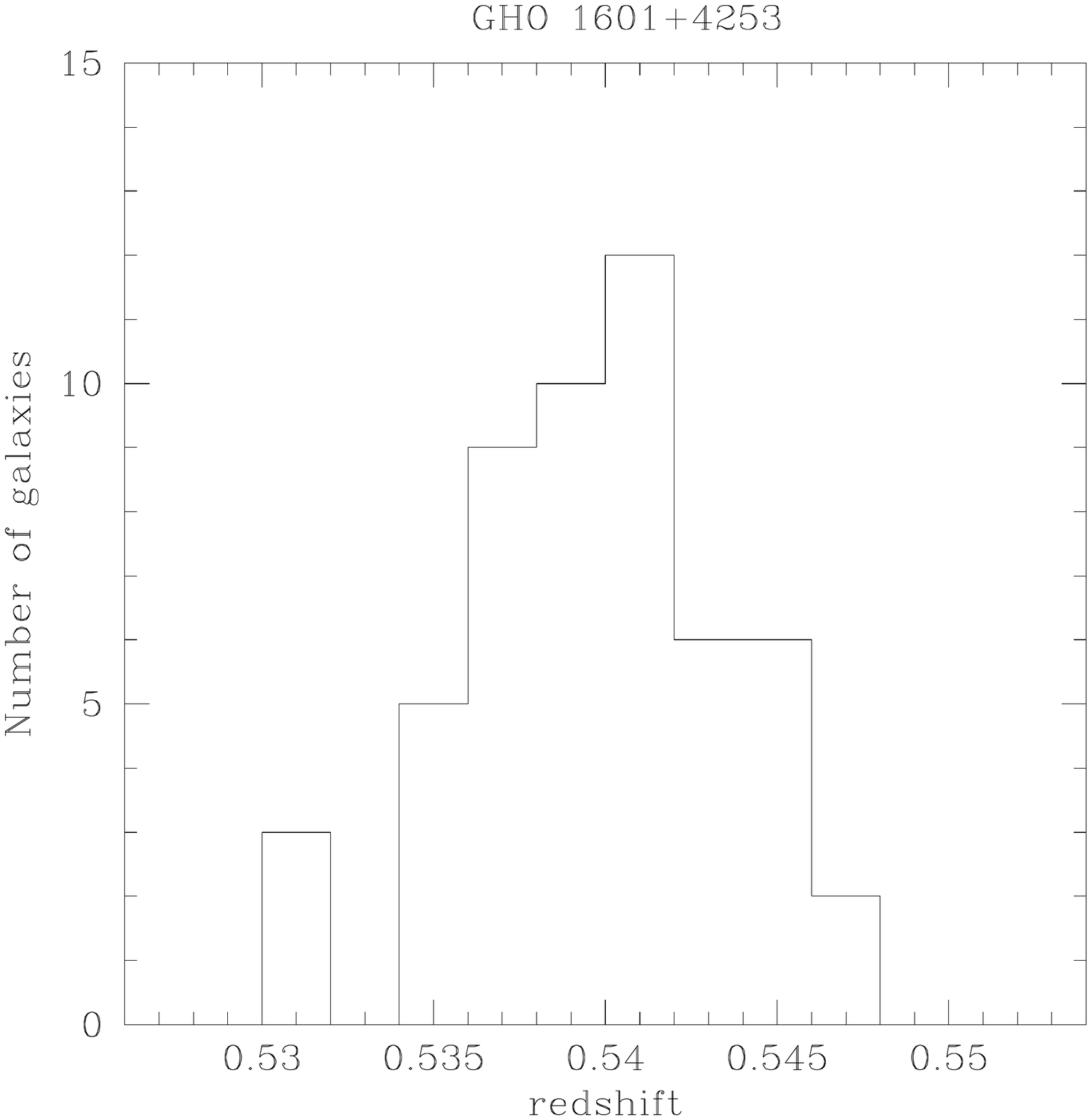} 
\includegraphics[width=6cm,angle=270]{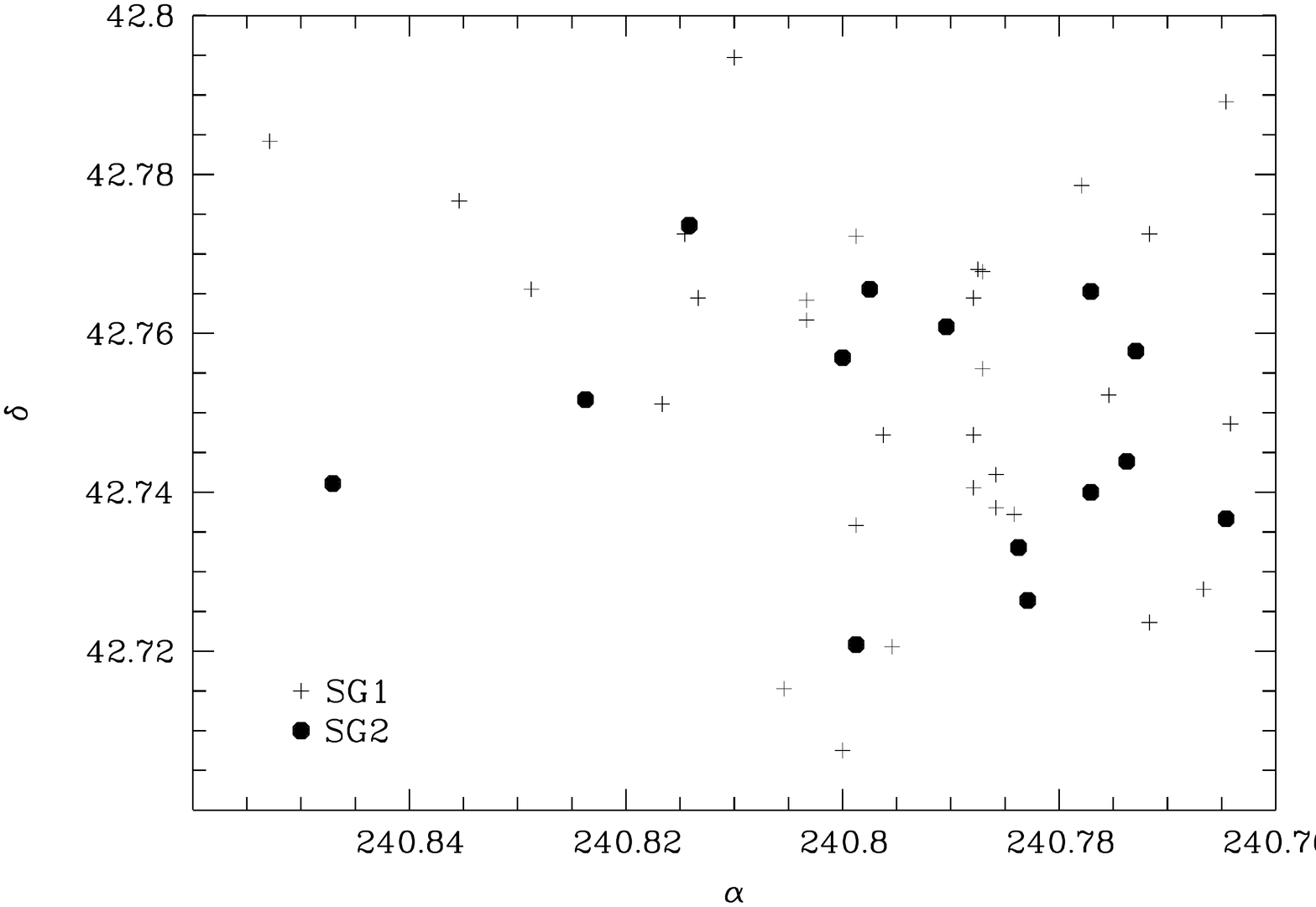} 
\caption{Upper figure: redshift histogram for GHO~1601+4253. Lower figure: 
SG1 and SG2 galaxy spatial distribution. }
\label{fig:gho1601_histoz}
\end{center}
\end{figure}

GHO~1601+4253 was discovered by Gunn et al. (1986). Its redshift
histogram  shows a strong and somewhat asymmetric peak
at z$\sim$0.54 (Fig.~\ref{fig:gho1601_histoz}).  There are 50 galaxies
in the [0.534,0.548] cluster range, and the SG method detects two 
substructures (see Table~\ref{tab:SG2}).

\subsection{RX J1716.4+6708 (259.20667$^o$, +67.1417$^o$, z=0.8130)} 

\begin{figure}
\begin{center}
\includegraphics[width=6cm]{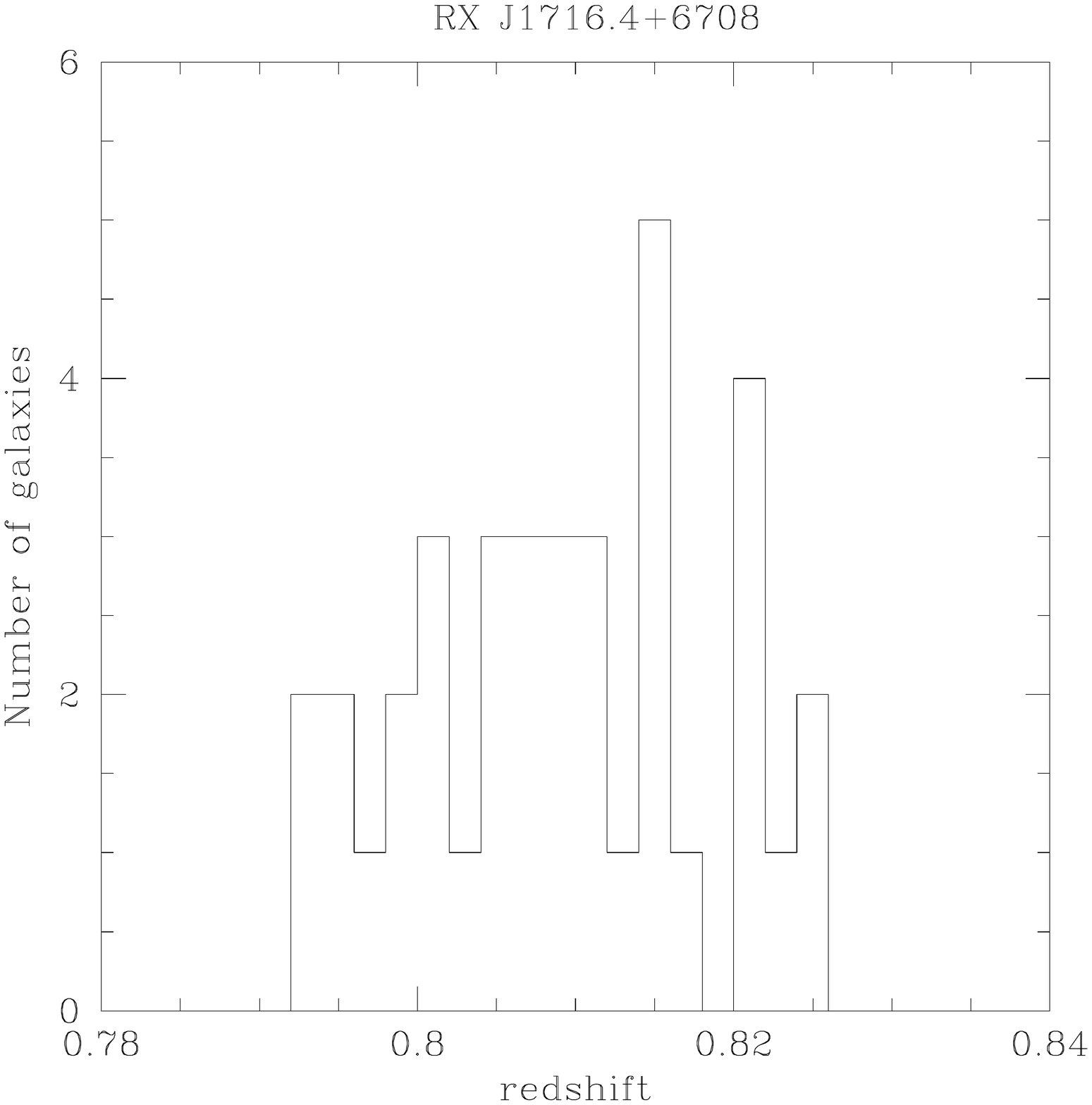}
\caption{Redshift histogram for RX J1716.4+6708. }
\label{fig:RX1716_histoz}
\end{center}
\end{figure}

RX J1716.4+6708 was discovered by Henry et al. (1997), who gave a
redshift z=0.8130 for this cluster and measured 12 galaxy redshifts in
the [0.7945,0.8266] range, spreading along a filament. Gioia et
al. (1999) increased this sample and published a list of 37 galaxies
with spectroscopic redshifts in the [0.7924,0.8256] range.  The
velocity dispersion calculated from these 37 redshifts is rather
large: 1522~km~s$^{-1}$ and agrees with the suggestion by Henry et
al. (1997) that RX J1716.4+6708 is instead a protocluster still in the
process of forming. This was confirmed by the weak lensing analysis of
Clowe et al. (1998), who stated that RX J1716.4+6708 is not a
well-formed cluster.  The SG analysis finds a large structure (the
cluster) and a few small substructures too poorly sampled to be
characterized.

The XMM-Newton image of this cluster is too faint to be usable.

\clearpage
\section{Notes on clusters with no usable XMM-Newton data and no possible 
SG analysis} 

We give below a few notes on clusters with no usable XMM-Newton data
(see Table~\ref{tab:fewz}) and too few galaxy redshifts for an SG analysis, but they may be of
some interest for various reasons, principally because in a number of
cases the redshift does not correspond to what is given by NED.

\begin{table*}[h!]
  \caption{Clusters with no usable XMM-Newton data. Notes: (1)~name (as in NED), 
    (2)~right ascension in degrees (J2000.0), (3)~declination in degrees (J2000.0), 
    (4)~redshift (in parentheses, redshift given by NED, if different), 
    (5)~approximate number of galaxies with redshifts in the cluster range. }
\begin{tabular}{lrrrr}
\hline
\hline
Name                 & RA & DEC & z      & Nz   \\ 
\hline
Abell 2843           &  14.15573 & -27.51298 & 0.215 (0.560)   &  11 \\ 
RX J0848.8+4455      & 132.20542 &  44.92944 & 0.5430          &   6 \\ 
GHO~0940+4819        & 145.92105 &  48.08725 & 0.4700?         &  4 \\ 
RX~J0957.8+6534      & 149.47167 &  65.57500 & 0.5300          &  1 \\ 
SEXCLASS 12          & 163.15917 &  57.51369 & 0.64 (0.6100)   &  4 \\ 
SEXCLASS 13          & 163.22583 &  57.53600 & 0.5800          &  3 \\ 
RX~J1540.8+1445      & 235.22208 &  14.75944 & 0.442           & 11 \\
CL J1604+4314        & 241.10750 &  43.23972 & 0.925 (0.8652)  & 11 \\ 
OC02 J1701+6412      & 255.34583 &  64.23583 & 0.4518  &  1 \\ 
\hline
\end{tabular}
\label{tab:fewz}
\end{table*}

\subsection{Abell 2843 (14.15573$^o$, --27.5130$^o$, z=0.215/0.5600?)} 

\begin{figure}
\begin{center}
\includegraphics[width=6cm]{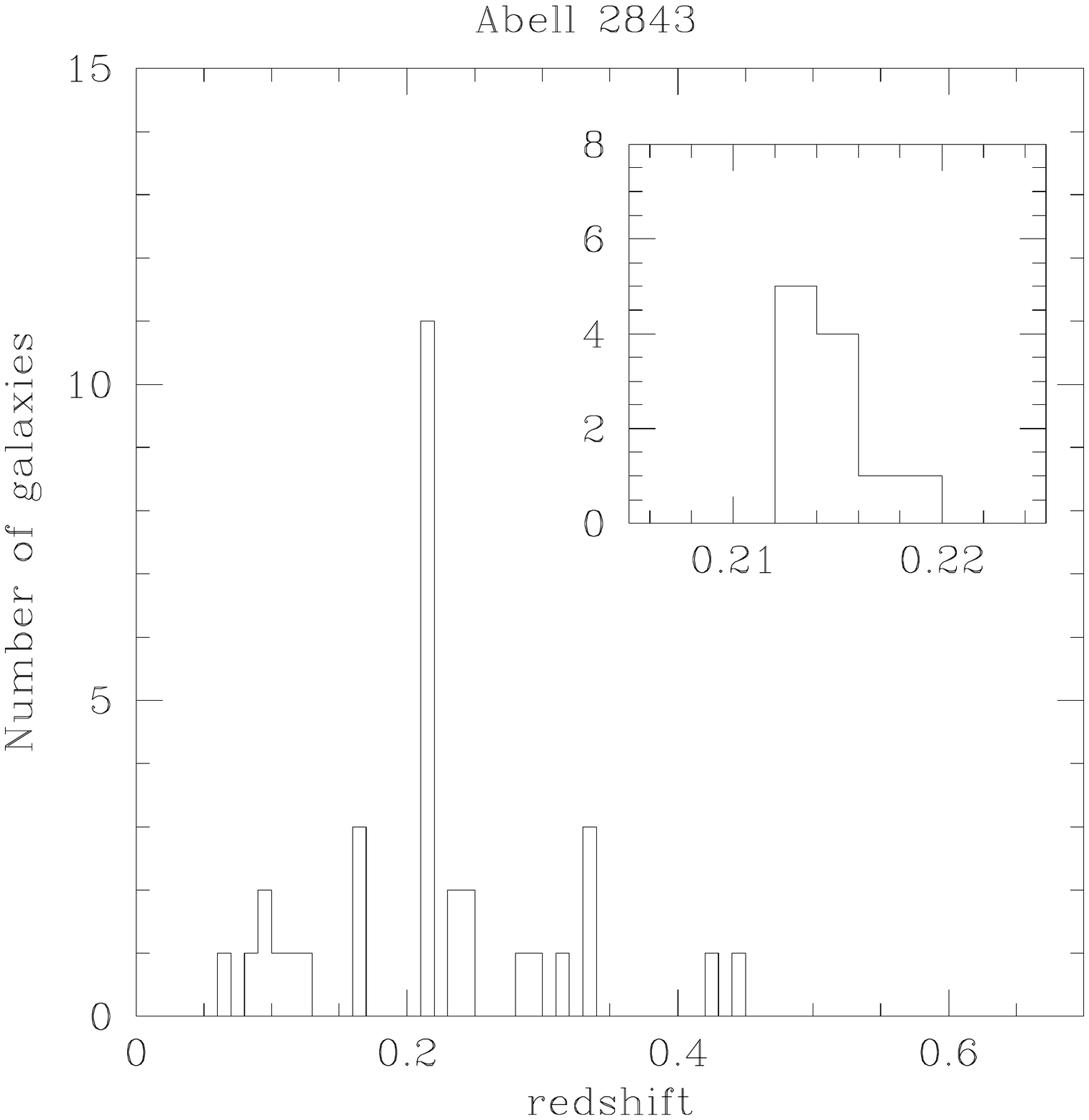}
\caption{Redshift histogram for Abell~2843. The insert shows a zoom around
  the possible alternative cluster redshift.}
\label{fig:abell2843_histoz}
\end{center}
\end{figure}

Little is known on Abell~2843 (Abell et al. 1989). According to NED
and Simbad, its redshift is 0.5600, but the redshift histogram shows no
galaxy at this redshift (Fig.~\ref{fig:abell2843_histoz}). On the
other hand, a peak is detected at z$\sim 0.215$, with 11 galaxies in
the [0.21,0.22] redshift range.

\subsection{RX J0848.8+4455 (132.20542$^o$, +44.9294$^o$, z=0.5430?)} 

According to Holden et al. (2001), there are two objects in this zone:
RX J0848+4456, an X-ray emitting cluster of galaxies at a redshift 
z = 0.570, and a group at a slightly lower redshift, z = 0.543.  These
authors state that the lower redshift group has, at most, one-fifth
and, more likely, 1/10$^{th}$, of the X-ray luminosity of RX J0848+4456.

\begin{figure}
\begin{center}
\includegraphics[width=6cm]{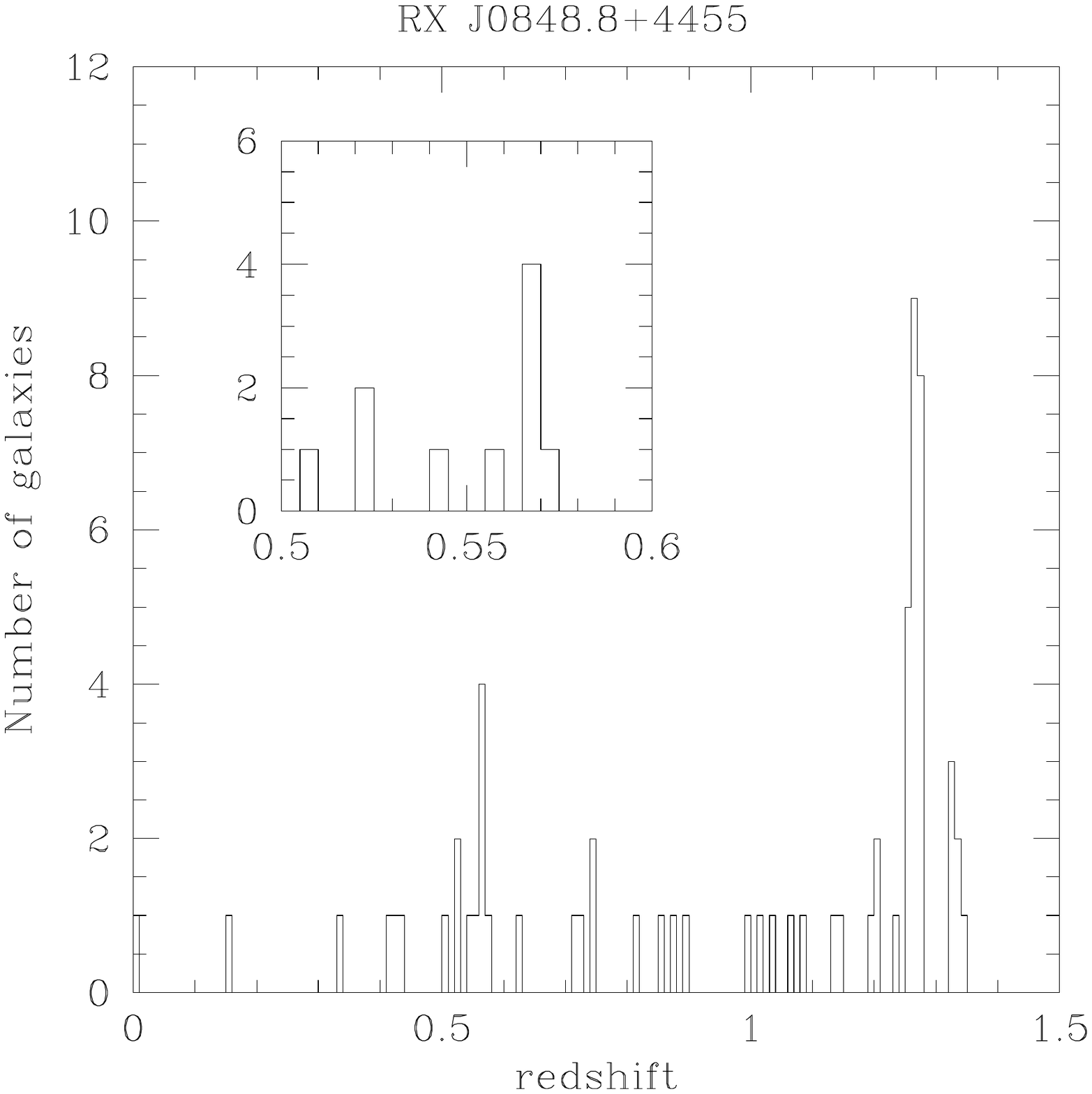}
\caption{Redshift histogram for RX J0848.8+4455. }
\label{fig:rx0848_histoz}
\end{center}
\end{figure}

The redshift histogram in the direction of this cluster only shows a
small peak in the region of the cluster redshift given by NED
(z=0.5430) (Fig.~\ref{fig:rx0848_histoz}), with 6 galaxies in the
[0.54,0.57] range. The redshift histogram shows a large
peak at redshift at z$\sim$1.26, due to the existence of a background
cluster in the Lynx field (Holden et al. 2001).

\subsection{GHO~0940+4819 (145.92105$^o$, +48.0873$^o$, z=0.4700?)} 

This cluster was first identified by Gunn et al. (1986). 
NED gives four different redshifts for the same cluster at the same coordinates, 
so the redshift of this cluster remains uncertain.

\subsection{RX~J0957.8+6534 (149.47167$^o$, +65.5750$^o$, z=0.5300?)}

There are only three galaxies in NED around this cluster position, and only one
at the cluster redshift (z=0.5300).

\subsection{SEXCLAS 12 (163.15917$^o$, +57.5137$^o$, z=0.6100?)} 

\begin{figure}
\begin{center}
\includegraphics[width=6cm]{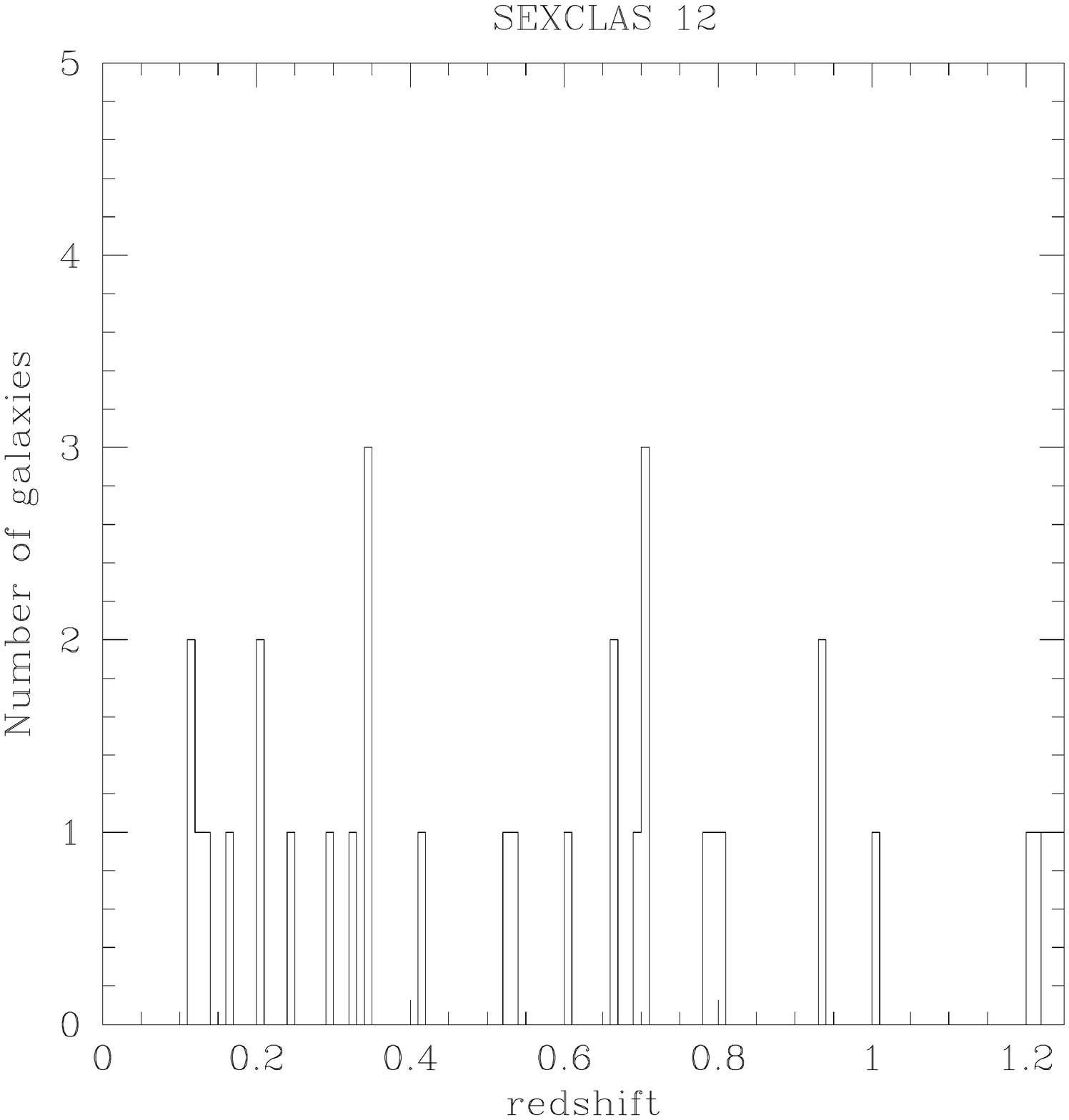}
\caption{Redshift histogram for SEXCLAS 12. }
\label{fig:sex12_histoz}
\end{center}
\end{figure}

SEXCLAS~12 is not significantly detected with XMM-Newton following our
criteria.

This cluster and the following one (they are about 3~arcmin apart)
were taken from the sample of Kolkotronis et al. (2006), who used
XMM-Newton pointings and five-band optical data to serendipitously
identify X-ray selected clusters.

The full redshift histogram of SEXCLAS~12 shows no prominent peak, in
particular at the redshift z=0.6100 given by NED
(Fig.~\ref{fig:sex12_histoz}). Four galaxies at z$\sim$0.64 could
indicate that this is the cluster redshift, but more spectra are
obviously needed to securely assign a redshift to these two potential
structures.

\subsection{SEXCLAS 13  (163.22583$^o$, +57.5360$^o$, z=0.5800)} 

\begin{figure}
\begin{center}
\includegraphics[width=6cm]{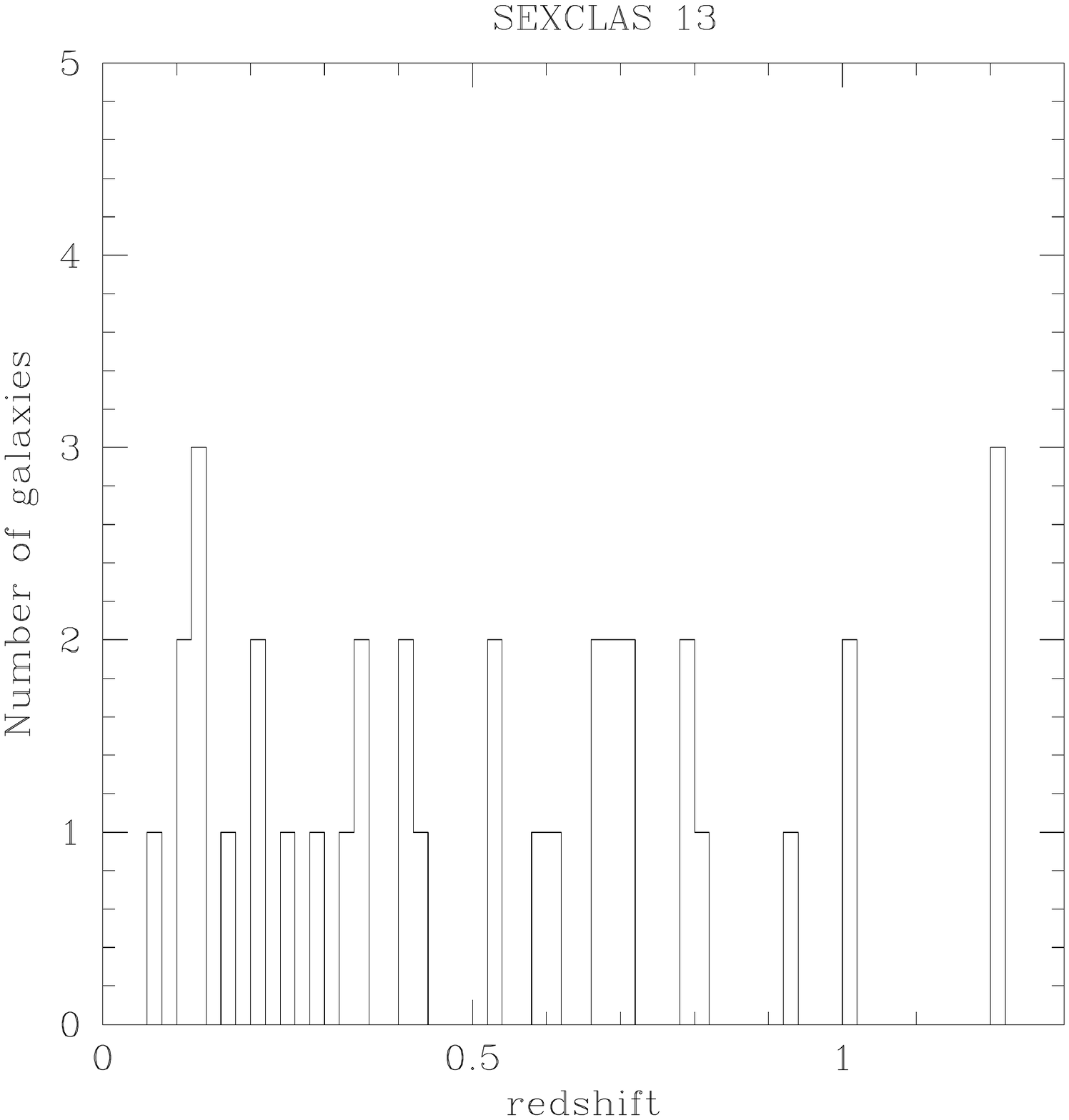}
\caption{Redshift histogram for SEXCLASS 13. }
\label{fig:sex13_histoz}
\end{center}
\end{figure}

SEXCLAS~13 is very close to SEXCLAS~12, and is not detected with
XMM-Newton either with our criteria.

The full redshift histogram of SEXCLAS~13 shows no prominent peak
either, in particular at the redshift z=0.5800 given by NED, where
there are only three galaxies (Fig.~\ref{fig:sex12_histoz}).

\subsection{RX~J1540.8+1445 (235.22208$^o$, +14.7594$^o$, z=0.4410)} 

\begin{figure}
\begin{center}
\includegraphics[width=6cm]{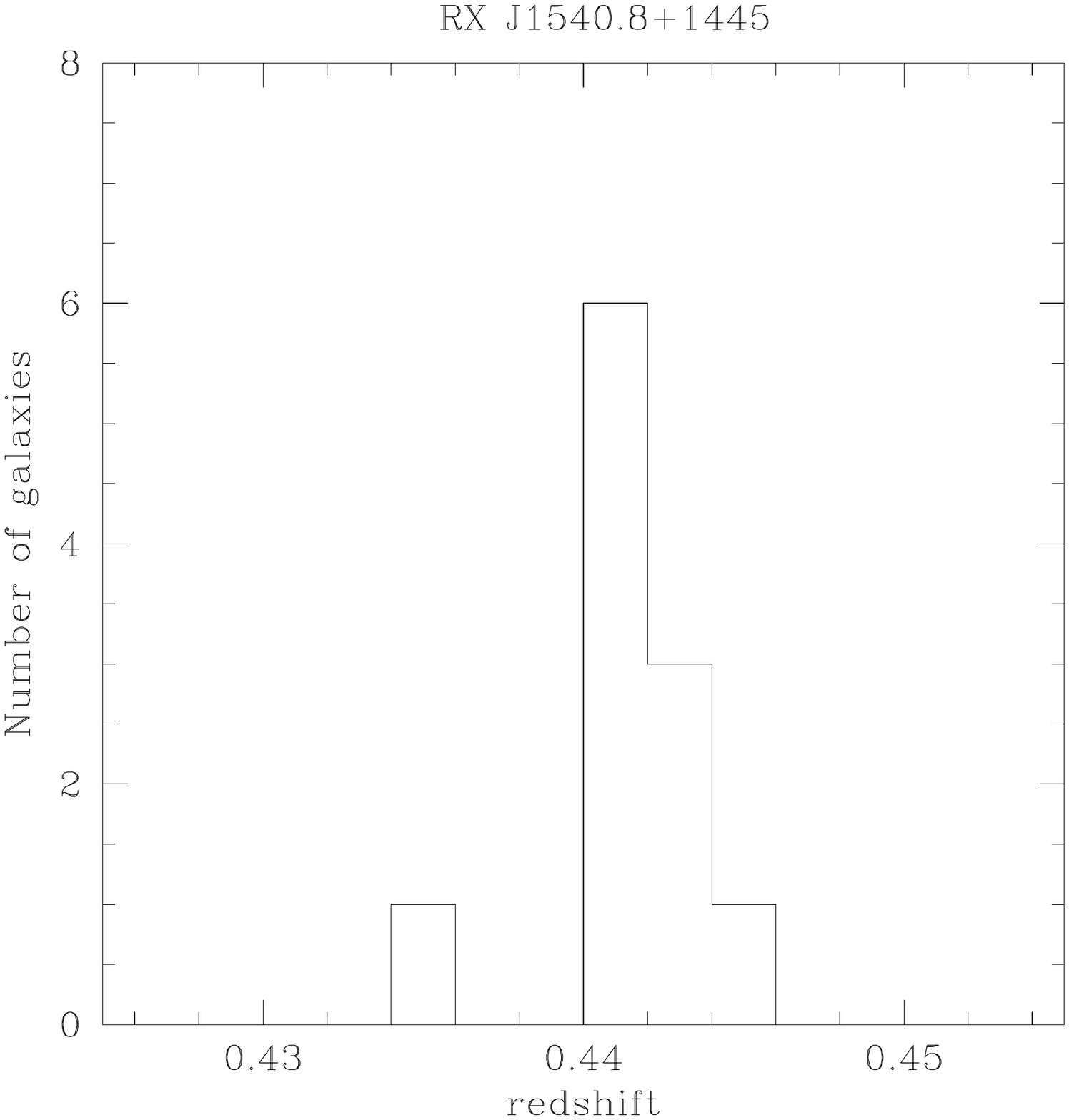}
\caption{Redshift histogram for RX~J1540.8+1445. }
\label{fig:rxj1540_histoz}
\end{center}
\end{figure}

Little is known about RX~J1540.8+1445, which was discovered in X-rays by
Vikhlinin et al. (1998).  We have obtained spectroscopic redshifts for
galaxies in this cluster with VLT/FORS2 (see Table~\ref{tabzspecus}).
However, only 11 redshifts are in the [0.43,0.45] interval (see
Fig.~\ref{fig:rxj1540_histoz}) so we cannot apply an SG.

\subsection{CL J1604+4314 (241.10750$^o$, +43.2397$^o$, z=0.8652/0.925?)} 


CL J1604+4314 is part of a supercluster made of several clusters at
redshift $\sim 0.9$ analysed in detail by Gal \& Lubin (2004), and by
Gal et al. (2005, 2008), with several hundred spectroscopically
confirmed members in a very large spatial area. Unfortunately, these
redshifts are not publicly available, so we cannot give a redshift
histogram or apply the SG method to this cluster. NED gives
z=0.8652 but the peak in the 11 redshifts available in NED rather
indicates z$\sim$0.925 (based on 11 galaxies), in agreement with Gal
\& Lubin (2004).


\subsection{OC02 J1701+6412 (255.34583$^o$, +64.2358$^o$, z=0.4518?)} 

\begin{figure}
\begin{center}
\includegraphics[width=6cm]{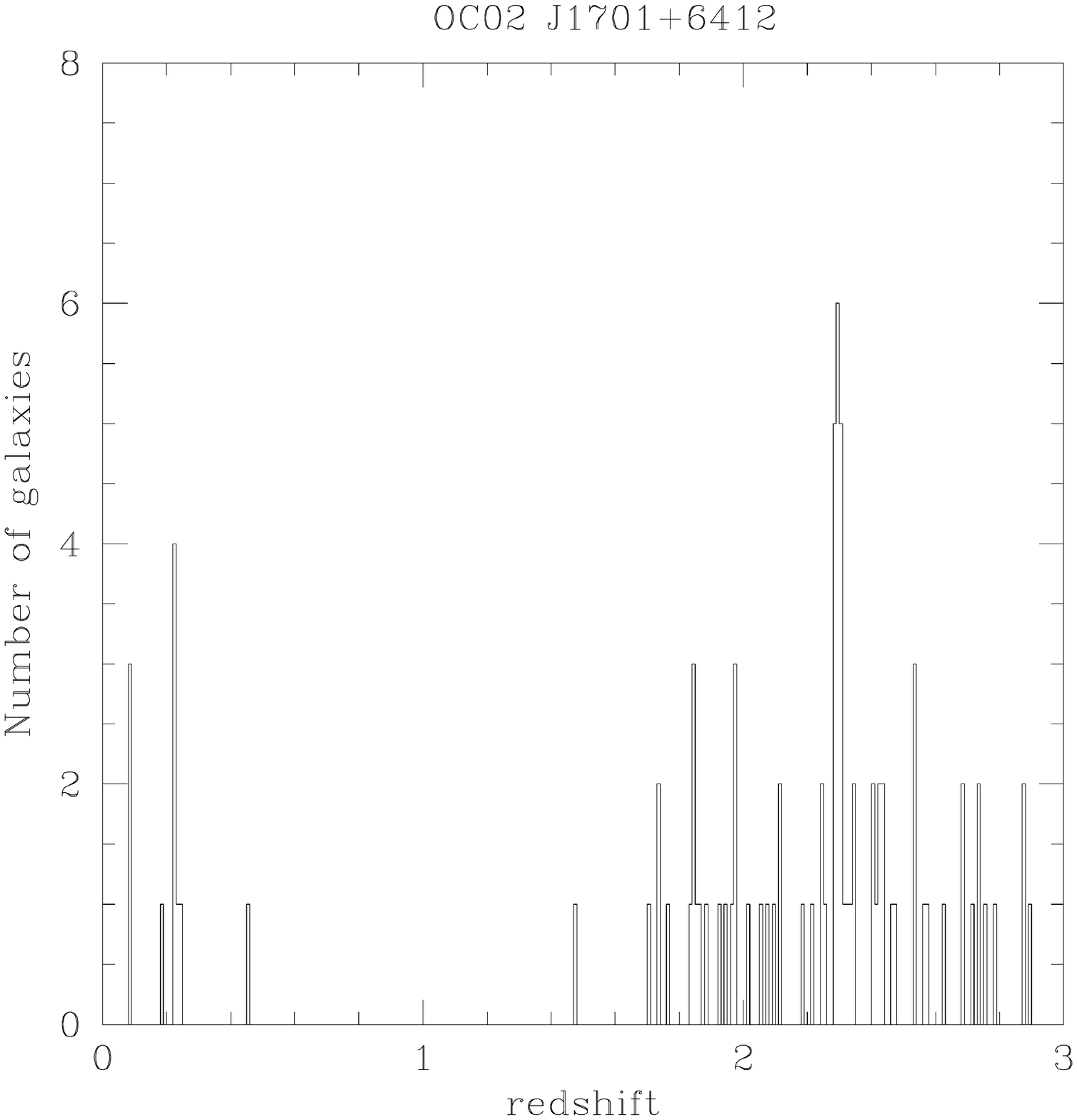}
\caption{Redshift histogram for OC02 J1701+6412. }
\label{fig:oc02_histoz}
\end{center}
\end{figure}

OC02 J1701+6412 was discovered by Vikhlinin et al. (1998) in their
ROSAT PSPC galaxy cluster survey, where they give for this cluster a
redshift z=0.4530 (the redshift of the cD galaxy). Out of the 88
redshifts gathered in an area of 5~arcmin radius around this cluster
there is only one galaxy at z=0.4518, most the other redshifts being
distant galaxies (Fig.~\ref{fig:oc02_histoz}). There are in particular
several galaxies at z$\sim$0.22 and a large peak at z$\sim$2.3 (17
galaxies with redshifts in the [2.28,2.31] redshift range).

\begin{table*}[]
  \caption{New spectroscopic redshift measurements obtained with VLT/FORS2 
    for galaxies located in the RX~J1540.8+1445 and GHO~2143+0408 clusters. 
The  columns are: right ascension, declination, and redshift.}
\label{tabzspecus}
\begin{tabular}{|lll|lll|}
\hline
\hline
 & RX~J1540.8+1445& & & GHO~2143+0408 & \\
\hline
 RA & DEC & z & RA & DEC & z\\
\hline
  235.1656134 &14.7352182 &0.4427  &  326.4801    &4.4216     &0.5562 \\ 
  235.1680243 &14.8065969 &0.4699  &  326.4843    &4.4264     &0.5259 \\ 
  235.1693535 &14.7595151 &0.19192 &  326.4862    &4.3711     &0.7385 \\ 
  235.1727369 &14.8025443 &0.5621  &  326.487     &4.4145     &0.4237 \\ 
  235.1751678 &14.7894211 &1.0626  &  326.4872    &4.4158     &0.0    \\ 
  235.1784827 &14.7920345 &0.411   &  326.4904    &4.4179     &0.1736 \\ 
  235.1815337 &14.8060684 &0.4418  &  326.4945    &4.3659     &0.4029 \\ 
  235.1896165 &14.752223  &0.4415  &  326.4949    &4.3676     &0.4015 \\ 
  235.1899269 &14.7878709 &0.0     &  326.4963    &4.413      &0.657  \\ 
  235.1969677 &14.77182   &0.19516 &  326.5008    &4.4087     &0.5171 \\ 
  235.1989185 &14.7371372 &0.8428  &  326.5008    &4.4087     &0.5171 \\ 
  235.1991002 &14.7479399 &0.6009  &  326.5058    &4.4166     &0.6034 \\ 
  235.2003501 &14.7968349 &0.4423  &  326.508     &4.41       &0.0    \\ 
  235.2007918 &14.7198888 &0.4407  &  326.5084    &4.351      &0.235  \\ 
  235.2034364 &14.7284524 &0.4444  &  326.5103    &4.4317     &0.601  \\ 
  235.204883  &14.7792275 &0.5638  &  326.5126    &4.3905     &0.5251 \\ 
  235.206242  &14.7412534 &0.442   &  326.5131    &4.4235     &0.5584 \\ 
  235.2062595 &14.7961091 &0.605   &  326.5136    &4.3634     &0.809  \\ 
  235.2103902 &14.7983381 &0.19469 &  326.5152    &4.3999     &0.6641 \\ 
  235.2158803 &14.7617361 &0.7473  &  326.5197    &4.3671     &0.0    \\ 
  235.2181524 &14.7638333 &0.1925  &  326.5198    &4.3587     &0.6479 \\ 
  235.2191101 &14.7743322 &0.4428  &  326.5211    &4.3734     &0.5137 \\ 
  235.2191998 &14.7723376 &0.4346  &  326.5215    &4.3511     &0.595  \\ 
  235.2217389 &14.7770748 &0.4419  &  326.5217    &4.3566     &0.78   \\ 
  235.2244809 &14.7584889 &0.5703  &  326.5234    &4.4028     &0.9993 \\ 
  235.2246155 &14.7657406 &0.44033 &  326.5235    &4.376      &0.0    \\ 
  235.227457  &14.7573185 &0.7356  &  326.5256    &4.3886     &0.6517 \\ 
  235.2332704 &14.7323585 &0.7659  &  326.527     &4.3871     &0.0374 \\ 
  235.2406444 &14.7849472 &1.9096  &  326.5425    &4.3626     &1.0079 \\ 
  235.24072   &14.7824836 &0.2831  & & &\\ 
  235.2611534 &14.7588384 &0.0     & & &\\ 
\hline
\end{tabular}
\end{table*}

\end{document}